\documentclass[prb,aps,nofootinbib,amssymb,twocolumn,superscriptaddress,10pt]{revtex4-2}

\usepackage[T1]{fontenc}
\usepackage{graphicx}
\usepackage{xcolor}
\usepackage{dcolumn}
\usepackage{comment}
\usepackage{amsmath,amssymb,bbold,bm}
\usepackage{hyperref}
\usepackage{amsfonts}
\usepackage{float,latexsym}
\usepackage{multirow}
\usepackage{color}
\usepackage[normalem]{ulem}
\usepackage{cleveref}
\usepackage{physics}
\usepackage{tabularx}
\usepackage{array}
\usepackage[caption=false]{subfig}
\usepackage{siunitx}
\usepackage{gensymb}
\usepackage{pifont}
\usepackage{placeins}
\usepackage{mathtools}
\let\vec\mathbf
\def\bSigma{\boldsymbol{\sigma}}
\def\ie{{\it i.e.}}
\def\eg{{\it e.g.}}

\def\omegaSLG{\Omega_{\mathrm{SLG}}}
\def\omegaTBG{\Omega_{\mathrm{TBG}}}

\def\MBZ{\mathrm{MBZ}}
\def\vk{\vec{k}}
\def\vQ{\vec{Q}}
\def\vq{\vec{q}}
\def\vG{\vec{G}}
\def\intD{\mathrm{d}}
\def\Q0{\mathcal{Q}_0}
\def\hU{\hat{U}}
\def\hN{\hat{N}}
\def\rthree{\sqrt{3} \times \sqrt{3}}
\def\sus{\mathrm{SU}_{\mathrm{S}} \left( 2\right)}
\def\suv{\mathrm{SU}_{\mathrm{V}} \left( 2\right)}
\def\susv{\sus \times \suv}
\def\uV{U_{\mathrm{V}}}
\def\uS{U_{\mathrm{S}}}
\def\huV{\hU_{\mathrm{V}}}
\def\huS{\hU_{\mathrm{S}}}
\def\Uncf{ \mathrm{U}_{\mathrm{nc}} \left( 4 \right) }

\newlength{\nsubht}

\newsavebox{\nsubbox}
\allowdisplaybreaks

\DeclarePairedDelimiterX{\setDef}[2]\{\}{%
  \, #1 \;\delimsize\vert\; #2 \,
}

\crefname{appendix}{Appendix}{Appendices}
\crefname{equation}{Eq.}{Eqs.}
\crefname{figure}{Fig.}{Figs.}
\crefname{table}{Table}{Tables}
\crefname{section}{Section}{Sections}
\crefname{enumi}{Case}{Cases}
\creflabelformat{appendix}{[#2#1#3]}
\crefrangelabelformat{figure}{#3#1#4--(#5\crefstripprefix{#1}{#2}#6}
\crefmultiformat{figure}{Figs.~#2#1\xdef\mycreffirstarg{#1}#3}%
 { and~#2#1#3}{, #2#1#3}{ and~#2#1#3}

\makeatletter     
\renewcommand\onecolumngrid{
\do@columngrid{one}{\@ne}%
\def\set@footnotewidth{\onecolumngrid}
\def\footnoterule{\kern-6pt\hrule width 1.5in\kern6pt}%
}

\makeatletter
\AtBeginDocument{\let\LS@rot\@undefined}
\makeatother

\newcommand{\citeInteracting}{~\cite{KAN19,BUL20,BER21a} (see \cref{app:sec:notation:ih})}
\newcommand{\citeSymmetries}{ (see \cref{app:sec:notation:sp})}
\newcommand{\citeSpectralFunction}{ (see \cref{app:sec:spectralf})}
\newcommand{\citeCharge}{ (see \cref{app:sec:charge})}
\newcommand{\citeME}{ (see \cref{app:sec:computingME})}
\newcommand{\citeExperimentalM}{ (see \cref{app:sec:experimental})}
\newcommand{\citeExperimentAdditional}{Details about the experimental measurements are provided in \cref{app:sec:experimental}, while the signal normalization based on tip height is explained in \cref{app:sec:spectralf}}
\newcommand{\citeExperimentResolution}{computed in \cref{app:sec:experimental}}
\newcommand{\citeSFSymmetries}{ (see \cref{app:sec:spFuncSym})}
\newcommand{\citeSFSymmetriesChiral}{ (see \cref{app:sec:spFuncSym:nonchiral})}
\newcommand{\citeSFSymmetriesPHBreak}{ (see \cref{app:sec:spFuncSym:breakingPH})}
\newcommand{\citeSimpleRules}{ (see \cref{app:sec:spFuncSym:add_examples:intuitive})}
\newcommand{\citeAdditionalExamples}{ in \cref{app:sec:results}}

\newcommand{\siSection}{appendix}
\newcommand{\SiSection}{Appendix}
\allowdisplaybreaks
\begin{document}

\title{Spectroscopy of Twisted Bilayer Graphene Correlated Insulators}

\author{Dumitru C\u{a}lug\u{a}ru}
\affiliation{Department of Physics, Princeton University, Princeton, New Jersey 08544, USA}

\author{Nicolas  Regnault}
\affiliation{Department of Physics, Princeton University, Princeton, New Jersey 08544, USA}

\author{Myungchul Oh}
\affiliation{Department of Physics, Princeton University, Princeton, New Jersey 08544, USA}

\author{Kevin P. Nuckolls}
\affiliation{Department of Physics, Princeton University, Princeton, New Jersey 08544, USA}

\author{Dillon  Wong}
\affiliation{Department of Physics, Princeton University, Princeton, New Jersey 08544, USA}

\author{Ryan L. Lee}
\affiliation{Department of Physics, Princeton University, Princeton, New Jersey 08544, USA}

\author{Ali Yazdani}
\affiliation{Department of Physics, Princeton University, Princeton, New Jersey 08544, USA}

\author{Oskar Vafek}
\affiliation{National High Magnetic Field Laboratory, Tallahassee, Florida, 32310, USA}
\affiliation{Department of Physics, Florida State University, Tallahassee, Florida 32306, USA}

\author{B. Andrei Bernevig}
\email{bernevig@princeton.edu}
\affiliation{Department of Physics, Princeton University, Princeton, New Jersey 08544, USA}
\affiliation{Donostia International Physics Center, P. Manuel de Lardizabal 4, 20018 Donostia-San Sebastian, Spain}
\affiliation{IKERBASQUE, Basque Foundation for Science, 48009 Bilbao, Spain}

\begin{abstract}

We analytically compute the scanning tunneling microscopy (STM) signatures of integer-filled correlated ground states of the magic angle twisted bilayer graphene (TBG) narrow bands. After experimentally validating the strong-coupling approach at $\pm 4$ electrons/moir\'e unit cell, we consider the spatial features of the STM signal for 14 different many-body correlated states and assess the possibility of Kekulé distortion (KD) emerging at the graphene lattice scale. Remarkably, we find that coupling the two opposite graphene valleys in the intervalley-coherent (IVC) TBG insulators does not always result in KD. As an example, we show that the Kramers IVC state and its nonchiral $\mathrm{U} \left( 4 \right)$ rotations do not exhibit any KD, while the time-reversal-symmetric IVC state does. Our results, obtained over a large range of energies and model parameters, show that the STM signal and Chern number of a state can be used to uniquely determine the nature of the TBG ground state. 
\end{abstract}
\maketitle

\emph{Introduction}.~Near the first magic angle~\cite{BIS11,LOP07,SUA10}, both transport~\cite{CAO18,YAN19,LU19,POL19,CAO20,SER20,CHE20a,STE20,LIU21e,PAR21c,SAI20,WU21a,LU21,SAI21,SAI21a,CAO21} and spectroscopy~\cite{KER19,XIE19,JIA19,CHO19,WON20,CHO20,NUC20,CHO21a,CHO21,TSC21} experiments have uncovered a wealth of superconducting and correlated insulating phases in twisted bilayer graphene (TBG), sparking considerable theoretical effort towards their understanding~\cite{YUA18,OCH18,THO18,DOD18,XU18b,KOS18,PO18,VEN18,KEN18,HUA19,LIU19,WU19,CLA19,KAN19,SEO19,DA19,WU20b,XIE20a,BUL20a,DAV22,CHI20b,CHA20,REP20,HEJ21,CEA20,ZHA20,KAN20a,FER20a,BRI20,BUL20,XIE21a,SOE20,CHR20,EUG20,VAF20,LIU21a,DA21,LIU21,KAN21,LIA21,XIE21,PAR21a,KWA21,CHE21,WAG22}. The physics of TBG near integer fillings with $\nu$ electrons per moir\'e unit cell was argued to be in the strong coupling regime, dictated by the interaction-only Hamiltonian projected onto its almost-flat bands~\cite{KAN19,SEO19,BUL20,VAF20,LIA21}. The enlarged continuous spin-valley symmetries thereof~\cite{KAN19,SEO19,BUL20,BER21a} have rendered a low-energy manifold of its many-body eigenstates~\cite{KAN19,SEO19,BUL20,VAF20,LIA21} and few-particle excitations~\cite{VAF20,BER21b} exactly solvable at integer fillings. Following numerically validated~\cite{KAN20a,BUL20,XIE21} analytical arguments~\cite{KAN19,BUL20,BER21a,LIA21}, the resulting eigenstates of the projected interaction Hamiltonian were shown to be energetically competitive ground-state candidates, if not \emph{the} actual ground states of the system, for a large range of parameters. 

Building on the aforementioned theoretical advances, this \textit{Letter} identifies spectroscopic signatures of the various competing states. For a given insulator, the differential conductance measured in scanning tunneling microscopy (STM) experiments is proportional to its spectral function~\cite{COL15}, which can be computed analytically from the readily available many-body electron and hole excitations~\cite{VAF20,BER21b}. We find that the STM features of the proposed correlated states -- particularly the presence or absence of a Kekul\'e distortion (KD) at the graphene lattice scale (\ie{} the modulation of the STM signal at wave vectors connecting the two graphene valleys) -- together with the knowledge of their Chern number, can distinguish among the candidate many-body states. Recent experiments~\cite{LIU22,COI22,LI19} demonstrating the ability of STM to visualize symmetry-broken states with KD arising from many-body interactions in the zeroth Landau level of monolayer graphene indicate that similar techniques can be employed to discriminate between the correlated insulators of TBG.

The competing correlated states of TBG at an integer filling $\nu$ can be characterized by their Chern number $\mathcal{C}$ and valley polarization, being either valley polarized (VP) or intervalley coherent (IVC). Additionally, even for $\mathcal{C}=0$, IVC states may either spontaneously break time-reversal symmetry ($T$), as in the Kramers IVC (K-IVC) state, or preserve it, as in the $T$-symmetric IVC (T-IVC) state~\cite{KAN19,SEO19,BUL20,BER21a}. Some of these states can be stabilized by magnetic field~\cite{NUC20,WU21a,DAS21,LIU21e,SAI21,CHO21a}. In this \textit{Letter}, we analyze numerically 14 different TBG correlated insulators, and show analytically that \emph{all} VP states together with the K-IVC states at $\nu = \pm 2,0$ exhibit no KD, while generic IVC states do display KD.
Furthermore, we show that the strong- versus weak-coupling nature of the system can be uniquely inferred from the spectral function of the $\nu=\pm 4$ band insulator. The correct asymmetric peak structure obtained in the strong-coupling approach differs significantly from the weak-coupling result and shows dramatic variations as the STM tip moves from the AA to AB moir\'e regions. While the experimental data display large sample-to-sample variation in the local density of states (LDOS), some datasets are uniquely compatible with the strong-coupling description.

\emph{Model}.~The physics of magic-angle TBG is dominated by the repulsive Coulomb interaction Hamiltonian projected in the almost-flat bands near charge neutrality\citeInteracting{}
\begin{equation}
	\label{eqn:proj_int}
	H_I = \frac{1}{2 \omegaTBG} \sum_{\vq \in \MBZ} \sum_{\vG \in \mathcal{Q}_0} O_{-\vq-\vG} O_{\vq+\vG}.
\end{equation}
where $\omegaTBG$ is the area of the TBG sample, $\MBZ$ and $\mathcal{Q}_0$, respectively, denote the moir\'e Brillouin zone (MBZ) and reciprocal lattice, while
\begin{align}
	\label{eqn:int_o_operators}
	O_{\vq+\vG} =& \sum_{\substack{\vk,\eta,s \\ n,m = \pm 1}} \sqrt{V (\vq + \vG)} M^{\eta}_{mn} \left(\vk,\vq+\vG \right) \nonumber \\
	&\times \left( \hat{c}^\dagger_{\vk+\vq,m,\eta,s} \hat{c}_{\vk,n,\eta,s} - \frac{1}{2} \delta_{\vq,\vec{0}}\delta_{m,n} \right)
\end{align}   
are proportional to the flat-band-projected density operators. In particular, $\hat{c}^\dagger_{\vk,n,\eta,s}$ is the electron creation operator for the TBG conduction ($n=+1$) and valence ($n=-1$) flat bands from valley $\eta = \pm$ and spin $s = \uparrow, \downarrow$, while $M^{\eta}_{mn} \left( \vk, \vq + \vG \right)$ are the TBG form factors. The Fourier-transformed screened Coulomb potential $V \left( \vq \right) = 2 \pi U_{\xi} \xi^{2} \left( 1-e^{-2 \xi q}\right)/ \left( \xi q \right)$ (with $U_{\xi}$ and $\xi$, respectively, denoting the interaction energy scale and the screening length) corresponds to the typical single-gate arrangement of the TBG sample in a STM experiment~\cite{NUC20}. 

The TBG single-particle Hamiltonian features a series of discrete symmetries\citeSymmetries{}: the $C_{2z}$, $C_{3z}$, $T$, and $C_{2x}$ \emph{commuting} symmetries, as well as an approximate unitary particle-hole $P$ \emph{anticommuting} symmetry~\cite{SON19,KAN18,SON21,BER21a,PO19}. The latter enlarges the valley-spin-charge $\mathrm{U} \left( 2 \right) \times \mathrm{U} \left( 2 \right)$ rotation symmetry of $H_{I}$ to the so-called nonchiral-flat $\mathrm{U} \left( 4 \right)$ symmetry~\cite{KAN19,SEO19,BUL20,BER21a}, henceforth denoted by $\Uncf{}$. Additionally, when the interlayer tunneling amplitude at the AA stacking centers ($w_0$) is neglected compared to the one at AB stacking centers ($w_1 = \SI{110}{\milli\eV}$) -- in the so-called chiral limit ($w_0 / w_1 = 0$) -- the single-particle Hamiltonian enjoys an additional \emph{anticommuting} chiral $C$ symmetry~\cite{TAR19,BER21a}, which further enlarges the symmetry group of $H_I$ to the chiral-flat $\mathrm{U} \left( 4 \right) \times \mathrm{U} \left( 4 \right)$ group~\cite{BER21a,BUL20}. Recombining the active TBG bands into Chern-number $e_Y$ bands with operators $\hat{d}^\dagger_{\vk,e_Y,\eta,s} = \frac{1}{\sqrt{2}} \left( \hat{c}^\dagger_{\vk,+1,\eta,s} + i e_Y \hat{c}^\dagger_{\vk,-1,\eta,s} \right)$, the 32 generators of the chiral-flat $\mathrm{U} \left( 4 \right) \times \mathrm{U} \left( 4 \right)$ group correspond to independent valley-spin rotations within each Chern sector. Away from the chiral limit the $\mathrm{U} \left( 4 \right) \times \mathrm{U} \left( 4 \right)$ generators get combined into the 16 $\Uncf{}$ generators such that $\Uncf{}$ intervalley (intravalley) rotations act on the two Chern sectors in the same (opposite) way~\cite{BER21a}.

The presence of enlarged symmetries renders some of the eigenstates of $H_I$ exactly solvable at integer fillings. Up to rotations $\hU$ belonging to the symmetry group of $H_I$, the TBG ground states have been shown to be Slater determinants obtained by populating the active TBG bands one Chern-valley-spin sector $\left( e_{Y_{j}},\eta_{j},s_{j} \right)$ at a time~\cite{KAN19,BUL20,KAN20a,BER21a,BER21b,LIA21,XIE21}   
\begin{equation}
	\label{eqn:ground_state}
	\ket{\varphi} = \hU \prod_{\vk} \prod_{j=1}^{4 + \nu} \hat{d}^\dagger_{\vk,e_{Y_{j}},\eta_{j},s_{j}}  \ket{0}. 
\end{equation}
In the chiral limit, $\ket{\varphi}$ is an \emph{exact} eigenstate of $H_I$ for any choice of the filled Chern-valley-spin sectors and $\hU \in \mathrm{U} \left( 4 \right) \times \mathrm{U} \left( 4 \right)$~\cite{LIA21}. Away from the chiral limit, only the insulators from \cref{eqn:ground_state} with fully filled or fully empty valley-spin flavors and $\hU \in \Uncf$ are \emph{exact}, with the rest being \emph{perturbative} eigenstates of $H_I$~\cite{LIA21}.

\emph{Spectral function.}~For a given state $\ket{\varphi}$ from \cref{eqn:ground_state}, the differential conductance as a function of bias voltage measured in STM experiments is proportional to its spectral function~\cite{COL15}
\begin{equation}
	\label{eqn:spec_func}
	\begin{split}
		A \left( \vec{r},\omega \right) = \sum_{\xi,s}  &\left[
			\abs{\mel**{\xi}{\hat{\psi}^\dagger_{s} \left( \vec{r} \right)}{\varphi}}^2 \delta \left( \omega - E_{\xi} + E_{\varphi} \right) \right. \\ 
			+ &\left.\abs{\mel**{\xi}{\hat{\psi}_{s} \left( \vec{r} \right)}{\varphi}}^2 \delta \left( \omega + E_{\xi} - E_{\varphi} \right) \right].
	\end{split}
\end{equation}
where $\psi_{s} \left( \vec{r} \right)$ denotes the electron field annihilation operator corresponding to spin $s = \uparrow,\downarrow$, and a summation is performed over all the \emph{many-body} eigenstates $\ket{\xi}$ of $H_I$ with energy $E_{\xi}$\citeSpectralFunction{}. Expressing the field operators in the TBG energy-band basis $\hat{\psi}^\dagger_{s} \left( \vec{r} \right) = \sum_{\eta,n} V_{\vec{r}, \vk n \eta} \hat{c}^\dagger_{\vk,n,\eta,s}$ (where the factors $V_{\vec{r}, \vk n \eta}$ depend on the carbon $p_z$ orbitals and the TBG flat band wave functions and include contributions from both graphene layers), we find that
\begin{equation}
	\label{eqn:spec_func_enbas_temp}
	A \left( \vec{r},\omega \right) = \hspace{-1 em} \sum_{\substack{ \vk \in \MBZ, \\ n,\eta, n',\eta' \\ c=\pm}} \left[\mathcal{M}_{\varphi}^{c} \left(\omega\right) \right]_{\vk n \eta,\vk n' \eta'} \left[ \mathcal{B} \left(\vec{r} \right) \right]_{\vk n \eta,\vk n' \eta'}. 
\end{equation}
In \cref{eqn:spec_func_enbas_temp}, we have introduced the spatial factor matrix $\left[\mathcal{B} \left( \vec{r} \right) \right]_{\vec{k} n \eta, \vec{k}' n' \eta'} = V_{\vec{r},\vec{k} n \eta} V^{*}_{\vec{r},\vec{k}' n' \eta'}$ (which depends only on the TBG single-particle Hamiltonian) and the spectral function matrices (which depend on the state $\ket{\varphi}$)
\begin{equation}
    \label{eqn:spec_func_mat_elem}
	\begin{split}
		\left[\mathcal{M}_{\varphi}^{c} \left(\omega\right) \right]_{\vk n \eta,\vk n' \eta'} = \sum_{\lambda_{-c},s} \mel**{\lambda_+}{\hat{c}_{\vk,n',\eta',s}}{\lambda_-} \\
		\times \mel**{\lambda_-}{\hat{c}^\dagger_{\vk,n,\eta,s}}{\lambda_+} \delta \left( \omega - E_{\lambda_-} + E_{\lambda_+} \right),
	\end{split}
\end{equation}
where $ \ket{\lambda_{c}} = \ket{\varphi}$ and we assumed no breaking of the moir\'e translation symmetry. Since a $\hat{c}^\dagger_{\vk,n,\eta,s}$ operator acts in \emph{one} single-layer graphene (SLG) valley, $\left[\mathcal{B} \left( \vec{r} \right) \right]_{\vec{k} n \eta, \vec{k} n' \eta}$ is only modulated at the level of the SLG and TBG lattices. In contrast, $\left[\mathcal{B} \left( \vec{r} \right) \right]_{\vec{k} n \eta, \vec{k} n' (-\eta)}$ contains an additional modulation corresponding to wave vectors linking the two Dirac points of the same graphene layer, which manifests in real space as a KD of the SLG.

As shown in \cref{eqn:spec_func_enbas_temp,eqn:spec_func_mat_elem}, computing the TBG spectral function requires the exact eigenstates of $H_I$ containing an extra electron or hole compared to $\ket{\varphi}$ (\ie{} the charge-one excitations). Despite $H_I$ being a quartic Hamiltonian, the exact charge-($\pm$)one excitations on top of $\ket{\varphi}$ can be computed as a zero-body problem using the \emph{charge-one commutation relations}~\cite{VAF20,BER21b}. For example, the electron commutation relation reads as
\begin{equation}
	\label{eqn:r_mat_commut}
		\left[H_I -\mu \hN , \hat{c}^\dagger_{\vk,n,\eta,s} \right] \ket{\varphi} = \sum_{m} R^{\eta}_{mn} \left( \vk \right) \hat{c}^\dagger_{\vk,m,\eta,s} \ket{\varphi} ,
\end{equation}
where $\mu$ denotes the chemical potential, $\hN$ is the total fermion number operator and the matrix $R \left( \vk \right)$ depends on $\nu$, the active TBG wave functions and the Coulomb repulsion potential\citeCharge. As such, the $\hat{c}^\dagger_{\vec{k},n,\eta,s}$ and $\hat{c}_{\vk,n,\eta,s}$ operators can be recombined into exact electron and hole excitations, allowing for the analytical calculation of the spectral function of $\ket{\varphi}$\citeME{}.

\begin{figure*}[!t]
	\includegraphics[width=\textwidth]{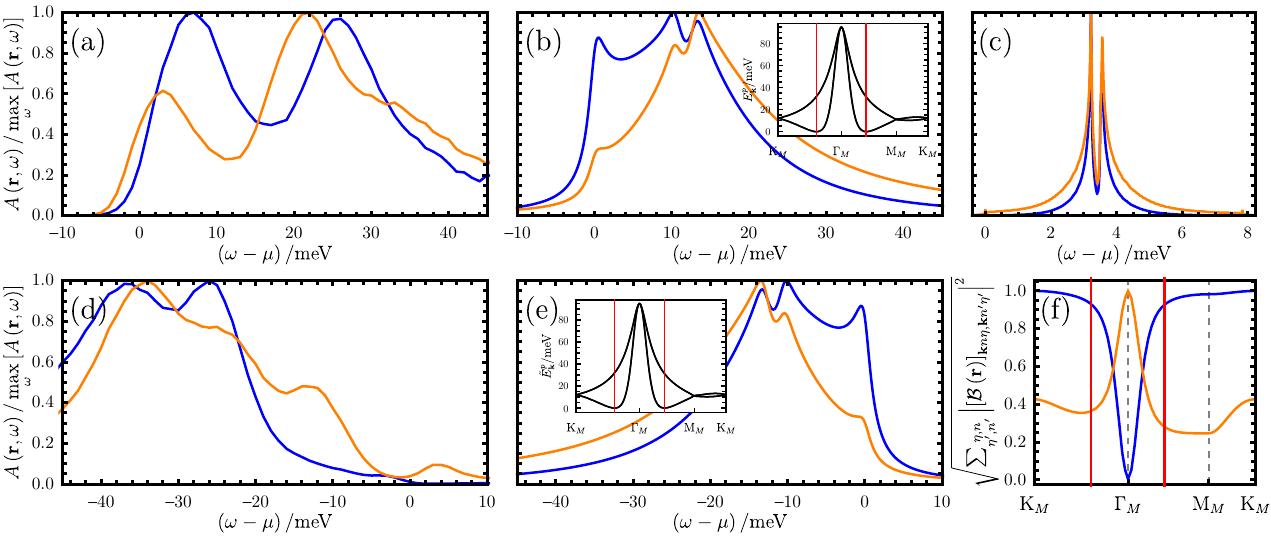}\subfloat{\label{fig:expVStheory:a}}\subfloat{\label{fig:expVStheory:b}}\subfloat{\label{fig:expVStheory:c}}\subfloat{\label{fig:expVStheory:d}}\subfloat{\label{fig:expVStheory:e}}\subfloat{\label{fig:expVStheory:f}}\caption{The TBG spectral function for the $\nu = \pm 4$ insulator. For $\nu=-4$ ($\nu = +4$), we compare the experimentally measured STM signal in (a) [(d)] with the spectral function computed from the charge-one excitation of $H_I$ in (b) [(e)]. We use $U_{\xi} =\SI{24}{\milli\eV}$, $\xi = \SI{300}{\nano\meter}$, $w_0/w_1 = 0.8$. For reference, the TBG spectral function at $\nu=-4$ derived from the single-particle TBG Hamiltonian is given in (c). The signal at the center of the AA (AB) site is shown in blue (orange) and normalized by its maxima in the energy range at that particular location. The theoretically computed spectral function is averaged over three SLG unit cells. The inset in (b) [(e)] shows the electron [hole] excitation dispersion $E^{p}_{\vk}$ ($\tilde{E}^{p}_{\vk}$) computed from \cref{eqn:r_mat_commut} (the red vertical lines are a guide for eye pointing the minima of the dispersion). (f) provides the spatial factor at the AA and AB sites along the high symmetry line of the MBZ (the red lines indicate the position of minima).}
	\label{fig:expVStheory}
\end{figure*}

\emph{Signatures of strong correlation}.~We first analyze the spectral function of the $\nu = -4$ ($\nu = +4$) TBG insulator from \cref{eqn:ground_state}, for which the active TBG bands are fully empty (fully filled) and no ambiguity in the choice of ground state arises. As shown in \cref{fig:expVStheory}, the strong-coupling and weak-coupling spectral functions at $\nu= \pm 4$ are markedly different as a result of the large interaction-induced dispersion of single-particle excitations in the strong-coupling regime [insets in \cref{fig:expVStheory:b,fig:expVStheory:e}] compared to the almost-flat dispersion in the weak-coupling (\ie{} noninteracting) regime, as well as from different van Hove singularities and Dirac points~\cite{VAF20,BER21b}. 

We will discuss the $\nu = -4$ insulator from \cref{fig:expVStheory:a,fig:expVStheory:b,fig:expVStheory:c}, with the $\nu = +4$ insulator [\cref{fig:expVStheory:d,fig:expVStheory:e}] following analogously from the many-body charge-conjugation symmetry of TBG~\cite{BER21a}. \citeExperimentAdditional{}. For $\nu = -4$, we focus on positive-energy biases ($\omega - \mu \geq 0$), such that the electrons tunnel into the fermion states of the active TBG bands, recombined into the electron excitations according to \cref{eqn:r_mat_commut}. The electron excitation energies $E^{p}_{\vk}$ [inset of \cref{fig:expVStheory:b}], obtained by diagonalizing the charge-one commutation matrices from \cref{eqn:r_mat_commut}\citeME{}, are comprised of four sets of twofold [spin $\mathrm{SU} \left( 2 \right)$] degenerate bands, which are further paired by the approximate $C_{2z} P$ symmetry of $H_I$ into two sets of almost fourfold degenerate bands~\cite{SON19,SON21,BER21b,VAF20}. For small biases, the electrons start tunneling into the regions at the bottom of the excitation bands away from any high-symmetry points (\eg{} halfway between the $\Gamma_M$ and $M_M$ points of the MBZ), giving rise to the peak near $\omega - \mu \approx \SI{0}{\milli\eV}$ in the spectral function at both the AA and AB centers. Upon increasing the bias to $\omega - \mu \approx \SI{20}{\milli\eV}$, the electrons tunnel into the almost-flat regions near the boundary of the MBZ, giving rise to two close peaks in the theoretical spectral function, merging into one in the experimental STM signal. For larger biases, the spectral function decreases as the electron tunnel in the strongly dispersive bands near $\Gamma_M$. 

The variation of the magnitude of $\left[ \mathcal{B} \left( \vec{r} \right) \right]_{\vec{k} n \eta, \vec{k} n' \eta'}$ with $ \vec{k} \in \MBZ$ at the AA and AB sites [\cref{fig:expVStheory:f}] \emph{qualitatively} explains the change in the STM signal between the two stacking centers: at the AA site, the spatial factor has roughly the same magnitude in the MBZ for the two almost-flat regions of the excitation bands, resulting in similar magnitudes for the LDOS peaks at $\omega-\mu \approx \SI{0}{\milli\eV},\SI{20}{\milli\eV}$. At the AB site, the spatial factor has a larger amplitude on the boundary of the MBZ, diminishing the peak at $\omega - \mu \approx \SI{0}{\milli\eV}$ compared to the one at $\omega - \mu \approx \SI{20}{\milli\eV}$. A similar decrease is also present in the experimental data [\cref{fig:expVStheory:a}], while clearly absent in the noninteracting LDOS [\cref{fig:expVStheory:c}]. Moreover, the half-maximum width of the spectral function ($\Delta \omega$) is much smaller in the noninteracting case ($\Delta \omega < \SI{1}{\milli\eV}$, comparable to the active TBG bandwidth) than in the experiment and the strong-coupling prediction ($\Delta \omega = \SIrange{20}{30}{\milli\eV}$, comparable to $U_{\xi}=\SI{24}{\milli\eV}$ and much larger than the resolution of the experiment $\delta \omega \approx \SI{3}{\milli\eV}$ \citeExperimentResolution{}). While the STM signal is sample dependent and may vary from different AA or AB sites\citeExperimentalM{}, this dataset indicates evidence of strong correlations governing the physics of TBG near charge neutrality.

\begin{figure*}[!t]
	\includegraphics[width=\textwidth]{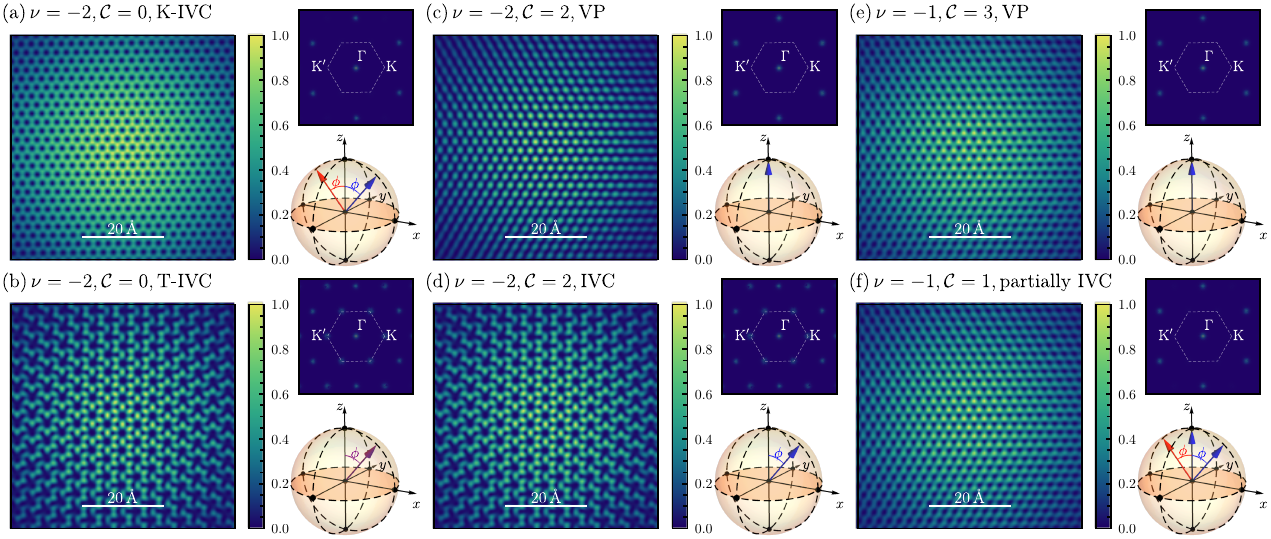}\subfloat{\label{fig:realSpaceExamples:a}}\subfloat{\label{fig:realSpaceExamples:b}}\subfloat{\label{fig:realSpaceExamples:c}}\subfloat{\label{fig:realSpaceExamples:d}}\subfloat{\label{fig:realSpaceExamples:e}}\subfloat{\label{fig:realSpaceExamples:f}}\caption{Kekul\'e distortion and intervalley coherence. For each insulator [(a)-(f)], we show the real-space spectral function centered at the AA site (left panel), its Fourier transformation (top-right panel), as well as the valley polarizations of the occupied Chern bands as blue ($e_Y = +1$) or red ($e_Y = -1$) unit vectors in the valley Bloch sphere. The valley polarization can be oriented parallel to the $\hat{\vec{z}}$ axis or at an angle $\phi = \pi/2$ to it. We consider the $\nu = -2$, $\mathcal{C} = 0$, K-IVC [(a)] and T-IVC [(b)] states, the $\nu = -2$, $\mathcal{C} = 2$, VP [(c)] and IVC [(d)] Chern insulators, as well as the $\nu = -1$, $\mathcal{C} = 3$ fully VP [(e)] and $\nu = -1$, $\mathcal{C} = 1$ partially IVC [(f)] Chern insulators. The presence of KD in (b) and (d) appears as threefold enlargement of the SLG unit cell and as a signal at the $\mathrm{K}$ and $\mathrm{K}'$ points of the SLG Brillouin zone. We use $U_{\xi} =\SI{8}{\milli\eV}$, $\xi = \SI{300}{\nano\meter}$, $w_0/w_1 = 0.8$. }
	\label{fig:realSpaceExamples}
\end{figure*}

\emph{Discriminating correlated insulating phases}.~We now investigate the effects of intervalley coherence on the \emph{spatial} variation of $A \left(\vec{r},\omega \right)$ for various $\abs{\nu}<4$ insulating states. Naively, coupling the two graphene valleys in an IVC insulator results in IVC charge-one excitations and should lead to the emergence of KD in the corresponding STM signal. However, due to the discrete symmetries of TBG, breaking the valley $\mathrm{U} \left( 1 \right)$ symmetry does \emph{not} guarantee the emergence of KD in $A \left(\vec{r},\omega \right)$\citeSFSymmetries{}. For instance, \cref{fig:realSpaceExamples:a,fig:realSpaceExamples:b} show the simulated STM patterns for two \emph{fully} IVC TBG insulators at $\nu = -2$~\cite{BUL20,LIA21}: 
\begin{equation}
    \begin{split}
        \ket{\text{K-IVC}} &= \prod_{\vk} \prod_{e_Y = \pm 1} \frac{\hat{d}^\dagger_{\vk,e_Y,+,\uparrow} + e_Y \hat{d}^\dagger_{\vk,e_Y,-,\uparrow}}{\sqrt{2}} \ket{0},\\
        \ket{\text{T-IVC}} &= \prod_{\vk} \prod_{e_Y = \pm 1} \frac{\hat{d}^\dagger_{\vk,e_Y,+,\uparrow} + \hat{d}^\dagger_{\vk,e_Y,-,\uparrow}}{\sqrt{2}} \ket{0}.
    \end{split}
\end{equation}
The K-IVC (T-IVC) state is obtained from a fully filled valley-spin flavor by rotating the two Chern bands in the $xz$ valley plane in opposite (identical) directions, as shown in \cref{fig:realSpaceExamples:a,fig:realSpaceExamples:b}. Remarkably, while the STM patterns of the T-IVC state show clear signs of KD, no KD emerges for the K-IVC state. 

The counterintuitive absence of KD in the $\nu = -2$ K-IVC state is part of a more general \emph{exact} result, relying on the $C_{2z}$, $T$, and $P$ symmetries of TBG\citeSFSymmetriesChiral{}: a VP even-$\nu$ insulator with only fully filled and fully empty valley-spin flavors and all its $\Uncf{}$ rotations have \emph{identical} spectral functions, without exhibiting KD. Note that these are precisely the theoretically proposed exact ground states of $H_I$ at even filling and away from the chiral limit~\cite{KAN19,BUL20,LIA21,XIE21}. Moreover, even when the $P$ symmetry is broken, we find that $C_{2z}$ and $T$ are enough to guarantee the \emph{exact} absence of KD in the K-IVC state, although not necessarily in its general $\Uncf{}$ rotations\citeSFSymmetriesPHBreak{}.   

When $\ket{\varphi}$ is \emph{not} a $\Uncf{}$ rotation of an insulator with only fully filled or fully empty valley-spin flavors, intervalley coherence \emph{can} lead to KD, but fine-tuned counter-examples do exist. For maximally spin polarized states, we have derived simple rules governing the presence of KD\citeSimpleRules{}: (1) Filling a single IVC Chern band gives rise to KD; (2) An exact cancellation of the KD signal occurs upon filling a pair of Chern bands with \emph{opposite} Chern numbers whose valley polarization projections in the valley $xy$ plane of the Bloch sphere are nonzero and cancel out, \eg{} the K-IVC from \cref{fig:realSpaceExamples:a}. In \cref{fig:realSpaceExamples:c,fig:realSpaceExamples:d,fig:realSpaceExamples:e,fig:realSpaceExamples:f}, we illustrate these rules for $\mathcal{C} \neq 0$ states, which at odd filling are the theoretical ground states, while at even filling are the ground states in-field~\cite{NUC20,WU21a,DAS21,LIU21e,SAI21,CHO21a}. The $\nu=-2$, $\mathcal{C}=2$ and $\nu=-1$, $\mathcal{C}=3$ VP Chern insulators trivially harbor no KD. The $\nu=-2 $, $\mathcal{C}=2$ IVC insulator does exhibit KD, since the valley polarizations of the two bands projected in the $xy$ plane of the valley Bloch sphere do not cancel out. Finally, the $\nu=-1$, $\mathcal{C}=1$ partially IVC insulator has one VP filled Chern band and a pair of filled IVC Chern bands, satisfying rule 2 and therefore not displaying any KD. Further examples are presented\citeAdditionalExamples{}. Between states showing no KD, further LDOS differences exist because the Chern band operators $\hat{d}^\dagger_{\vk,e_Y,\eta,s}$ are primarily located on a single SLG sublattice, depending on $e_Y \eta = \pm 1$. In the $\nu=-2$, $\mathcal{C}=2$ state, two Chern bands with $e_Y=+1$ are occupied in the same valley $\eta = +$ (and two spin sectors), leading to the appearance of a triangular lattice in the STM signal from \cref{fig:realSpaceExamples:c}. To obtain the VP $\nu =-1 $, $\mathcal{C}=3$ (IVC $\nu = -1$, $\mathcal{C}=1$) state, a Chern band with $\eta = -$, $e_Y=+1$ ($\eta = +$, $e_Y = -1$) is added, which is polarized primarily on the other graphene sublattice. Hence two interpenetrating triangular lattices of weight $2:1$ appear in the LDOS patterns from \cref{fig:realSpaceExamples:e,fig:realSpaceExamples:f}.

\emph{Conclusions}.~We analyzed the STM signal of a multitude of the predicted candidates for the correlated insulators in TBG and showed that it can be used to differentiate between the ground states. Both numerically and analytically, we found that the $\mathcal{C} = 0$ celebrated K-IVC states at $\nu = \pm 2, 0$ do not exhibit a KD, while other IVC states, including the $\nu=\pm 2$ T-IVC or $\mathcal{C} = 2$ states do display KD. We experimentally measured the STM signal for the $\nu = \pm 4$ band insulator, and used it to validate the the strong-coupling regime. The broadening of the signal and the specific variations of the signal from the AA to the AB sites are unique signatures of the strong-coupling regime. Our work paves the road toward the unambiguous identification of the TBG correlated insulators.

\emph{Acknowledgments}.~We thank Fang Xie for fruitful discussions. 
The simulations presented in this work were performed using the Princeton Research Computing resources at Princeton University, which is a consortium of groups led by the Princeton Institute for Computational Science and Engineering (PICSciE) and Office of Information Technology's Research Computing.
B.A.B. and N.R. were supported by the Office of Naval Research (ONR Grant No. N00014-20-1-2303), the National Science Foundation (EAGER Grant No. DMR 1643312), a Simons Investigator grant (No. 404513), the BSF Israel US foundation (Grant No. 2018226), the Gordon and Betty Moore Foundation through Grant No. GBMF8685 towards the Princeton theory program, the Gordon and Betty Moore Foundation's EPiQS Initiative (Grant No. GBMF11070), and a Guggenheim Fellowship from the John Simon Guggenheim Memorial Foundation.
B.A.B. and N.R. were also supported by the NSF-MRSEC (Grant No. DMR-2011750), and the Princeton Global Network Funds.
B.A.B. and N.R. gratefully acknowledge financial support from the Schmidt DataX Fund at Princeton University made possible through a major gift from the Schmidt Futures Foundation.
This work is also partly supported by a project that has received funding from the European Research Council (ERC) under the European Union's Horizon 2020 research and innovation programme (Grant Agreement No. 101020833).
O. V. is supported by NSF DMR-1916958 and by the Gordon and Betty Moore Foundation's EPiQS Initiative Grant No. GBMF11070.
This work was also supported by the Gordon and Betty Moore Foundation’s EPiQS initiative grants GBMF9469 and DOE-BES Grant No. DE-FG02-07ER46419 to A.Y. Other support for the experimental work was provided by NSF-MRSEC through the Princeton Center for Complex Materials NSF-DMR- 2011750, NSF-DMR-1904442, ExxonMobil through the Andlinger Center for Energy and the Environment at Princeton, and the Princeton Catalysis Initiative. We are grateful to K. Watanabe and T. Taniguchi for providing high-quality hexagonal boron nitride crystals used for the experimental work. A.Y. acknowledges the hospitality of the Aspen Center for Physics, which is supported by National Science Foundation Grant No. PHY-1607611. 
{}

\emph{Note added}.~
During the later stages of preparation of our manuscript, we became aware of Ref.~\cite{HON21} posted on the same date on arXiv, which also computes the STM signal of various TBG states using Hartree-Fock methods. Where they overlap, our conclusions (\ie{} the vanishing of KD in the K-IVC state) agree with the ones presented in Ref.~\cite{HON21}.
{}

\let\oldaddcontentsline\addcontentsline
\renewcommand{\addcontentsline}[3]{}
\bibliographystyle{apsrev4-2}
\bibliography{STMRefs,TBGRefs}

\let\addcontentsline\oldaddcontentsline

\appendix
\onecolumngrid
\newpage
\tableofcontents
\newpage

\section{Review of notation}\label{app:sec:notation}
This \siSection{} provides a short review of the single-particle and interacting twisted bilayer graphene (TBG) Hamiltonians. We will follow the same conventions as those employed in Refs.~\cite{BER21,SON21,BER21a,LIA21,BER21b,XIE21}, to which the interested reader is referred for more details. After briefly discussing the Bistritzer-MacDonald model of the single-particle Hamiltonian~\cite{BIS11}, we outline the discrete symmetries of TBG, which were extensively discussed in Refs.~\cite{SON19,TAR19,PO19,SON21,BER21a}, as well as the gauge-fixing procedure used throughout this work~\cite{BER21a}. We then review the TBG interaction Hamiltonian~\cite{KAN19,SEO19,BUL20,BER21a} and its enlarged continuous symmetries arising under various limits~\cite{KAN19,SEO19,BUL20,VAF20,BER21a}. These will be used extensively in our analytical proofs concerning the scanning tunneling microscopy (STM) signal of TBG.   

\subsection{Single-particle Hamiltonian}\label{app:sec:notation:sp}

\subsubsection{Fermion operators on the moir\'e lattice}\label{app:sec:notation:sp:fermion_ops}

\begin{figure}[!t]
	\includegraphics[width=0.3\textwidth]{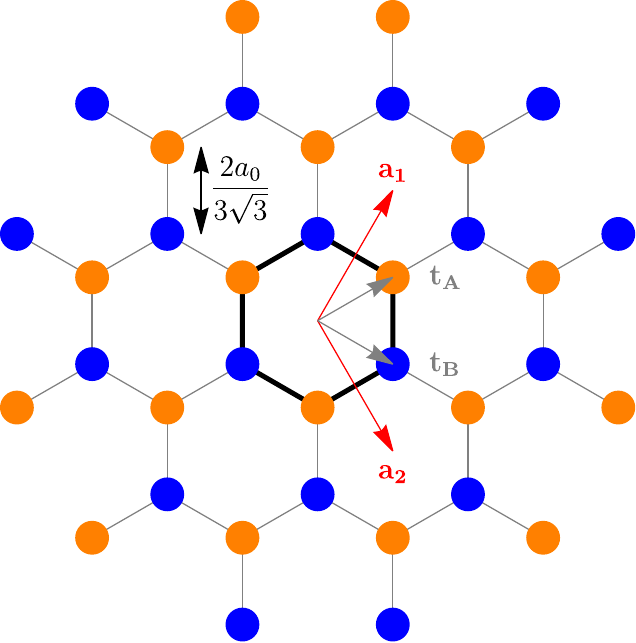}
	\caption{The reference single layer graphene (SLG) lattice. The corresponding triangular lattice is generated by the lattice vectors $\vec{a}_{1}$ and $\vec{a}_{2}$. Within each unit cell, we denote the carbon atoms belonging to the two sublattices $A$ and $B$ by orange and blue circles, respectively. The nearest-neighbor distance between two carbon atoms is given by $\frac{2 a_0}{3 \sqrt{3}}$. Additionally, $\vec{t}_A$ and $\vec{t}_B$ respectively denote the displacements corresponding to sublattices $A$ and $B$. The top ($l=+1$) and bottom ($l=-1$) graphene layers of TBG are obtained by rotating the reference lattice by an angle $l \theta /2$.}
	\label{app:fig:slg_reference_lattice}
\end{figure}

TBG consists of two graphene layers rotated at an angle $\theta$ relative to one another. We define $\mathcal{R}_{\theta, l}$ to be the matrix implementing the rotation transformation corresponding to the graphene layer $l = \pm$ (where $l=+$ denotes the top layer, while $l=-$ denotes the bottom layer) relative to a \emph{reference} (\ie{} unrotated) coordinate system
\begin{equation}
	\label{app:eqn:notRrotation}
	\mathcal{R}_{\theta, l} =
	\begin{pmatrix}
		\cos \left(\frac{\theta  l}{2}\right) & -\sin \left(\frac{\theta  l}{2}\right) \\
		\sin \left(\frac{\theta  l}{2}\right) & \cos \left(\frac{\theta  l}{2}\right) \\	
	\end{pmatrix}.
\end{equation}
Within layer $l$, we define $\hat{a}^\dagger_{\vec{R},\alpha,s,l}$ to be the microscopic fermion operator creating an electron of spin $s=\uparrow,\downarrow$ in the unit cell indexed by $\vec{R}$ (but located at $\mathcal{R}_{\theta,l} \vec{R}$) and graphene sublattice $\alpha=A,B$. $\vec{R}$ belongs to the reference single layer graphene (SLG) lattice, \ie{} $\vec{R} \in \mathbb{Z} \vec{a}_1 + \mathbb{Z} \vec{a}_2$, where 
\begin{equation}
	\label{app:eqn:grapheneLat}
	\vec{a}_1 = a_0 \left( \frac{1}{3}, \frac{1}{\sqrt{3}} \right)^T, \qquad
	\vec{a}_2 = a_0 \left( \frac{1}{3}, -\frac{1}{\sqrt{3}} \right)^T
\end{equation}
are the primitive SLG lattice vectors, with $\frac{2 a_0}{3 \sqrt{3}}$ being the length of a carbon-carbon bond, as shown in \cref{app:fig:slg_reference_lattice}. We also define the displacement vectors for the two graphene sublattices relative to the SLG unit cell origin according to 
\begin{equation}
	\label{app:eqn:def_graphene_displ_vecs}
	\vec{t}_{A} = \frac{2}{3} \vec{a}_1 + \frac{1}{3} \vec{a}_2, \qquad
	\vec{t}_{B} = \frac{1}{3} \vec{a}_1 + \frac{2}{3} \vec{a}_2.
\end{equation}
Using \cref{app:eqn:notRrotation,app:eqn:grapheneLat,app:eqn:def_graphene_displ_vecs}, we can write the Fourier transformation of the $\hat{a}^\dagger_{\vec{R},\alpha,s,l}$ operators over the SLG Brillouin Zone $\mathrm{BZ}_{l}$ corresponding to layer $l$ as
\begin{equation}
	\label{app:eqn:ft_a_ops}
	\hat{a}^\dagger_{\vec{R},\alpha,s,l} = \frac{1}{\sqrt{N_0}} \sum_{\vec{p} \in \mathrm{BZ}_l} \hat{a}^\dagger_{\vec{p},\alpha,s,l} e^{-i \vec{p} \cdot \mathcal{R}_{\theta,l} \left( \vec{R} + \vec{t}_\alpha \right)}.
\end{equation}
In \cref{app:eqn:ft_a_ops}, $N_0$ represents the number of unit cells in each graphene layer, and the momentum $\vec{p}$ is measured from the $\Gamma$ point of $\mathrm{BZ}_{l}$. It is also useful to introduce the reciprocal lattices corresponding to the two graphene layers. More precisely, we let $\mathcal{G}_{l} = \mathbb{Z} \vec{g}_{l,1} + \mathbb{Z} \vec{g}_{l,2}$ be the reciprocal lattice corresponding to layer $l$, generated by the reciprocal vectors $\vec{g}_{l,i} = \mathcal{R}_{\theta,l} \vec{g}_i$, where the reference or unrotated reciprocal lattice vectors $\vec{g}_i$ are given by
\begin{equation}
	\label{app:eqn:gen_rec_slg}
	\vec{g}_1 = \frac{2 \pi}{a_0} \left( \frac{3}{2}, \frac{\sqrt{3}}{2} \right)^T, \qquad
	\vec{g}_2 = \frac{2 \pi}{a_0} \left( \frac{3}{2}, -\frac{\sqrt{3}}{2} \right)^T.
\end{equation}

Focusing on TBG, we define $\vec{K}_{l} = \frac{1}{3} \left(\vec{g}_{l,1} + \vec{g}_{l,2} \right)$ as the $K$ point of the SLG Brillouin Zone $\mathrm{BZ}_l$. $\vec{K}_{+}$ and $\vec{K}_{-}$ differ by a twist angle $\theta$. We take $\mathbf{K}_{\pm}$ to be along the direction with an angle $ \pm \theta/2$ to the $\hat{\vec{x}}$ axis. Each graphene layer contains two valleys $K$ and $K'$, labeled by $\eta=\pm$ and located at momenta $\eta \vec{K}_{\pm}$, corresponding to the two (decoupled) valleys of the moir\'e single-particle Hamiltonian.

Introducing the two-dimensional momenta
\begin{equation}
	\vq_{1}=\left(\vec{K}_{+}-\vec{K}_{-}\right)=k_\theta \left( 0,1 \right)^T, \qquad 
	\vq_{2}=C_{3z}\vq_{1}=k_\theta \left(-\frac{\sqrt{3}}{2},-\frac{1}{2} \right)^T,\qquad
	\vq_{3}=C_{3z}^2\vq_{1}=k_\theta \left( \frac{\sqrt{3}}{2},-\frac{1}{2} \right)^T,
\end{equation}
where $k_\theta=|\vec{K}_{-}-\vec{K}_{+}|=2|\mathbf{K}_{+}|\sin(\theta/2)$, we can define the moir\'e Brillouin Zone (MBZ) for the (triangular) TBG moir\'e lattice $\mathcal{Q}_{0}=\mathbb{Z}\mathbf{b}_{M1}+\mathbb{Z}\mathbf{b}_{M2}$, which is generated by the reciprocal vectors
\begin{equation}
	\vec{b}_{M1}=\vq_3-\vq_1\ ,\qquad  \vec{b}_{M2}=\vq_3-\vq_2 .
\end{equation}
Additionally, we introduce two shifted momentum lattices $\mathcal{Q}_{+}=\mathbf{q}_{1}+\mathcal{Q}_{0}$ and $\mathcal{Q}_{-}=-\mathbf{q}_{1}+\mathcal{Q}_{0}$, which together form a honeycomb lattice. This allows us to define the low-energy TBG fermion operators as 
\begin{equation}	
		\label{app:eqn:low_en_fermions_c}
		\hat{c}^\dagger_{\vk,\vQ,\eta,\alpha,s} \equiv \hat{a}^\dagger_{\eta \vec{K}_l + \vk - \vQ, \alpha, s, l}, \qquad \text{for} \qquad \vQ \in \mathcal{Q}_{\eta l},
\end{equation}
with $\eta l$ denoting the product between the valley and layer indices, $\vk \in \text{MBZ}$, and $\vk=\vec{0}$ representing the $\Gamma_{M}$ point (\ie{} the $\Gamma$ point of the MBZ).

\subsubsection{Hamiltonian and the energy band basis}\label{app:sec:notation:sp:hamiltonian}

The Bistrizer-MacDonald model~\cite{BIS11} for the single-particle TBG Hamiltonian reads as~\cite{BER21,SON21,BER21a}
\begin{equation}
	\label{app:eqn:spHamiltonian}
	\hat{H}_{0} = \sum_{\vk \in \text{MBZ}} \sum_{\eta, \alpha, \beta, s} \sum_{\vQ,\vQ' \in \mathcal{Q}_{\pm}} \left[h^{\left(\eta\right)}_{\vQ,\vQ'} \left( \vk \right) \right]_{\alpha \beta} \hat{c}^\dagger_{\vk,\vQ,\eta,\alpha,s} \hat{c}_{\vk,\vQ',\eta,\beta,s},
\end{equation}
where the first-quantized TBG Hamiltonian in valley $\eta$, $h^{\left(\eta\right)}_{\vQ,\vQ'} \left( \vk \right)$, is independent on spin as a consequence of the absence of spin-orbit coupling in SLG. In valley $\eta = +$, it takes the form
\begin{equation}
	\label{app:eqn:firstQuantHamiltonian}
	h^{\left(+\right)}_{\vQ,\vQ'} \left( \vk \right) = v_F \delta_{\vQ,\vQ'} \left( \vk - \vQ \right) \cdot \bSigma - \lambda v_F\frac{\theta}{2} \zeta_{\vQ} \delta_{\vQ,\vQ'} \left( \vk - \vQ \right) \times \bSigma + \sum_{j=1}^{3} T_{j} \delta_{\vQ,\vQ' \pm \vq_j}, \\
\end{equation}
where $v_F$ is SLG Fermi velocity, $\theta$ is the twist angle, $\bSigma = \left( \sigma_x,\sigma_y \right)$, $\zeta_{\vQ}$ is the sublattice factor
\begin{equation}
	\label{app:eqn:defZeta}
	\zeta_{\vQ} = \begin{cases}
		+1 \quad \text{if} \quad \vQ \in \mathcal{Q}_{+} \\
		-1 \quad \text{if} \quad \vQ \in \mathcal{Q}_{-} 
	\end{cases},
\end{equation}
and the $T_j$'s represent the interlayer tunneling matrices defined according to 
\begin{equation}
	\label{app:eqn:interlayerT}
	T_j = w_0 \sigma_0 + w_1 \left[ \sigma_x \cos \frac{2 \pi \left( j-1 \right)}{3} + \sigma_y \sin \frac{2 \pi \left( j-1 \right)}{3} \right], \quad \text{for} \quad j=1,2,3
\end{equation}
For small angles $\theta$, the second term in \cref{app:eqn:firstQuantHamiltonian} is suppressed in magnitude with respect to the first one and is usually neglected~\cite{SON19,SON21}. We will investigate its effects through the parameter $\lambda=0,1$. For $\lambda = 0$ the TBG Hamiltonian features a unitary particle-hole symmetry, which is only slightly broken in the case $\lambda = 1$~\cite{SON19,SON21}. The tunneling matrices $T_j$ ($1 \leq j \leq 3$) depend on two parameters, $w_0 \geq 0$ and $w_1 \geq 0$, which are the interlayer hoppings at the AA and AB/BA stacking centers of the two graphene sheets, respectively. Generically, in realistic systems $w_0 < w_1$ due to lattice relaxation and corrugation effects~\cite{UCH14,WIJ15,DAI16,JAI16,SON21}. For the numerical calculations in this paper, we will use $\theta = 1.05 \degree$, $w_1 = \SI{110}{\milli\eV}$, $v_F = \SI{5.944}{\eV \angstrom}$, $\abs{\vec{K}_+} = \SI{1.703}{\angstrom^{-1}}$, and explore multiple values of the ratio $w_0 / w_1 \leq 1$. Finally, we note that the first-quantized Hamiltonian in valley $\eta = -$ is obtained through
\begin{equation}
	h^{\left(-\right)}_{\vQ,\vQ'} \left( \vk \right) = \sigma_x h^{\left(+\right)}_{-\vQ,-\vQ'} \left( -\vk \right) \sigma_x .
\end{equation}  

The TBG Hamiltonian in \cref{app:eqn:spHamiltonian} can be diagonalized as follows
\begin{equation}
	\label{app:eqn:diag_sp_Hamiltonian}
	\hat{H}_{0} = \sum_{\vk \in \text{MBZ}} \sum_{\eta, n, s} \epsilon_{n,\eta} \left( \vk \right) \hat{c}^\dagger_{\vk,n,\eta,s} \hat{c}_{\vk,n,\eta,s},
\end{equation}
where 
\begin{equation}
	\label{app:eqn:en_band}
	\hat{c}^\dagger_{\vk,n,\eta,s} = \sum_{\vQ\in \mathcal{Q}_{\pm},\alpha} u_{\vQ \alpha; n \eta} \left( \vk \right) \hat{c}^\dagger_{\vk,\vQ,\eta,\alpha,s},	
\end{equation}
are the energy band operators. We define $u_{\vQ \alpha; n \eta} \left( \vk \right)$ to be the eigenstate wave functions of energy band $n$ of the first quantized single-particle TBG Hamiltonian~\cite{BER21a}
\begin{equation}
	\sum_{\beta} \sum_{\vQ' \in \mathcal{Q}_{\pm}} \left[ h^{\left(\eta\right)}_{\vQ,\vQ'} \right]_{\alpha \beta} u_{\vQ' \beta; n \eta} \left( \vk \right) = \epsilon_{n,\eta} u_{\vQ \alpha; n \eta} \left( \vk \right). 
\end{equation} 
For each valley and spin, we will use the integer $n>0$ to denote the $n$-th conduction band and use the integer $n<0$ to label the $\abs{n}$-th valence band. Throughout this paper, we will be concerned exclusively with the active TBG bands (corresponding to $n= \pm 1$). Finally, we note that the completeness of the TBG eigenstate wave functions allows us to express the moir\'e lattice operators from \cref{app:eqn:low_en_fermions_c} in the energy band basis as
\begin{equation}
	\label{app:eqn:low_en_to_eig}
	\hat{c}^\dagger_{\vk,\vQ,\eta,\alpha,s} = \sum_{n} u^{*}_{\vQ\alpha;n \eta} \left(\vk \right) \hat{c}^\dagger_{\vk,n,\eta,s}.
\end{equation}

\subsubsection{Discrete symmetries of TBG}\label{app:sec:notation:sp:symmetries}
We now briefly review the discrete symmetries of TBG which have been derived and extensively discussed in Refs.~\cite{SON19,TAR19,PO19,SON21,BER21a}. Since graphene has zero spin-orbit coupling, we can define a set of spinless symmetry transformations for TBG: the spinless unitary discrete symmetries $C_{2z}$, $C_{3z}$, $C_{2x}$, and the spinless anti-unitary time-reversal symmetry $T$. In addition to the above symmetry operators which commute with the many-body projected Hamiltonian of TBG ($H$), one can also define a unitary particle-hole transformation $P$~\cite{SON19,SON21}, as well as a chiral transformation $C$~\cite{TAR19}. The particle-hole transformation is an anticommuting symmetry of the single-particle TBG Hamiltonian from \cref{app:eqn:spHamiltonian} for $\lambda = 0$, and remains an approximate symmetry of the model for $\lambda = 1$~\cite{SON19,SON21}. In the first chiral limit, when the interlayer AA-hopping $w_0$ can be neglected ($w_0 = 0$), the chiral transformation $C$ also denotes an anticommuting symmetry of $\hat{H}_0$~\cite{TAR19,BER21a}. It is worth noting, that the first chiral limit \emph{always} implies the presence of particle-hole symmetry, as the second term in \cref{app:eqn:firstQuantHamiltonian} can be gauged away~\cite{TAR19,WAN21a}. 

We denote the action of any symmetry transformation operator $g$ on the moir\'e lattice fermions to be
\begin{equation}
	\label{app:eqn:action_of_g_moire_fermions}
	g \hat{c}^\dagger_{\vk,\vQ,\eta,\alpha,s} g^{-1} = \sum_{\vQ', \eta', \beta}\left[ D \left( g \right) \right]_{\vQ' \eta' \beta, \vQ \eta \alpha}\hat{c}^\dagger_{g\vk,\vQ',\eta',\beta,s},
\end{equation}
where $D \left( g \right)$ is the representation matrix of the symmetry operator $g$ in the space of indices $\left\lbrace \vQ,\eta,\alpha \right\rbrace$ of the $\hat{c}^\dagger_{\vk,\vQ,\eta,\alpha,s}$ fermion operators. We denote $g \vk$ to be the momentum obtained after acting the transformation $g$ on momentum $\vk$. In particular $T \vk= P \vk = - \vk$, while $C\vk = \vk$. The representation matrices for the discrete symmetries of TBG are given by~\cite{SON19, SON21, BER21a}
\begin{align}
	\left[D \left(C_{2z}\right)\right]_{\vQ' \eta' \beta, \vQ \eta \alpha} &= \delta_{\vQ', -\vQ} \delta_{\eta',-\eta} \left(\sigma_x\right)_{\beta\alpha}, \label{app:eqn:c2z_rep} \\
	\left[D \left(C_{3z}\right)\right]_{\vQ' \eta' \beta, \vQ \eta \alpha} &= \delta_{\vQ', C_{3z} \vQ} \delta_{\eta', \eta} \left(e^{i\eta  \frac{2\pi}{3}\sigma_{z}} \right)_{\beta\alpha}, \label{app:eqn:c3z_rep} \\ 
	\left[D \left(C_{2x}\right)\right]_{\vQ' \eta' \beta, \vQ \eta \alpha} &= \delta_{\vQ', C_{2x} \vQ} \delta_{\eta', \eta} \left(\sigma_x\right)_{\beta\alpha}, \label{app:eqn:c2x_rep} \\ 
	\left[D \left(T\right)\right]_{\vQ' \eta' \beta, \vQ \eta \alpha} &= \delta_{\vQ', -\vQ} \delta_{\eta',-\eta} \delta_{\beta,\alpha}, \label{app:eqn:t_rep}  \\	
	\left[D \left( P \right) \right]_{\vQ' \eta' \beta, \vQ \eta \alpha} &= \delta_{\vQ', -\vQ} \delta_{\eta', \eta} \delta_{\beta,\alpha} \zeta_{\vQ}, \label{app:eqn:p_rep} \\
	\left[D \left( C \right) \right]_{\vQ' \eta' \beta, \vQ \eta \alpha} &= \delta_{\vQ', \vQ} \delta_{\eta', \eta} \left( \sigma_z \right)_{\beta\alpha},  \label{app:eqn:c_rep} 
\end{align}
where in \cref{app:eqn:p_rep}, we have employed the sublattice factor $\zeta_{\vQ}$ defined in \cref{app:eqn:defZeta}.

\subsubsection{Gauge-fixing the single-particle spectrum}\label{app:sec:notation:sp:gauge}
The symmetries presented in \cref{app:sec:notation:sp:symmetries} yield certain relations between single-particle TBG eigenstates, which will prove instrumental in deriving the properties of the TBG spectral function. Here, we briefly review the gauge-fixing conditions for the TBG single-particle eigenstates defined in \cref{app:eqn:en_band} which were introduced in Refs.~\cite{SON21,BER21a}.

For brevity, we will consider the wave function $u_{\mathbf{Q} \alpha; n \eta}\left( \vk \right)$ as a column vector $u_{n\eta}\left( \vk \right)$ in the space of indices $\left\lbrace \vQ,\alpha \right\rbrace$. Furthermore, when a representation matrix $D \left(g \right)$ of an operation $g$ defined in \cref{app:eqn:action_of_g_moire_fermions} acts on a wave function $u_{n\eta'}\left( \vk \right)$, we denote the resulting wave function in valley $\eta$ for short as $\sum_{\eta'} \left[D(g)\right]_{\eta\eta'} u_{n\eta'}\left( \vk \right)$, the components of which are given by $\sum_{\vQ'\beta\eta'}\left[D\left(g\right)\right]_{\vQ\alpha\eta,\vQ'\beta\eta'}u_{\vQ'\beta;n\eta'}\left( \vk \right)$. Namely, we suppress the indices $\left\lbrace \vQ,\alpha \right\rbrace$ of the representation matrix $D\left(g\right)$ to streamline notation.

When $g$ is a commuting (anticommuting) symmetry operator of the single-particle TBG Hamiltonian, if $u_{n\eta'}( \vk )$ is an eigenstate wave function at momentum $\vk$, the wave function $\sum_{\eta} \left[ D \left(g\right)\right]_{\eta\eta'} u_{n\eta'}\left(\vk\right)$ (an additional complex conjugation is needed if $g$ is anti-unitary) must also be an eigenstate wave function at momentum $g\vk$ at the same  (opposite) single-particle energy. This allows us to define a sewing matrix corresponding to the symmetry operator $g$ and the eigenstates $u_{n\eta'}( \vk )$
\begin{equation}
	\label{app:eqn:symmetry_sewing_definition}
	\left[ D \left(g\right)\right]_{\eta\eta'} u_{n\eta'}\left(\vk\right) = \sum_{m} \left[ B^g \left( \vk \right) \right]_{m \eta, n\eta'} u_{m\eta} \left(g \vk \right).
\end{equation}
In the energy band basis defined in \cref{app:eqn:en_band}, a symmetry $g$ acts as 
\begin{equation}
	\label{app:eqn:symmetry_action_sewing}
	g \hat{c}^\dagger_{\vk, n, \eta', s} g^{-1} = \sum_{m,\eta} \left[ B^g \left( \vk \right) \right]_{m \eta, n\eta'}\hat{c}^\dagger_{g\vk, m, \eta, s}.
\end{equation}

The gauge-fixing of the TBG energy band operators was discussed at length in Refs.~\cite{SON21,BER21a}. We will first consider the particle-hole symmetric case ($\lambda = 0$). We will only summarize the results here and refer the reader to Refs.~\cite{SON21, BER21a} for complete proofs. All sewing matrices are closed within the pair of bands $n= \pm 1$. Therefore, for the bands with band index $n= \pm 1$, we will use $\zeta^a$ and $\tau^a$ ($a=0,x,y,z$) to denote the identity and Pauli matrices in the energy band $n=\pm 1$ and the valley spaces, respectively. For all the symmetries that leave $\vk$ invariant, the following $\vk$-independent gauge-fixings will be adopted in this paper
\begin{equation}
	\label{app:eqn:sewing_mats}
	B^{C_{2z} T} \left( \vk \right) = \zeta^0 \tau^0, \qquad
	B^{C_{2z} P} \left( \vk \right) = \zeta^y \tau^y, \qquad
	B^{C} \left( \vk \right) = \zeta^y \tau^z,
\end{equation}
where the sewing matrix corresponding to the chiral symmetry operator $C$ is only applicable in the first chiral limit ($w_0 / w_1 = 0$)~\cite{TAR19,BER21a}. Additionally, we will further fix the relative gauge between wave functions at momenta $\vk$ and $-\vk$ by fixing the sewing matrices of $C_{2z}$ and $P$. 
\begin{equation}
	\label{app:eqn:sewing_mats_kDep}
	B^{C_{2z}}\left(\vk\right)=\begin{cases}
		\zeta^0\tau^x &\vk \neq \vk_{M_M} \\
		-\zeta^0\tau^x &\vk = \vk_{M_M} \\ 
	\end{cases} \quad
	B^{T}\left(\vk\right)=\begin{cases}
		\zeta^0\tau^x &\vk \neq \vk_{M_M} \\
		-\zeta^0\tau^x &\vk = \vk_{M_M} \\ 
	\end{cases} \quad
	B^{P}\left(\vk\right)=\begin{cases}
		-i\zeta^y\tau^z &\vk \neq \vk_{M_M} \\
		i\zeta^y\tau^z &\vk = \vk_{M_M} \\ 
	\end{cases},
\end{equation}
where $\vk_{M_M}$ denotes any one of the three equivalent $M_M$ points in the MBZ
\begin{equation}
	\label{app:eqn:three_trim_TBG}
	\vk_{M_M} \in \left\lbrace \frac{1}{2} \vec{b}_{M1}, \frac{1}{2} \vec{b}_{M2}, \frac{1}{2} \vec{b}_{M1} + \frac{1}{2} \vec{b}_{M2} \right\rbrace .
\end{equation}
The reason for the additional minus sign of the sewing matrix $B^{P}\left(\vk\right)$ at $\vk = \vk_{M_M}$ was explained in Ref.~\cite{BER21a}. In addition to the gauge-fixing conditions given in \cref{app:eqn:sewing_mats,app:eqn:sewing_mats_kDep}, we fix the relative sign between the single-particle wave functions $u_{+, \eta}\left( \vk \right)$ and $u_{-, \eta}\left( \vk \right)$ imposing~\cite{BER21}
\begin{equation}
	\label{app:eqn:c_continuous}
	\lim_{\vq \rightarrow \mathbf{0}}\abs{u^{\dagger}_{n \eta} \left( \vk + \vq \right) u_{n \eta}\left( \vk \right)- u^{\dagger}_{-n \eta} \left(\vk +\vq \right) u_{-n \eta} \left(\vk \right)}=0.
\end{equation}
By fixing the sewing matrix of the $C_{2z}T$ transformation according to \cref{app:eqn:sewing_mats_kDep}, as well as the continuous gauge condition from \cref{app:eqn:c_continuous}, we can introduce the Chern band basis~\cite{AHN19,SON21,BER21a} within the two active bands in each valley-spin flavor  
\begin{equation}
	\label{app:eqn:chern_band}
	\hat{d}^\dagger_{\vk,e_Y,\eta,s} = \frac{1}{\sqrt{2}} \left( \hat{c}^\dagger_{\vk,+1,\eta,s} + i e_Y \hat{c}^\dagger_{\vk,-1,\eta,s} \right) = \sum_{n} W_{e_Y,n} \hat{c}^\dagger_{\vk,n,\eta,s},
\end{equation}
where $e_Y=\pm1$ and the unitary matrix $W$ is given by 
\begin{equation}
	\label{app:eqn:chern_W_matrix}
	W = \frac{1}{\sqrt{2}}
	\begin{pmatrix}
		1 & i \\
		1 & -i \\
	\end{pmatrix}
\end{equation} As proven in Refs.~\cite{SON21,BER21a}, the operator $\hat{d}^\dagger_{\vk,e_Y,\eta,s}$ for $\vk \in \MBZ$ and fixed $e_Y$, $\eta$, and $s$ corresponds to a Chern band carrying Chern number $e_Y$. To numerically determine the gauge-fixed TBG wave functions, we follow the procedure employed in Ref.~\cite{XIE21}.

In the case $\lambda = 1$, $P$ is no longer an exact symmetry of $\hat{H}_{0}$ in \cref{app:eqn:firstQuantHamiltonian}, and a different approach is needed. We start by diagonalizing the first-quantized TBG Hamiltonian from \cref{app:eqn:firstQuantHamiltonian} in valley $\eta = +$, fix the sewing matrix of $C_{2z}T$ according to \cref{app:eqn:sewing_mats}, and impose the continuous gauge condition from \cref{app:eqn:c_continuous}. These conditions are enough to guarantee the existence of the Chern band basis as defined in \cref{app:eqn:chern_band}~\cite{SON21} for valley $\eta = +$. At the same time, for each $\vk$, we are free to transform the wave functions according to 
\begin{equation}
	u_{n +} \left( \vk \right) = \sum_{m=\pm 1} U_{nm} \left( \vk \right) u_{m +} \left( \vk \right),
\end{equation}
where the $2 \times 2$, momentum-dependent matrix $U \left( \vk \right)$ obeys
\begin{equation}
	\begin{split}
		U \left( \vk \right) &= \pm \mathbb{1} \quad \text{if} \quad \epsilon_{+1,\eta} \left( \vk \right) \neq \epsilon_{-1,\eta} \left( \vk \right), \\
		U \left( \vk \right) &\in \mathrm{SO}\left( 2 \right) \quad \text{if} \quad \epsilon_{+1,\eta} \left( \vk \right) = \epsilon_{-1,\eta} \left( \vk \right)
	\end{split}.
\end{equation}
Ref.~\cite{SON21} showed that \emph{even} when $\lambda = 1$, one still has
\begin{equation}
	1-\abs{u^{\dagger}_{-n\eta} \left( - \vk \right) \left[D(P)\right]_{\eta\eta} u_{n\eta}\left( \vk \right)}^2 \lesssim 0.04,
\end{equation}
meaning that the particle-hole symmetry breaking is small \emph{even} when $\lambda = 1$. We choose $U \left( \vk \right)$ in such a way as to ensure that
\begin{equation}
	\label{app:eqn:approx_p_sewing}
	\norm{\left[D(P)\right]_{++} u_{n+}\left( \vk \right) - \left[ B^P \left( \vk \right) \right]_{-n +, n+} u_{-n+}\left( \vk \right)} \quad \text{is minimized}.
\end{equation}
To find the wave functions in the $\eta = -$ valley, we then use the $C_{2z}$ symmetry of TBG and impose the corresponding sewing matrix from \cref{app:eqn:sewing_mats_kDep}. By \emph{approximately} fixing the sewing matrix of the particle-hole transformation in the $\lambda = 1$ case according to \cref{app:eqn:approx_p_sewing}, we ensure that this case is smoothly connected to the $\lambda = 0$ one. At the same time, \cref{app:eqn:approx_p_sewing} guarantees that even when $\lambda = 1$, \cref{app:eqn:symmetry_action_sewing} still holds \emph{approximately} for $g=P$.

\subsection{Interaction Hamiltonian}\label{app:sec:notation:ih}

\subsubsection{Form factor matrices}\label{app:sec:notation:ih:formFactor}

After discussing the single-particle wave functions of TBG, we now briefly review the Coulomb interaction Hamiltonian, which has been derived and discussed at length in Refs.~\cite{BER21a,LIA21}. We let $V ( \vec{r} )$ denote the interaction potential between two electrons within the TBG sample, and define $V ( \vq )$ as its Fourier transformation. For the time being, we will keep $V(\vec{r})$ generic and only require that it denotes a repulsive interaction ($V ( \vq ) \geq 0 $). It was shown in Ref.~\cite{KAN19,BUL20,BER21a} that the interaction Hamiltonian \emph{projected} in the active TBG bands is a positive semi-definite operator and reads as 
\begin{equation}
	\label{app:eqn:proj_int}
	H_I = \frac{1}{2 \omegaTBG} \sum_{\vq \in \MBZ} \sum_{\vG \in \mathcal{Q}_0} O_{-\vq-\vG} O_{\vq+\vG},
\end{equation}
where $\omegaTBG$ is the area of the TBG sample, and we have introduced the operators
\begin{equation}
	\label{app:eqn:int_o_operators}
	O_{\vq+\vG} = \sum_{\vk,\eta,s} \sum_{n,m = \pm 1} \sqrt{V (\vq + \vG)} M^{\eta}_{mn} \left(\vk,\vq+\vG \right) \left( \hat{c}^\dagger_{\vk+\vq,m,\eta,s} \hat{c}_{\vk,n,\eta,s} - \frac{1}{2} \delta_{\vq,\vec{0}}\delta_{m,n} \right).
\end{equation}
In \cref{app:eqn:int_o_operators} we have employed the wave function overlap matrix (also known as the form factor matrix) which is defined in terms of the active TBG wave functions as~\cite{BER21a}
\begin{equation}
	\label{app:eqn:ff_def}
	M^{\eta}_{mn} \left( \vk, \vq + \vG \right) = \sum_{\alpha} \sum_{\vQ \in \mathcal{Q}_{\pm}} u^*_{\vQ - \vG \alpha; m \eta} \left( \vk + \vq \right) u_{\vQ \alpha; n\eta} \left( \vk \right).
\end{equation}

The symmetries of the single-particle TBG Hamiltonian impose a series of constraints on the form factors matrices, through the gauge-fixing conditions from \cref{app:sec:notation:sp:gauge}~\cite{BER21a}. Letting $g$ be a symmetry of $\hat{H}_0$ as defined in \cref{app:sec:notation:sp:symmetries}, \cref{app:eqn:symmetry_sewing_definition} implies that 
\begin{align}
	&\sum_{\eta,n',m'} \left[ B^{g} (\vk) \right]_{n'\eta',n\eta} \left[ B^{g} (\vk + \vq) \right]^*_{m'\eta',m\eta} M^{\eta'}_{m'n'} \left( g \vk, g \vq + g \vG \right) \nonumber \\
	=& \sum_{\eta,n',m'}	\sum_{\alpha} \sum_{\vQ \in \mathcal{Q}_{\pm}} \left[ B^{g} (\vk + \vq) \right]^*_{m'\eta',m\eta} u^*_{g\vQ - g\vG \alpha; m' \eta'} \left( g\vk + g\vq \right) \left[ B^{g} (\vk) \right]_{n'\eta',n\eta} u_{g\vQ \alpha; n'\eta'} \left( g\vk \right) \nonumber \\
	=& \sum_{\eta}	\sum_{\alpha} \sum_{\vQ \in \mathcal{Q}_{\pm}} \left[\left[D(g)\right]_{\eta' \eta} u_{m \eta} \left(\vk + \vq \right) \right]^{*}_{\vQ - \vG \alpha} \left[\left[D(g)\right]_{\eta' \eta} u_{n \eta} \left(\vk \right) \right]_{\vQ\alpha} \nonumber \\
	=& \sum_{\alpha} \sum_{\vQ \in \mathcal{Q}_{\pm}} u^*_{\vQ - \vG \alpha; m \eta} \left( \vk + \vq \right) u_{\vQ \alpha; n\eta} \left( \vk \right),
	\label{app:eqn:ff_sym_index}
\end{align}
with an additional complex conjugation when $g$ is anti-unitary. Written in matrix form, \cref{app:eqn:ff_sym_index}  reads as 
\begin{equation}
	B^{\dagger g} (\vk + \vq)  M^{(*)} \left( g \vk, g \vq + g \vG \right) B^{g} (\vk) = M \left( \vk, \vq + \vG \right),
	\label{app:eqn:ff_sym_matrix}
\end{equation}
where ${}^{(*)}$ denotes an additional complex conjugation when $g$ is anti-unitary. Additionally, the form factor matrix obeys the following Hermiticity condition
\begin{equation}
	\label{app:eqn:ff_hermiticity}
	M^{*\eta}_{mn} \left( \vk, \vq + \vG \right) = M^{\eta}_{nm} \left( \vk + \vq, - \vq - \vG \right),
\end{equation}
which can be readily checked from its definition from \cref{app:eqn:ff_def}.

Under the gauge-fixing conditions outlined in \cref{app:sec:notation:sp:gauge} and as a consequence of \cref{app:eqn:ff_sym_matrix}, the $C_{2z}T$ symmetry imposes a reality condition for both the $\lambda = 0$ and $\lambda = 1$ cases~\cite{BER21a}. As such, the form factor matrix can be generically parameterized as 
\begin{align}
	M \left( \vk, \vq + \vG \right) = &
	\zeta^0 \tau^0 \alpha_{0} \left( \vk,\vq+\vG \right)+
	\zeta^x \tau^z \alpha_{1} \left( \vk,\vq+\vG \right)+
	i\zeta^y \tau^0 \alpha_{2} \left( \vk,\vq+\vG \right)+
	\zeta^z \tau^z \alpha_{3} \left( \vk,\vq+\vG \right) \nonumber \\
	+ &\zeta^0 \tau^z \alpha_{4} \left( \vk,\vq+\vG \right)+
	\zeta^x \tau^0 \alpha_{5} \left( \vk,\vq+\vG \right)+
	i\zeta^y \tau^z \alpha_{6} \left( \vk,\vq+\vG \right)+
	\zeta^z \tau^0 \alpha_{7} \left( \vk,\vq+\vG \right).
	\label{app:eqn:ff_param_1}
\end{align} 
When $\lambda = 0$, the presence of the anticommuting $C_{2z}P$ symmetry enforces $\alpha_{i} \left( \vk, \vq + \vG \right) = 0$ for all $4 \leq i \leq 7$, leading to the following parameterization in the band and valley subspaces~\cite{BER21a}
\begin{equation}
	\label{app:eqn:ff_param_2}
	M \left( \vk, \vq + \vG \right) = 
	\zeta^0 \tau^0 \alpha_{0} \left( \vk,\vq+\vG \right)+
	\zeta^x \tau^z \alpha_{1} \left( \vk,\vq+\vG \right)+
	i\zeta^y \tau^0 \alpha_{2} \left( \vk,\vq+\vG \right)+
	\zeta^z \tau^z \alpha_{3} \left( \vk,\vq+\vG \right). 
\end{equation} 
It is worth noting that, because the $P$ anticommuting symmetry (and hence the $C_{2z}P$ anticommuting symmetry) is only slightly broken even in the $\lambda = 1$ case~\cite{SON21}, we generically find that in \cref{app:eqn:ff_param_1}
\begin{equation}
	\abs{\alpha_i \left( \vk, \vq + \vG \right)} \ll \abs{\alpha_j \left( \vk, \vq + \vG \right)}, \quad \text{for} \quad 0 \leq i \leq 3 \quad \text{and} \quad 4 \leq j \leq 7,
\end{equation}
provided that \cref{app:eqn:approx_p_sewing} is imposed. Finally, we note that in the first chiral limit ($w_0 = 0$), the single-particle wave functions at a given momentum are additionally constrained by the chiral symmetry operator $C$. As shown in Ref.~\cite{BER21a}, this implies that the form factors are further restricted to the parameterization 
\begin{equation}
	\label{app:eqn:ff_param_3}
	M \left( \vk, \vq + \vG \right) = 
	\zeta^0 \tau^0 \alpha_{0} \left( \vk,\vq+\vG \right)+
	i\zeta^y \tau^0 \alpha_{2} \left( \vk,\vq+\vG \right). 
\end{equation} 

\subsubsection{Coulomb interaction potential}\label{app:sec:notation:ih:coulomb}

\begin{figure}[!t]
	\includegraphics[width=0.5\textwidth]{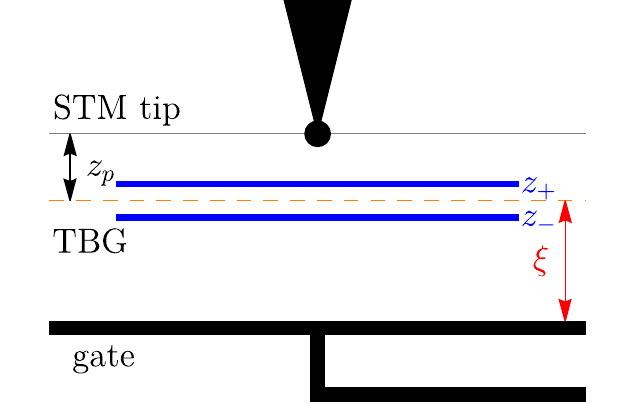}
	\caption{Schematic of the typical experimental setup considered. The TBG sample is is located in between a gate plate and the scanning tunneling microscopy (STM) tip, with $\xi$ denoting the distance between the gate and the sample and $z_p$ being the height of the STM tip. The two graphene monolayers are located at heights $z_l$ ($l = \pm$) with the interlayer TBG separation obeying $\abs{z_{+} - z_{-}} \ll \xi$.}
	\label{app:fig:schematicExperimentSetup}
\end{figure}
With the general form of the projected interaction TBG Hamiltonian at hand, we now turn to the electron-electron repulsion potential. We consider the experimental setup shown schematically in \cref{app:fig:schematicExperimentSetup}, corresponding to a single-gate arrangement for STM experiments. The potential between two electrons separated by $\vec{r}$ in the plane of the TBG sample includes a contribution from the image charge formed on the gate located at a distance $\xi$ below the sample
\begin{equation}
	V \left( \vec{r} \right) = \frac{e^2}{\epsilon} \left( \frac{1}{r} - \frac{1}{\sqrt{r^2 + (2\xi)^2}} \right).
\end{equation}
Using the identity
\begin{equation}
	\int \frac{\intD^2 q}{(2\pi)^2} \frac{e^{-\alpha q + i \vq \cdot \vec{r}}}{q} = \frac{1}{2\pi} \frac{1}{\sqrt{r^2 + \alpha^2}},
\end{equation}
we find that the Fourier transformation of the single-gate potential reads
\begin{equation}
	\label{app:eqn:potSingle}
	V(\vq) = \frac{2\pi e^2}{\epsilon} \frac{1-e^{-2 \xi q}}{q} = 2\pi U_{\xi} \xi^2 \frac{1-e^{-2 \xi q}}{\xi q}.
\end{equation}
In \cref{app:eqn:potSingle}, $e$ represents the electron charge, $\epsilon$ is the dielectric constant, $q = \abs{\vq}$, and $U_{\xi} = e^2/\left( \epsilon \xi \right)$. In this work and unless stated otherwise, we use $U_{\xi} =\SI{8}{\milli\eV}$ and $\xi = \SI{30}{\nano\meter}$. 

\subsubsection{Continuous symmetries of the projected interaction Hamiltonian}\label{app:sec:notation:ih:symmetries}

Throughout this paper, we will work exclusively in the flat limit, as defined by Refs.~\cite{BER21a,LIA21}, implying that the TBG Hamiltonian $H$ is given solely by the projected interaction Hamiltonian from \cref{app:eqn:proj_int}, \ie{} $H = H_I$. Refs.~\cite{LIA21,XIE21} have found that owing to its enlarged continuous symmetries (which will be discussed below), $H_I$ features a large manifold of degenerate ground states. This degeneracy is broken by the finite dispersion of the single-particle Hamiltonian which acts perturbatively within the ground state manifold~\cite{LIA21,XIE21}. Instead of focusing on a single ground state at each integer filling as predicted by perturbation theory~\cite{BUL20,LIA21}, we will instead explore the properties of the degenerate manifold in an effort to identify the experimentally-relevant ground state. 

We will now briefly review the symmetries of the interaction Hamiltonian $H_I$, which were derived and extensively discussed in Refs.~\cite{KAN19,SEO19,BUL20,BER21a}. Following the notation of Ref.~\cite{BER21a}, we shall use $\zeta^a$, $\tau^a$, and $s^a$ to denote the identity matrix ($a=0$), and Pauli matrices ($a=x,y,z$) in the band ($n = \pm 1$), valley ($\eta = \pm$), and spin ($s= \uparrow,\downarrow$) subspaces, respectively. 
\begin{itemize}
	\item \emph{$\mathrm{U} \left( 2 \right) \times \mathrm{U} \left( 2 \right)$ symmetry in the non-particle-hole-symmetric case ($\lambda = 1$)}. Due to the absence of spin-orbit coupling in TBG, as well as the suppression of intervalley scattering processes in \cref{app:eqn:proj_int}, $H_I$, enjoys a $\mathrm{U} \left( 2 \right) \times \mathrm{U} \left( 2 \right)$ rotation symmetry corresponding to independent spin-charge rotation within each valley. The corresponding generators are given by 
	\begin{equation}
		\label{app:eqn:generatorsu2u2}
		S^{ab} = \sum_{\vk}\sum_{\substack{n,\eta,s \\ n', \eta', s'}} \left[ s^{ab} \right]_{n \eta s, n' \eta' s'} \hat{c}^\dagger_{\vk,n,\eta,s} \hat{c}_{\vk,n',\eta',s'}, \quad \text{for} \quad a=0,z \quad b=0,x,y,z,
	\end{equation}
	where the matrices $s^{ab}$ read as
	\begin{equation}
		\label{app:eqn:generatorsu2u2:matrix}
		s^{0b} = \zeta^0 \tau^0 s^{b}, \quad 
		s^{zb} = \zeta^0 \tau^z s^{b}.
	\end{equation}
	\item \emph{$\mathrm{U} \left( 4 \right)$ symmetry in the particle-hole-symmetric case ($\lambda = 0$)}. The presence of the anticommuting particle-hole symmetry restricts the parameterization of the form factors, as shown in \cref{app:eqn:ff_param_2}. In turn, this enlarges the $\mathrm{U} \left( 2 \right) \times \mathrm{U} \left( 2 \right)$ symmetry of $H_I$ to the $\mathrm{U} \left( 4 \right)$ group~\cite{KAN19,SEO19,BUL20,BER21a} (corresponding to the so-called \emph{nonchiral-flat} limit, as defined by Ref.~\cite{BER21a}). In what follows, we will always denote the nonchiral-flat $\mathrm{U} \left( 4 \right)$ group as $\Uncf$ to distinguish it from other $\mathrm{U} \left( 4 \right)$ groups. The generators of $\Uncf$ read as 
	\begin{equation}
		\label{app:eqn:generatorsu4}
		S^{ab} = \sum_{\vk}\sum_{\substack{n,\eta,s \\ n', \eta', s'}} \left[ s^{ab} \right]_{n \eta s, n' \eta' s'} \hat{c}^\dagger_{\vk,n,\eta,s} \hat{c}_{\vk,n',\eta',s'}, \quad \text{for} \quad a,b=0,x,y,z,
	\end{equation}
	with the matrices $s^{ab}$ being given by
	\begin{equation}
		\label{app:eqn:generatorsu4:matrix}
		s^{0b} = \zeta^0 \tau^0 s^{b}, \quad
		s^{xb} = \zeta^y \tau^x s^{b}, \quad
		s^{yb} = \zeta^y \tau^y s^{b}, \quad 
		s^{zb} = \zeta^0 \tau^z s^{b}. 
	\end{equation}
	For future reference, we also define the set
	\begin{equation}	
		\label{app:eqn:def_N_u4}
		\mathcal{N}_{\Uncf} = \setDef*{ \sum_{a,b \in \left\lbrace 0,x,y,z \right\rbrace} \phi^{ab} s^{ab} }{\phi^{ab} \in \mathbb{C}, \phi^{00} = 0},
	\end{equation}
	which contains all eight-dimensional matrices which can be expressed as a linear combinations of the generators $s^{ab}$ from \cref{app:eqn:generatorsu4:matrix} for $a,b \in \left\lbrace 0,x,y,z \right\rbrace$ and $ab \neq 0$. For any $M \in \mathcal{N}_{\Uncf}$, $\left[M,s^{ab}\right] \in \mathcal{N}_{\Uncf}$, implying that the set $\mathcal{N}_{\Uncf}$ is closed under any $\Uncf$ group transformation.  
	\item \emph{$\mathrm{U} \left( 4 \right) \times \mathrm{U} \left( 4 \right)$ symmetry in the first chiral limit}. Finally, in the first chiral limit ($w_0 = 0 $)~\cite{TAR19,BER21a}, the anticommuting $C$ symmetry further restricts the parameterization of the form factors, given in \cref{app:eqn:ff_param_3}. As a consequence, the projected interaction Hamiltonian enjoys a large $\mathrm{U} \left( 4 \right) \times \mathrm{U} \left( 4 \right)$ symmetry~\cite{BUL20,BER21a} (corresponding to the \emph{chiral-flat} limit, as defined in Ref.~\cite{BER21a}) which is generated by the 32 operators
	\begin{equation}
		\label{app:eqn:generatorsu4u4}
		S_{\pm}^{ab} = \sum_{\vk}\sum_{\substack{n,\eta,s \\ n', \eta', s'}} \left[ s_{\pm}^{ab} \right]_{n \eta s, n' \eta' s'} \hat{c}^\dagger_{\vk,n,\eta,s} \hat{c}_{\vk,n',\eta',s'}, \quad \text{for} \quad a,b=0,x,y,z,
	\end{equation}
	with the matrices $s_{\pm}^{ab}$ being given by
	\begin{equation}
		\label{app:eqn:generatorsu4u4:matrix}
		s_{\pm}^{ab} = \frac{1}{2} \left( \zeta^0 \pm \zeta^y \right)\tau^a s^{b}. 
	\end{equation}
	In analogy with \cref{app:eqn:def_N_u4}, we also introduce the set 
	\begin{equation}	
		\label{app:eqn:def_N_u4u4}
		\mathcal{N}_{\mathrm{U} \left( 4 \right) \times  \mathrm{U} \left( 4 \right)} = \setDef*{ \sum_{a,b \in \left\lbrace 0,x,y,z \right\rbrace} \left( \phi_+^{ab} s_+^{ab} + \phi_-^{ab} s_-^{ab} \right) }{\phi_{\pm}^{ab} \in \mathbb{C}, \phi_{+}^{00} = \phi_{-}^{00} = 0},
	\end{equation}
	which is closed under any $\mathrm{U} \left( 4 \right) \times \mathrm{U} \left( 4 \right)$ transformation. 
\end{itemize}

The $\Uncf$ group in the nonchiral-flat limit is a subgroup of the $\mathrm{U} \left( 4 \right) \times \mathrm{U} \left( 4 \right)$ group in the chiral-flat limit, but is not one of the tensor-producted $\mathrm{U} \left( 4 \right)$ subgroups of the later~\cite{BER21a}. To better understand the group-subgroup relation between the two groups, is is instructive to recast the generators from \cref{app:eqn:generatorsu4,app:eqn:generatorsu4u4} in the Chern band basis defined in \cref{app:eqn:chern_band}. In the Chern band basis, the chiral-flat generators from \cref{app:eqn:generatorsu4u4} correspond to independent $\mathrm{U} \left( 4 \right)$ charge-valley-spin rotations within each Chern sector
\begin{equation}
	S_{\pm}^{ab} = \sum_{\vk}\sum_{\eta,s,\eta', s'} \left[ \tau^{a} s^{b} \right]_{\eta s, \eta' s'} \hat{d}^\dagger_{\vk,\pm 1,\eta,s} \hat{d}_{\vk,\pm 1,\eta',s'}, \quad \text{for} \quad a,b=0,x,y,z.
\end{equation}
When the chiral anticommuting symmetry is broken, the generators of the $\mathrm{U} \left( 4 \right) \times \mathrm{U} \left( 4 \right)$ group get combined into the generators of the $\Uncf$ group from \cref{app:eqn:generatorsu4}
\begin{equation}
	S^{ab} = \begin{cases}
		S_{+}^{ab} + S_{-}^{ab}, &\quad \text{if} \quad a = 0,z \\
		S_{+}^{ab} - S_{-}^{ab}, &\quad \text{if} \quad a = x,y \\
	\end{cases}, \quad \text{for} \quad a,b=0,x,y,z.
\end{equation}
As such, we find that away from the chiral limit, the fermions belonging to the two Chern sectors can no longer be rotated independently in the charge-valley-spin space. Instead, the $\Uncf$ rotations which do not mix the two valleys rotate the two Chern sectors in the same directions, while the generators corresponding to valley rotations in the $xz$ and $yz$ planes rotate the two Chern sectors in in opposite directions.

\section{Charge-one excitations at integer fillings}\label{app:sec:charge}
The spectral function of the TBG insulators which is measured by STM experiments depends on the eigenstates of the TBG Hamiltonian obtained by adding or removing one electron from the ground state -- the charge-one excitations. The latter constitute the main focus of this \siSection{}. We start with a brief review of the method devised by Ref.~\cite{BER21b} for obtaining the charge-one excitations above the ground states of TBG obtained in Ref.~\cite{LIA21}: the charge-one excitation can be found by diagonalizing the so-called charge-one excitation matrices. We then investigate the consequences of the various discrete symmetries of TBG from \cref{app:sec:notation:sp:symmetries} on the charge-one excitation matrices and obtain their parameterizations under different limits. Finally, by focusing on a specific state from the approximately degenerate ground state manifold of TBG derived in Ref.~\cite{LIA21}, we explicitly work out the charge-one excitation spectrum.  

\subsection{Charge-one excitation matrices}\label{app:sec:charge:matrices}

Ref.~\cite{BER21b} has shown that for the different ground states of the TBG Hamiltonian at integer fillings derived in Ref.~\cite{LIA21}, the charge-one excitations can be computed directly as a zero-body problem. Here, we will briefly review the procedure introduced in Ref.~\cite{BER21b} for obtaining them.

Let $\ket{\varphi}$ be one of the exact eigenstates of $H_I$ at filling $\nu$ introduced in Ref.~\cite{LIA21} (which will be specified in detail below). The energies and wave functions of the charge-one excitations can be determined using the following commutation relations~\cite{BER21b}
\begin{align}
	\left[H_I -\mu \hN , \hat{c}^\dagger_{\vk,n,\eta,s} \right] \ket{\varphi} &= \sum_{m} R^{\eta}_{mn} \left( \vk \right) \hat{c}^\dagger_{\vk,m,\eta,s} \ket{\varphi}, \label{app:eqn:r_mat_commut_+} \\
	\left[H_I -\mu \hN, \hat{c}_{\vk,n,\eta,s} \right] \ket{\varphi} &= \sum_{m} \tilde{R}^{\eta}_{mn} \left( \vk \right) \hat{c}_{\vk,m,\eta,s} \ket{\varphi}, \label{app:eqn:r_mat_commut_-}
\end{align}
where $\mu$ denotes the chemical potential and $\hN$ is the total fermion number operator
\begin{equation}
	\hN = \sum_{\vk,n,\eta,s} \hat{c}^\dagger_{\vk,n,\eta,s} \hat{c}_{\vk,n,\eta,s}.
\end{equation}
The charge-one excitation matrices $R^{\eta}_{mn} \left( \vk \right)$ and $\tilde{R}^{\eta}_{mn} \left( \vk \right)$ depend only on the filling $\nu$ of the ground state $\ket{\varphi}$ and are given in terms of the form factors introduced in \cref{app:eqn:ff_def}
\begin{align}
	R^{\eta}_{mn} \left( \vk \right) &= \frac{1}{2 \omegaTBG} \sum_{\vG \in \Q0} \left[ \left( \sum_{\vq \in \MBZ, m'} V\left( \vq + \vG \right) M^{* \eta}_{m'm} \left(\vk, \vq + \vG \right) M^{\eta}_{m'n} \left(\vk, \vq + \vG \right) \right) \right. \nonumber \\
	&\left.+ 2 A_{-\vG} \sqrt{V \left(\vG \right)} M^{\eta}_{mn} \left( \vk,\vG \right) \right] - \mu \delta_{mn}, \label{app:eqn:def_r_+}\\
	\tilde{R}^{\eta}_{mn} \left( \vk \right) &= \frac{1}{2 \omegaTBG} \sum_{\vG \in \Q0} \left[ \left( \sum_{\vq \in \MBZ, m'} V\left( \vq + \vG \right) M^{\eta}_{m'm} \left(\vk, \vq + \vG \right) M^{*\eta}_{m'n} \left(\vk, \vq + \vG \right) \right) \right. \nonumber \\
	&\left.- 2 A_{-\vG} \sqrt{V \left(\vG \right)} M^{*\eta}_{mn} \left( \vk,\vG \right) \right] + \mu \delta_{mn}, \label{app:eqn:def_r_-}
\end{align}
with the factor $A_{\vG}$ depending on the filling $\nu$ according to~\cite{BER21b}
\begin{equation}
	\label{app:eqn:def_Afact}
	A_{\vG} = \sqrt{V \left( \vG \right)} \sum_{\substack{\vk \in \MBZ \\ n,\eta}} \frac{\nu}{4} M^{\eta}_{nn} \left( \vk, \vG \right) = \sqrt{V \left( \vG \right)} \sum_{\vk \in \MBZ} \frac{\nu}{4} \Tr \left[ M \left( \vk, \vG \right)\right] .
\end{equation} 
The charge-one excitation matrices defined in \cref{app:eqn:def_r_+} are valid for any gauge choice. Under the gauge-fixing conditions defined in \cref{app:sec:notation:sp:gauge}, the form factors are real, so the complex-conjugation can be dropped. 

Refs.~\cite{BER21b, VAF20} derived the charge-one commutation relations from \cref{app:eqn:r_mat_commut_+,app:eqn:r_mat_commut_-} in the particle-hole symmetric case ($\lambda = 0$), for $\ket{\varphi}$ being one of the ground states\footnote{Strictly speaking, Ref.~\cite{LIA21} showed that in the nonchiral- and chiral-flat cases, respectively, the states from \cref{app:eqn:nonchiralGS} and \cref{app:eqn:chiralGS} are \emph{eigenstates} of $H_I$. Additionally, they were shown to be the \emph{ground states} of $H_I$ only for $\nu = 0$, or for $\nu \neq 0$ assuming that the flat metric condition holds~\cite{BER21,LIA21}. However, Ref.~\cite{XIE21} proved the validity of the flat metric approximation by offering compelling numerical evidence that the states from \cref{app:eqn:nonchiralGS} and \cref{app:eqn:chiralGS} are indeed the ground states of $H_I$ in the corresponding limits.} of $H_I$ derived in Ref.~\cite{LIA21}. Away from the first chiral limit ($w_0 \neq 0$), \cref{app:eqn:r_mat_commut_+,app:eqn:r_mat_commut_-} are valid provided $\ket{\varphi}$ is one of ground states $\ket{\Psi_{\nu}}$ of $H_I$ at even filling $\nu$ defined by~\cite{LIA21}
\begin{equation}
	\label{app:eqn:nonchiralGS}
	\ket{\Psi_{\nu}} = \prod_{\vk} \left( \prod_{j=1}^{(\nu + 4)/2} \hat{c}^\dagger_{\vk,+1,\eta_j,s_j} \hat{c}^\dagger_{\vk,-1,\eta_j,s_j} \right) \ket{0}, \quad \text{for} \quad \nu =0, \pm 2, \pm 4,
\end{equation}
as well as any $\Uncf$ rotation thereof given by the generators from \cref{app:eqn:generatorsu4}. In \cref{app:eqn:nonchiralGS}, $\left\lbrace \eta_j,s_j \right\rbrace$ denote the occupied valley-spin flavors of $\ket{\Psi_{\nu}}$ and the ``vacuum state'' $\ket{0}$ corresponds to filling $\nu = -4$ (\ie{} unoccupied TBG active bands). The states $\ket{\Psi_{\nu}}$ and their $\Uncf$ rotations carry zero Chern number.

In the first chiral limit ($w_0 = 0$), \cref{app:eqn:r_mat_commut_+,app:eqn:r_mat_commut_-} are valid for any of the Chern number $\mathcal{C}$ ground states $\ket{\Psi^{\nu_+, \nu_-}_{\nu}}$ of $H_I$ at integer filling $\nu$ defined by~\cite{LIA21}
\begin{equation}
	\label{app:eqn:chiralGS}
	\ket{\Psi^{\nu_+, \nu_-}_{\nu}} = \prod_{\vk} \left( \prod_{j_1=1}^{\nu_+} \hat{d}^\dagger_{\vk,+1,\eta_{j_1},s_{j_1}} \prod_{j_2=1}^{\nu_-} \hat{d}^\dagger_{\vk,-1,\eta_{j_2},s_{j_2}} \right )\ket{0}, \quad \text{for} \quad -4 \leq \nu \leq 4
\end{equation}
and any of their $\mathrm{U} \left( 4 \right) \times \mathrm{U} \left( 4 \right)$ rotations generated by the operators from \cref{app:eqn:generatorsu4u4}. In \cref{app:eqn:chiralGS}, the occupancies of the two Chern sectors are given by
\begin{equation}
	\nu_+ = \frac{\nu+4+\mathcal{C}}{2} \quad \text{and} \quad
	\nu_- = \frac{\nu+4-\mathcal{C}}{2},
\end{equation}
with $\left\lbrace \eta_{j_1},s_{j_1} \right\rbrace$ and $\left\lbrace \eta_{j_2},s_{j_2} \right\rbrace$ denoting the (arbitrarily chosen) occupied valley-spin flavors of the two Chern sectors.  

Additionally, the charge-one commutation relations from \cref{app:eqn:r_mat_commut_+,app:eqn:r_mat_commut_-} are also valid in the $\lambda = 1$ case for any of the ground states of $H_I$ from \cref{app:eqn:nonchiralGS}, along with any $\mathrm{U} \left( 2 \right) \times \mathrm{U} \left( 2 \right)$ rotation thereof. In what follows, we will assume the charge-one commutation relations to hold for any $\mathrm{U} \left( 4 \right) \times \mathrm{U} \left( 4 \right)$ rotation of the integer filling states from \cref{app:eqn:chiralGS}, even \emph{away} from the first chiral limit ($w_0 \neq 0$) \emph{and} in the \emph{absence} of exact particle-hole symmetry ($\lambda = 1$). To see why this approximation is justified, we first note that in moving away from the chiral limit ($w_0 = 0$ and $\lambda = 0$) to the nonchiral, but particle-hole symmetric case ($w_0 \neq 0$ and $\lambda = 0$), the states of the form in \cref{app:eqn:chiralGS} are still perturbatively the ground states of the TBG Hamiltonian, even at odd integer fillings~\cite{LIA21,XIE21}. Moreover, through a renormalization group approach, Ref.~\cite{VAF20} has shown that by successively integrating the remote TBG bands in the nonchiral case ($w_0 \neq 0$), the system flows towards the chiral limit, thus approaching the $\mathrm{U} \left( 4 \right) \times \mathrm{U} \left( 4 \right)$-symmetric case. It is therefore justified to use the same charge-one commutation relation away from the chiral limit for $\mathrm{U} \left( 4 \right) \times \mathrm{U} \left( 4 \right)$ rotations of the states in \cref{app:eqn:chiralGS}. Finally, moving away from the case with \emph{exact} particle-hole symmetry can be justified by noting that even in the $\lambda = 1$ case, particle-hole is still and excellent \emph{approximate} symmetry~\cite{SON21}. 

\subsection{Symmetry properties of the charge-one excitation matrices}\label{app:sec:charge:symmetries}

Under the gauge-fixing conditions outlined in \cref{app:sec:notation:sp:gauge}, the symmetries of the single-particle TBG Hamiltonian impose a series of constraints on the charge-one excitation matrices. This \siSection{} aims to derive these constraints with the goal of parameterizing the $R \left( \vk \right)$ and $\tilde{R} \left( \vk \right)$ matrices within the band and valley subspaces. This will then used for analytic approximations of the LDOS. Part of this parametrization was first derived in \cite{KAN21}. 

Let $g$ be one of the symmetries of the single-particle TBG Hamiltonian from \cref{app:sec:notation:sp:symmetries}. Using the symmetry properties of the form factor matrix from \cref{app:eqn:ff_sym_matrix}, \cref{app:eqn:def_Afact} implies that 
\begin{align}
	\label{app:eqn:trafo_Afact}
	A_{\vG} =& \sqrt{V \left( \vG \right)} \sum_{\vk \in \MBZ} \frac{\nu}{4} \Tr \left[ M \left( \vk, \vG \right)\right]
	= \sqrt{V \left( \vG \right)} \sum_{\vk \in \MBZ} \frac{\nu}{4} \Tr \left[ B^{\dagger g} (\vk)  M \left( g \vk, g \vG \right) B^{g} (\vk) \right] \nonumber \\
	=& \sqrt{V \left( g\vG \right)} \sum_{\vk \in \MBZ} \frac{\nu}{4} \Tr \left[ M \left( g \vk, g \vG \right) \right] = A_{g\vG},
\end{align} 
where we have used the reality of the form factor matrix, the invariance of the interaction potential under two-dimensional spatial rotations, as well as the unitarity of the sewing matrices. Applying \cref{app:eqn:ff_sym_matrix,app:eqn:trafo_Afact} in the definition from \cref{app:eqn:def_r_+}, we find that 
\begin{align}
	R^{\eta}_{mn} \left( \vk \right) &= \frac{1}{2 \omegaTBG} \sum_{\vG \in \Q0} \left[ \left( \sum_{\vq \in \MBZ, m'} V\left( g\vq + g\vG \right) 
	\left[B^{\dagger g} (\vk + \vq)  M \left( g \vk, g \vq + g \vG \right) B^{g} (\vk)\right]^{*\eta}_{m'm} \nonumber \right. \right. \\
	&\times \left.
	\left[B^{\dagger g} (\vk + \vq)  M \left( g \vk, g \vq + g \vG \right) B^{g} (\vk)\right]^{\eta}_{m'n} \right)  \nonumber \\
	&\left.+ 2 A_{-g\vG} \sqrt{V \left(g\vG \right)} \left[B^{\dagger g} (\vk)  M \left( g \vk, g \vG \right) B^{g} (\vk)\right]^{\eta}_{mn} \right] - \mu \delta_{mn} \nonumber \\
	&= \frac{1}{2 \omegaTBG} \sum_{\vG \in \Q0} \left[ \left( \sum_{\vq \in \MBZ, m'} V\left( \vq + \vG \right) 
	\left[B^{\dagger g} (\vk + \vq)  M \left( g \vk, \vq + \vG \right) B^{g} (\vk)\right]^{*\eta}_{m'm} \nonumber \right. \right. \\
	&\times \left.\left.
	\left[B^{\dagger g} (\vk + \vq)  M \left( g \vk, \vq + \vG \right) B^{g} (\vk)\right]^{\eta}_{m'n} \right) + 2 A_{-\vG} \sqrt{V \left(\vG \right)} \left[B^{\dagger g} (\vk)  M \left( g \vk, \vG \right) B^{g} (\vk)\right]^{\eta}_{mn} \right] - \mu \delta_{mn}. \label{app:eqn:trafo_rMat_1}
\end{align}
We now proceed to simplify the first term in \cref{app:eqn:trafo_rMat_1}, 
\begin{align}
	&\sum_{m'} \left[B^{\dagger g} (\vk + \vq)  M \left( g \vk, \vq + \vG \right) B^{g} (\vk)\right]^{*\eta}_{m'm} 
	\left[B^{\dagger g} (\vk + \vq)  M \left( g \vk, \vq + \vG \right) B^{g} (\vk)\right]^{\eta}_{m'n} \nonumber \\
	=& \sum_{m'} \sum_{\substack{\eta_1, n_1, m_1 \\ \eta_2, n_2, m_2}} \left[ B^{g} (\vk) \right]^*_{n_1 \eta_1, m\eta} \left[ B^{g} (\vk + \vq) \right]_{m_1 \eta_1,m'\eta} M^{*\eta_1}_{m_1 n_1} \left( g \vk, \vq + \vG \right) \nonumber \\
	\times & \left[ B^{g} (\vk) \right]_{n_2 \eta_2, n\eta} \left[ B^{g} (\vk + \vq) \right]^*_{m_2 \eta_2,m'\eta} M^{\eta_2}_{m_2 n_2} \left( g \vk, \vq + \vG \right) \nonumber \\
	=& \sum_{\substack{\eta_1, n_1, m_1 \\ \eta_2, n_2, m_2}} \left[ B^{g} (\vk) \right]^*_{n_1 \eta_1, m\eta}  M^{*\eta_1}_{m_1 n_1} \left( g \vk, \vq + \vG \right) \left[ B^{g} (\vk) \right]_{n_2 \eta_2, n\eta} M^{\eta_2}_{m_2 n_2} \left( g \vk, \vq + \vG \right) \delta_{m_1 m_2} \delta_{\eta_1 \eta_2} \nonumber \\
	=& \sum_{\substack{\eta',m'\\ n_1, n_2}} \left[ B^{g} (\vk) \right]^*_{n_1 \eta', m\eta}  M^{*\eta'}_{m' n_1} \left( g \vk, \vq + \vG \right) \left[ B^{g} (\vk) \right]_{n_2 \eta', n\eta} M^{\eta'}_{m' n_2} \left( g \vk, \vq + \vG \right). \label{app:eqn:trafo_rMat_2}
\end{align}

Finally, combining \cref{app:eqn:trafo_rMat_1,app:eqn:trafo_rMat_2}, it is straightforward to show that 
\begin{equation}
	R^{\eta}_{mn} \left( \vk \right) = \sum_{\eta', n_1,n_2} \left[ B^{g} (\vk) \right]^*_{n_1 \eta', m\eta} \left[ B^{g} (\vk) \right]_{n_2 \eta', n\eta} R^{\eta'}_{n_1 n_2} \left( g \vk \right),
\end{equation}
or alternatively, in matrix notation 
\begin{equation}
	\label{app:eqn:trafo_temp_r_+}
	R \left( \vk \right) =  B^{\dagger g} (\vk) R \left( g \vk \right)  B^{g} (\vk). \\
\end{equation}
Note that the $C_{2z}T$ symmetry of TBG, through the gauge-fixing conditions from \cref{app:sec:notation:sp:gauge}, imposes a reality condition on the form-factor matrix, and consequently on the matrix $R \left( \vk \right)$. As such, we have not included a complex conjugation when $g$ is anti-unitary. Strictly speaking, for a different gauge choice (\ie{} when the charge-excitation matrices are \emph{not} real), an additional complex conjugation could be required in \cref{app:eqn:trafo_temp_r_+} when $g$ is anti-unitary.

Obtaining the symmetry transformation of the $\tilde{R} \left( \vk \right)$ matrix proceeds analogously with the derivation of \cref{app:eqn:trafo_temp_r_+}, as $R \left( \vk \right)$ and $\tilde{R} \left( \vk \right)$ only differ by the sign of the second term and of the chemical potential
\begin{equation}
	\label{app:eqn:trafo_temp_r_-}
	\tilde{R} \left( \vk \right) =  B^{T g} (\vk) \tilde{R} \left( g \vk \right)  B^{*g} (\vk).
\end{equation}
Under the gauge-fixing conditions from \cref{app:sec:notation:sp:gauge}, all sewing matrices are real, \cref{app:eqn:trafo_temp_r_+,app:eqn:trafo_temp_r_-} can be written equivalently as
\begin{align}
	R \left( \vk \right) =  B^{\dagger g} (\vk) R \left( g \vk \right)  B^{g} (\vk), \label{app:eqn:trafo_r_+} \\
	\tilde{R} \left( \vk \right) =  B^{\dagger g} (\vk) \tilde{R} \left( g \vk \right)  B^{g} (\vk). \label{app:eqn:trafo_r_-}
\end{align}

Finally, we note that the charge-one excitation matrices are Hermitian
\begin{align}
	R^\dagger \left( \vk \right) = R \left( \vk \right) \label{app:eqn:herm_r_+} \\
	\tilde{R}^\dagger \left( \vk \right) = \tilde{R} \left( \vk \right) \label{app:eqn:herm_r_-}
\end{align}
 We will prove this for $R \left( \vk \right)$ using the definition from \cref{app:eqn:trafo_temp_r_+}, as well as the Hermiticity of the form factor matrix from \cref{app:eqn:ff_hermiticity}
\begin{align}
	R^{*\eta}_{nm} \left( \vk \right) &= \frac{1}{2 \omegaTBG} \sum_{\vG \in \Q0} \left[ \left( \sum_{\vq \in \MBZ, m'} V\left( \vq + \vG \right) M^{\eta}_{m'n} \left(\vk, \vq + \vG \right) M^{*\eta}_{m'm} \left(\vk, \vq + \vG \right) \right) \right. \nonumber \\
	&\left.+ 2 A^{*}_{-\vG} \sqrt{V \left(\vG \right)} M^{*\eta}_{nm} \left( \vk,\vG \right) \right] - \mu \delta_{mn} \nonumber \\
	&= \frac{1}{2 \omegaTBG} \sum_{\vG \in \Q0} \left[ \left( \sum_{\vq \in \MBZ, m'} V\left( \vq + \vG \right) M^{*\eta}_{m'm} \left(\vk, \vq + \vG \right) M^{\eta}_{m'n} \left(\vk, \vq + \vG \right) \right) \right. \nonumber \\
	&\left.+ 2 A_{\vG} \sqrt{V \left(\vG \right)} M^{\eta}_{mn} \left( \vk,-\vG \right) \right] - \mu \delta_{mn} \nonumber \\
	&= R^{\eta}_{mn} \left( \vk \right),
\end{align} 
with the Hermiticity of $\tilde{R} \left( \vk \right)$ following analogously.
\subsection{Parameterization of the charge-one excitation matrices}\label{app:sec:charge:parameterization}

With the transformation properties of the charge-one excitation matrices at hand, we now analyze the consequence of each discrete symmetry of TBG from \cref{app:sec:notation:ih:symmetries}.

Firstly, as shown in \cref{app:sec:charge:symmetries}, the charge-one excitation matrices are Hermitian and diagonal in valley space. Moreover, as a consequence of the $C_{2z}T$ symmetry, they are real, and hence symmetric. In the most general case, they can be parameterized as 
\begin{align}
	R \left( \vk \right) &= 
	\zeta^{0} \tau^{z} d_1 \left( \vk \right) + 
	\zeta^{0} \tau^{0} d_2 \left( \vk \right) + 
	\zeta^{x} \tau^{z} d_3 \left( \vk \right) + 
	\zeta^{x} \tau^{0} d_4 \left( \vk \right) + 
	\zeta^{z} \tau^{z} d_5 \left( \vk \right) + 
	\zeta^{z} \tau^{0} d_6 \left( \vk \right), \label{app:eqn:param_r_+}\\
	\tilde{R} \left( \vk \right) &= 
	\zeta^{0} \tau^{z} \tilde{d}_1 \left( \vk \right) + 
	\zeta^{0} \tau^{0} \tilde{d}_2 \left( \vk \right) + 
	\zeta^{x} \tau^{z} \tilde{d}_3 \left( \vk \right) + 
	\zeta^{x} \tau^{0} \tilde{d}_4 \left( \vk \right) + 
	\zeta^{z} \tau^{z} \tilde{d}_5 \left( \vk \right) + 
	\zeta^{z} \tau^{0} \tilde{d}_6 \left( \vk \right),\label{app:eqn:param_r_-}
\end{align}
where $d_i \left( \vk \right)$ and $\tilde{d}_i \left( \vk \right)$ ($1 \leq i \leq 6$) are real functions of the crystalline momentum $\vk$. Moreover, as a consequence of \cref{app:eqn:trafo_r_+,app:eqn:trafo_r_-} for $g=C_{2z}$, we find that the parity of $d_i \left( \vk \right)$ and $\tilde{d}_i \left( \vk \right)$ ($1 \leq i \leq 6$) with respect to $\vk$ is given by the parity of $i$, \ie{}
\begin{equation}
	\label{app:eqn:parity_d_c2z}
	d_i \left( - \vk \right) = (-1)^{i} d_i \left( \vk \right),
	\qquad
	\tilde{d}_i \left( - \vk \right) = (-1)^{i} \tilde{d}_i \left( \vk \right), \quad
	\text{for} \quad 1\leq i \leq 6.
\end{equation}
No additional restrictions are imposed by the time-reversal symmetry $T$. 

In the particle-hole symmetric case ($\lambda = 0$), we find that the particle-hole transformation $P$ additionally imposes 
\begin{equation}
	\label{app:eqn:parity_d_p}
	d_i \left(-\vk \right) = \begin{cases}
		d_i \left( \vk \right) & \quad \text{if} \quad 1 \leq i \leq 2 \\
		- d_i \left( \vk \right) & \quad \text{if} \quad 3 \leq i \leq 6
	\end{cases},
	\qquad
	\tilde{d}_i \left(-\vk \right) = \begin{cases}
		\tilde{d}_i \left( \vk \right) & \quad \text{if} \quad 1 \leq i \leq 2 \\
		- \tilde{d}_i \left( \vk \right) & \quad \text{if} \quad 3 \leq i \leq 6
	\end{cases},
\end{equation}
which together with \cref{app:eqn:parity_d_c2z} requires that $d_i \left(\vk \right) =\tilde{d}_i \left(\vk \right) = 0$ for $i=1,4,6$. In the particle-hole symmetric case ($\lambda = 0$), the parameterization of the charge-one excitation matrices reads as
\begin{align}
	R \left( \vk \right) &= 
	\zeta^{0} \tau^{0} d_2 \left( \vk \right) + 
	\zeta^{x} \tau^{z} d_3 \left( \vk \right) + 
	\zeta^{z} \tau^{z} d_5 \left( \vk \right), \label{app:eqn:param_P_r_+}\\
	\tilde{R} \left( \vk \right) &= 
	\zeta^{0} \tau^{0} \tilde{d}_2 \left( \vk \right) + 
	\zeta^{x} \tau^{z} \tilde{d}_3 \left( \vk \right) + 
	\zeta^{z} \tau^{z} \tilde{d}_5 \left( \vk \right),\label{app:eqn:param_P_r_-}
\end{align}
where the momentum parity of the functions $d_i \left(\vk \right)$ and $\tilde{d}_i \left(\vk \right)$ for $i=2,3,5$ is given by \cref{app:eqn:parity_d_c2z}. 

In the $\lambda = 1$ case, although not exact, the particle-hole transformation $P$ is an an excellent \emph{approximate} symmetry~\cite{SON21}. Provided the gauge-fixing condition from \cref{app:eqn:approx_p_sewing} is imposed, we find that the parameterization of the charge-one excitation matrices from  \cref{app:eqn:param_r_+,app:eqn:param_r_-} obeys
\begin{equation}
	\label{app:eqn:parity_d_p_approx}
	\abs{d_i \left( \vk \right)} \ll \abs{d_j \left( \vk \right)} \quad \text{and} \quad
	\abs{\tilde{d}_i \left( \vk \right)} \ll \abs{\tilde{d}_j \left( \vk \right)}, \quad \text{for $ i=1,4,6$ and $j=2,3,5$},
\end{equation}
meaning that \cref{app:eqn:param_P_r_+,app:eqn:param_P_r_-} still hold \emph{approximately}. Nevertheless, in the absence of \emph{exact} particle-hole symmetry, we will employ the exact parameterizations from \cref{app:eqn:param_r_+,app:eqn:param_r_-} and only then explore the consequences of \cref{app:eqn:parity_d_p_approx}.

Finally, in the first chiral limit ($w_0 = 0$), the presence of the anticommuting $C$ symmetry implies that the charge-one excitation matrices are diagonal
\begin{align}
	R \left( \vk \right) &= 
	\zeta^{0} \tau^{0} d_2 \left( \vk \right), \label{app:eqn:param_C_r_+}\\
	\tilde{R} \left( \vk \right) &= 
	\zeta^{0} \tau^{0} \tilde{d}_2 \left( \vk \right) ,\label{app:eqn:param_C_r_-}
\end{align}
where the real functions $d_{2} \left( \vk \right)$ and $\tilde{d}_{2} \left( \vk \right)$ are even with respect to momentum inversion.

\subsection{Charge-one excitation above specific ground states}\label{app:sec:charge:specific}

As written in \cref{app:eqn:r_mat_commut_+,app:eqn:r_mat_commut_-}, the charge-one commutation relations are cumbersome to apply for any choice of TBG ground states apart from the states $\ket{\Psi_{\nu}}$ defined in \cref{app:eqn:nonchiralGS}. This is because for a generic $\mathrm{U} \left( 4 \right) \times \mathrm{U} \left( 4 \right)$ rotation of the state $\ket{\Psi^{\nu_+, \nu_-}_{\nu}}$ introduced in \cref{app:eqn:chiralGS} and for a given momentum $\vk$, the states $\hat{c}^\dagger_{\vk,n,\eta,s} \ket{\Psi^{\nu_+, \nu_-}_{\nu}}$ ($\hat{c}_{\vk,n,\eta,s} \ket{\Psi^{\nu_+, \nu_-}_{\nu}}$) are not necessarily linearly independent, and therefore provide a redundant basis for the electron (hole) excitations above the ground state. As a simple example, consider the $\nu = -3$ valley-polarized ground state with $\mathcal{C} = 1$
\begin{equation}
	\ket{\varphi} = \prod_{\vk} \hat{d}^\dagger_{\vk,+1,+,\uparrow} \ket{0} = \prod_{\vk} \frac{1}{\sqrt{2}} \left( \hat{c}^\dagger_{\vk,+1,+,\uparrow} + i \hat{c}^\dagger_{\vk,-1,+,\uparrow} \right) \ket{0}.
\end{equation}
With only one occupied Chern band, the state $\ket{\varphi}$ admits a single hole excitation with momentum $\vk$. On the other hand, both $\hat{c}_{\vk,+1,+,\uparrow} \ket{\varphi}$ and $\hat{c}_{\vk,-1,+,\uparrow} \ket{\varphi}$ are non-vanishing. The solution to this apparent contradiction is that $\hat{c}_{\vk,+1,+,\uparrow} \ket{\varphi} = i\hat{c}_{\vk,-1,+,\uparrow} \ket{\varphi}$, meaning that the two hole excitations are in fact one and the same. The situation becomes even worse when considering generic rotations of the state $\ket{\varphi}$, where a coherent superposition of potentially \emph{all} the TBG active bands is filled: despite having only \emph{one} filled band and hence a single hole excitation for a given momentum, acting with \emph{any} of the energy band operators $\hat{c}_{\vk,n,\eta,s}$ leads to a non-vanishing state. 

Since the states $\ket{\Psi^{\nu_+, \nu_-}_{\nu}}$ from \cref{app:eqn:chiralGS} are obtained by filling Chern bands of different valley-spin flavors, we will find it useful perform a basis change and recast \cref{app:eqn:r_mat_commut_+,app:eqn:r_mat_commut_-} in terms of the Chern band operators from \cref{app:eqn:chern_band} as 
\begin{align}
	\left[H_I -\mu \hN , \hat{d}^\dagger_{\vk,e_{Y_1},\eta,s} \right] \ket{\varphi} &= \sum_{e_{Y_2}} R^{\prime \eta}_{e_{Y_2} e_{Y_1}} \left( \vk \right) \hat{d}^\dagger_{\vk,e_{Y_2},\eta,s} \ket{\varphi}, \label{app:eqn:r_mat_commut_ch_+} \\
	\left[H_I -\mu \hN, \hat{d}_{\vk,e_{Y_1},\eta,s} \right] \ket{\varphi} &= \sum_{e_{Y_2}} \tilde{R}^{\prime \eta}_{e_{Y_2} e_{Y_1}} \left( \vk \right) \hat{d}_{\vk,e_{Y_2},\eta,s} \ket{\varphi}, \label{app:eqn:r_mat_commut_ch_-}
\end{align}
where the charge-one excitation matrices expressed in the Chern-band basis are given by 
\begin{equation}
	\begin{split}
		R^{\prime \eta}_{e_{Y_2} e_{Y_1}} \left( \vk \right)&=  \sum_{m,n} W^*_{e_{Y_2},m} W_{e_{Y_1},n} R^{\eta}_{m n} \left( \vk \right),\\
		\tilde{R}^{\prime \eta}_{e_{Y_2} e_{Y_1}} \left( \vk \right) &= \sum_{m,n} W_{e_{Y_2},m} W^*_{e_{Y_1},n} \tilde{R}^{\eta}_{m n} \left( \vk \right).\\
	\end{split}
\end{equation}
Here and in what follows, we will use the same symbol to represent a matrix or a vector in both the Chern ($e_Y = \pm 1$) and energy band ($n=\pm 1$) bases. To avoid confusion, we will employ a ``prime'' (${}'$) symbol to denote that a matrix or vector is expressed in the Chern band basis, rather than the energy band basis. For generic $\mathrm{U} \left( 4 \right) \times \mathrm{U} \left( 4 \right)$ rotations of the states $\ket{\Psi^{\nu_+, \nu_-}_{\nu}}$, we will also define \emph{rotated} Chern and energy band fermion operators. By doing so, we avoid the problems associated with redundant bases for the charge-one excitations. 

\subsubsection{Rotated fermion operators}\label{app:sec:charge:specific:rotated_fermions}
Consider a specific $\mathrm{U} \left( 4 \right) \times \mathrm{U} \left( 4 \right)$ rotation (henceforth denoted by $\hU)$ of the state from \cref{app:eqn:chiralGS} 
\begin{equation}
	\label{app:eqn:chosen_ground_state}
	\ket{\varphi} = \hU \ket{\Psi^{\nu_+, \nu_-}_{\nu}}.
\end{equation}
We employ $\rho_{e_Y,\eta,s} = 0,1$ to explicitly show which Chern-valley-spin flavors are occupied in the unrotated state $\ket{\Psi^{\nu_+, \nu_-}_{\nu}}$, such that
\begin{equation}
	\label{app:eqn:chosen_ground_state_rho}
	\rho_{e_Y,\eta,s} = \begin{cases} 
		1, \quad & $if the Chern-valley-spin flavor $ \left\lbrace e_Y,\eta,s \right\rbrace$ band is occupied in $\ket{\Psi^{\nu_+, \nu_-}_{\nu}} \\
		0, \quad & $if the Chern-valley-spin flavor $ \left\lbrace e_Y,\eta,s \right\rbrace$ band is empty in $\ket{\Psi^{\nu_+, \nu_-}_{\nu}}
	\end{cases}.
\end{equation}
We are interested in explicitly deriving the charge-one excitation states above $\ket{\varphi}$, by particularizing \cref{app:eqn:r_mat_commut_+,app:eqn:r_mat_commut_-}. To find a non-redundant basis for the charge-one excitations, we define the \emph{rotated} Chern band basis 
\begin{equation}
	\label{app:eqn:chern_band_rot}
	\hat{g}^\dagger_{\vk,e_Y,\eta,s} \equiv \hU \hat{d}^\dagger_{\vk,e_Y,\eta,s} \hU^{\dagger} = \sum_{e_Y',\eta',s'} U'_{e_Y \eta s, e_Y' \eta' s'} \hat{d}^\dagger_{\vk,e_Y',\eta',s'},
\end{equation} 
as well as the \emph{rotated} energy band basis
\begin{equation}
	\label{app:eqn:en_band_rot}
	\hat{f}^\dagger_{\vk,n,\eta,s} \equiv \hU \hat{c}^\dagger_{\vk,n,\eta,s} \hU^{\dagger} = \sum_{n',\eta',s'} U_{n \eta s, n' \eta' s'} \hat{c}^\dagger_{\vk,n',\eta',s'}.
\end{equation} 
In \cref{app:eqn:chern_band_rot} and \cref{app:eqn:en_band_rot}, $U'_{e_Y \eta s, e_Y' \eta' s'}$ and $U_{n \eta s, n' \eta' s'}$ respectively denote eight-dimensional unitary matrices implementing the rotation $\hU$ within the original Chern band ($\hat{d}^\dagger_{\vk,e_Y,\eta,s}$) and energy band ($\hat{c}^\dagger_{\vk,n,\eta,s}$) bases. The rotated Chern band operator $\hat{g}^\dagger_{\vk,e_Y,\eta,s}$ creates a fermion with a Chern number $e_Y = \pm 1$, as the $\mathrm{U} \left( 4 \right) \times \mathrm{U} \left( 4 \right)$ rotations generated by \cref{app:eqn:generatorsu4u4} do not mix the different Chern sectors. On the other hand, $\hat{g}^\dagger_{\vk,e_Y,\eta,s}$ generally denotes a coherent superposition of fermions at momentum $\vk$ from \emph{all} valley and spin sectors, with the indices $\eta$ and $s$ merely indicating that $\hat{g}^\dagger_{\vk,e_Y,\eta,s}$ is obtained from $\hat{d}^\dagger_{\vk,e_Y,\eta,s}$ by acting with the transformation $\hU$. For the rotated energy band fermion $\hat{f}^\dagger_{\vk,n,\eta,s}$, the band ($n$), valley ($\eta$), and spin ($s$) indices are simply an indication that it was obtained by rotating the original energy band fermion $\hat{c}^\dagger_{\vk,n,\eta,s}$ according to the transformation $\hU$.

\begin{figure}[!t]
	\includegraphics[width=\textwidth]{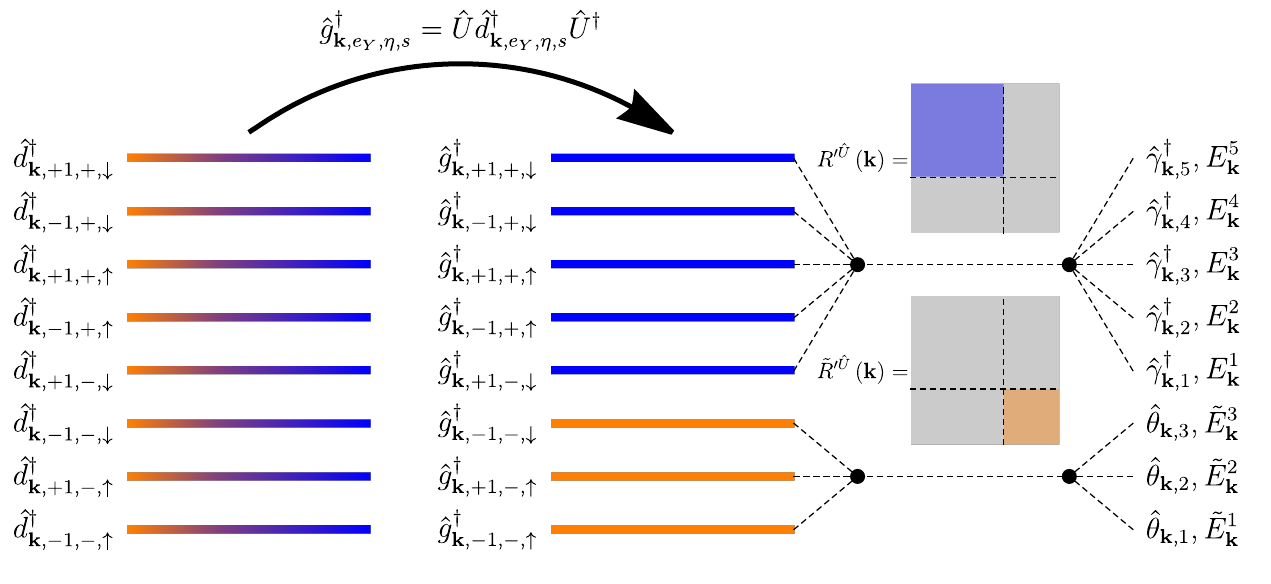}
	\caption{Defining a non-redundant basis for the charge-one excitations. We consider the state $\ket{\varphi} = \hU \ket{\Psi^{1, 2}_{-1}}$, where $\hU$ denotes a $\mathrm{U} \left( 4 \right) \times \mathrm{U} \left( 4 \right)$ rotation. For generic rotations $\hU$ the three occupied bands of $\ket{\varphi}$ represent coherent superpositions of the original TBG Chern bands, implying that the operators $\hat{d}^\dagger_{\vk,e_Y,\eta,s}$ ($\hat{d}_{\vk,e_Y,\eta,s}$) acting on the state $\ket{\varphi}$ constitute a redundant basis for the electron (hole) excitations. By defining a \emph{rotated} Chern band basis according to \cref{app:eqn:chern_band_rot}, the state $\ket{\varphi}$ can be rewritten simply as $\prod_{\vk} \left( \hat{g}^\dagger_{\vk,-1,-,\uparrow} \hat{g}^\dagger_{\vk,+1,-,\uparrow} \hat{g}^\dagger_{\vk,-1,-,\downarrow} \right) \ket{0}$. The rotated operators $\hat{g}^\dagger_{\vk,e_Y,\eta,s}$ ($\hat{g}_{\vk,e_Y,\eta,s}$) corresponding to the empty (filled) bands in $\ket{\varphi}$ provide a linearly independent basis for \emph{all} the electron (hole) excitations on top of $\ket{\varphi}$. The rotated charge-one excitation matrices $R^{\prime \hU} \left( \vk \right)$ and $\tilde{R}^{\prime \hU} \left( \vk \right)$ defined in \cref{app:eqn:rot_rmat_idx_+,app:eqn:rot_rmat_idx_-} are then diagonalized in the space of empty and filled bands of $\ket{\varphi}$, respectively. For a given momentum $\vk$, the rotated operators $\hat{g}^\dagger_{\vk,e_Y,\eta,s}$ ($\hat{g}_{\vk,e_Y,\eta,s}$) corresponding to the empty (filled) bands in $\ket{\varphi}$ can be recombined into the operators $\hat{\gamma}^\dagger_{\vk,p}$ for $1 \leq p \leq 4-\nu$ ($\hat{\theta}_{\vk,p}$ for $1 \leq p \leq 4+\nu$), which create an electron (hole) excitation above $\ket{\varphi}$ with energy $E^{p}_{\vk}$ ($\tilde{E}^{p}_{\vk}$).}
	\label{app:fig:rotatedbases}
\end{figure}

The main benefit of using the rotated Chern basis from \cref{app:eqn:chern_band_rot} is that the ground state $\ket{\varphi}$ of \cref{app:eqn:chosen_ground_state} has a particularly simple expression 
\begin{equation}
	\ket{\varphi} = \prod_{\vk} \left( \prod_{\substack{e_Y,\eta,s \\ \rho_{e_Y,\eta,s} = 1}} \hat{g}^\dagger_{\vk,e_Y,\eta,s} \right)\ket{0},
\end{equation}
where the product runs over those values $ \left\lbrace e_Y,\eta,s \right\rbrace$ for which $\rho_{e_Y,\eta,s} = 1$. When written in this form, it becomes clear that a non-redundant basis for the electron excitations above $\ket{\varphi}$ with a definite momentum $\vk$ is given by the $4-\nu$ operators $\hat{g}^\dagger_{\vk,e_Y,\eta,s}$ for which $\rho_{e_Y,\eta,s} = 0$. Similarly, a linearly-independent basis for the hole excitations is given by the $4+\nu$ operators $\hat{g}_{\vk,e_Y,\eta,s}$ for which $\rho_{e_Y,\eta,s} = 0$. We illustrate this schematically for a generic $\mathrm{U} \left( 4 \right) \times \mathrm{U} \left( 4 \right)$ rotation of the state $\ket{\Psi^{1,2}_{-1}}$ in \cref{app:fig:rotatedbases}.

\subsubsection{Rotated charge-one excitation matrices}\label{app:sec:charge:specific:rotated_matrix}

As discussed in \cref{app:sec:charge:specific:rotated_fermions} and shown schematically in \cref{app:fig:rotatedbases}, the Chern band operators $\hat{d}^\dagger_{\vk,e_Y,\eta,s}$ represent coherent superpositions of the occupied bands in $\ket{\varphi}$. As such, we will rewrite \cref{app:eqn:r_mat_commut_ch_+,app:eqn:r_mat_commut_ch_-} in terms of the rotated Chern band basis defined in \cref{app:eqn:chern_band_rot}
\begin{align}
	\left[H_I -\mu \hN , \hat{g}^\dagger_{\vk,e_{Y_1},\eta_1,s_1} \right] \ket{\varphi} &= \sum_{\substack{e_{Y_2},\eta_2,s_2 \\ \rho_{e_{Y_2},\eta_2,s_2} = 0}} \left[ R^{\prime \hU}\left( \vk \right) \right]_{e_{Y_2} \eta_2 s_2, e_{Y_1} \eta_1 s_1} \hat{g}^\dagger_{\vk,e_{Y_2},\eta_2,s_2} \ket{\varphi}, \quad \text{for} \quad \rho_{e_{Y_1},\eta_1,s_1} = 0, \label{app:eqn:r_mat_commut_b_ch_+} \\
	\left[H_I -\mu \hN , \hat{g}_{\vk,e_{Y_1},\eta_1,s_1} \right] \ket{\varphi} &= \sum_{\substack{e_{Y_2},\eta_2,s_2 \\ \rho_{e_{Y_2},\eta_2,s_2} = 1}} \left[ \tilde{R}^{\prime \hU}\left( \vk \right) \right]_{e_{Y_2} \eta_2 s_2, e_{Y_1} \eta_1 s_1} \hat{g}_{\vk,e_{Y_2},\eta_2,s_2} \ket{\varphi}, \quad \text{for} \quad \rho_{e_{Y_1},\eta_1,s_1} = 1. \label{app:eqn:r_mat_commut_b_ch_-}
\end{align}
where we have also introduced the \emph{rotated} charge-one excitation matrices given by 
\begin{align}
	\left[ R^{\prime \hU}\left( \vk \right) \right]_{e_{Y_2} \eta_2 s_2, e_{Y_1} \eta_1 s_1} &= \sum_{\substack{e_{Y_3},\eta_3,s_3 \\ e_{Y_4},\eta_4,s_4}} U'_{e_{Y_1} \eta_1 s_1,e_{Y_3} \eta_3 s_3} R^{\prime \eta_3}_{e_{Y_4},e_{Y_3}} \left( \vk \right) \delta_{\eta_3 \eta_4} \delta_{s_3 s_4} U^{\prime *}_{e_{Y_2} \eta_2 s_2,e_{Y_4} \eta_4 s_4}, \label{app:eqn:rot_rmat_idx_+} \\
	\left[ \tilde{R}^{\prime  \hU}\left( \vk \right) \right]_{e_{Y_2} \eta_2 s_2, e_{Y_1} \eta_1 s_1} &= \sum_{\substack{e_{Y_3},\eta_3,s_3 \\ e_{Y_4},\eta_4,s_4}} U^{\prime *}_{e_{Y_1} \eta_1 s_1,e_{Y_3} \eta_3 s_3} \tilde{R}^{\prime \eta_3}_{e_{Y_4},e_{Y_3}} \left( \vk \right) \delta_{\eta_3 \eta_4} \delta_{s_3 s_4} U'_{e_{Y_2} \eta_2 s_2,e_{Y_4} \eta_4 s_4}, \label{app:eqn:rot_rmat_idx_-}
\end{align}
or equivalently, in matrix form, by
\begin{align}
	R^{\prime \hU}\left( \vk \right) &=  U^{\prime *} \left( R' \left( \vk \right) \otimes s^0 \right) U^{\prime T}, \label{app:eqn:rot_rmat_mat_+} \\
	\tilde{R}^{\prime \hU}\left( \vk \right) &=  U' \left( \tilde{R}' \left( \vk \right) \otimes s^0 \right) U^{\prime \dagger}. \label{app:eqn:rot_rmat_mat_-}
\end{align}
Note that in the electron (hole) commutation relation, we have restricted to only those operators $\hat{g}^\dagger_{\vk,e_Y,\eta,s}$ ($\hat{g}_{\vk,e_Y,\eta,s}$) which create (destroy) fermions belonging to the empty (filled) rotated Chern bands of $\ket{\varphi}$. In the chiral limit ($w_0 = 0$), \cref{app:eqn:param_C_r_+,app:eqn:param_C_r_-} imply that the charge-one excitation matrices $R \left(\vk \right)$ and $\tilde{R} \left(\vk \right)$, and hence the rotated charge-one excitation matrices are proportional to identity. As such, they do not include any off-diagonal elements between the fermions belonging to the empty and filled rotated Chern bands in $\ket{\varphi}$. Away from the chiral limit ($w_0 \neq 0$), without exact particle-hole symmetry ($\lambda = 1$), and for general $\mathrm{U} \left( 4 \right) \times \mathrm{U} \left( 4 \right)$ rotations, the $R^{\prime \hU}\left( \vk \right)$ and $\tilde{R}^{\prime \hU}\left( \vk \right)$ matrices will generically contain non-vanishing off-diagonal elements between the fermions belonging to the empty and filled bands of $\ket{\varphi}$. This is a consequence of $\ket{\varphi}$ being a perturbative rather than exact eigenstate of $H_I$. 

The charge commutation relations from \cref{app:eqn:r_mat_commut_b_ch_+,app:eqn:r_mat_commut_b_ch_-} written in the rotated Chern band basis allow us to find the electron and hole excitation states above the ground state $\ket{\varphi}$. To see this, we first note that $R^{\prime \hU}\left( \vk \right)$ and $\tilde{R}^{\prime  \hU}\left( \vk \right)$ are Hermitian, being related by a unitary transformations to the Hermitian matrices $R \left( \vk \right) \otimes s^0$ and $\tilde{R} \left( \vk \right) \otimes s^0$, respectively. Therefore, their restrictions into the filled or empty rotated Chern bands from \cref{app:eqn:r_mat_commut_b_ch_+,app:eqn:r_mat_commut_b_ch_-} are Hermitian and can be diagonalized. We define the electron ($\Xi^{\prime p}_{\vk,e_Y \eta s}$ for $1\leq p \leq 4-\nu$) and hole ($\tilde{\Xi}^{\prime p}_{\vk,e_Y \eta s}$  for $1\leq p \leq 4+\nu$) excitation wave functions in the rotated Chern basis from \cref{app:eqn:chern_band_rot} (where $p$ indexes the excitation), with support only on the empty and occupied rotated Chern bands, respectively, \ie{}
\begin{align}
	\Xi^{p}_{\vk,e_Y \eta s} &= 0, \quad \text{if} \quad \rho_{e_Y,\eta,s} = 1, \\
	\tilde{\Xi}^{p}_{\vk,e_Y \eta s} &= 0, \quad \text{if} \quad \rho_{e_Y,\eta,s} = 0.
\end{align}
The charge excitation wave functions diagonalize the restriction of $R^{\prime \hU}\left( \vk \right)$ [$\tilde{R}^{\prime \hU}\left( \vk \right)$] in the empty (occupied) rotated Chern bands
\begin{align}
	\sum_{\substack{e_{Y_1},\eta_1,s_1 \\ \rho_{e_{Y_1},\eta_1,s_1} = 0}} \left[ R^{\prime \hU}\left( \vk \right) \right]_{e_{Y_2} \eta_2 s_2, e_{Y_1} \eta_1 s_1} \Xi^{\prime p}_{\vk,e_{Y_1} \eta_1 s_1} &= E^p_{\vk} \Xi^{\prime p}_{\vk,e_{Y_2} \eta_2 s_2}, \text{ for } 1\leq p \leq 4-\nu, \text{ if } \rho_{e_{Y_2},\eta_2,s_2} = 0 \label{app:eqn:ch_exc_xi_wavf+} \\
	\sum_{\substack{e_{Y_1},\eta_1,s_1 \\ \rho_{e_{Y_1},\eta_1,s_1} = 1}} \left[ \tilde{R}^{\prime \hU}\left( \vk \right) \right]_{e_{Y_2} \eta_2 s_2, e_{Y_1} \eta_1 s_1} \tilde{\Xi}^{\prime p}_{\vk,e_{Y_1} \eta_1 s_1} &= \tilde{E}^p_{\vk} \tilde{\Xi}^{\prime p}_{\vk,e_{Y_2} \eta_2 s_2}, \text{ for } 1\leq p \leq 4+\nu, \text{ if } \rho_{e_{Y_2},\eta_2,s_2} = 1, \label{app:eqn:ch_exc_xi_wavf-}
\end{align}
where $E^{p}_{\vk}$ and $\tilde{E}^{p}_{\vk}$ denote the electron and hole excitation energies\footnote{The excitation energies are defined with respect to the grand canonical Hamiltonian $H_{I} - \mu \hN$.}, respectively. Assuming that $\ket{\varphi}$ is a ground state of the TBG interaction Hamiltonian, the excitation energies must be positive $E^{p}_{ \vk }, \tilde{E}^{p}_{ \vk } \geq 0$~\cite{BER21b}. Defining the charge-one excitation operators
\begin{align}
	\hat{\gamma}^\dagger_{\vk,p} &= \sum_{e_Y,\eta,s} \Xi^{\prime p}_{\vk,e_Y \eta s} \hat{g}^\dagger_{\vk,e_Y,\eta,s} = \sum_{n,\eta,s} \Xi^{p}_{\vk,n \eta s} \hat{f}^\dagger_{\vk,n,\eta,s},\text{ for } 1\leq p \leq 4-\nu, \label{app:eqn:ch_exc_op_+} \\
	\hat{\theta}_{\vk,p} &= \sum_{e_Y,\eta,s} \tilde{\Xi}^{\prime p}_{\vk,e_Y \eta s} \hat{g}_{\vk,e_Y,\eta,s} = \sum_{n,\eta,s} \tilde{\Xi}^{p}_{\vk,n \eta s} \hat{f}_{\vk,n,\eta,s},\text{ for } 1\leq p \leq 4+\nu, \label{app:eqn:ch_exc_op_-}
\end{align}
corresponding respectively to the electron and hole excitations above the $\ket{\varphi}$ state, we find that 
\begin{align}
	\left[H_I -\mu \hN , \hat{\gamma}^\dagger_{\vk,p} \right] \ket{\varphi} &= E^p_{\vk} \hat{\gamma}^\dagger_{\vk,p} \ket{\varphi},\text{ for } 1\leq p \leq 4-\nu, \label{app:eqn:ch_exc_com_operators+}  \\
	\left[H_I -\mu \hN, \hat{\theta}_{\vk,p} \right] \ket{\varphi} &= \tilde{E}^p_{\vk} \hat{\theta}_{\vk,p} \ket{\varphi},\text{ for } 1\leq p \leq 4+\nu, \label{app:eqn:ch_exc_com_operators-}
\end{align}
implying that $\hat{\gamma}^\dagger_{\vk,p} \ket{\varphi}$ ($\hat{\theta}_{\vk,p} \ket{\varphi}$) is an electron (hole) excited eigenstate of $H_{I}$ above the ground state $\ket{\varphi}$ with eigenvalue $E_{\varphi} + E^p_{\vk} + \mu$ ($E_{\varphi} + \tilde{E}^p_{\vk} - \mu$), where $E_{\varphi}$ is the energy of $\ket{\varphi}$, \ie{} $H_{I} \ket{\varphi} = E_{\varphi} \ket{\varphi}$. In \cref{app:eqn:ch_exc_op_+,app:eqn:ch_exc_op_-}, we have also introduced the charge-one excitation wave functions in the rotated energy band basis 
\begin{align}
	\Xi^{p}_{\vk,n \eta s} &= \sum_{e_Y} W_{e_Y,n} \Xi^{\prime p}_{\vk,e_Y \eta s} \label{app:eqn:ch_exc_wavf_rot_-} \\
	\tilde{\Xi}^{p}_{\vk,n \eta s} &= \sum_{e_Y} W^*_{e_Y,n} \tilde{\Xi}^{\prime p}_{\vk,e_Y \eta s} \label{app:eqn:ch_exc_wavf_rot_+}
\end{align}

Having obtained the charge-one excitation operators from \cref{app:eqn:ch_exc_op_+,app:eqn:ch_exc_op_-}, we now perform one last basis transformation to express the corresponding wave functions in the original TBG fermion basis. Defining
\begin{align}
	Z^{\prime p}_{\vk,e_Y \eta s} &= \sum_{e_Y',\eta',s'} U'_{e_Y'\eta's',e_Y \eta s} \Xi^{\prime p}_{\vk, e_Y' \eta' s'},\text{ for } 1\leq p \leq 4-\nu, \label{app:eqn:ch_exc_wavf_trafo_+}  \\
	\tilde{Z}^{\prime p}_{\vk,e_Y \eta s} &= \sum_{e_Y',\eta',s'} U^{\prime *}_{e_Y'\eta's',e_Y \eta s} \tilde{\Xi}^{\prime p}_{\vk, e_Y' \eta' s'},\text{ for } 1\leq p \leq 4+\nu, \label{app:eqn:ch_exc_wavf_trafo_-}
\end{align}
we can rewrite the charge-one excitation operators as
\begin{align}
	\hat{\gamma}^\dagger_{\vk,p} &= \sum_{e_Y,\eta,s} Z^{\prime p}_{\vk,e_Y \eta s} \hat{d}^\dagger_{\vk,e_Y,\eta,s} = \sum_{n,\eta,s} Z^{p}_{\vk,n \eta s} \hat{c}^\dagger_{\vk,n,\eta,s} ,\text{ for } 1\leq p \leq 4-\nu, \label{app:eqn:ch_exc_op_orig_+} \\
	\hat{\theta}_{\vk,p} &= \sum_{e_Y,\eta,s} \tilde{Z}^{\prime p}_{\vk,e_Y \eta s} \hat{d}_{\vk,e_Y,\eta,s} = \sum_{n,\eta,s} \tilde{Z}^{p}_{\vk,n \eta s} \hat{c}_{\vk,n,\eta,s},\text{ for } 1\leq p \leq 4+\nu, \label{app:eqn:ch_exc_op_orig_-}
\end{align}
where we have also defined the charge-one excitation wave functions $Z^{p}_{\vk,n \eta s}$ and $\tilde{Z}^{p}_{\vk,n \eta s}$ in the TBG energy band basis. Finally, we can define two projector matrices which project in the empty and occupied Chern bands of the unrotated state $\ket{\Psi^{\nu_+, \nu_-}_{\nu}}$, whose components in the Chern band basis read as
\begin{align}
	\Pi'_{e_{Y_1} \eta_1 s_1, e_{Y_2} \eta_2 s_2} &= \delta_{e_{Y_1}, e_{Y_2}} \delta_{\eta_1, \eta_2} \delta_{s_1, s_2} - \rho_{e_{Y_1},\eta_1,s_1}  \rho_{e_{Y_2},\eta_2,s_2}, \label{app:eqn:def_proj_1} \\
	\tilde{\Pi}'_{e_{Y_1} \eta_1 s_1, e_{Y_2} \eta_2 s_2} &= \rho_{e_{Y_1},\eta_1,s_1}  \rho_{e_{Y_2},\eta_2,s_2}, \label{app:eqn:def_proj_2}
\end{align}
respectively. \Cref{app:eqn:def_proj_1,app:eqn:def_proj_2} can be employed to obtain a set of important relations which relate the $\mathrm{U} \left( 4 \right) \times \mathrm{U} \left( 4 \right)$ transformation $\hU$, the occupied bands of the unrotated state $\ket{\Psi^{\nu_+, \nu_-}_{\nu}}$, the charge-one excitation matrices from \cref{app:eqn:r_mat_commut_+,app:eqn:r_mat_commut_-}, and the charge-one excitation spectra for $\ket{\varphi}$
\begin{alignat}{3}
	&\sum_{p=1}^{4-\nu} E^{p}_{\vk} \Xi^{*p}_{\vk,n \eta s} \Xi^{p}_{\vk,n' \eta' s'} \quad &&= \left[ \Pi R^{*\hU} \left(\vk \right) \Pi \right]_{n \eta s, n' \eta' s'} \quad &&= \left[ \Pi U \left( R \left( \vk \right) \otimes s^0 \right) U^{\dagger} \Pi  \right]_{n \eta s, n' \eta' s'}, \label{app:eqn:ch_exc_unrot_eigendecomp_+} \\
	&\sum_{p=1}^{4+\nu} \tilde{E}^{p}_{\vk} \tilde{\Xi}^{p}_{\vk,n \eta s} \tilde{\Xi}^{*p}_{\vk,n' \eta' s'} \quad &&= \left[ \tilde{\Pi} \tilde{R}^{\hU} \left(\vk \right) \tilde{\Pi} \right]_{n \eta s, n' \eta' s'} \quad &&= \left[ \tilde{\Pi} U \left( \tilde{R} \left( \vk \right) \otimes s^0 \right) U^{\dagger} \tilde{\Pi} \right]_{n \eta s, n' \eta' s'}, \label{app:eqn:ch_exc_unrot_eigendecomp_-} \\
	&\sum_{p=1}^{4-\nu} E^{p}_{\vk} Z^{*p}_{\vk,n \eta s} Z^{p}_{\vk,n' \eta' s'} \quad &&= \left[ U^{\dagger} \Pi R^{*\hU} \left(\vk \right) \Pi U \right]_{n \eta s, n' \eta' s'} \quad &&= \left[ U^{\dagger} \Pi U \left( R \left( \vk \right) \otimes s^0 \right) U^{\dagger} \Pi U \right]_{n \eta s, n' \eta' s'}, \label{app:eqn:ch_exc_eigendecomp_+} \\
	&\sum_{p=1}^{4+\nu} \tilde{E}^{p}_{\vk} \tilde{Z}^{p}_{\vk,n \eta s} \tilde{Z}^{*p}_{\vk,n' \eta' s'} \quad &&= \left[ U^{\dagger} \tilde{\Pi} \tilde{R}^{\hU} \left(\vk \right) \tilde{\Pi} U \right]_{n \eta s, n' \eta' s'} \quad &&= \left[ U^{\dagger} \tilde{\Pi} U \left( \tilde{R} \left( \vk \right) \otimes s^0 \right) U^{\dagger} \tilde{\Pi} U \right]_{n \eta s, n' \eta' s'}. \label{app:eqn:ch_exc_eigendecomp_-}
\end{alignat}
In \cref{app:sec:spFuncSym}, we will employ these relations in order to provide an analytical understanding of the symmetry properties of the spectral function of the TBG ground states.

\section{Spectral function}\label{app:sec:spectralf}

An STM experiment allows for the indirect determination of the spectral function $A \left( \vec{r}, \omega \right)$ of a certain quantum system. This \siSection{} is dedicated to defining and then deriving the spectral function for a TBG sample (for a discussion on the relation between the spectral function and STM measurements, see \cref{app:sec:experimental}). We show that the TBG spectral function can be written as a contraction between two (gauge-dependent) tensors: the spectral function matrix (which depends on the specific ground state that we consider and on the many-body TBG Hamiltonian), and the spatial factor (which depends on the active band TBG wave functions). We then investigate the properties of the spatial factor arising from the single-particle symmetries of the single-particle TBG Hamiltonian outlined in \cref{app:sec:notation:sp:symmetries}, as well as on the gauge-fixing conditions from \cref{app:sec:notation:sp:gauge}. Finally, we will provide an approximation to the spatial factor and compute it at key momenta within the SLG Brillouin zone.

\subsection{Derivation}\label{app:sec:spectralf:derivation}

\subsubsection{Fermionic field operators}\label{app:sec:spectralf:derivation:fieldOps}

\begin{table*}[!t]
	\begin{tabular}{c | c | c | c} 
		$i$ & $\xi_i$ & $C_i$ & $N_i$ \\ 
  		$1$ & $0.9807$ & $0.2824$ & $1.100$ \\
  		$2$ & $1.4436$ & $0.5470$ & $2.891$ \\
  		$3$ & $2.6005$ & $0.2320$ & $12.59$ \\
  		$4$ & $6.5100$ & $0.0103$ & $124.9$ \\
	\end{tabular} 
	\caption{Parameters of the valence electron wave function of the carbon atom $p_z$ orbital from \cref{app:eqn:pz_orb_wavf}. These parameters were tabulated by Ref.~\cite{RAD85}, and were obtained by a series expansion in the basis set of atomic Slater orbitals.}
	\label{app:tab:param_pz_orbitals}
\end{table*}

We start by introducing the fermionic field operator $\hat{\Psi}_{s} \left( \vec{r} \right)$ which annihilates an electron of spin $s$ at position $\vec{r}$. Letting $\phi \left( \vec{r} \right)$ denote the carbon $p_z$ orbital wave function, we can obtain the following anti-commutation relation between the field operator $\hat{\Psi}_{s} \left( \vec{r} \right)$ and the microscopic fermion operators $\hat{a}^\dagger_{\vec{R},\alpha,s,l}$ introduced in \cref{app:sec:notation:sp:fermion_ops}
\begin{equation}
	\label{app:eqn:anticomm_field_a}
	\left\lbrace \hat{\Psi}_{s} \left( \vec{r} \right), \hat{a}^\dagger_{\vec{R},\alpha,s,l} \right\rbrace = \bra{0} \hat{\Psi}_{s} \left( \vec{r} \right) \hat{a}^\dagger_{\vec{R},\alpha,s,l} \ket{0} = \phi \left[ \vec{r} - \mathcal{R}_{\theta,l} \left( \vec{R} + \vec{t}_\alpha \right) - \vec{z}_l \right],
\end{equation}
with $\vec{z}_l = z_l \hat{\vec{z}}$, where $z_l$ is the height of the layer $l$ (see \cref{app:fig:schematicExperimentSetup}). Note that $\hat{a}^\dagger_{\vec{R},\alpha,s,l}$ creates an electron of spin $s$ in a carbon $p_z$ orbital located at position $\mathcal{R}_{\theta,l} \left( \vec{R} + \vec{t}_\alpha \right)$. Throughout this work, we take the $p_z$ orbital wave function to be given by an analytic approximation of the carbon atom $p_z$ orbital~\cite{RAD85,IJA13}
\begin{equation}
	\label{app:eqn:pz_orb_wavf}
	\phi\left( {\vec{r}} \right) = \frac{1}{2} \sqrt{\frac{3}{\pi a_B^{5}}} z \sum_{i=1}^{4} C_i N_i e^{-\frac{\xi_i r}{a_B}},
\end{equation}
where $\vec{r} = x \hat{\vec{x}} + y \hat{\vec{y}} + z \hat{\vec{z}}$, $a_{B}$ is the Bohr radius, and the dimensionless parameters $C_i$, $N_i$ and $\xi_i$ (for $1 \leq i \leq 4$) were defined and tabulated by Ref.~\cite{RAD85} and are also provided in \cref{app:tab:param_pz_orbitals}. In what follows, we will assume $p_z$ orbitals belonging to different carbon atoms to be orthogonal and thus
\begin{equation}
	\label{app:eqn:anticomm_a_a}
	\left\lbrace \hat{a}^\dagger_{\vec{R},\alpha,s,l},\hat{a}_{\vec{R}',\beta,s',l'} \right\rbrace = \delta_{\vec{R},\vec{R}'} \delta_{\alpha,\beta} \delta_{s,s'} \delta_{l,l'}.
\end{equation}
Strictly speaking, this is not true, as we are assuming atomic orbitals as opposed to Wannier ones. Nevertheless, for visualizing STM patterns, \cref{app:eqn:pz_orb_wavf} provides a good enough approximation~\cite{IJA13}. Using the anti-commutation relations from \cref{app:eqn:anticomm_field_a,app:eqn:anticomm_a_a} we can express the fermionic field operator at the STM tip position in terms of the microscopic graphene orbitals as
\begin{equation}
	\label{app:eqn:real_space_field_ops}
	\hat{\Psi}^\dagger_{s} \left( \vec{r} + \vec{z}_p \right) = \sum_{\alpha,l} \sum_{\vec{R}} \phi \left[ \vec{r} + \vec{z}_p - R_{\theta,l} \left( \vec{R} + \vec{t}_\alpha \right) - \vec{z}_l \right] \hat{a}^\dagger_{\vec{R},\alpha,s,l} + \dots,
\end{equation}
where $\vec{z}_p = z_p \hat{\vec{z}}$, with $z_p$ being the height of the STM tip (see \cref{app:fig:schematicExperimentSetup}). The dots at the end imply that \cref{app:eqn:real_space_field_ops} does not provide a full expansion of the fermionic field, as the $p_z$ Carbon atoms of the TBG sample do not form a complete basis set. For \cref{app:eqn:real_space_field_ops} to be complete, one would need to include the all the (infinitely many) orbitals \emph{orthogonal} to the ones created by the set of operators $\hat{a}^\dagger_{\vec{R},\alpha,s,l}$. As they are not relevant for the physics of TBG near charge neutrality, we leave them unspecified in the expansion from \cref{app:eqn:real_space_field_ops} and omit them completely henceforth.

We assume the height of the STM tip to remain constant throughout an experiment\footnote{For the theoretically predicted STM signal, we assume the height of the STM tip to remain constant, whereas for experiments, its height changes so as to keep the tunneling current constant. As such, we normalize the experimentally-measured STM signals and the theoretically-predicted spectral functions in \cref{fig:expVStheory} of the main text according to their maxima.}. As such, we will employ a convention in which $\vec{r}$ denotes a strictly two-dimensional vectors in the plane of the TBG sample, and use the shorthand notation where
\begin{equation}
	\label{app:eqn:shorthand_real_space_field_ops}
	\hat{\psi}^\dagger_{s} \left( \vec{r} \right) \equiv \hat{\Psi}^\dagger_{s} \left( \vec{r} + \vec{z}_p \right)
\end{equation}
denotes the Fermionic field operator corresponding to the STM tip position and
\begin{equation}
	\label{app:eqn:shorthand_pz_orbitals}
	\phi_{l} \left(\vec{r} \right) \equiv \phi \left(\vec{r} + \vec{z}_p - \vec{z}_l \right)
\end{equation}
is the orbital wave function for a $p_z$ orbital located at the origin, within layer $l$.

Using the notation in \cref{app:eqn:shorthand_pz_orbitals,app:eqn:shorthand_real_space_field_ops}, \cref{app:eqn:real_space_field_ops} becomes
\begin{equation}
	\label{app:eqn:real_space_field_ops_2D}
	\hat{\psi}^\dagger_{s} \left( \vec{r} \right) = \sum_{\alpha,l} \sum_{\vec{R}} \phi_l \left[ \vec{r} - R_{\theta,l} \left( \vec{R} + \vec{t}_\alpha \right) \right] \hat{a}^\dagger_{\vec{R},\alpha,s,l}.
\end{equation}
We will now relate the Fermion field operator to the TBG energy band operators introduced in \cref{app:eqn:en_band}. We start by introducing the two-dimensional Fourier transformation of the $p_z$ orbital wave function over the SLG Brillouin Zone $\mathrm{BZ}_l$
\begin{equation}
	\label{app:eqn:ft_orbs}
	\phi_l \left( \vec{r} \right) = \frac{1}{N_0\omegaSLG}\sum_{\substack{\vec{p} \in \mathrm{BZ}_l \\ \vG \in \mathcal{G}_l}} \phi_{l} \left( \vec{p} + \vG \right) e^{-i \left( \vec{p} + \vG \right) \cdot \vec{r} },
\end{equation}
where $\mathcal{G}_l$ is the reciprocal lattice of the graphene layer $l$, as defined in the text surrounding \cref{app:eqn:gen_rec_slg}, $\omegaSLG$ denotes the surface area of the SLG unit cell, and $N_0$ denotes the number of SLG unit cells. The total area of the TBG sample is thus given by $\omegaTBG = N_0 \omegaSLG$. Owing to the rotational symmetry of the $p_z$ orbital around the $\hat{\vec{z}}$ axis, we have that
\begin{equation}
	\label{app:eqn:rot_sym_of_phi_l}
	\phi_{l} \left( \vec{p} + \vG \right) =  \phi_{l} \left( \abs{\vec{p} + \vG} \right)
\end{equation}  
and $ \phi_{l} \left( \vec{p} + \vG \right)$ is also real, \ie{}
\begin{equation}
	\label{app:eqn:reality_of_phi}
	\phi_{l} \left( \vec{p} + \vG \right) = \phi^{*}_{l} \left( \vec{p} + \vG \right).
\end{equation}
Employing the Fourier transformations introduced in \cref{app:eqn:ft_orbs,app:eqn:ft_a_ops}, we can express the Fermionic field operators $\hat{\psi}^\dagger_{s} \left( \vec{r} \right)$ in terms of the moir\'e lattice low-energy operators defined in \cref{app:eqn:low_en_fermions_c}
\begin{align}
	\hat{\psi}^\dagger_{s} \left( \vec{r} \right) 
	&= \frac{1}{N_0 \omegaSLG \sqrt{N_0}} \sum_{\alpha,l} \sum_{\vec{R}} \sum_{\substack{\vec{p}_1,\vec{p}_2 \in \mathrm{BZ}_l \\ \vG \in \mathcal{G}_{l}}} \phi_{l} \left( \vec{p}_1 + \vG \right) \hat{a}^\dagger_{\vec{p}_2,\alpha,s,l}  e^{-i \left( \vec{p}_1 + \vG \right) \cdot \vec{r}} e^{i \left( \vec{p}_1 - \vec{p}_2 \right)\cdot R_{\theta,l} \left( \vec{R} + \vec{t}_{\alpha} \right)} e^{i \vG \cdot R_{\theta,l} \vec{t}_{\alpha}} \nonumber \\
	&= \frac{1}{\omegaSLG \sqrt{N_0}} \sum_{\alpha,l} \sum_{\substack{\vec{p} \in \mathrm{BZ}_l \\ \vG \in \mathcal{G}_{l}}} \phi_{l} \left( \vec{p} + \vG \right) \hat{a}^\dagger_{\vec{p},\alpha,s,l} e^{i \vG \cdot R_{\theta,l} \vec{t}_{\alpha}} e^{-i \left( \vec{p} + \vG \right) \cdot \vec{r}} \nonumber \\
	&= \frac{1}{\omegaSLG \sqrt{N_0}} \sum_{\alpha,l,\eta} \sum_{\substack{\vk \in \mathrm{MBZ}_l \\ \vQ \in \mathcal{Q}_{\eta l} \\ \vG \in \mathcal{G}_{l}}} \phi_{l} \left( \eta \vec{K}_l + 
	\vk - \vQ + \vG \right) e^{i \vG \cdot R_{\theta,l} \vec{t}_{\alpha}} e^{-i \left( \eta \vec{K}_l + 
	\vk - \vQ + \vG \right) \cdot \vec{r}} \hat{c}^\dagger_{\vk,\vQ,\eta,\alpha,s}.  \label{app:eqn:fieldOps_moire_fermions}
\end{align}
Using \cref{app:eqn:low_en_to_eig}, we can also write \cref{app:eqn:fieldOps_moire_fermions} in the energy band basis as 
\begin{equation}
	\hat{\psi}^\dagger_{s} \left( \vec{r} \right) = \sum_{\eta,n} V_{\vec{r},\vk n \eta} \hat{c}^\dagger_{\vk,n,\eta,s},   \label{app:eqn:fieldOps_enBand}
\end{equation}
where we have defined
\begin{equation}
	V_{\vec{r},\vk n \eta} = \frac{1}{\omegaSLG \sqrt{N_0}} \sum_{\alpha,l} \sum_{\substack{\vQ \in \mathcal{Q}_{\eta l} \\ \vG \in \mathcal{G}_{l}}} \phi_{l} \left( \eta \vec{K}_l + 
	\vk - \vQ + \vG \right) e^{i \vG \cdot R_{\theta,l} \vec{t}_{\alpha}} e^{-i \left( \eta \vec{K}_l + 
	\vk - \vQ + \vG \right) \cdot \vec{r}} u^{*}_{\vQ\alpha;n \eta} \left(\vk \right).  \label{app:eqn:fieldOps_enBand_vMat}
\end{equation}

\subsubsection{The TBG spectral function}\label{app:sec:spectralf:derivation:tbgSpectralF}

As mentioned previously, the key quantity which is measured in an STM experiment is the real-space spectral function $A \left(\vec{r},\omega \right)$ which can be written as~\cite{PER09,COL15,ERV17}
\begin{equation}
	\label{app:eqn:spec_func}
	A \left( \vec{r},\omega \right) = \sum_{\lambda, \xi} \sum_{s} P_{\lambda} \left[
		\abs{\mel**{\xi}{\hat{\psi}^\dagger_{s} \left( \vec{r} \right)}{\lambda}}^2 \delta \left( \omega - E_{\xi} + E_{\lambda} \right) + 
		\abs{\mel**{\xi}{\hat{\psi}_{s} \left( \vec{r} \right)}{\lambda}}^2 \delta \left( \omega + E_{\xi} - E_{\lambda} \right) 
	\right],
\end{equation}
where $\ket{\lambda}$ and $\ket{\xi}$ denote the exact many-body eigenstates of TBG, with the corresponding energies being given by $E_{\xi}$ and $E_{\lambda}$, respectively. In \cref{app:eqn:spec_func} $P_{\lambda}$ represents the thermodynamic probability of the system being in state $\lambda$. In what follows, we will focus on low-temperature measurements, and so we will assume that $P_{\lambda}=1$ if $\ket{\lambda}$ is one of the gound states of TBG, and zero otherwise~\cite{KAN19,BUL20,LIA21}. In \cref{app:eqn:spec_func}, the state $\hat{\psi}^\dagger_{s} \left( \vec{r} \right)\ket{\lambda}$ ($\hat{\psi}_{s} \left( \vec{r} \right)\ket{\lambda}$) is formed by creating (destroying) an electron from the TBG many-body ground state, and thus represents a superposition of charge-one excitations~\cite{BER21b}. It follows that the only states $\ket{\xi}$ which give a non-zero contribution to the spectral function in \cref{app:eqn:spec_func} are the charge-one excitations above the many-body ground states of TBG, whose properties were analytically derived in Ref.~\cite{BER21b} and summarized and extended in \cref{app:sec:charge}. 

As discussed in \cref{app:sec:charge}, Ref.~\cite{BER21b} has shown that the low-energy charge-one excitation above a given TBG ground state at integer fillings $-4 \leq \nu \leq 4$~\cite{LIA21} are obtained by acting with the energy band operators from \cref{app:eqn:en_band} with $n = \pm 1$, or alternatively, with the Chern band basis operators from \cref{app:eqn:chern_band} on the many-body ground states of TBG. As such, we can employ \cref{app:eqn:fieldOps_enBand} to write the real-space spectral function in terms of the energy-band operators as
\begin{align}
	A \left( \vec{r},\omega \right) &= \sum_{\vk',\vk \in \mathrm{MBZ}} \sum_{\substack{n,\eta\\ n', \eta'}} \left\lbrace \left[\mathcal{M}^{+}\left(\omega\right) \right]_{\vk n \eta,\vk' n' \eta'} V_{\vec{r},\vk n \eta} V^{*}_{\vec{r},\vk' n' \eta'} + \left[\mathcal{M}^{-}\left(\omega\right) \right]_{\vk n \eta,\vk' n' \eta'} V_{\vec{r},\vk n \eta} V^{*}_{\vec{r},\vk' n' \eta'} \right\rbrace = \nonumber \\
		&= \sum_{\vk',\vk \in \mathrm{MBZ}} \sum_{\substack{n,\eta \\ n',\eta'}} \left[\mathcal{M}^{+} \left(\omega\right) + \mathcal{M}^{-} \left(\omega\right) \right]_{\vk n \eta,\vk' n' \eta'} \left[ \mathcal{B} \left(\vec{r} \right) \right]_{\vk n \eta,\vk' n' \eta'} \label{app:eqn:spec_func_enbas_temp}
\end{align}
where here, and in what follows, all summations over the TBG energy bands will be restricted to the \emph{active} TBG bands (\ie{} $n,n'=\pm 1$). \Cref{app:eqn:spec_func_enbas_temp} represents the central result of this \siSection{}: the spectral function of TBG is the tensor contraction between the spectral function matrices $\left[\mathcal{M}^{\pm}\left(\omega\right) \right]_{\vk n \eta,\vk' n' \eta'}$, which depend on the particular TBG ground state that we consider and whose elements are given by
\begin{align}
	\left[\mathcal{M}^{+}\left(\omega\right) \right]_{\vk n \eta,\vk' n' \eta'} = & \sum_{\lambda, \xi} \sum_{ s,s'} P_{\lambda} \mel**{\lambda}{\hat{c}_{\vk',n',\eta',s'}}{\xi} \mel**{\xi}{\hat{c}^\dagger_{\vk,n,\eta,s}}{\lambda} \delta_{s, s'} \delta \left( \omega - E_{\xi} + E_{\lambda} \right),  \label{app:eqn:spec_func_mat_elem+} \\
	\left[\mathcal{M}^{-}\left(\omega \right) \right]_{\vk n \eta, \vk' n' \eta'}  = & \sum_{\lambda, \xi} \sum_{ s,s'} P_{\lambda} \mel**{\lambda}{\hat{c}^\dagger_{\vk,n,\eta,s}}{\xi} \mel**{\xi}{\hat{c}_{\vk',n',\eta',s'}}{\lambda} \delta_{s, s'} \delta \left( \omega + E_{\xi} - E_{\lambda} \right), \label{app:eqn:spec_func_mat_elem-}
\end{align}
respectively for the so called ``electron'' and ``hole'' contributions, and the spatial factor matrix $\mathcal{B} \left(\vec{r} \right)$ whose elements are given by $\left[ \mathcal{B} \left(\vec{r} \right) \right]_{\vk n \eta,\vk' n' \eta'} = V^{*}_{\vec{r},\vk' n' \eta'} V_{\vec{r},\vk n \eta}$, or more precisely by
\begin{align}
	\left[ \mathcal{B} \left(\vec{r} \right) \right]_{\vk n \eta,\vk' n' \eta'}=\frac{1}{\omegaTBG \omegaSLG } \sum_{\substack{\alpha_1,l_1 \\ \alpha_2,l_2 }} 
	\sum_{\substack{\vQ_1 \in \mathcal{Q}_{\eta l_1} \\ \vQ_2 \in \mathcal{Q}_{\eta' l_2}}} 
	\sum_{\substack{ \vG_1 \in \mathcal{G}_{l_1} \\ \vG_2 \in \mathcal{G}_{l_2} }} 
	\phi_{l_1} \left( \eta \vec{K}_{l_1} + \vk - \vQ_1 + \vG_1 \right)
	\phi^{*}_{l_2} \left( \eta' \vec{K}_{l_2} + 	\vk' - \vQ_2 + \vG_2 \right) \nonumber \\ 
	\times u^{*}_{\vQ_1 \alpha_1;n \eta} \left(\vk \right)
	u_{\vQ_2 \alpha_2;n' \eta'} \left(\vk' \right)
	e^{i \left( \vG_1 \cdot R_{\theta,l_1} \vec{t}_{\alpha_1} - \vG_2 \cdot R_{\theta,l_2} \vec{t}_{\alpha_2} \right)} 
	e^{-i \left( \eta \vec{K}_{l_1} + \vk - \vQ_1 + \vG_1 \right) \cdot \vec{r}}
	e^{i \left( \eta' \vec{K}_{l_2} + \vk' - \vQ_2 + \vG_2 \right) \cdot \vec{r}} .  \label{app:eqn:spatial_fact_def}
\end{align}
It is important to note that neither the spectral function matrix elements, nor the spatial factor are separately gauge-invariant quantities. By fixing the gauge according to \cref{app:sec:notation:sp:gauge}, we can however discuss their properties individually. We also note that provided that the moir\'e translation symmetry is not broken by the TBG ground state (which will always assume to be the case in this work), the spectral function matrix is diagonal in momentum space
\begin{equation}
	\left[\mathcal{M}^{\pm}\left(\omega \right) \right]_{\vk n \eta, \vk' n' \eta'} = 0, \quad \text{if} \quad \vk' \neq \vk.
\end{equation}  
Finally, the spectral function matrix elements can be used to determine the density of states $\rho \left( \omega \right)$ by tracing over all degrees of freedom
\begin{equation}
	\rho \left( \omega \right) = \sum_{\vk \in \MBZ} \sum_{n,\eta} \left[\mathcal{M}^{+} \left(\omega\right) + \mathcal{M}^{-} \left(\omega\right) \right]_{\vk n \eta,\vk n \eta}
\end{equation}

Alternatively, one can also adopt a momentum-space description and define the Fourier transformation of the spectral function and spatial factor
\begin{equation}
	A \left( \vec{r},\omega \right) = \int \frac{d^2 \vq}{\left(2 \pi \right)^2} A \left( \vq,\omega \right) e^{i \vq \cdot \vec{r}}, \quad 
	\mathcal{B}\left( \vec{r} \right) = \int \frac{d^2 \vq}{\left(2 \pi \right)^2} \mathcal{B}\left( \vq \right) e^{i \vq \cdot \vec{r}}.
\end{equation}
By analogy with \cref{app:eqn:spec_func_enbas_temp} $A \left( \vq,\omega \right)$ can also be expressed in terms of the spectral function matrix elements as
\begin{equation}
	A \left( \vq,\omega \right) = \sum_{\vk',\vk \in \mathrm{MBZ}} \sum_{\substack{n,\eta\\ n', \eta'}} 
		\left[\mathcal{M}^{+} \left(\omega\right) + \mathcal{M}^{-} \left(\omega\right) \right]_{\vk n \eta,\vk' n' \eta'} \left[ \mathcal{B} \left(\vq \right) \right]_{\vk n \eta,\vk' n' \eta'},	\label{app:eqn:spec_func_enbas_temp_qspace}
\end{equation}
where the Fourier-transformed spatial factor is given explicitly by 
\begin{align}
	\left[ \mathcal{B} \left( \vq \right) \right]_{\vk n \eta, \vk' n' \eta'} = 
	\frac{\left(2 \pi \right)^2}{\omegaTBG \omegaSLG } \sum_{\substack{\alpha_1,l_1 \\ \alpha_2,l_2 }} 
	\sum_{\substack{\vQ_1 \in \mathcal{Q}_{\eta l_1} \\ \vQ_2 \in \mathcal{Q}_{\eta' l_2}}} 
	\sum_{\substack{\vG_1 \in \mathcal{G}_{l_1} \\ \vG_2 \in \mathcal{G}_{l_2}}} 
	\phi_{l_1} \left( \eta \vec{K}_{l_1} + 	\vk - \vQ_1 + \vG_1 \right) 
	\phi^{*}_{l_2} \left( \eta' \vec{K}_{l_2} + \vk' - \vQ_2 + \vG_2 \right) \nonumber \\ 
	\times u^{*}_{\vQ_1 \alpha_1;n \eta} \left(\vk \right)
	u_{\vQ_2 \alpha_2;n' \eta'} \left(\vk' \right)
	e^{i \left( \vG_1 \cdot R_{\theta,l_1} \vec{t}_{\alpha_1} - \vG_2 \cdot R_{\theta,l_2} \vec{t}_{\alpha_2}\right)} 
	\delta \left( \eta \vec{K}_{l_1} + \vk - \vQ_1 + \vG_1 - \eta' \vec{K}_{l_2} - \vk' + \vQ_2 - \vG_2 + \vq \right).  \label{app:eqn:spatial_fact_def_qspace}
\end{align}
From a computational standpoint, we evaluate and store the Fourier-transformed spatial factor. After contracting it with the spectral function matrix elements for a given insulator ground state, we employ the FINUFFT package~\cite{BAR19,BAR21} for the efficient inverse (nonuniform) Fourier transformation back to real space. For the evaluation of the spectral function matrix elements, we approximate the $\delta$-functions from \cref{app:eqn:spec_func_mat_elem+,app:eqn:spec_func_mat_elem-} by Lorentzians according to 
\begin{equation}
	\delta(\omega) \to \frac{1}{\pi} \frac{\varepsilon}{x^2 + \varepsilon^2},
\end{equation}
where the broadening width $\varepsilon$ is chosen adaptively throughout the MBZ, depending on the local gradient of the charge-one excitation dispersion~\cite{YAT07}.

\subsection{Symmetries and the spatial factor}\label{app:sec:spectralf:sp_fact_sym}
The discrete spatial symmetries of the single-particle TBG Hamiltonian summarized in \cref{app:sec:notation:sp:symmetries} impose several constraints on the spatial factor, as a consequence of the gauge-fixing conventions from \cref{app:sec:notation:sp:gauge}. In this section, we discuss the consequences of the $C_{2z}T$ and $T$ symmetries of single-particle TBG Hamiltonian on the spatial factor. The discussion of the particle-hole symmetry of $\hat{H}_0$ (in the $\lambda = 0$ case), which does not impose any further constraints on the spatial factor will be relegated to \cref{app:sec:spectralf:sp_fact_approx:Psym}.  

\subsubsection{$C_{2z}T$ symmetry}\label{app:sec:spectralf:sp_fact_sym:c2zt}
Due to the $C_{2z}T$ symmetry implemented by \cref{app:eqn:c2z_rep,app:eqn:t_rep}, as well as the gauge-fixing from \cref{app:eqn:sewing_mats}, we find that the single-particle TBG wave functions obey
\begin{equation}
	u_{\vQ \alpha;n \eta} \left(\vk \right) = u^{*}_{\vQ (-\alpha);n \eta} \left(\vk \right),
\end{equation} 
where $-\alpha=B,A$ for $\alpha=A,B$. Additionally, from \cref{app:eqn:def_graphene_displ_vecs}, we have that 
\begin{equation}
	e^{-i \vG \cdot R_{\theta,l}\vec{t}_{\alpha}} = e^{i \vG \cdot R_{\theta,l} \vec{t}_{-\alpha}}, \quad \text{for} \quad \vG \in \mathcal{G}_{l}, 
\end{equation}
where $\vec{t}_{\alpha}$ is the sublattice displacement vector for the graphene sublattice $\alpha$. Using also the reality of the Fourier-transformed $p_z$ orbital wave functions from \cref{app:eqn:reality_of_phi}, one can show from \cref{app:eqn:spatial_fact_def_qspace} that  
\begin{align}
	\left[ \mathcal{B} \left( \vq \right) \right]^{*}_{\vk n \eta, \vk' n' \eta'} = 
	\frac{\left(2 \pi \right)^2}{\omegaTBG \omegaSLG } \sum_{\substack{\alpha_1,l_1 \\ \alpha_2,l_2 }} 
	\sum_{\substack{\vQ_1 \in \mathcal{Q}_{\eta l_1} \\ \vQ_2 \in \mathcal{Q}_{\eta' l_2}}} 
	\sum_{\substack{\vG_1 \in \mathcal{G}_{l_1} \\ \vG_2 \in \mathcal{G}_{l_2}}} 
	\phi^{*}_{l_1} \left( \eta \vec{K}_{l_1} + 	\vk - \vQ_1 + \vG_1 \right) 
	\phi_{l_2} \left( \eta' \vec{K}_{l_2} + \vk' - \vQ_2 + \vG_2 \right) \nonumber \\ 
	\times u_{\vQ_1 \alpha_1;n \eta} \left(\vk \right)
	u^{*}_{\vQ_2 \alpha_2;n' \eta'} \left(\vk' \right)
	e^{-i \left( \vG_1 \cdot R_{\theta,l_1} \vec{t}_{\alpha_1} - \vG_2 \cdot R_{\theta,l_2} \vec{t}_{\alpha_2}\right)} 
	\delta \left( \eta \vec{K}_{l_1} + \vk - \vQ_1 + \vG_1 - \eta' \vec{K}_{l_2} - \vk' + \vQ_2 - \vG_2 + \vq \right)  \nonumber \\
	= 	\frac{\left(2 \pi \right)^2}{\omegaTBG \omegaSLG } \sum_{\substack{\alpha_1,l_1 \\ \alpha_2,l_2 }} 
	\sum_{\substack{\vQ_1 \in \mathcal{Q}_{\eta l_1} \\ \vQ_2 \in \mathcal{Q}_{\eta' l_2}}} 
	\sum_{\substack{\vG_1 \in \mathcal{G}_{l_1} \\ \vG_2 \in \mathcal{G}_{l_2}}} 
	\phi_{l_1} \left( \eta \vec{K}_{l_1} + 	\vk - \vQ_1 + \vG_1 \right) 
	\phi^{*}_{l_2} \left( \eta' \vec{K}_{l_2} + \vk' - \vQ_2 + \vG_2 \right) \nonumber \\ 
	\times u^{*}_{\vQ_1 -\alpha_1;n \eta} \left(\vk \right)
	u_{\vQ_2 -\alpha_2;n' \eta'} \left(\vk' \right)
	e^{i \left( \vG_1 \cdot R_{\theta,l_1} \vec{t}_{-\alpha_1} - \vG_2 \cdot R_{\theta,l_2} \vec{t}_{-\alpha_2}\right)} 
	\delta \left( \eta \vec{K}_{l_1} + \vk - \vQ_1 + \vG_1 - \eta' \vec{K}_{l_2} - \vk' + \vQ_2 - \vG_2 + \vq \right)  \nonumber \\
	= 	\frac{\left(2 \pi \right)^2}{\omegaTBG \omegaSLG } \sum_{\substack{\alpha_1,l_1 \\ \alpha_2,l_2 }} 
	\sum_{\substack{\vQ_1 \in \mathcal{Q}_{\eta l_1} \\ \vQ_2 \in \mathcal{Q}_{\eta' l_2}}} 
	\sum_{\substack{\vG_1 \in \mathcal{G}_{l_1} \\ \vG_2 \in \mathcal{G}_{l_2}}} 
	\phi_{l_1} \left( \eta \vec{K}_{l_1} + 	\vk - \vQ_1 + \vG_1 \right) 
	\phi^{*}_{l_2} \left( \eta' \vec{K}_{l_2} + \vk' - \vQ_2 + \vG_2 \right) \nonumber \\ 
	\times u^{*}_{\vQ_1 \alpha_1;n \eta} \left(\vk \right)
	u_{\vQ_2 \alpha_2;n' \eta'} \left(\vk' \right)
	e^{i \left( \vG_1 \cdot R_{\theta,l_1} \vec{t}_{\alpha_1} - \vG_2 \cdot R_{\theta,l_2} \vec{t}_{\alpha_2}\right)} 
	\delta \left( \eta \vec{K}_{l_1} + \vk - \vQ_1 + \vG_1 - \eta' \vec{K}_{l_2} - \vk' + \vQ_2 - \vG_2 + \vq \right).
\end{align}
Otherwise stated, the spatial factor has a real Fourier transformation
\begin{equation}
	\label{app:eqn:prop_B_from_C2zT}
	\mathcal{B} \left(\vec{r} \right) = \mathcal{B}^{*} \left(-\vec{r} \right), \qquad 
	\mathcal{B} \left(\vq \right) = \mathcal{B}^{*} \left( \vq \right),
\end{equation}
which greatly simplifies all numerical computations, by halving the memory usage and required processing power.

\subsubsection{$T$ symmetry}\label{app:sec:spectralf:sp_fact_sym:T}

Another important property can be obtained by considering the $T$ symmetry of TBG. Under the gauge-fixing conditions from \cref{app:eqn:sewing_mats}, \cref{app:eqn:t_rep} implies that the single-particle TBG wave functions obey
\begin{equation}
	u^{*}_{-\vQ \alpha; n (-\eta)} \left( -\vk \right) = \left[ B^{T} \left( \vk \right) \right]_{n \eta, n-\eta} u_{\vQ \alpha, n \eta} \left( \vk \right) = \left(-1 \right)^{\vk} u_{\vQ \alpha; n \eta} \left( \vk \right),
\end{equation} 
where we have introduced the factor 
\begin{equation}
	\label{app:eqn:trim_factor}
	(-1)^{\vk} = \begin{cases}
		1 &\vk \neq \vk_{M_M} \\
		-1 &\vk = \vk_{M_M} \\ 
	\end{cases},
\end{equation}
with $\vk_{M_M}$ denoting any one of the three equivalent $M_M$ points in the MBZ from \cref{app:eqn:three_trim_TBG}.

Additionally, using the rotational invariance, as well as the reality of the $p_z$ orbital wave functions from \cref{app:eqn:rot_sym_of_phi_l,app:eqn:reality_of_phi}, respectively, one can show from \cref{app:eqn:spatial_fact_def_qspace} that
\begin{align}
	\left[ \mathcal{B} \left(-\vq \right) \right]^*_{\vk n (-\eta),\vk' n' (-\eta')} = 
	\frac{\left(2 \pi \right)^2}{\omegaTBG \omegaSLG } \sum_{\substack{\alpha_1,l_1 \\ \alpha_2,l_2 }} 
	\sum_{\substack{\vQ_1 \in \mathcal{Q}_{-\eta l_1} \\ \vQ_2 \in \mathcal{Q}_{-\eta' l_2}}} 
	\sum_{\substack{\vG_1 \in \mathcal{G}_{l_1} \\ \vG_2 \in \mathcal{G}_{l_2}}} 
	\phi^{*}_{l_1} \left( -\eta \vec{K}_{l_1} + \vk - \vQ_1 + \vG_1 \right)
	\phi_{l_2} \left( -\eta' \vec{K}_{l_2} + 	\vk' - \vQ_2 + \vG_2 \right) \nonumber \\ 
	\times u_{\vQ_1 \alpha_1;n (-\eta)} \left(\vk \right)
	u^{*}_{\vQ_2 \alpha_2;n' (-\eta')} \left(\vk' \right)
	e^{-i \left( \vG_1 \cdot R_{\theta,l_1} \vec{t}_{\alpha_1} - \vG_2 \cdot R_{\theta,l_2} \vec{t}_{\alpha_2} \right)} 
	\delta \left( \eta \vec{K}_{l_1} - \vk + \vQ_1 - \vG_1 - \eta' \vec{K}_{l_2} + \vk' - \vQ_2 + \vG_2 + \vq \right)  \nonumber \\
	= \frac{\left(2 \pi \right)^2}{\omegaTBG \omegaSLG }\sum_{\substack{\alpha_1,l_1 \\ \alpha_2,l_2 }} 
	\sum_{\substack{\vQ_1 \in \mathcal{Q}_{\eta l_1} \\ \vQ_2 \in \mathcal{Q}_{\eta' l_2}}} 
	\sum_{\substack{\vG_1 \in \mathcal{G}_{l_1} \\ \vG_2 \in \mathcal{G}_{l_2}}} 
	\phi_{l_1} \left( -\eta \vec{K}_{l_1} + \vk + \vQ_1 - \vG_1 \right) 
	\phi^{*}_{l_2} \left( -\eta' \vec{K}_{l_2} + \vk' + \vQ_2 - \vG_2 \right)  \nonumber \\ 
	\times 
	(-1)^{\vk} (-1)^{\vk'} u^{*}_{\vQ_1 \alpha_1; n \eta} \left(-\vk \right)
	u_{\vQ_2 \alpha_2;n' \eta'} \left(-\vk' \right)
	e^{i \left( \vG_1 \cdot R_{\theta,l_1} \vec{t}_{\alpha_1} - \vG_2 \cdot R_{\theta,l_2} \vec{t}_{\alpha_2} \right)} \nonumber \\
	\times	 
	\delta \left( \eta \vec{K}_{l_1} - \vk - \vQ_1 + \vG_1 - \eta' \vec{K}_{l_2} + \vk' + \vQ_2 - \vG_2 + \vq \right).
\end{align}
This implies using the definition from \cref{app:eqn:spatial_fact_def_qspace} that
\begin{equation}
	\label{app:eqn:trivial_prop_t_directly}
	\left[ \mathcal{B} \left(-\vq \right) \right]^*_{\vk n (-\eta),\vk' n' (-\eta')} = (-1)^{\vk} (-1)^{\vk'} \left[ \mathcal{B} \left(\vq \right) \right]_{-\vk n \eta,-\vk' n' \eta'}.
\end{equation}
Additionally, the spectral function matrix obeys the trivial property
\begin{equation}
	\label{app:eqn:trivial_prop_B}
	\left[ \mathcal{B} \left(- \vq \right) \right]^{*}_{\vk n \eta,\vk' n' \eta'} = \left[ \mathcal{B} \left( \vq \right) \right]_{\vk' n' \eta', \vk n \eta},
\end{equation}
which can be readily verified using the definition from \cref{app:eqn:spatial_fact_def_qspace}. Combining
\cref{app:eqn:trivial_prop_t_directly,app:eqn:trivial_prop_B}, we obtain
\begin{equation}
	\label{app:eqn:prop_B_from_T}
	\left[ \mathcal{B} \left(\vq \right) \right]_{\vk' n' (-\eta'),\vk n (-\eta)} = (-1)^{\vk} (-1)^{\vk'} \left[ \mathcal{B} \left(\vq \right) \right]_{-\vk n \eta,-\vk' n' \eta'}.
\end{equation}
\Cref{app:eqn:prop_B_from_T} can be simplified for the case without translation symmetry breaking (for which the condition $\vk = \vk'$ is imposed from the spectral function matrix elements)
\begin{equation}
	\label{app:eqn:prop_B_R3_cancelation}
	\left[ \mathcal{B} \left(\vec{r} \right) \right]_{\vk n' \eta',\vk n \eta} = \left[ \mathcal{B} \left(\vec{r} \right) \right]_{-\vk n (-\eta),-\vk n' (-\eta')}, \quad
	\left[ \mathcal{B} \left(\vq \right) \right]_{\vk n' \eta',\vk n \eta} = \left[ \mathcal{B} \left(\vq \right) \right]_{-\vk n (-\eta),-\vk n' (-\eta')}.
\end{equation}
\Cref{app:eqn:prop_B_R3_cancelation} will prove instrumental in providing a general understanding of the TBG spectral function. In anticipation of the results from \cref{app:sec:spFuncSym}, we briefly mention that because of \cref{app:eqn:prop_B_R3_cancelation}, the spectral function can be alternatively computed as 
\begin{equation}
	A \left( \vq,\omega \right) = \hspace{-1em} \sum_{\substack{\vk \in \mathrm{MBZ} \\ n,\eta, n', \eta'}} 
		\left[\mathcal{M}^{S+} \left(\omega\right) + \mathcal{M}^{S-} \left(\omega\right) \right]_{\vk n \eta,\vk n' \eta'} 
		\left[ \mathcal{B} \left(\vq \right) \right]_{\vk n \eta,\vk n' \eta'},	\label{app:eqn:spec_func_enbas_temp_qspace_symmetrized}
\end{equation}
where we have introduced the \emph{symmetrized} spectral function matrices 
\begin{equation}
	\label{app:eqn:symmetrized_m_tensor}
	\left[\mathcal{M}_{\varphi}^{S \pm}\left(\omega\right) \right]_{\vk n \eta,\vk n' \eta'} = \frac{1}{2}\left( \left[\mathcal{M}_{\varphi}^{\pm}\left(\omega\right) \right]_{\vk n \eta,\vk n' \eta'} + \left[\mathcal{M}_{\varphi}^{\pm}\left(\omega\right) \right]_{-\vk n' -\eta',-\vk n -\eta}\right).
\end{equation}
\Cref{app:eqn:spec_func_enbas_temp_qspace_symmetrized} implies that any components of $\mathcal{M}^{\pm} \left(\omega\right)$ which are anti-symmetric with respect to the transformation $\left[ \vk n \eta, \vk n' \eta' \right] \to \left[ -\vk n' (-\eta'), -\vk n (-\eta) \right]$ will necessarily vanish upon contracting with the spatial factor. 

\subsection{Approximations of the spatial factor}
\label{app:sec:spectralf:sp_fact_approx}
The goal of this section is to formulate a series of approximations to the spatial factor matrix for the purpose of obtaining some analytical intuition on the STM patterns. After deriving a general approximation to the spatial factor, we briefly review the consequences of the anticommuting particle-hole symmetry of $\hat{H}_{0}$ on the approximations of the spatial factor. Finally, we compute the approximate spatial factor at key momenta at the SLG graphene scale. 
\subsubsection{General approximations}
\label{app:sec:spectralf:sp_fact_approx:general}
In deriving approximations to the spatial factor matrix, we will assume that the translation symmetry (at the level of the moir\'e lattice) is not broken and focus on the properties of $\left[\mathcal{B} \left( \vq \right) \right]_{\vk n \eta, \vk n' \eta'}$. The generalization to the case when translation symmetry is broken is straightforward. 

First, we note that $\vk \in \mathrm{MBZ}$ and therefore $\abs{\vk} \ll \abs{\eta \vec{K}_l + \vG}$ for any $\eta = \pm$, $l=\pm$, and $\vG\in \mathcal{G}_{l}$. The Fourier transformations of $p_z$ orbital wave functions do not change significantly on the scale of the $\mathrm{MBZ}$, justifying the approximation $\phi_{l} \left(\eta \vec{K}_{l} + \vk - \vQ + \vG \right) \approx \phi_{l} \left(\eta \vec{K}_{l} - \vQ + \vG \right)$ for any $l = \pm$, $\vQ \in \mathcal{Q}_{\pm}$, $\vk \in \mathrm{MBZ}$, and $\vG \in \mathcal{G}_{\vec{l}}$. Thus a first simplification of the spatial factor matrix reads
\begin{align}
	\left[ \mathcal{B}^{(1)} \left( \vq \right) \right]_{\vk n \eta,\vk n' \eta'}=\frac{\left(2 \pi \right)^2}{\omegaTBG \omegaSLG } \sum_{\substack{\alpha_1,l_1 \\ \alpha_2,l_2 }} 
	\sum_{\substack{\vQ_1 \in \mathcal{Q}_{\eta l_1} \\ \vQ_2 \in \mathcal{Q}_{\eta' l_2}}} 
	\sum_{\substack{ \vG_1 \in \mathcal{G}_{l_1} \\ \vG_2 \in \mathcal{G}_{l_2} }} 
	\phi_{l_1} \left( \eta \vec{K}_{l_1} - \vQ_1 + \vG_1 \right)
	\phi^{*}_{l_2} \left( \eta' \vec{K}_{l_2} - \vQ_2 + \vG_2 \right) \nonumber \\ 
	\times u^{*}_{\vQ_1 \alpha_1;n \eta} \left(\vk \right)
	u_{\vQ_2 \alpha_2;n' \eta'} \left(\vk' \right)
	e^{i \left( \vG_1 \cdot R_{\theta,l_1} \vec{t}_{\alpha_1} - \vG_2 \cdot R_{\theta,l_2} \vec{t}_{\alpha_2} \right)} 
	\delta \left( \eta \vec{K}_{l_1} - \vQ_1 + \vG_1 - \eta' \vec{K}_{l_2} + \vQ_2 - \vG_2 + \vq \right) .  \label{app:eqn:spatial_fact_def_qspace_approx1}
\end{align}

Additionally, the TBG wave functions $u_{\vQ \alpha; n \eta} \left( \vk \right)$ decay exponentially in magnitude for large values of $\vQ$ as shown in Ref.~\cite{BER21}. As such, there is a natural cutoff $Q_{\mathrm{max}}$ for the magnitude of the vectors in $\mathcal{Q}_{\pm}$. More precisely, one has $\abs{\vQ} \leq Q_{\mathrm{max}}$ for any $\vQ \in \mathcal{Q}_{\pm}$. Crucially, Ref.~\cite{BER21} found that $Q_{\mathrm{max}} \lesssim 2 k_{\theta}$, implying that $\abs{\vQ} \ll \abs{\eta \vec{K}_l + \vG}$ for any $\eta = \pm$, $l=\pm$, $\vQ \in \mathcal{Q}_{\pm}$, and $\vG\in \mathcal{G}_{l}$. As the $p_z$ orbital wave functions do not change significantly on the scale $k_{\theta}$, we can approximate $\phi_{l} \left(\eta \vec{K}_{l} - \vQ + \vG \right) \approx \phi_{l} \left(\eta \vec{K}_{l} + \vG \right)$ for any $l = \pm$, $\vQ \in \mathcal{Q}_{\pm}$, $\vG \in \mathcal{G}_{\vec{l}}$. Thus, we can construct a second approximation to the spatial factor matrix
\begin{align}
	\left[ \mathcal{B}^{(2)} \left( \vq \right) \right]_{\vk n \eta,\vk n' \eta'}=\frac{\left(2 \pi \right)^2}{\omegaTBG \omegaSLG } \sum_{\substack{\alpha_1,l_1 \\ \alpha_2,l_2 }} 
	\sum_{\substack{\vQ_1 \in \mathcal{Q}_{\eta l_1} \\ \vQ_2 \in \mathcal{Q}_{\eta' l_2}}} 
	\sum_{\substack{ \vG_1 \in \mathcal{G}_{l_1} \\ \vG_2 \in \mathcal{G}_{l_2} }} 
	\phi_{l_1} \left( \eta \vec{K}_{l_1} + \vG_1 \right)
	\phi^{*}_{l_2} \left( \eta' \vec{K}_{l_2} + \vG_2 \right) \nonumber \\ 
	\times u^{*}_{\vQ_1 \alpha_1;n \eta} \left(\vk \right)
	u_{\vQ_2 \alpha_2;n' \eta'} \left(\vk' \right)
	e^{i \left( \vG_1 \cdot R_{\theta,l_1} \vec{t}_{\alpha_1} - \vG_2 \cdot R_{\theta,l_2} \vec{t}_{\alpha_2} \right)} 
	\delta \left( \eta \vec{K}_{l_1} - \vQ_1 + \vG_1 - \eta' \vec{K}_{l_2} + \vQ_2 - \vG_2 + \vq \right) .  \label{app:eqn:spatial_fact_def_qspace_approx2}
\end{align}

Although the $p_z$ orbital wave functions are roughly constant at the MBZ scale, they do decay significantly at the graphene reciprocal lattice scale. In fact, it can be checked numerically that
\begin{equation}
	\abs{\frac{\phi_{l} \left( \eta \vec{K}_{l} + \vG \right)}{\phi_{l} \left( \eta \vec{K}_{l} \right)}} \lesssim 0.03, \text{ for any } \eta = \pm \text{ and } \vG \in \mathcal{G}_l \text{ with } \abs{\eta \vec{K}_{l} + \vG} \neq \abs{\eta \vec{K}_{l}}.  
\end{equation}  
This implies that in \cref{app:eqn:spatial_fact_def_qspace_approx2}, the only $\vG_1$ and $\vG_2$ vectors that contribute significantly to $\left[ \mathcal{B}^2 \left( \vq \right) \right]_{\vk n \eta, \vk n' \eta'}$ are $\vG_1 \in \mathcal{G}^{*}_{\eta,l_1}$ and $\vG_2 \in \mathcal{G}^{*}_{\eta',l_2}$, where we have defined
\begin{equation}
	\mathcal{G}^{*}_{\eta,l} = \left \lbrace \vec{0}, -\eta \vec{g}_{l,1},-\eta \vec{g}_{l,2} \right \rbrace.
\end{equation}
Moreover, $\abs{\eta \vec{K}_l + \vG} = \abs{\eta \vec{K}_l}$ for any $\vG \in \mathcal{G}^{*}_{\eta,l}$, and as a consequence of \cref{app:eqn:rot_sym_of_phi_l}, we have that $\phi_{l} \left( \eta \vec{K}_l + \vG \right) = \phi_{l} \left( \vec{K}_+ \right)$ for any $l=\pm$, $\eta=\pm$, and $\vG \in \mathcal{G}^{*}_{\eta,l}$. Therefore, we can write a third approximation for the spatial factor matrix as 
\begin{align}
	\left[ \mathcal{B}^{(3)} \left( \vq \right) \right]_{\vk n \eta,\vk n' \eta'}=\frac{\left(2 \pi \right)^2}{\omegaTBG \omegaSLG } \sum_{\substack{\alpha_1,l_1 \\ \alpha_2,l_2 }} 
	\sum_{\substack{\vQ_1 \in \mathcal{Q}_{\eta l_1} \\ \vQ_2 \in \mathcal{Q}_{\eta' l_2}}} 
	\sum_{\substack{ \vG_1 \in \mathcal{G}_{l_1} \\ \vG_2 \in \mathcal{G}_{l_2} }} 
	\phi_{l_1} \left( \vec{K}_{+} \right)
	\phi^{*}_{l_2} \left( \vec{K}_{+} \right) \nonumber \\ 
	\times u^{*}_{\vQ_1 \alpha_1;n \eta} \left(\vk \right)
	u_{\vQ_2 \alpha_2;n' \eta'} \left(\vk' \right)
	e^{i \left( \vG_1 \cdot R_{\theta,l_1} \vec{t}_{\alpha_1} - \vG_2 \cdot R_{\theta,l_2} \vec{t}_{\alpha_2} \right)} 
	\delta \left( \eta \vec{K}_{l_1} - \vQ_1 + \vG_1 - \eta' \vec{K}_{l_2} + \vQ_2 - \vG_2 + \vq \right) .  \label{app:eqn:spatial_fact_def_qspace_approx3}
\end{align}

Finally, the bottom layer $l=-$ is about twice as far away from the STM tip than the top layer $l=+$ is ($z_p - z_+ \approx \SI{3}{\angstrom}$ and $z_p - z_- \approx \SI{6}{\angstrom}$). Mathematically, this implies that   
\begin{equation}
	\abs{\frac{\phi_{-}\left(\vq \right)}{\phi_{+}\left(\vq \right)}} \lesssim 0.01	,
\end{equation}
allowing us to further neglect the bottom layer contribution in \cref{app:eqn:spatial_fact_def_qspace_approx3}, and write a fourth approximation to the spatial factor matrix
\begin{align}
	\left[ \mathcal{B}^{(4)} \left( \vq \right) \right]_{\vk n \eta,\vk n' \eta'}=
	\frac{\left(2 \pi \right)^2}{\omegaTBG \omegaSLG } \abs{\phi_{+} \left( \vec{K}_{+} \right)}^{2} \sum_{\substack{\alpha_1, \alpha_2 \\ \vG_1,\vG_2 \in \mathcal{G}^{*}_{+} }} 
	\sum_{\substack{\vQ_1 \in \mathcal{Q}_{\eta} \\ \vQ_2 \in \mathcal{Q}_{\eta'}}} 
	u^{*}_{\vQ_1 \alpha_1;n \eta} \left(\vk \right)
	u_{\vQ_2 \alpha_2;n' \eta'} \left(\vk' \right) \nonumber \\
	\times e^{i \left( \vG_1 \cdot R_{\theta,l_1} \vec{t}_{\alpha_1} - \vG_2 \cdot R_{\theta,l_2} \vec{t}_{\alpha_2} \right)}
	\delta \left[ \left(\eta - \eta'\right) \vec{K}_{+} - \vQ_1 + \vQ_2 + \vG_1 - \vG_2 + \vq \right] .  \label{app:eqn:spatial_fact_def_qspace_approx4}
\end{align}

Defining the two order-three complex non-real roots of unity $\omega_{\alpha}$ with $\alpha=A,B$ and 
\begin{equation}
	\omega_{A} = e^{i\frac{2 \pi}{3}} \quad \text{and} \quad \omega_{B} = e^{-i \frac{2\pi}{3}},
\end{equation} 
one can rewrite the fourth approximation from \cref{app:eqn:spatial_fact_def_qspace_approx4} as
\begin{align}
	\left[ \mathcal{B}^{(4)} \left( \vq \right) \right]_{\vk n \eta, \vk n' \eta'} = 
	\frac{\left(2 \pi \right)^2}{\omegaTBG \omegaSLG } \abs{\phi_{+} \left( \vec{K}_{+} \right)}^{2} \sum_{\substack{\alpha_1, \alpha_2 \\ 0 \leq j_1,j_2 <3 }} 
	\sum_{\substack{\vQ_1 \in \mathcal{Q}_{\eta} \\ \vQ_2 \in \mathcal{Q}_{\eta'}}}  
	u^{*}_{\vQ_1 \alpha_1;n \eta} \left(\vk \right)
	u_{\vQ_2 \alpha_2;n' \eta'} \left(\vk \right) \nonumber \\ 
	\times
	\omega^{\eta' j_2}_{\alpha_2} \omega^{-\eta j_1}_{\alpha_1}
	\delta \left[ C^{j_2}_{3z}\eta' \vec{K}_{+} - C^{j_1}_{3z} \eta \vec{K}_{+} - \vQ_2 + \vQ_1 - \vq \right],   \label{app:eqn:spatial_fact_def_qspace_approx4_simple}
\end{align}
where $C_{3z}$ denotes a threefold rotation around the $\hat{\vec{z}}$ axis. It is important to note that \emph{all} the approximations to the spatial factor matrix obey the properties derived in \cref{app:sec:spectralf:sp_fact_sym}. Specifically, as a consequence of the $C_{2z}T$ symmetry of the TBG single-particle Hamiltonian, the approximated spatial factor matrix has a real Fourier transformation, meaning that $\mathcal{B}^{(i)} \left( \vq \right)$ (for $1 \leq i \leq 4$) is real. Additionally, because of the $T$ symmetry, the approximated spatial factor matrices obey
\begin{equation}
	\left[ \mathcal{B}^{(i)} \left(\vq \right) \right]_{\vk n' -\eta',\vk n -\eta} = \left[ \mathcal{B}^{(i)} \left(\vq \right) \right]_{-\vk n \eta,-\vk n' \eta'}, \quad \text{for} \quad 1 \leq i \leq 4.
\end{equation}
Finally, we note that the fourth approximation from \cref{app:eqn:spatial_fact_def_qspace_approx4} agrees with the unappoximated spatial factor to a relative error smaller than $5 \%$.  

\subsubsection{The particle-hole symmetry and the spatial factor}\label{app:sec:spectralf:sp_fact_approx:Psym}
In \cref{app:sec:spectralf:sp_fact_sym}, we have argued that the anticommuting $P$ symmetry of $\hat{H}_0$ in the $\lambda = 0$ does not provide any additional constraints on the spatial factors. Here, we show that the particle-hole symmetry allows us to relate the spatial factor in the fourth approximation to the TBG form factors defined in \cref{app:eqn:ff_def}. Under the gauge-fixing conditions from \cref{app:eqn:sewing_mats}, \cref{app:eqn:p_rep} implies that the single-particle TBG wave functions obey
\begin{equation}
	\label{app:eqn:explicit_P_requirement}
	u_{-\vQ \alpha, -n \eta} \left( -\vk \right) = - \zeta_{\vQ} \eta n (-1)^{\vk} u_{\vQ \alpha, n  \eta} \left( \vk \right),
\end{equation}
where the factor $(-1)^{\vk}$ was defined in \cref{{app:eqn:trim_factor}}. We now consider the fourth approximation of the spatial factor from \cref{app:eqn:spatial_fact_def_qspace_approx4_simple}, where as a consequence of \cref{app:eqn:explicit_P_requirement}, we have 
\begin{align}
	\left[ \mathcal{B}^{(4)} \left( \vq \right) \right]_{-\vk -n \eta, -\vk -n' \eta'}  &= 
	\frac{\left(2 \pi \right)^2}{\omegaTBG \omegaSLG } \abs{\phi_{+} \left( \vec{K}_{+} \right)}^{2} \sum_{\substack{\alpha_1, \alpha_2 \\ 0 \leq j_1,j_2 <3 }} 
	\sum_{\substack{\vQ_1 \in \mathcal{Q}_{-\eta} \\ \vQ_2 \in \mathcal{Q}_{-\eta'}}}  
	u^{*}_{-\vQ_1 \alpha_1;-n \eta} \left( -\vk \right)
	u_{-\vQ_2 \alpha_2;-n' \eta'} \left( -\vk \right) \nonumber \\ 
	&\times
	\omega^{\eta' j_2}_{\alpha_2} \omega^{-\eta j_1}_{\alpha_1}
	\delta \left[ C^{j_2}_{3z}\eta' \vec{K}_{+} - C^{j_1}_{3z} \eta \vec{K}_{+} + \vQ_2 - \vQ_1 - \vq \right] \nonumber \\
	&= \frac{\left(2 \pi \right)^2}{\omegaTBG \omegaSLG } \abs{\phi_{+} \left( \vec{K}_{+} \right)}^{2} \eta \eta' n n' \sum_{\substack{\alpha_1, \alpha_2 \\ 0 \leq j_1,j_2 <3 }} 
	\sum_{\substack{\vQ_1 \in \mathcal{Q}_{-\eta} \\ \vQ_2 \in \mathcal{Q}_{-\eta'}}}  
	u^{*}_{\vQ_1 \alpha_1;n \eta} \left( \vk \right)
	u_{\vQ_2 \alpha_2;n' \eta'} \left( \vk \right) \nonumber \\ 
	&\times
	\omega^{\eta' j_2}_{\alpha_2} \omega^{-\eta j_1}_{\alpha_1}
	\delta \left[ C^{j_2}_{3z}\eta' \vec{K}_{+} - C^{j_1}_{3z} \eta \vec{K}_{+} + \vQ_2 - \vQ_1 - \vq \right] \label{app:eqn:explicit_P_consequence_1}.   
\end{align}
In the neighborhood of $\vq \approx \vG \in \mathcal{Q}_0$, \cref{app:eqn:explicit_P_consequence_1} is dominated by $\vq = \vG$, leading to\footnote{Strictly speaking, the Fourier-transformed spatial factor matrix is not a function, but rather a distribution. Because of the Dirac $\delta$-function located at $\vq = \vG$, $\abs{\left[ \mathcal{B}^{(4)} \left( \vG \right) \right]_{\vk n \eta, \vk n' \eta'}} = + \infty$. We therefore use the notation in \cref{app:eqn:explicit_P_consequence_2} to specify the amplitude of the Dirac $\delta$-function at $\vq = \vG$.}
\begin{align}
	\left[ \mathcal{B}^{(4)} \left( \vq \right) \right]_{-\vk -n \eta, -\vk -n' \eta'} \biggr\lvert_{\vq \approx \vG} &= 
	\frac{3 \left(2 \pi \right)^2 \delta_{\eta, \eta'}}{\omegaTBG \omegaSLG } \abs{\phi_{+} \left( \vec{K}_{+} \right)}^{2} n n' \sum_{\alpha} 
	\sum_{\vQ_1,\vQ_2 \in \mathcal{Q}_{-\eta}}  
	u^{*}_{\vQ_1 \alpha;n \eta} \left( \vk \right)
	u_{\vQ_2 \alpha;n' \eta} \left( \vk \right) \nonumber \\ 
	&\times
	\delta \left(\vQ_2 - \vQ_1 - \vq\right) \nonumber \\
	&= 
	\frac{3 \left(2 \pi \right)^2 \delta_{\eta, \eta'}}{\omegaTBG \omegaSLG } \abs{\phi_{+} \left( \vec{K}_{+} \right)}^{2} n n' \sum_{\alpha} 
	\sum_{\vQ \in \mathcal{Q}_{-\eta}}  
	u^{*}_{\vQ \alpha;n \eta} \left( \vk \right)
	u_{\vQ + \vG \alpha;n' \eta} \left( \vk \right) \delta \left(\vq - \vG\right) \label{app:eqn:explicit_P_consequence_2}.   
\end{align}
At the same time, using the approximation from \cref{app:eqn:spatial_fact_def_qspace_approx4_simple} directly for $\vq \approx \vG$
\begin{align}
	\left[ \mathcal{B}^{(4)} \left( \vq \right) \right]_{\vk n \eta, \vk n' \eta'} \biggr\lvert_{\vq \approx \vG} &= 
	\frac{3 \left(2 \pi \right)^2 \delta_{\eta, \eta'}}{\omegaTBG \omegaSLG } \abs{\phi_{+} \left( \vec{K}_{+} \right)}^{2} \sum_{\alpha} 
	\sum_{\vQ_1,\vQ_2 \in \mathcal{Q}_{\eta}}  
	u^{*}_{\vQ_1 \alpha;n \eta} \left(\vk \right)
	u_{\vQ_2 \alpha;n' \eta} \left(\vk \right) \nonumber \\ 
	&\times	\delta \left( - \vQ_2 + \vQ_1 - \vq \right) \nonumber \\
	&= \frac{3 \left(2 \pi \right)^2 \delta_{\eta, \eta'}}{\omegaTBG \omegaSLG } \abs{\phi_{+} \left( \vec{K}_{+} \right)}^{2} \sum_{\alpha} 
	\sum_{\vQ \in \mathcal{Q}_{\eta}}  
	u^{*}_{\vQ + \vG \alpha;n \eta} \left(\vk \right)
	u_{\vQ \alpha;n' \eta} \left(\vk \right)  \delta \left( \vG - \vq \right). \label{app:eqn:explicit_P_consequence_3}     
\end{align}
Combining \cref{app:eqn:explicit_P_consequence_2,app:eqn:explicit_P_consequence_3} with the definition of the TBG form factors from \cref{app:eqn:ff_def}, we find that 
\begin{equation}
	\left\lbrace nn'\left[ \mathcal{B}^{(4)} \left( \vq \right) \right]^{*}_{-\vk -n' \eta, -\vk -n \eta'} + \left[ \mathcal{B}^{(4)} \left( \vq \right) \right]_{\vk n \eta, \vk n' \eta'} \right\rbrace \biggr\lvert_{\vq \approx \vG} =  M^{\eta}_{nn'} \left(\vec{k,\vG} \right) \frac{3 \left(2 \pi \right)^2 \delta_{\eta, \eta'}}{\omegaTBG \omegaSLG } \abs{\phi_{+} \left( \vec{K}_{+} \right)}^{2} \delta \left(\vq - \vG \right).
\end{equation}

\subsubsection{Computing the spectral function at key momenta in the SLG BZ}
\label{app:sec:spectralf:sp_fact_approx:computed}
As discussed in \cref{app:sec:spectralf:sp_fact_approx:general}, the active TBG wave functions $u_{\vQ \alpha; n \eta} \left( \vk \right)$ decay exponentially in magnitude for large values of $\vQ$~\cite{BER21}. As such, there is a natural cutoff $Q_{\mathrm{max}}$ for the magnitude of the vectors in $\mathcal{Q}_{\pm}$. More precisely, one has $\abs{\vQ} \leq Q_{\mathrm{max}} \lesssim 2 k_{\theta}$ for any $\vQ \in \mathcal{Q}_{\pm}$~\cite{BER21}, implying that $\abs{\vQ} \ll \abs{\vec{K}_+}$ for any $\vQ \in \mathcal{Q}_{\pm}$. Considering the fourth approximation from \cref{app:eqn:spatial_fact_def_qspace_approx4_simple}, we find that the contributions to the spatial factor matrix are clustered around four types of momenta $\vq$:
\begin{equation}
	\label{app:eqn:spectralf_relevant_mom}
	\abs{\vq} \approx 0, \quad
	\abs{\vq} \approx \abs{\vec{K}_+}, \quad
	\abs{\vq} \approx 2\abs{\vec{K}_+}, \quad
	\abs{\vq} \approx \abs{\vec{g}_{+,1}}. \quad
\end{equation}
For each of the four types of clusters, we will chose \emph{one} relevant momentum $\vq$ and explicitly work out the spatial factor matrix within the fourth approximation from \cref{app:eqn:spatial_fact_def_qspace_approx4_simple}. 
It is also important to note that any non-vanishing amplitude of the spectral function with momenta $\abs{\vq} \approx \abs{\vec{K}_+}$ or $\abs{\vq} \approx 2\abs{\vec{K}_+}$ signals the presence of so-called \emph{$\rthree$ ordering} at the level of the SLG. Without $\rthree$ ordering, the STM patterns of TBG display an apparent periodicity corresponding to the usual SLG lattice up to an additional modulation at the moir\'e scale. In the presence of $\rthree$ ordering, the apparent periodicity of the graphene mono-layer lattice is broken, with the SLG unit cell becoming enlarged from two to six carbon atoms per unit cell~\cite{CHA00,GUS07,HER09,NOM09,KOS14,GUT16,LI19a,BAO21,LIU22,COI22}.

In fact, inspecting the (unapproximated) spatial factor from \cref{app:eqn:spatial_fact_def_qspace} and remembering that $\abs{\vQ} \ll \abs{\vec{K}_+}$ for any $\vQ \in \mathcal{Q}_{\pm}$ as a consequence of the exponential decay of the TBG wave functions $u_{\vQ \alpha, n \eta} \left( \vk \right)$ with $\abs{\vQ}$~\cite{BER21}, we see that the contributions to the spatial factor $\left[ \mathcal{B} \left( \vq \right) \right]_{\vk n \eta,\vk' n' \eta'}$ cluster around three types of momenta, depending on the valley indices $\eta$ and $\eta'$:
\begin{equation}
	\label{app:eqn:clustering_sp_factor}
	\begin{split}
		\left[ \mathcal{B} \left( \vq \right) \right]_{\vk n \eta,\vk' n' \eta} \neq 0 \quad &\text{if and only if} \quad \vq = \Delta \vq + \vG,  \\
		\left[ \mathcal{B} \left( \vq \right) \right]_{\vk n \eta,\vk' n' (-\eta)} \neq 0 \quad &\text{if and only if} \quad \vq = \Delta \vq + 2 \eta \vec{K}_+ + \vG,
	\end{split}
\end{equation}
where $\vG \in \mathcal{G}_+$, and $\abs{\Delta \vq} \sim k_\theta \ll \abs{\vec{K}_+}$. One important consequence of the second line in \cref{app:eqn:clustering_sp_factor} is that $\rthree$ ordering arises from the valley-off-diagonal elements of the spatial factor, a fact upon which we will rely heavily in \cref{app:sec:spFuncSym} for understanding the real-space STM patterns of the various TBG insulators. 

We will now consider each of the four types of momenta from \cref{app:eqn:spectralf_relevant_mom} individually and compute the spatial factor using the fourth approximation:
\begin{enumerate}
	\item \emph{$\mathcal{B}\left( \vq \right)$ for $\abs{\vq} \approx 0$}. 
	
	For the contributions with $\abs{\vq} \approx 0$, we choose to compute the spatial factor matrix at $\vq = \vec{0}$, where we must have that $C^{j_2}_{3z}\eta' \vec{K}_{+} - C^{j_1}_{3z} \eta \vec{K}_{+} - \vQ_2 + \vQ_1 = \vec{0}$ in \cref{app:eqn:spatial_fact_def_qspace_approx4_simple}. Since $\abs{\vQ_1},\abs{\vQ_2} \ll \abs{\vec{K}_+}$, this is only true if $j_1 = j_2$, $\vQ_1 = \vQ_2$ and $\eta = \eta'$. As such, we find that 
	\begin{align} 
		\left[ \mathcal{B}^{(4)} \left( \vq \right) \right]_{\vk n \eta, \vk n' \eta'} \biggr\lvert_{\vq \approx \vec{0}} 
		&= \frac{\left(2 \pi \right)^2 \delta_{\eta,\eta'}}{\omegaTBG \omegaSLG } \abs{\phi_{+} \left( \vec{K}_{+} \right)}^{2} \sum_{\substack{\alpha_1, \alpha_2 \\ 0 \leq j <3 }} 
		\sum_{\vQ \in \mathcal{Q}_{\eta} }  
		u^{*}_{\vQ \alpha_1;n \eta} \left(\vk \right)
		u_{\vQ \alpha_2;n' \eta} \left(\vk \right) 
		\omega^{\eta j}_{\alpha_2} \omega^{-\eta j}_{\alpha_1}
		\delta \left( \vq \right) \nonumber \\
		&= \frac{3 \left(2 \pi \right)^2 \delta_{\eta,\eta'}}{\omegaTBG \omegaSLG } \abs{\phi_{+} \left( \vec{K}_{+} \right)}^{2} \sum_{\alpha} 
		\sum_{\vQ \in \mathcal{Q}_{\eta} }  
		u^{*}_{\vQ \alpha;n \eta} \left(\vk \right)
		u_{\vQ \alpha;n' \eta} \left(\vk \right) 
		\delta \left( \vq \right).
		\label{app:eqn:spatial_fact_cluster_1}
	\end{align}
	Notice that in \cref{app:eqn:spatial_fact_cluster_1} the summation is only over the $\vQ \in \mathcal{Q}_\eta$ and not over the entire $\mathcal{Q}_{\pm}$ lattice. Had \cref{app:eqn:spatial_fact_cluster_1} included a summation over the entire $\mathcal{Q}_{\pm}$ lattice, this contribution would have been proportional to $\delta_{nn'}$. Nevertheless, we have checked numerically that to an error smaller than $1 \%$,
 	\begin{equation} 
		\left[ \mathcal{B}^{(4)} \left( \vq \right) \right]_{\vk n \eta, \vk n' \eta'} \biggr\lvert_{\vq \approx \vec{0}} 
		\approx \frac{3 \left(2 \pi \right)^2 \delta_{\eta,\eta'}}{\omegaTBG \omegaSLG } \abs{\phi_{+} \left( \vec{K}_{+} \right)}^{2} \frac{\delta_{n,n'}}{2} \delta \left(\vq \right).
		\label{app:eqn:spatial_fact_cluster_1_approx}
	\end{equation}
	\item \emph{$\mathcal{B}\left( \vq \right)$ for $\abs{\vq} \approx \abs{\vec{K}_+}$}. 
	
	When $\abs{\vq} \approx \abs{\vec{K}_+}$, we choose to focus on the contribution at $\vq  \approx \vec{K}_+$. As such, we must have that $C^{j_2}_{3z}\eta' \vec{K}_{+} - C^{j_1}_{3z} \eta \vec{K}_{+} - \vQ_2 + \vQ_1 \approx \vec{K}_+$ in \cref{app:eqn:spatial_fact_def_qspace_approx4_simple}. This necessarily means that $\eta = -\eta'=+$ and $j_1=1$, $j_2 = 2$, or $j_1=2$, $j_2 = 1$. Since this implies that $\vQ_1 \in \mathcal{Q}_{+}$ and $\vQ_2 \in \mathcal{Q}_{-}$, it follows that $\vQ_1 \neq \vQ_2$, meaning that we cannot chose $\vq$ to be exactly at $\vec{K}_+$. However, we can still make $\vq$ as close as possible to $\vec{K}_+$, by letting $\abs{\vQ_1 - \vQ_2}$ be as small as possible. Here, we will consider $\vQ_1 - \vQ_2 = -\vq_1$ to find that 
	\begin{align} 
		\left[ \mathcal{B}^{(4)} \left( \vq \right) \right]_{\vk n \eta, \vk n' \eta'} \biggr\lvert_{\vq \approx \vec{K}_+ - \vq_1} 
		=& \frac{\left(2 \pi \right)^2 \delta_{\eta,+} \delta_{\eta',-}}{\omegaTBG \omegaSLG } \abs{\phi_{+} \left( \vec{K}_{+} \right)}^{2} \sum_{\alpha_1, \alpha_2} 
		\sum_{\vQ \in \mathcal{Q}_{+}}  
		u^{*}_{\vQ \alpha_1;n +} \left(\vk \right)
		u_{\vQ + \vq_1 \alpha_2;n' -} \left(\vk \right) \nonumber \\ 
		& \times \left( \omega^{-1}_{\alpha_2} \omega^{-2}_{\alpha_1} + \omega^{-2}_{\alpha_2} \omega^{-1}_{\alpha_1} \right)
		\delta \left(\vec{K}_{+} - \vq_1 - \vq \right) \nonumber \\
		=& \frac{\left(2 \pi \right)^2 \delta_{\eta,+} \delta_{\eta',-}}{\omegaTBG \omegaSLG } \abs{\phi_{+} \left( \vec{K}_{+} \right)}^{2} \delta \left(\vec{K}_{+} - \vq_1 - \vq \right)  \nonumber \\
		&\times \sum_{\alpha} 
		\sum_{\vQ \in \mathcal{Q}_{+}}  
		2 u^{*}_{\vQ \alpha;n +} \left(\vk \right)
		u_{\vQ + \vq_1 \alpha;n' -} \left(\vk \right)
		- u^{*}_{\vQ \alpha;n +} \left(\vk \right)
		u_{\vQ + \vq_1 (-\alpha);n' -} \left(\vk \right). \label{app:eqn:spatial_fact_cluster_2}
	\end{align}
	\item \emph{$\mathcal{B}\left( \vq \right)$ for $\abs{\vq} \approx 2  \abs{\vec{K}_+}$}. 
	
	When $\abs{\vq} \approx 2\abs{\vec{K}_+}$, we choose to focus on the contribution at $\vq \approx 2\vec{K}_+$. As such, we must have that $C^{j_2}_{3z}\eta' \vec{K}_{+} - C^{j_1}_{3z} \eta \vec{K}_{+} - \vQ_2 + \vQ_1 \approx 2\vec{K}_+$ in \cref{app:eqn:spatial_fact_def_qspace_approx4_simple}. This necessarily means that $\eta = -\eta'=+$ and $j_1=j_2=0$. Since this implies that $\vQ_1 \in \mathcal{Q}_{+}$ and $\vQ_2 \in \mathcal{Q}_{-}$, it follows that $\vQ_1 \neq \vQ_2$, meaning that we cannot chose $\vq$ to be exactly at $2\vec{K}_+$. However, we can still make $\vq$ as close as possible to $2\vec{K}_+$, by letting $\abs{\vQ_1 - \vQ_2}$ be as small as possible. Here, we will consider $\vQ_1 - \vQ_2 = -\vq_1$ and find that 
	\begin{align} 
		\left[ \mathcal{B}^{(4)} \left( \vq \right) \right]_{\vk n \eta, \vk n' \eta'} \biggr\lvert_{\vq \approx 2\vec{K}_+ - \vq_1} 
		=& \frac{\left(2 \pi \right)^2 \delta_{\eta,+} \delta_{\eta',-}}{\omegaTBG \omegaSLG } \abs{\phi_{+} \left( \vec{K}_{+} \right)}^{2} \sum_{\alpha_1, \alpha_2} 
		\sum_{\vQ \in \mathcal{Q}_{+}}  
		u^{*}_{\vQ \alpha_1;n +} \left(\vk \right)
		u_{\vQ + \vq_1 \alpha_2;n' -} \left(\vk \right) \delta \left(2\vec{K}_{+} - \vq_1 - \vq \right) \nonumber \\
		=& \frac{\left(2 \pi \right)^2 \delta_{\eta,+} \delta_{\eta',-}}{\omegaTBG \omegaSLG } \abs{\phi_{+} \left( \vec{K}_{+} \right)}^{2} \delta \left(2 \vec{K}_{+} - \vq_1 - \vq \right)  \nonumber \\
		&\times \sum_{\alpha} 
		\sum_{\vQ \in \mathcal{Q}_{+}}  
		u^{*}_{\vQ \alpha;n +} \left(\vk \right)
		u_{\vQ + \vq_1 \alpha;n' -} \left(\vk \right)
		+ u^{*}_{\vQ \alpha;n +} \left(\vk \right)
		u_{\vQ + \vq_1 (-\alpha);n' -} \left(\vk \right).	
		\label{app:eqn:spatial_fact_cluster_3}
	\end{align}
	\item \emph{$\mathcal{B}\left( \vq \right)$ for $\abs{\vq} \approx \abs{\vec{g}_{+,1}}$}. 
	
	Finally, for the contribution with $\abs{\vq} \approx \abs{\vec{g}_{+,1}}$, we chose to focus on $\vq = \vec{g}_{+,1}$. As such, we must have that $C^{j_2}_{3z}\eta' \vec{K}_{+} - C^{j_1}_{3z} \eta \vec{K}_{+} - \vQ_2 + \vQ_1 \approx \vec{g}_{+,1}$ in \cref{app:eqn:spatial_fact_def_qspace_approx4_simple}. This necessarily means that $\eta = \eta' = +$,  $j_1 = 0$ and $j_2 = 2$, or $\eta = \eta' = -$, and $j_1 = 2$ and $j_2 = 0$. Since $\vQ_1,\vQ_2 \in \mathcal{Q}_{+}$, it follows that we can chose $\vq$ to be exactly at $\vec{g}_{+,1}$. We find that 
\begin{align} 
	\left[ \mathcal{B}^{(4)} \left( \vq \right) \right]_{\vk n \eta, \vk n' \eta'} \biggr\lvert_{\vq \approx \vec{g}_{+,1}} 
	=& \frac{\left(2 \pi \right)^2 \delta_{\eta,+} \delta_{\eta',+}}{\omegaTBG \omegaSLG } \abs{\phi_{+} \left( \vec{K}_{+} \right)}^{2} \sum_{\alpha_1, \alpha_2} 
	\sum_{\vQ \in \mathcal{Q}_{+}}  
	u^{*}_{\vQ \alpha_1;n +} \left(\vk \right)
	u_{\vQ \alpha_2;n' +} \left(\vk \right)  \omega^{2}_{\alpha_2}
	\delta \left(\vec{g}_{+,1} - \vq \right) \nonumber \\
	+& \frac{\left(2 \pi \right)^2 \delta_{\eta,-} \delta_{\eta',-}}{\omegaTBG \omegaSLG } \abs{\phi_{+} \left( \vec{K}_{+} \right)}^{2} \sum_{\alpha_1, \alpha_2} 
	\sum_{\vQ \in \mathcal{Q}_{-}}  
	u^{*}_{\vQ \alpha_1;n -} \left(\vk \right)
	u_{\vQ \alpha_2;n' -} \left(\vk \right)  \omega^{2}_{\alpha_1}
	\delta \left(\vec{g}_{+,1} - \vq \right). 
	\label{app:eqn:spatial_fact_cluster_4}
\end{align}
\end{enumerate}

\section{The spectral function matrix elements}\label{app:sec:computingME}

In \cref{app:sec:spectralf}, we have shown that the spectral function for TBG sample can be obtained by contracting two (gauge-dependent) tensors: the spatial factor -- which depends on the TBG active band wave functions -- and the spectral function matrices -- which depend on the TBG ground state under consideration, as well as on the TBG Hamiltonian. With the properties of the spatial factor derived and discussed in \cref{app:sec:spectralf}, this \siSection{} turns to the spectral function matrices $\mathcal{M}^{\pm} \left( \omega \right)$ and their evaluation. We start by focusing on the experimentally-relevant flat limit~\cite{BER21a}, in which the physics of the system is governed by the TBG interaction Hamiltonian $H_I$ from \cref{app:sec:notation:ih}. We show how the charge-one excitations derived in Ref.~\cite{BER21b} and reviewed in \cref{app:sec:charge} can be used to compute the spectral function matrices for the TBG correlated insulators found in Ref.~\cite{LIA21}. Under the assumption of rigidly filling the charge-one excitation bands, we additionally consider the effects of small doping away from integer filling on the TBG spectral function. Finally, the spectral function matrices are also computed in the noninteracting limit (\ie{} when the electron-electron interactions are ignored). The direct comparison presented in the main paper between the computed spectral functions in both the noninteracting and interacting regimes and the STM experimental data at $\nu = \pm 4$ offers compelling evidence for the validity of the strongly-coupled limit in TBG at the magic angle.

\subsection{Computing the spectral function matrices at integer fillings in the interacting limit}\label{app:sec:computingME:integer}
In this section, we derive the spectral function matrix $\mathcal{M}^{\pm} \left( \omega \right)$ using the exact charge-one excitations of TBG found by Ref.~\cite{BER21b}, and briefly reviewed in \cref{app:sec:charge}. We will always assume to be in the so-called flat limit~\cite{BER21a}, and therefore ignore the projected single-particle Hamiltonian $H_0$ with respect to the interaction one $H_I$. 

As in \cref{app:sec:charge:specific} (whose notation we will follow), we assume that the ground state is given by $\ket{\varphi}$ from \cref{app:eqn:chosen_ground_state}, which represents a certain $\mathrm{U} \left( 4 \right) \times \mathrm{U} \left( 4 \right)$ rotation $\hU$ of the integer-filling ground state $\ket{\Psi^{\nu_+, \nu_-}_{\nu}}$ from \cref{app:eqn:chiralGS}. Using \cref{app:eqn:spec_func_mat_elem+,app:eqn:spec_func_mat_elem-}, we find the corresponding spectral function matrices $\mathcal{M}_{\varphi}^{\pm} \left( \omega \right)$ to be
\begin{align}
	\left[\mathcal{M}_{\varphi}^{+}\left(\omega\right) \right]_{\vk n \eta,\vk' n' \eta'} = & \sum_{\xi} \sum_{ s,s'} \mel**{\varphi}{\hat{c}_{\vk',n',\eta',s'}}{\xi} \mel**{\xi}{\hat{c}^\dagger_{\vk,n,\eta,s}}{\varphi} \delta_{s, s'} \delta \left( \omega - E_{\xi} + E_{\varphi} \right),  \label{app:eqn:spec_func_mat_elem_varphi+} \\
	\left[\mathcal{M}_{\varphi}^{-}\left(\omega \right) \right]_{\vk n \eta, \vk' n' \eta'}  = & \sum_{\xi} \sum_{ s,s'} \mel**{\varphi}{\hat{c}^\dagger_{\vk,n,\eta,s}}{\xi} \mel**{\xi}{\hat{c}_{\vk',n',\eta',s'}}{\varphi} \delta_{s, s'} \delta \left( \omega + E_{\xi} - E_{\varphi} \right). \label{app:eqn:spec_func_mat_elem_varphi-}
\end{align}
\Cref{app:eqn:spec_func_mat_elem_varphi+,app:eqn:spec_func_mat_elem_varphi-} include a summation over all the eigenstates of the TBG interaction Hamiltonian, $\ket{\xi}$. Nevertheless, one of the central results of the charge-commutation relations from \cref{app:sec:charge:specific} was that acting with the $\hat{c}^\dagger_{\vk,n,\eta,s}$ ($\hat{c}_{\vk,n,\eta,s}$) operators on the ground state $\ket{\varphi}$ leads to a superposition of electron (hole) excitations, which can be readily computed as a zero-body problem. Focusing on the electron excitations, we have shown in \cref{app:sec:charge:specific} that the eight fermionic operators $\hat{c}^\dagger_{\vk,n,\eta,s} \ket{\varphi}$ (for $s=\uparrow,\downarrow$, $\eta=\pm$, $n=\pm 1$) can be recombined into $\hat{\gamma}^\dagger_{\vk,p}$ (for $1 \leq p \leq 4 - \nu$) and $\hat{\theta}^\dagger_{\vk,p}$ (for $1 \leq p \leq 4 + \nu$), which were respectively defined in \cref{app:eqn:ch_exc_op_orig_+,app:eqn:ch_exc_op_orig_-}. Thus, the Hillbert space of spanned by \emph{all} states of the form $\hat{c}^\dagger_{\vk,n,\eta,s} \ket{\varphi}$ is identical to the space spanned by $\hat{\gamma}^\dagger_{\vk,p} \ket{\varphi}$ for $1 \leq p \leq 4 - \nu$, \ie{} the eigenstates of $H_I$ corresponding to an electron excitation above the ground state $\ket{\varphi}$\footnote{Note that $\hat{\theta}^\dagger_{\vk,p}$ (for $1 \leq p \leq 4 + \nu$) are linear combinations of fermions corresponding to the occupied bands in $\ket{\varphi}$ and, therefore, $\hat{\theta}^\dagger_{\vk,p} \ket{\varphi} = 0$.}. Therefore, the \emph{only} many-body eigenstates of $H_I$ that have a non-vanishing overlap with $\hat{c}^\dagger_{\vk,n,\eta,s} \ket{\varphi}$ \emph{are} the electron excitations $\hat{\gamma}^\dagger_{\vk,p} \ket{\varphi}$, for $1 \leq p \leq 4 - \nu$\footnote{Strictly speaking, this relies on $\hat{\gamma}^\dagger_{\vk,p} \ket{\varphi}$ being \emph{exact} eigenstates of $H_I$. As argued in \cref{app:sec:charge:matrices}, this is only true in the chiral limit ($w_0 = 0$) for $\ket{\varphi}$ being any $\mathrm{U} \left( 4 \right) \times \mathrm{U} \left( 4 \right)$ rotation of the states from \cref{app:eqn:chiralGS}, or away from the chiral limit in the particle-hole-symmetric case ($w_0 \neq 0$ and $\lambda = 0$) for $\ket{\varphi}$ being any $\Uncf$ rotation of the states in \cref{app:eqn:nonchiralGS}. In the general case, we \emph{approximate} the spectral function by treating both $\ket{\varphi}$ and $\hat{\gamma}^\dagger_{\vk,p} \ket{\varphi}$ as being \emph{exact} eigenstates of $H_I$ and ignore any further perturbative corrections to the spectral function matrix elements.}, whose properties were given in \cref{app:sec:charge:specific}. As shown in the text surrounding \cref{app:eqn:ch_exc_com_operators+}, $\hat{\gamma}^\dagger_{\vk,p} \ket{\varphi}$ are eigenstates of the interaction Hamiltonian with eigenvalue $E_{\varphi} + E^p_{\vk} + \mu$, for $1 \leq p \leq 4 - \nu$, where $E_{\varphi}$ is the ground state energy and the addition of the chemical potential $\mu$ reflects the fact that the excitation energies $E^p_{\vk}$ were defined with respect to the grand canonical Hamiltonian ($H_I - \mu \hN$), as opposed to microcanonical one, $H_I$. As a result, one can replace the summation over all excited states $\ket{\xi}$ in \cref{app:eqn:spec_func_mat_elem_varphi+}, according to 
\begin{equation}
	\label{app:eqn:sum_replace+}
	\sum_{\xi} \ket{\xi} \delta \left(\omega - E_{\xi} + E_{\varphi} \right) \bra{\xi}  \to \sum_{\vk} \sum_{p=1}^{4-\nu} \hat{\gamma}^\dagger_{\vk,p} \ket{\varphi} \delta \left(\omega - \mu - E^p_{\vk}\right) \bra{\varphi} \hat{\gamma}_{\vk,p}  .
\end{equation}
By analogy, the sum over $\ket{\xi}$ in \cref{app:eqn:spec_func_mat_elem_varphi-} can be changed to include only the hole excitations on top of $\ket{\varphi}$ from \cref{app:eqn:ch_exc_op_orig_-} according to 
\begin{equation}
	\label{app:eqn:sum_replace-}
	\sum_{\xi} \ket{\xi} \delta \left(\omega + E_{\xi} - E_{\varphi} \right) \bra{\xi}  \to \sum_{\vk} \sum_{p=1}^{4+\nu} \hat{\theta}_{\vk,p} \ket{\varphi} \delta \left(\omega - \mu + \tilde{E}^p_{\vk}\right) \bra{\varphi} \hat{\theta}^\dagger_{\vk,p}  .
\end{equation}
Using \cref{app:eqn:sum_replace+,app:eqn:sum_replace-}, as well as the wave functions of the charge-one excitations from \cref{app:eqn:ch_exc_op_orig_+,app:eqn:ch_exc_op_orig_-}, we find that the spectral function matrices are given by
\begin{align}
	\left[\mathcal{M}_{\varphi}^{+}\left(\omega\right) \right]_{\vk n \eta,\vk' n' \eta'} = & \delta_{\vk,\vk'} \sum_{ s,s'} \left[ Y \left( \vk, \omega \right) \right]_{n \eta s, n' \eta' s'} \delta_{s, s'} ,  \label{app:eqn:spec_func_from_rMat_y+} \\
	\left[\mathcal{M}_{\varphi}^{-}\left(\omega \right) \right]_{\vk n \eta, \vk' n' \eta'}  = & \delta_{\vk,\vk'} \sum_{ s,s'} \left[ \tilde{Y} \left( \vk, \omega \right) \right]_{n \eta s, n' \eta' s'} \delta_{s, s'}.  \label{app:eqn:spec_func_from_rMat_y-}
\end{align}
where, for simplicity, we have defined
\begin{align}
	\left[ Y \left( \vk, \omega \right) \right]_{n \eta s, n' \eta' s'} = & \sum_{p=1}^{4-\nu} \delta \left(\omega - \mu - E^p_{\vk}\right) Z^{*p}_{\vk,n \eta s} Z^{p}_{\vk,n' \eta' s'}  , \label{app:eqn:def_ys+} \\
	\left[ \tilde{Y} \left( \vk, \omega \right) \right]_{n \eta s, n' \eta' s'} = & \sum_{p=1}^{4+\nu} \delta \left(\omega - \mu + \tilde{E}^p_{\vk}\right) \tilde{Z}^{p}_{\vk,n \eta s} \tilde{Z}^{*p}_{\vk,n' \eta' s'} . \label{app:eqn:def_ys-}
\end{align}
Alternatively, the spectral function can also be defined in terms of charge-one excitation wave functions in the rotated fermion basis, $\Xi^{p}_{\vk,n \eta s}$ and $\tilde{\Xi}^{p}_{\vk,n \eta s}$, which were defined in \cref{app:eqn:ch_exc_op_+,app:eqn:ch_exc_op_-}, respectively,
\begin{align}
	\left[\mathcal{M}_{\varphi}^{+}\left(\omega\right) \right]_{\vk n \eta,\vk' n' \eta'} = & \delta_{\vk,\vk'} \sum_{ s,s'} \left[U^{\dagger} \Upsilon \left( \vk, \omega \right) U \right]_{n \eta s, n' \eta' s'} \delta_{s, s'},  \label{app:eqn:spec_func_from_rMat_ups+} \\
	\left[\mathcal{M}_{\varphi}^{-}\left(\omega \right) \right]_{\vk n \eta, \vk' n' \eta'}  = & \delta_{\vk,\vk'} \sum_{ s,s'} \left[U^{\dagger} \tilde{\Upsilon} \left( \vk, \omega \right) U \right]_{n \eta s, n' \eta' s'} \delta_{s, s'}, \label{app:eqn:spec_func_from_rMat_ups-}
\end{align}
where $U$ is the eight-dimensional unitary matrix defined in \cref{app:eqn:en_band_rot} implementing the $\mathrm{U} \left( 4 \right) \times \mathrm{U} \left( 4 \right)$ rotation $\hU$ from \cref{app:eqn:chosen_ground_state} and 
\begin{align}
	\left[ \Upsilon \left( \vk, \omega \right) \right]_{n \eta s, n' \eta' s'} &= \sum_{p=1}^{4-\nu} \delta \left(\omega - \mu - E^p_{\vk}\right) \Xi^{*p}_{\vk,n \eta s} \Xi^{p}_{\vk,n' \eta' s'}, \label{app:eqn:def_upsilons+}\\
	\left[  \tilde{\Upsilon} \left( \vk, \omega \right)\right]_{n \eta s, n' \eta' s'} &= \sum_{p=1}^{4+\nu} \delta \left(\omega - \mu + \tilde{E}^p_{\vk}\right) \tilde{\Xi}^{p}_{\vk,n \eta s} \tilde{\Xi}^{*p}_{\vk,n' \eta' s'}. \label{app:eqn:def_upsilons-}
\end{align}
Finally, from \cref{app:eqn:ch_exc_wavf_trafo_+,app:eqn:ch_exc_wavf_trafo_-}, we note that 
\begin{equation}
	U^{\dagger} \Upsilon \left( \vk, \omega \right) U =  Y \left( \vk, \omega \right)
	\quad \text{and} \quad
	U^{\dagger} \tilde{\Upsilon} \left( \vk, \omega \right) U =  \tilde{Y} \left( \vk, \omega \right),
\end{equation}
and therefore, the matrices $\Upsilon \left( \vk, \omega \right)$ and $\tilde{\Upsilon} \left( \vk, \omega \right)$ are related to $Y \left( \vk, \omega \right)$ and $\tilde{Y} \left( \vk, \omega \right)$ through a $\mathrm{U} \left( 4 \right) \times \mathrm{U} \left( 4 \right)$ rotation. 

The advantage of using the notation from \cref{app:eqn:spec_func_from_rMat_y+,app:eqn:spec_func_from_rMat_y-} becomes apparent when comparing \cref{app:eqn:def_ys+,app:eqn:def_ys-} with \cref{app:eqn:ch_exc_eigendecomp_+,app:eqn:ch_exc_eigendecomp_-}: up to a Dirac $\delta$-function being applied on the (non-zero) eigenvalues, the matrices $Y \left( \vk, \omega \right)$ and $\tilde{Y} \left( \vk, \omega \right)$ share identical eigenspectra with the matrices $U^{\dagger} \Pi U \left( R \left( \vk \right) \otimes s^0 \right) U^{\dagger} \Pi U$ and $U^{\dagger} \Pi U \left( \tilde{R} \left( \vk \right) \otimes s^0 \right) U^{\dagger} \Pi U$, respectively. Similarly, the matrices $\Upsilon \left( \vk, \omega \right)$ and $\tilde{\Upsilon} \left( \vk, \omega \right)$ share identical eigenvectors with the matrices $\Pi U \left( R \left( \vk \right) \otimes s^0 \right) U^{\dagger} \Pi$ and $\Pi U \left( \tilde{R} \left( \vk \right) \otimes s^0 \right) U^{\dagger} \Pi$, respectively. In \cref{app:sec:spFuncSym}, we will leverage this connection extensively for deriving parameterizations for the $\mathcal{M}_{\varphi}^{\pm}\left(\omega\right)$ matrices and for obtaining an \emph{analytical} understanding of the TBG spectral function patterns. 

\subsection{Influence of small doping away from integer fillings in the interacting limit}\label{app:sec:computingME:doping}
Having derived the spectral function matrix elements of the insulator $\ket{\varphi}$ (corresponding to an \emph{exact} integer filling $\nu$), we now investigate the influence of small doping \emph{away} from integer filling. As a zeroth order approximation, we will assume that the ground state of the interacting Hamiltonian away from exact integer filling can be found by ``rigidly'' filling or depleting the charge-one excitation bands.  In other words, we assume that for small doping away from integer fillings, the charge-one excitations can be approximated as noninteracting fermions and, as such, the charge-one excitation bands can be filled (or depleted) according to the Pauli exclusion principle. Such an approximation can be justified by noting that the steep dispersion of the charge-one excitations above $\ket{\varphi}$ is enough to stabilize a Fermi liquid for small doping away from integer filling~\cite{KAN21}.

To start with, we let $\ket{\varphi_{\Delta \nu}}$ be the ground state of the interaction TBG Hamiltonian at filling $\nu + \Delta \nu$, obtained by doping the insulator $\ket{\varphi}$
\begin{equation}
	\label{app:eqn:doped_gs_def}
	\ket{\varphi_{\Delta \nu}} = \left( \prod_{p=1}^{4-\nu}
\prod_{\substack{\vk \\ \left( \mu_{\Delta \nu} - \mu -E^p_{\vk} \right) > 0}}	\hat{\gamma}^\dagger_{\vk,p} \right)
	\left( \prod_{p=1}^{4+\nu}
\prod_{\substack{\vk \\ \left( \mu -\mu_{\Delta \nu} - \tilde{E}^p_{\vk} \right) > 0}}	\hat{\theta}_{\vk,p} \right)
	\ket{\varphi},
\end{equation}
where $\mu_{\Delta \nu}$ ($\mu$) is the chemical potential of the doped (un-doped) ground state. Because $\ket{\varphi}$ is a ground state of the TBG interaction Hamiltonian, the charge-one excitation energies are strictly positive. Therefore, \cref{app:eqn:doped_gs_def} is valid for both electron doping (where $\mu_{\Delta \nu}>\mu$ and the second product is identity) and hole doping (where $\mu_{\Delta \nu} < \mu$ and the first product is identity). The chemical potential of the doped ground state $\mu_{\Delta \nu}$ satisfies the self-consistent condition 
\begin{equation}
	\label{app:eqn:doped_chem_pot_1}
	\Delta \nu = \frac{1}{N_M} \begin{cases}
		\sum_{\vk} \sum_{p=1}^{4-\nu} \Theta \left( \mu_{\Delta \nu} - \mu - E^p_{\vk} \right), \quad &\text{if} \quad \Delta \nu >0 \\
		- \sum_{\vk} \sum_{p=1}^{4+\nu} \Theta \left( \mu - \mu_{\Delta \nu} - \tilde{E}^p_{\vk} \right), \quad &\text{if} \quad \Delta \nu <0
	\end{cases},
\end{equation}
where $N_M$ denotes the number of moir\'e unit cells in the system and $\Theta (x)$ is the Heaviside step function 
\begin{equation}
	\Theta(x) = \begin{cases}
		1, \quad x>0 \\
		0, \quad x<0
	\end{cases}.
\end{equation} 
\Cref{app:eqn:doped_chem_pot_1} can also be written more succinctly as 
\begin{equation}
	\Delta \nu = \frac{1}{N_M} \left[ \sum_{\vk} \sum_{p=1}^{4-\nu} \Theta \left( \mu_{\Delta \nu} - \mu - E^p_{\vk} \right) - \sum_{\vk} \sum_{p=1}^{4+\nu} \Theta \left( \mu - \mu_{\Delta \nu} - \tilde{E}^p_{\vk} \right) \right].
\end{equation}

To obtain the spectral function matrix elements
\begin{align}
	\left[\mathcal{M}_{\varphi_{\Delta \nu}}^{+}\left(\omega\right) \right]_{\vk n \eta,\vk' n' \eta'} = & \sum_{\xi} \sum_{ s,s'} \mel**{\varphi_{\Delta \nu}}{\hat{c}_{\vk',n',\eta',s'}}{\xi} \mel**{\xi}{\hat{c}^\dagger_{\vk,n,\eta,s}}{\varphi_{\Delta \nu}} \delta_{s, s'} \delta \left( \omega - E_{\xi} + E_{\varphi_{\Delta \nu}} \right),  \label{app:eqn:spec_func_mat_elem_varphi_dop+} \\
	\left[\mathcal{M}_{\varphi_{\Delta \nu}}^{-}\left(\omega \right) \right]_{\vk n \eta, \vk' n' \eta'}  = & \sum_{\xi} \sum_{ s,s'} \mel**{\varphi_{\Delta \nu}}{\hat{c}^\dagger_{\vk,n,\eta,s}}{\xi} \mel**{\xi}{\hat{c}_{\vk',n',\eta',s'}}{\varphi_{\Delta \nu}} \delta_{s, s'} \delta \left( \omega + E_{\xi} - E_{\varphi_{\Delta \nu}} \right). \label{app:eqn:spec_func_mat_elem_varphi_dop-}
\end{align}
we still need to compute the eigenstates of the interaction Hamiltonian that contain one additional electron or hole compared to $\ket{\varphi_{\Delta \nu}}$. Focusing on the electron contribution from \cref{app:eqn:spec_func_mat_elem_varphi_dop+}, we first note that the Hilbert space spanned by $\hat{c}^\dagger_{\vk,n,\eta,s} \ket{\varphi_{\Delta \nu}}$ for $s=\uparrow,\downarrow$, $\eta=\pm$, $n=\pm 1$, is identical to the Hilbert space spanned by $\hat{\gamma}^\dagger_{\vk,p} \ket{\varphi_{\Delta \nu}}$, for $1 \leq p \leq 4-\nu$, and $\hat{\theta}^\dagger_{\vk,p} \ket{\varphi_{\Delta \nu}}$, for $1 \leq p \leq 4+\nu$, because the corresponding operators, $\hat{\gamma}^\dagger_{\vk,p}$ and $\hat{\theta}^\dagger_{\vk,p}$, are just linear combinations of the energy band operators $\hat{c}^\dagger_{\vk,n,\eta,s}$. Therefore, the \emph{only} states containing one additional electron compared to $\ket{\varphi_{\Delta \nu}}$ that have a non-vanishing overlap with $\hat{c}^\dagger_{\vk,n,\eta,s} \ket{\varphi_{\Delta \nu}}$ are of the form $\hat{\gamma}^\dagger_{\vk,p} \ket{\varphi_{\Delta \nu}}$ or $\hat{\theta}^\dagger_{\vk,p} \ket{\varphi_{\Delta \nu}}$, provided that these states do themselves not vanish. Under the approximation that the charge-one excitations do not interact, these states are also \emph{eigenstates} of the interaction Hamiltonian, and so we find that $\ket{\xi}$ from \cref{app:eqn:spec_func_mat_elem_varphi_dop+} can be either $\hat{\gamma}^\dagger_{\vk,p} \ket{\varphi_{\Delta \nu}}$ (with energy $E_{\xi} = E_{\varphi_{\Delta}} + E^p_{\vk} + \mu$) or $\hat{\theta}^\dagger_{\vk,p} \ket{\varphi_{\Delta \nu}}$ (with energy $E_{\xi} = E_{\varphi_{\Delta}} - \tilde{E}^p_{\vk} + \mu$). As such, the summation over the states $\ket{\xi}$ in \cref{app:eqn:spec_func_mat_elem_varphi_dop+} can be changed according to 
\begin{align}
	\sum_{\xi}  \ket{\xi} \delta \left(\omega - E_{\xi} + E_{\varphi_{\Delta \nu}} \right) \bra{\xi}   \to \sum_{\vk} \left( \sum_{p=1}^{4-\nu} \hat{\gamma}^\dagger_{\vk,p} \ket{\varphi_{\Delta \nu}} \delta \left(\omega - \mu - E^p_{\vk}\right) \bra{\varphi_{\Delta \nu}} \hat{\gamma}_{\vk,p}  \right. \nonumber\\
	\left.+ \sum_{p=1}^{4+\nu} \hat{\theta}^\dagger_{\vk,p} \ket{\varphi_{\Delta \nu}} \delta \left(\omega - \mu + \tilde{E}^p_{\vk}\right) \bra{\varphi_{\Delta \nu}} \hat{\theta}_{\vk,p} \right). \label{app:eqn:sum_replace_dope+}
\end{align}
Using a similar argument for the hole contribution from \cref{app:eqn:spec_func_mat_elem_varphi_dop-}, we find that the summation over the states $\ket{\xi}$ in \cref{app:eqn:spec_func_mat_elem_varphi_dop-} can also be replaced as
\begin{align}
	\sum_{\xi}  \ket{\xi} \delta \left(\omega + E_{\xi} - E_{\varphi_{\Delta \nu}} \right) \bra{\xi}   \to \sum_{\vk} \left( \sum_{p=1}^{4-\nu} \hat{\gamma}_{\vk,p} \ket{\varphi_{\Delta \nu}} \delta \left(\omega - \mu - E^p_{\vk}\right) \bra{\varphi_{\Delta \nu}} \hat{\gamma}^\dagger_{\vk,p}  \right. \nonumber\\
	\left.+ \sum_{p=1}^{4+\nu} \hat{\theta}_{\vk,p} \ket{\varphi_{\Delta \nu}} \delta \left(\omega - \mu + \tilde{E}^p_{\vk}\right) \bra{\varphi_{\Delta \nu}} \hat{\theta}^\dagger_{\vk,p} \right). \label{app:eqn:sum_replace_dope-}
\end{align}

By substituting \cref{app:eqn:sum_replace_dope+,app:eqn:sum_replace_dope-,app:eqn:doped_gs_def} in the expressions for the spectral function matrices from \cref{app:eqn:spec_func_mat_elem_varphi_dop+,app:eqn:spec_func_mat_elem_varphi_dop-}, we find
\begin{align}
	\left[\mathcal{M}_{\varphi_{\Delta \nu}}^{+}\left(\omega\right) \right]_{\vk n \eta,\vk' n' \eta'} = & \delta_{\vk,\vk'} \left \lbrace \sum_{p=1}^{4-\nu} \sum_{ s,s'} Z^{*p}_{\vk,n \eta s} Z^{p}_{\vk,n' \eta' s'} \delta_{s, s'} \delta \left(\omega - \mu - E^p_{\vk}\right) \left[1- \Theta \left( \mu_{\Delta \nu} - \mu - E^p_{\vk} \right) \right] \right. \nonumber \\
	+& \left. \sum_{p=1}^{4+\nu} \sum_{ s,s'} \tilde{Z}^{p}_{\vk,n \eta s} \tilde{Z}^{*p}_{\vk,n' \eta' s'} \delta_{s, s'} \delta \left(\omega - \mu + \tilde{E}^p_{\vk}\right) \Theta \left( \mu - \mu_{\Delta \nu} - \tilde{E}^p_{\vk} \right) \right\rbrace,   \label{app:eqn:spec_func_from_rMat_dope+} \\
	\left[\mathcal{M}_{\varphi_{\Delta \nu}}^{-}\left(\omega \right) \right]_{\vk n \eta, \vk' n' \eta'}  = & \delta_{\vk,\vk'} \left \lbrace \sum_{p=1}^{4+\nu} \sum_{ s,s'} \tilde{Z}^{p}_{\vk,n \eta s} \tilde{Z}^{*p}_{\vk,n' \eta' s'} \delta_{s, s'} \delta \left(\omega - \mu + \tilde{E}^p_{\vk}\right) \left[1- \Theta \left( \mu - \mu_{\Delta \nu} - \tilde{E}^p_{\vk} \right) \right] \right. \nonumber \\
	+& \left. \sum_{p=1}^{4-\nu} \sum_{ s,s'} Z^{*p}_{\vk,n \eta s} Z^{p}_{\vk,n' \eta' s'} \delta_{s, s'} \delta \left(\omega - \mu - E^p_{\vk}\right) \Theta \left( \mu_{\Delta \nu} - \mu - E^p_{\vk} \right) \right\rbrace . \label{app:eqn:spec_func_from_rMat_dope-}
\end{align}
At first glance, since the spectral function matrix elements of the doped insulator from \cref{app:eqn:spec_func_from_rMat_dope+,app:eqn:spec_func_from_rMat_dope-} differ from the ones of the un-doped state $\ket{\varphi}$ from \cref{app:eqn:spec_func_from_rMat_y+,app:eqn:spec_func_from_rMat_y-}, one might conclude that the spectral function itself also changes under small doping. However, as shown in \cref{app:eqn:spec_func_enbas_temp}, the spectral function depends on the \emph{sum} between the electron and hole contributions from \cref{app:eqn:spec_func_from_rMat_dope+,app:eqn:spec_func_from_rMat_dope-}, which can be seen to obey
\begin{equation}
	\label{app:eqn:doped_sp_func_matrices}
	\mathcal{M}_{\varphi_{\Delta \nu}}^{+}\left(\omega\right) + \mathcal{M}_{\varphi_{\Delta \nu}}^{-}\left(\omega\right) = 
	\mathcal{M}_{\varphi}^{+}\left(\omega\right) + \mathcal{M}_{\varphi}^{-}\left(\omega\right),
\end{equation}
implying that, under the assumption of rigidly filling the charge-one excitation bands, the spectral function does not change. This result is a direct consequence of treating the charge-one excitations as being noninteracting and represents a well-known result in noninteracting band theory~\cite{ERV17}. For example, one can consider doping the insulator $\ket{\varphi}$ with electrons: there are fewer available electron excitations in $\ket{\varphi_{\Delta \nu}}$. However, the missing electron excitations from $\ket{\varphi_{\Delta \nu}}$ now appear as hole excitations, meaning that although the individual hole and electron contributions to the spectral function change, their sum does not. 

\subsection{Computing the spectral function matrices in the absence of interactions}\label{app:sec:computingME:nonInteracting}

For comparison purposes, it is instructive to compute the TBG spectral function in the noninteracting limit at a certain (not necessarily integer) filling factor $\nu$. In this limit, the physics is governed by the single-particle TBG Hamiltonian from \cref{app:eqn:diag_sp_Hamiltonian}, as electron-electron interactions are ignored. The corresponding ground state $\ket{\phi_{\nu}}$ at filling $\nu$ is found by populating the active TBG bands according the Pauli exclusion principle
\begin{equation}
	\label{app:eqn:non_int_gs}
	\ket{\phi_{\nu}} = \prod_{n,\eta}
\prod_{\substack{\vk \\ \epsilon_{n,\eta} \left( \vk \right) < \mu}}	\hat{c}^\dagger_{\vk,n,\eta,\uparrow} \hat{c}^\dagger_{\vk,n,\eta,\downarrow}
	\ket{0}.
\end{equation}
In \cref{app:eqn:non_int_gs}, $\epsilon_{n,\eta}  \left( \vk \right)$ are the single-particle energies of the active TBG bands and $\mu$ is the chemical potential of $\ket{\phi_{\nu}}$ satisfying the self-consistent condition
\begin{equation}
	\label{app:eqn:nonint_chem_pot}
	\nu = -4 + \frac{2}{N_M}  \sum_{\vk} \sum_{n,\eta} \Theta \big( \mu - \epsilon_{n,\eta} \left( \vk \right) \big).
\end{equation}
Since interactions are ignored completely, the state $\hat{c}^\dagger_{\vk,n,\eta,s} \ket{\phi_{\nu}}$ [$\hat{c}_{\vk,n,\eta,s} \ket{\phi_{\nu}}$] is an exact eigenstate of the noninteracting TBG Hamiltonian with energy $E_{\phi_{\nu}} + \epsilon_{n,\eta} \left( \vk \right)$ [$E_{\phi_{\nu}} - \epsilon_{n,\eta} \left( \vk \right)$], provided that the state itself does not vanish. As a result, one can replace the summation over all excited states $\ket{\xi}$ in the definition of the spectral function matrices from \cref{app:eqn:spec_func_mat_elem+,app:eqn:spec_func_mat_elem-} according to 
\begin{align}
	\sum_{\xi}  \ket{\xi} \delta \left(\omega - E_{\xi} + E_{\varphi_{\Delta \nu}} \right) \bra{\xi} &\to \sum_{\vk} \sum_{n,\eta,s} \hat{c}^\dagger_{\vk,n,\eta,s} \ket{\phi_{\nu}} \delta \big(\omega - \epsilon_{n,\eta} \left( \vk \right) \big) \bra{\phi_{\nu}} \hat{c}_{\vk,n,\eta,s}, \label{app:eqn:sum_replace_nonint+} \\
	\sum_{\xi}  \ket{\xi} \delta \left(\omega + E_{\xi} - E_{\varphi_{\Delta \nu}} \right) \bra{\xi} &\to \sum_{\vk} \sum_{n,\eta,s} \hat{c}_{\vk,n,\eta,s} \ket{\phi_{\nu}} \delta \big(\omega - \epsilon_{n,\eta} \left( \vk \right) \big) \bra{\phi_{\nu}} \hat{c}^\dagger_{\vk,n,\eta,s}, \label{app:eqn:sum_replace_nonint-}
\end{align}
respectively. The corresponding matrix elements can be readily computed to be 
\begin{align}
	\left[\mathcal{M}_{\phi_{\nu}}^{+}\left(\omega\right) \right]_{\vk n \eta,\vk' n' \eta'} = & \sum_{s,s'} \delta_{\vk,\vk'} \delta_{n,n'} \delta_{\eta,\eta'} \delta_{s,s'}  \delta \big(\omega - \epsilon_{n,\eta} \left( \vk \right) \big) \left[ 1- \Theta \big( \mu - \epsilon_{n,\eta} \left( \vk \right) \big) \right], \label{app:eqn:spec_func_nonInt+} \\
	\left[\mathcal{M}_{\phi_{\nu}}^{-}\left(\omega\right) \right]_{\vk n \eta,\vk' n' \eta'} = & \sum_{s,s'} \delta_{\vk,\vk'} \delta_{n,n'} \delta_{\eta,\eta'} \delta_{s,s'}  \delta \big(\omega - \epsilon_{n,\eta} \left( \vk \right) \big) \Theta \big( \mu - \epsilon_{n,\eta} \left( \vk \right) \big). \label{app:eqn:spec_func_nonInt-}
\end{align}
The TBG spectral function depends on the sum between the electron and hole contribution,
\begin{equation}
	\left[\mathcal{M}_{\phi_{\nu}}^{+}\left(\omega\right) \right]_{\vk n \eta,\vk' n' \eta'} + \left[\mathcal{M}_{\phi_{\nu}}^{-}\left(\omega\right) \right]_{\vk n \eta,\vk' n' \eta'} = 2\delta_{\vk,\vk'} \delta_{n,n'} \delta_{\eta,\eta'} \delta \big(\omega - \epsilon_{n,\eta} \left( \vk \right) \big),
\end{equation}
which is independent on the filling $\nu$. Unlike the interacting spectral function matrices computed in \cref{app:sec:computingME:integer}, the spectral function matrices in the noninteracting case are necessarily diagonal in the band and valley subspaces.

\section{Symmetry properties of the spectral function matrices in the interacting case}\label{app:sec:spFuncSym}

In \cref{app:sec:computingME:integer}, we showed how the exact charge-one excitations from \cref{app:sec:charge:specific} can be employed to compute the spectral function matrices for the correlated TBG insulating states found by Ref.~\cite{LIA21}. Since evaluating the charge-one excitation spectrum for a given ground state $\ket{\varphi}$ in the interacting case is equivalent to a zero-body problem~\cite{BER21b}, the direct numerical calculation of the corresponding spectral function is computationally feasible even for relatively large system sizes (\ie{} much higher than can be achieved with techniques such as exact diagonalization). Nevertheless, the various ways of populating the eight Chern bands in the unrotated state $\ket{\Psi^{\nu_+, \nu_-}_{\nu}}$, coupled with the thermodynamically extensive number of possible $\mathrm{U} \left( 4 \right) \times \mathrm{U} \left( 4 \right)$ rotations thereof, results in a virtually infinite number of choices for the correlated TBG ground states $\ket{\varphi}$. As such, the numerical evaluation of the spectral functions for \emph{all} possible insulating ground states found by Ref.~\cite{LIA21} becomes neither feasible, nor useful.

With the goal of building a more complete understanding of the spectral functions for the various insulators found by Ref.~\cite{LIA21}, this \siSection{} analytically derives the symmetry properties of the spectral functions matrices. To do so, we will leverage the connection outlined in \cref{app:sec:charge:specific:rotated_matrix,app:sec:computingME:integer} between the spectral function and charge-one excitation matrices. For a given ground state $\ket{\varphi}$ defined in \cref{app:eqn:chosen_ground_state}, comparing the spectral function matrices from \cref{app:eqn:spec_func_from_rMat_y+,app:eqn:spec_func_from_rMat_y-} with \cref{app:eqn:ch_exc_eigendecomp_+,app:eqn:ch_exc_eigendecomp_-} reveals that, up to a rescaling of the eigenvalues and a trace over the spin degrees of freedom, $\mathcal{M}_{\varphi}^{+}\left(\omega\right)$ and $\mathcal{M}_{\varphi}^{-}\left(\omega\right)$ have the same 
eigendecomposition as the matrices $U^{\dagger} \Pi U \left( R \left( \vk \right) \otimes s^0 \right) U^{\dagger} \Pi U$ and $U^{\dagger} \tilde{\Pi} U \left( \tilde{R} \left( \vk \right) \otimes s^0 \right) U^{\dagger} \tilde{\Pi} U$, respectively, where the unitary matrix $U$ defined in \cref{app:eqn:en_band_rot} specifies the $\mathrm{U} \left( 4 \right) \times \mathrm{U} \left( 4 \right)$ transformation defining $\ket{\varphi}$, and the projectors $\Pi$ and $\tilde{\Pi}$ are related to the occupied bands of $\ket{\varphi}$ according to \cref{app:eqn:def_proj_1,app:eqn:def_proj_2}. Since the charge-one excitation matrices $R \left( \vk \right)$ and $\tilde{R} \left( \vk \right)$ defined in \cref{app:sec:charge:matrices} have been explicitly parameterized in \cref{app:sec:charge:parameterization}, this connection allows us to derive parameterizations of the spectral function matrices, which take into consideration both the discrete and continuous symmetries of TBG. Coupled with the properties of the spatial factor derived in \cref{app:sec:spectralf:sp_fact_sym}, these will enable us to obtain a general understanding of the real-space STM patterns of the integer-filled TBG insulators.

As argued in \cref{app:sec:spectralf:sp_fact_approx:computed}, the valley-off-diagonal elements of the spatial factor correspond to the emergence of $\rthree$ symmetry-breaking at the level of the SLG lattice. In turn, for intervalley-coherent TBG ground states, the spectral function matrices also contain valley-off-diagonal terms. As such, one might conclude that $\rthree$ ordering represents a generic feature of intervalley-coherent TBG ground states. In this \siSection{}, we will show that this naive picture is, in fact, incorrect: $\rthree$ symmetry-breaking is \emph{not} generic \emph{even} in the presence of intervalley coherence, and therefore its presence or absence in the STM patterns can serve as a robust means of experimentally discriminating between the various theoretically proposed TBG ground state at a given filling. Consequently, deriving the conditions under which $\rthree$ ordering emerges within the TBG insulators found by Ref.~\cite{LIA21} constitutes the main focus of this \siSection{}. 

We start by investigating the STM patterns of the insulators $\ket{\Psi_{\nu}}$, defined in \cref{app:eqn:nonchiralGS}, together with their $\Uncf$ rotations. The resulting states $\ket{\varphi}$ correspond to an even filling and zero Chern number, and were analytically shown~\cite{KAN19,BUL20,LIA21} and numerically verified~\cite{BUL20,XIE21} to be exact ground states of the projected interaction TBG Hamiltonian $H_I$ (in the $\lambda = 0$ case), thus making our analysis exact in the nonchiral-flat limit~\cite{BER21a}. We show that the spectral function of these insulators is independent of the particular $\Uncf$ rotation we consider, and prove that no $\rthree$ symmetry breaking is present in the corresponding STM patterns, \emph{even} in the presence of intervalley coherence. We find that the counter-intuitive absence of $\rthree$ ordering stems from occupying both Chern bands of a given valley-spin flavor, considering only $\Uncf$ rotations (as opposed to the more general $\mathrm{U} \left( 4 \right)\times \mathrm{U} \left( 4 \right)$ rotations), and the discrete symmetries of TBG -- $C_{2z}$, $T$, and $P$.  

We then consider the more general states $\ket{\Psi^{\nu_+, \nu_-}_{\nu}}$, alongside their $\mathrm{U} \left( 4 \right) \times \mathrm{U} \left( 4 \right)$ rotations, which were shown to be exact ground states of the interaction Hamiltonian in the chiral limit~\cite{LIA21,XIE21}. Working similarly in the chiral limit, our analysis is also exact. We find that $\rthree$ ordering \emph{can} arise for generic intervalley-coherent $\mathrm{U} \left( 4 \right) \times \mathrm{U} \left( 4 \right)$ rotations of the states $\ket{\Psi^{\nu_+ ,\nu_-}_{\nu}}$, provided that they are not $\Uncf$ rotations of the insulators $\ket{\Psi_{\nu}}$, defined in \cref{app:eqn:nonchiralGS}. The latter does not however constitute the only exception. Indeed, without being exhaustive, we show that there are possible choices (to be defined in \cref{app:sec:spFuncSym:add_examples}) for the states $\ket{\Psi^{\nu_+ ,\nu_-}_{\nu}}$ (different from $\ket{\Psi_{\nu}}$), which, when rotated according to a certain subgroup (to be defined in \cref{app:sec:spFuncSym:add_examples}) of $\mathrm{U} \left( 4 \right) \times \mathrm{U} \left( 4 \right)$ (different from the $\Uncf$ group), give rise to intervalley-coherent insulators \emph{without} any $\rthree$ ordering. On the other hand, we argue that such examples are fine-tuned, as there is no \emph{a priori} reason for restricting to these $\mathrm{U} \left( 4 \right) \times \mathrm{U} \left( 4 \right)$ subgroups (unlike the $\Uncf$ group which is obtained from $\mathrm{U} \left( 4 \right) \times \mathrm{U} \left( 4 \right)$ as a result of breaking the chiral symmetry). We end the discussion with an intuitive picture explaining the presence or absence of $\rthree$ ordering for the \emph{maximally spin-polarized} TBG insulators.

Finally, we investigate the effects of breaking the exact particle-hole symmetry on the absence of $\rthree$ symmetry-breaking in the  the STM patterns of the insulators $\ket{\Psi_{\nu}}$, defined in \cref{app:eqn:nonchiralGS}, together with their $\Uncf$ rotations.

In the forthcoming \cref{app:sec:results}, we also provide a series of numerical results illustrating the main conclusions of this \siSection{}. 
 
\subsection{The spectral function of $\ket{\Psi_{\nu}}$ and its $\Uncf$ rotations}\label{app:sec:spFuncSym:nonchiral}

In this section we consider the spectral function of the even-integer-filled insulators $\ket{\Psi_{\nu}}$ defined in \cref{app:eqn:nonchiralGS}, and their $\Uncf$ rotations. We will assume that the single-particle TBG Hamiltonian has an exact particle-hole symmetry by focusing on the case $\lambda = 0$ (the $\lambda = 1$ case will be discussed in detail in \cref{app:sec:spFuncSym:breakingPH}). 

Following the notation in \cref{app:sec:charge:specific}, we let 
\begin{equation}
	\label{app:eqn:ncf_rot_U}
	\hU = \exp \left( i \sum_{a,b} \theta^{ab} S^{ab}  \right)
	\quad \text{and} \quad 
	U = \exp \left( i \sum_{a,b} \theta^{ab} \left(s^{ab}\right)^{T}  \right)
\end{equation}
respectively denote a nonchiral-flat rotation operator and the corresponding eight-dimensional unitary matrix $U$ implementing it in the energy band basis. 
In \cref{app:eqn:ncf_rot_U}, $\theta^{ab} \in \mathbb{R}$ (for $a,b=0,x,y,z$) denote the angles parameterizing the $\Uncf$ rotation, $S^{ab}$ are the $\Uncf$ generators from \cref{app:eqn:generatorsu4}, while $s^{ab}$ are the corresponding matrix generators given in \cref{app:eqn:generatorsu4:matrix}. The TBG ground state we consider can be written as 
\begin{equation}
	\label{app:eqn:rot_ncf_state}
	\ket{\varphi} = \hU \ket{\Psi_{\nu}},
\end{equation}
where the unrotated state $\ket{\Psi_{\nu}}$ was defined in \cref{app:eqn:nonchiralGS}. Note that the states of the form in \cref{app:eqn:rot_ncf_state} form a subset of the more general states considered in \cref{app:eqn:chosen_ground_state}. In what follows, we will employ $a_{\eta,s}$ to specify which valley-spin flavors of $\ket{\Psi_{\nu}}$ are occupied, such that
\begin{equation}
	\label{app:eqn:ncf_occupations}
	a_{\eta,s} = \begin{cases}
		1, \quad &\text{if the valley-spin flavor $\left\lbrace \eta,s \right\rbrace$ is filled in $\ket{\Psi_{\nu}}$} \\
		0, \quad &\text{if the valley-spin flavor $\left\lbrace \eta,s \right\rbrace$ is empty in $\ket{\Psi_{\nu}}$} \\  
	\end{cases}.
\end{equation}
In the notation of \cref{app:eqn:chosen_ground_state_rho}, this implies that $\rho_{+1,\eta,s} = \rho_{-1,\eta,s} = a_{\eta,s}$. The corresponding projectors in the empty and filled bands of $\ket{\Psi_{\nu}}$ defined in \cref{app:eqn:def_proj_1,app:eqn:def_proj_2}, respectively read as
\begin{align}
	\Pi =& \zeta^0 \tau^0 s^0 -\frac{1}{4} \zeta^{0} \otimes \sum_{\eta} \left( \tau^0 + \eta \tau^z \right) \otimes 
	\left[
		a_{\eta,\uparrow} \left( s^0 + s^z \right) +
		a_{\eta,\downarrow} \left( s^0 - s^z \right) 
	\right] \label{app:eqn:ncf_projectors_+} \\
	\tilde{\Pi} =& \frac{1}{4} \zeta^{0} \otimes \sum_{\eta} \left( \tau^0 + \eta \tau^z \right) \otimes 
	\left[
		a_{\eta,\uparrow} \left( s^0 + s^z \right) +
		a_{\eta,\downarrow} \left( s^0 - s^z \right) 
	\right] \label{app:eqn:ncf_projectors_-}
\end{align}

Our goal is to obtain parameterizations of the spectral function matrix elements of the insulator $\ket{\varphi}$. To do so, we note that the spectral function matrices are obtained by tracing over the spin degrees of freedom of the matrices $\Upsilon \left( \vk,\omega \right)$ and $\tilde{\Upsilon} \left( \vk,\omega \right)$, defined in \cref{app:eqn:def_upsilons+,app:eqn:def_upsilons-}. By comparing \cref{app:eqn:def_upsilons+,app:eqn:def_upsilons-} with \cref{app:eqn:ch_exc_unrot_eigendecomp_+,app:eqn:ch_exc_unrot_eigendecomp_-}, respectively, we find that, up to a Dirac $\delta$-function being applied on the (non-zero) eigenvalues, the matrices $\Upsilon \left( \vk, \omega \right)$ and $\tilde{\Upsilon} \left( \vk, \omega \right)$ share identical eigenspectra with the matrices $\Pi U \left( R \left( \vk \right) \otimes s^0 \right) U^{\dagger} \Pi$ and $\Pi U \left( \tilde{R} \left( \vk \right) \otimes s^0 \right) U^{\dagger} \Pi$, respectively. Therefore, parameterized forms of the former can be inferred from the parameterizations of the latter. In what follows, we will focus on the electron excitations, and obtain the parameterization of $\Upsilon \left( \vk, \omega \right)$, which we then use to parameterize $\mathcal{M}^{+} \left( \omega \right)$. The hole contribution can be computed analogously, which allows us to discuss the spectral function of the insulator $\ket{\varphi}$. 

\subsubsection{Parameterized form of $\Upsilon \left( \vk, \omega \right)$ }\label{app:sec:spFuncSym:nonchiral:upsilon}
To obtain $\Pi U \left( R \left( \vk \right) \otimes s^0 \right) U^{\dagger} \Pi$, we first note that \cref{app:eqn:param_P_r_+} implies for the $\lambda = 0$ case that  
\begin{equation}
	R \left( \vk \right) \otimes s^0 = 
	\zeta^{0} \tau^{0} s^0 d_2 \left( \vk \right) + 
	\zeta^{x} \tau^{z} s^0 d_3 \left( \vk \right) + 
	\zeta^{z} \tau^{z} s^0 d_5 \left( \vk \right),
\end{equation}
and, because $\zeta^{0} \tau^{0} s^0 $, $\zeta^{x} \tau^{z} s^0$, and $\zeta^{z} \tau^{z} s^0 $ commute with \emph{all} the $\Uncf$ generators $s^{ab}$ from \cref{app:eqn:generatorsu4:matrix},
\begin{equation}
	\label{app:eqn:ncf_rot_invariance_1}
	U \left( R \left( \vk \right) \otimes s^0 \right) U^{\dagger} = R \left( \vk \right) \otimes s^0.
\end{equation}
We then project \cref{app:eqn:ncf_rot_invariance_1} using the projector $\Pi$ defined in \cref{app:eqn:ncf_projectors_+} to obtain 
\begin{align}
	&\Pi U \left( R \left( \vk \right) \otimes s^0 \right) U^{\dagger} \Pi = \Pi \left( R \left( \vk \right) \otimes s^0 \right) \Pi = \nonumber \\
	&\frac{1}{4} \zeta ^0 \tau ^0 s^0\left(4-a_{-,\downarrow }-a_{-,\uparrow }-a_{+,\downarrow }-a_{+,\uparrow }\right) d_2\left( \vec{k} \right)+\frac{1}{4} \zeta ^0 \tau ^0 s^z\left(-a_{-,\downarrow }+a_{-,\uparrow }-a_{+,\downarrow }+a_{+,\uparrow }\right) d_2\left( \vec{k} \right)\nonumber \\ 
 +&\frac{1}{4} \zeta ^0 \tau ^z s^0\left(-a_{-,\downarrow }-a_{-,\uparrow }+a_{+,\downarrow }+a_{+,\uparrow }\right) d_2\left( \vec{k} \right)+\frac{1}{4} \zeta ^0 \tau ^z s^z\left(-a_{-,\downarrow }+a_{-,\uparrow }+a_{+,\downarrow }-a_{+,\uparrow }\right) d_2\left( \vec{k} \right)\nonumber \\ 
 +&\frac{1}{4} \zeta ^x \tau ^0 s^0\left(-a_{-,\downarrow }-a_{-,\uparrow }+a_{+,\downarrow }+a_{+,\uparrow }\right) d_3\left( \vec{k} \right)+\frac{1}{4} \zeta ^x \tau ^0 s^z\left(-a_{-,\downarrow }+a_{-,\uparrow }+a_{+,\downarrow }-a_{+,\uparrow }\right) d_3\left( \vec{k} \right)\nonumber \\ 
 +&\frac{1}{4} \zeta ^x \tau ^z s^0\left(4-a_{-,\downarrow }-a_{-,\uparrow }-a_{+,\downarrow }-a_{+,\uparrow }\right) d_3\left( \vec{k} \right)+\frac{1}{4} \zeta ^x \tau ^z s^z\left(-a_{-,\downarrow }+a_{-,\uparrow }-a_{+,\downarrow }+a_{+,\uparrow }\right) d_3\left( \vec{k} \right)\nonumber \\ 
 +&\frac{1}{4} \zeta ^z \tau ^0 s^0\left(-a_{-,\downarrow }-a_{-,\uparrow }+a_{+,\downarrow }+a_{+,\uparrow }\right) d_5\left( \vec{k} \right)+\frac{1}{4} \zeta ^z \tau ^0 s^z\left(-a_{-,\downarrow }+a_{-,\uparrow }+a_{+,\downarrow }-a_{+,\uparrow }\right) d_5\left( \vec{k} \right)\nonumber \\ 
 +&\frac{1}{4} \zeta ^z \tau ^z s^0\left(4-a_{-,\downarrow }-a_{-,\uparrow }-a_{+,\downarrow }-a_{+,\uparrow }\right) d_5\left( \vec{k} \right)+\frac{1}{4} \zeta ^z \tau ^z s^z\left(-a_{-,\downarrow }+a_{-,\uparrow }-a_{+,\downarrow }+a_{+,\uparrow }\right) d_5\left( \vec{k} \right) 	, \label{app:eqn:ncf_rot_invariance_2}
\end{align}
where we have employed $a_{\eta,s}^2 = a_{\eta,s}$. Using \cref{app:eqn:ncf_rot_invariance_2}, as well as the parity of the function $d_i$ ($i=2,3,5$) from \cref{app:eqn:parity_d_c2z}, we can find \emph{all} the symmetries of the matrix $\Pi U \left( R \left( \vk \right) \otimes s^0 \right) U^{\dagger} \Pi$
\begin{equation}
	\begin{split}
		\left[\zeta^0 \tau^0 s^z, \Pi U \left( R \left( \vk \right) \otimes s^0 \right) U^{\dagger} \Pi \right] &= \left[\zeta^0 \tau^z s^0, \Pi U \left( R \left( \vk \right) \otimes s^0 \right) U^{\dagger} \Pi \right] = 0, \\ 
		\left( \zeta^y \tau^0 s^0 \right) \Pi U \left( R \left( \vk \right) \otimes s^0 \right) U^{\dagger} \Pi \left( \zeta^y \tau^0 s^0 \right) &= \Pi U \left( R \left( - \vk \right) \otimes s^0 \right) U^{\dagger} \Pi.
	\end{split}
\end{equation}
Because $\Pi U \left( R \left( \vk \right) \otimes s^0 \right) U^{\dagger} \Pi$ and $\Upsilon \left( \vk, \omega \right)$ share the same eigenvectors, as can be seen from \cref{app:eqn:ch_exc_unrot_eigendecomp_+,app:eqn:def_upsilons+}, the symmetries of the former also represent symmetries of the latter, thus implying that
\begin{equation}
	\label{app:eqn:symmetriesUpsilon}
	\begin{split}
		\left[\zeta^0 \tau^0 s^z, \Upsilon \left( \vk, \omega \right) \right] &= \left[\zeta^0 \tau^z s^0, \Upsilon \left( \vk, \omega \right) \right] = 0,  \\ 
		\left( \zeta^y \tau^0 s^0 \right) \Upsilon \left( \vk, \omega \right) \left( \zeta^y \tau^0 s^0 \right) &= \Upsilon \left( - \vk, \omega \right).
	\end{split}
\end{equation}
Additionally, since $\Pi U \left( R \left( \vk \right) \otimes s^0 \right) U^{\dagger} \Pi$ is real, its eigenstate are real, implying that $\Upsilon \left( \vk, \omega \right)$ is also real. Moreover, from its definition in \cref{app:eqn:def_upsilons+}, $\Upsilon \left( \vk, \omega \right)$ is manifestly Hermitian. Coupled with the symmetries from \cref{app:eqn:symmetriesUpsilon}, we obtain the decomposition of $\Upsilon \left( \vk, \omega \right)$ in the energy band, valley, and spin subspaces as
\begin{equation}
	\label{app:eqn:param_upsilon+}
	\Upsilon \left( \vk, \omega \right) = \sum_{\substack{a \in \left\lbrace 0, x, z \right\rbrace \\ b,c \in \left\lbrace 0,z \right\rbrace}} \zeta^a \tau^b s^c \beta_{abc} \left( \vk, \omega \right),
\end{equation}
where $\beta_{abc} \left(\vk, \omega\right)$ are real functions whose parity with respect to momentum inversion is given by
\begin{equation}
	\label{app:eqn:parity_beta_three_params+}
	\beta_{abc} \left(\vk, \omega\right) = \begin{cases}
		\beta_{abc} \left(-\vk, \omega \right), \quad &\text{if } a=0 \\
		-\beta_{abc} \left(-\vk, \omega \right), \quad &\text{if } a=x,z \\
	\end{cases},
	\quad \text{for} \quad a \in \left\lbrace 0, x, z \right\rbrace \text{ and } b,c \in \left\lbrace 0,z \right\rbrace,
\end{equation}
and which only depend on the occupation of the valley-spin flavors $a_{\eta,s}$ defined in \cref{app:eqn:ncf_occupations}, but \emph{not} on the $\hU$ rotation. 

\subsubsection{Parameterized form of $\mathcal{M}_{\varphi}^{+}\left(\omega\right)$ }\label{app:sec:spFuncSym:nonchiral:m}
As shown in \cref{app:eqn:spec_func_from_rMat_ups+}, to obtain the electron spectral function matrix from \cref{app:eqn:parity_beta_three_params+}, we need to perform one final $\Uncf$ transformation on $\Upsilon \left( \vk, \omega \right)$, which results in
\begin{equation}
	\label{app:eqn:param_upsilonU+}
	Y \left( \vk, \omega \right) = U^{\dagger} \Upsilon \left( \vk, \omega \right) U = \zeta^0 \tau^0 s^0 \beta_{000} \left( \vk, \omega \right) + \zeta^x \tau^z s^0 \beta_{xz0} \left( \vk, \omega \right) + \zeta^z \tau^z s^0 \beta_{zz0} \left( \vk, \omega \right) 	+ \sum_{j=1}^{3} N^{j} \beta_j \left( \vec{k}, \omega \right),
\end{equation}
where $\beta_i \left( \vk,\omega \right)$ are real functions ($i=1,2,3$) obeying
\begin{equation}
	\label{app:eqn:parity_beta+}
	\beta_1 \left( \vk,\omega \right) = \beta_1 \left( - \vk,\omega \right), \quad
	\beta_2 \left( \vk,\omega \right) = -\beta_2 \left( - \vk,\omega \right), \quad
	\beta_3 \left( \vk,\omega \right) = -\beta_3 \left( - \vk,\omega \right),
\end{equation}
but which we otherwise leave unspecified. In \cref{app:eqn:param_upsilonU+}, we have also introduced the matrices $N^{i}$ for $i=1,2,3$, which depend on the specific $\Uncf$ rotation $\hU$ and belong to the sets $N^{i} \in \mathcal{N}^i$, with 
\begin{equation}
	\label{app:eqn:ncf_invariant_sets}
	\begin{split}
	\mathcal{N}^{1} &= \mathcal{N}_{\Uncf},  \\
	\mathcal{N}^{2} &= \setDef*{ \sum_{a,b \in \left\lbrace 0,x,y,z \right\rbrace} \phi^{ab} t^{ab} }{\phi^{ab} \in \mathbb{C}, \phi^{00} = 0, 
		t^{0b} = \zeta^x \tau^z s^{b}, 
		t^{xb} = \zeta^z \tau^x s^{b}, 
		t^{yb} = \zeta^z \tau^y s^{b}, 
		t^{zb} = \zeta^x \tau^0 s^{b}},  \\
	\mathcal{N}^{3} &= \setDef*{ \sum_{a,b \in \left\lbrace 0,x,y,z \right\rbrace} \phi^{ab} t^{ab} }{\phi^{ab} \in \mathbb{C}, \phi^{00} = 0, 
		t^{0b} = \zeta^z \tau^z s^{b}, 
		t^{xb} = \zeta^x \tau^x s^{b}, 
		t^{yb} = \zeta^x \tau^y s^{b}, 
		t^{zb} = \zeta^z \tau^0 s^{b} },
	\end{split}	
\end{equation}
where $\mathcal{N}_{\Uncf}$ was defined in \cref{app:eqn:def_N_u4}. To see how \cref{app:eqn:param_upsilonU+} was obtained from  \cref{app:eqn:param_upsilon+}, we start by observing that the sets from \cref{app:eqn:ncf_invariant_sets} share one important property: they form invariant spaces under the $\Uncf$ group. This can be seen by remarking that for any matrix $T\in \mathcal{N}^{i}$, $\left[ T, s^{ab} \right] \in \mathcal{N}^{i}$, and so, from \cref{app:eqn:ncf_rot_U}, $U^{\dagger} T U \in \mathcal{N}^{i}$, for any $i=1,2,3$. We then group all the matrices in the expansion from \cref{app:eqn:param_upsilon+} according to how they transform under a $\Uncf$ rotation. The first three terms of \cref{app:eqn:param_upsilonU+} are obtained by noting that the matrices $\zeta^0 \tau^0 s^0$, $\zeta^x \tau^z s^0$, $\zeta^z \tau^z s^0$ are invariant under any transformation $U$, as they commute with all the nonchrial-flat $\Uncf$ generators $s^{ab}$. The other matrices from \cref{app:eqn:param_upsilon+} can be grouped as follows
\begin{equation}
	\begin{split}
		\zeta^0 \tau^0 s^z, \zeta^0 \tau^z s^0, \zeta^0 \tau^z s^z &\in \mathcal{N}^{1}, \\
		\zeta^x \tau^0 s^0, \zeta^x \tau^x s^z, \zeta^x \tau^z s^z &\in \mathcal{N}^{2}, \\
		\zeta^z \tau^0 s^0, \zeta^z \tau^0 s^z, \zeta^z \tau^z s^z &\in \mathcal{N}^{3}.
	\end{split}
\end{equation}
Within each one of the three groups, the parity with respect to momentum inversion of the real functions $\beta_{abc} \left( \vk, \omega \right)$ multiplying each matrix is the same, thus leading to the expansion from \cref{app:eqn:param_upsilonU+}.

The parameterization of the electron spectral function matrix is obtained from \cref{app:eqn:param_upsilonU+}, by employing \cref{app:eqn:spec_func_from_rMat_ups+}
\begin{equation}
	\label{app:eqn:expansion_spec_func_ncf+}
	\left[\mathcal{M}_{\varphi}^{+}\left(\omega\right) \right]_{\vk n \eta,\vk' n' \eta'} = \delta_{\vk,\vk'} \left[ \zeta^0 \tau^0 \beta_{000} \left( \vk, \omega \right) + \zeta^x \tau^z \beta_{xz0} \left( \vk, \omega \right) + \zeta^z \tau^z \beta_{zz0} \left( \vk, \omega \right) + \sum_{j=1}^{3} \left(\tr_s N^{j} \right) \beta_j \left( \vec{k}, \omega \right)	 \right]_{n \eta, n' \eta'},
\end{equation}
where $\tr_s$ denotes the trace over the spin degrees of freedom, and the spin-traced matrices obey
\begin{equation}
	\begin{split}
		\tr_s N^{1} &\in \setDef*{ \phi_x \zeta^y \tau^x + \phi_y \zeta^y \tau^y + \phi_z \zeta^0 \tau^z }{\phi_x, \phi_y, \phi_z \in \mathbb{C} }, \\
		\tr_s N^{2} &\in \setDef*{ \phi_x \zeta^z \tau^x + \phi_y \zeta^z \tau^y + \phi_z \zeta^x \tau^0 }{\phi_x, \phi_y, \phi_z \in \mathbb{C} }, \\
		\tr_s N^{3} &\in \setDef*{ \phi_x \zeta^x \tau^x + \phi_y \zeta^x \tau^y + \phi_z \zeta^z \tau^0 }{\phi_x, \phi_y, \phi_z \in \mathbb{C} }. 
	\end{split}
\end{equation}

\subsubsection{Parameterized form of $\mathcal{M}_{\varphi}^{S\pm}$ }\label{app:sec:spFuncSym:nonchiral:mSymmetric}

As seen in \cref{app:eqn:expansion_spec_func_ncf+}, for general non-chiral flat $\Uncf$ rotations, $\mathcal{M}_{\varphi}^{+}\left(\omega\right)$ contains off-diagonal elements in the valley subspace. Following the discussion surrounding \cref{app:eqn:clustering_sp_factor}, one might conclude that this signals the presence of $\rthree$ ordering at the level of the SLG lattice. In fact, introducing intervalley coherence by means of a $\Uncf$ rotation in the ground state $\ket{\varphi}$ results to intervalley-coherent charge-one excitations, according to the discussion from \cref{app:sec:charge:specific}. In turn, intervalley-coherent charge-one excitations lead to scattering between the two $K$ points of the SLG Brillouin Zone and generically give rise to $\rthree$ ordering  at the level of the SLG lattice~\cite{GUT16,BAO21,LIU22}.

However, the above reasoning neglects the discrete symmetries of TBG, which impose several restrictions on both the spatial factor and the spectral function matrices. It was shown in \cref{app:sec:spectralf:sp_fact_sym:T} that owing to the time-reversal symmetry of $\hat{H}_0$, the components of $\left[ \mathcal{M}^{\pm} \left(\omega\right) \right]_{\vk n \eta, \vk n' \eta' }$ which are anti-symmetric with respect to the transformation $\left[ \vk n \eta, \vk n' \eta' \right] \to \left[ -\vk n' (-\eta'), -\vk n (-\eta) \right]$ vanish upon contracting with the spatial factor tensor. Following \cref{app:eqn:spec_func_enbas_temp_qspace_symmetrized}, we find that instead of studying $\mathcal{M}_{\varphi}^{\pm}\left(\omega\right)$ directly, one should instead focus on the symmetrized tensors $\mathcal{M}_{\varphi}^{S \pm}\left(\omega\right)$ defined in \cref{app:eqn:symmetrized_m_tensor}.

To compute $\mathcal{M}_{\varphi}^{S +}$, we first note that in \cref{app:eqn:expansion_spec_func_ncf+}, 
\begin{alignat}{4}
		& \left[ \zeta^0 \tau^0 \right]_{n \eta, n' \eta'} \quad &&=  \left[ \zeta^0 \tau^0 \right]_{n' (-\eta'), n (-\eta)}, \quad && \left[ \tr_s N^1 \right]_{n \eta, n' \eta'} \quad &&=  - \left[ \tr_s N^1 \right]_{n' (-\eta'), n (-\eta)}, \nonumber \\
		& \left[ \zeta^x \tau^z \right]_{n \eta, n' \eta'} \quad &&= -\left[ \zeta^x \tau^z \right]_{n' (-\eta'), n (-\eta)}, \quad && \left[ \tr_s N^2 \right]_{n \eta, n' \eta'} \quad &&=  \left[ \tr_s N^2 \right]_{n' (-\eta'), n (-\eta)}, \\
		& \left[ \zeta^z \tau^z \right]_{n \eta, n' \eta'} \quad &&= -\left[ \zeta^z \tau^z \right]_{n' (-\eta'), n (-\eta)}, \quad && \left[ \tr_s N^3 \right]_{n \eta, n' \eta'} \quad &&=  \left[ \tr_s N^3 \right]_{n' (-\eta'), n (-\eta)}, \nonumber
\end{alignat}
which coupled with \cref{app:eqn:parity_beta_three_params+,app:eqn:parity_beta+,app:eqn:symmetrized_m_tensor} leads to 
\begin{equation}
	\label{app:eqn:expansion_spec_func_ncf_sym+}
	\left[\mathcal{M}_{\varphi}^{S +}\left(\omega\right) \right]_{\vk n \eta,\vk n' \eta'} = \delta_{\vk,\vk'} \left[ \zeta^0 \tau^0 \beta_{000} \left( \vk, \omega \right) + \zeta^x \tau^z \beta_{xz0} \left( \vk, \omega \right) + \zeta^z \tau^z \beta_{zz0} \left( \vk, \omega \right) \right]_{n \eta, n' \eta'}.
\end{equation}
In an analogous fashion, it can be shown that the hole contribution to the spectral function matrix elements is 
\begin{equation}
	\label{app:eqn:expansion_spec_func_ncf_sym-}
	\left[\mathcal{M}_{\varphi}^{S -}\left(\omega\right) \right]_{\vk n \eta,\vk n' \eta'} = \delta_{\vk,\vk'} \left[ \zeta^0 \tau^0 \tilde{\beta}_{000} \left( \vk, \omega \right) + \zeta^x \tau^z \tilde{\beta}_{xz0} \left( \vk, \omega \right) + \zeta^z \tau^z \tilde{\beta}_{zz0} \left( \vk, \omega \right) \right]_{n \eta, n' \eta'},
\end{equation}
where the real functions $\tilde{\beta}_{abc} \left( \vk, \omega \right)$ have the same parity as the functions $\beta_{abc} \left( \vk, \omega \right)$ (for $abc = 000, xz0, zz0$) with respect to momentum inversion and, similarly, only depend on the occupation of the valley-spin flavors, $a_{\eta,s}$, but \emph{not} on the exact $\Uncf$ rotation. 

The parameterizations in \cref{app:eqn:expansion_spec_func_ncf_sym+,app:eqn:expansion_spec_func_ncf_sym-} imply that the spectral function matrices are diagonal in the valley-subspace. As a result of this and of \cref{app:eqn:spec_func_enbas_temp_qspace_symmetrized}, for the insulators $\ket{\Psi_{\nu}}$ defined in \cref{app:eqn:nonchiralGS}, and their $\Uncf$ rotations, there is no $\rthree$ ordering, \emph{despite} the possible presence of intervalley coherence. Equally surprising is the fact that the spectral function of the insulators of the form $\hU \ket{\Psi_{\nu}}$ is \emph{independent} on the particular $\Uncf$ rotation considered.  

The counter-intuitive absence of $\rthree$ ordering even in the presence of intervalley coherence in \emph{any} $\Uncf$ rotation of the states $\ket{\Psi_{\nu}}$ warrants a more careful explanation. At the level of the spatial factor, the presence of $T$ symmetry imposes \cref{app:eqn:prop_B_R3_cancelation}, which requires that we consider only the symmetrized spectral function matrices defined \cref{app:eqn:symmetrized_m_tensor}. For the spectral function matrices themselves, the vanishing of the valley-off-diagonal terms in the parameterizations from \cref{app:eqn:expansion_spec_func_ncf_sym+,app:eqn:expansion_spec_func_ncf_sym-} \emph{even} in the case of valley-coherent insulators $\hU\ket{\Psi_{\nu}}$ relies on three main peculiarities of this specific ground state and the TBG Hamiltonian: 
\begin{enumerate}
	\item Occupying \emph{both} Chern bands in each of the filled spin-valley sector of the unrotated state $\ket{\Psi_{\nu}}$.
	\item Restricting to the non-chiral flat $\Uncf$ rotations of the $\ket{\Psi_{\nu}}$ state, as opposed to the more general $\mathrm{U} \left( 4 \right) \times \mathrm{U} \left( 4 \right)$ rotations. 
	\item The presence of $C_{2z}$, $T$, and $P$ symmetries of the TBG single-particle Hamiltonian which constrain the forms of the charge-one excitation matrices, as derived in \cref{app:eqn:param_P_r_+,app:eqn:param_P_r_-}. In turn, the parameterizations of the charge-one excitation matrices directly determine the parameterization of the spectral function matrices.
\end{enumerate}
In \cref{app:sec:spFuncSym:chiral}, we will investigate the effects of relaxing the first two assumptions in the first chiral-flat limit~\cite{TAR19,BER21a}. Additionally, in \cref{app:sec:spFuncSym:breakingPH} we will relax the third assumption and explore the spectral function of $\ket{\Psi_{\nu}}$ as defined in \cref{app:eqn:nonchiralGS} and some of its $\Uncf$ rotations using the full-fledged TBG interaction Hamiltonian away from the first chiral-flat limit and without assuming \emph{exact} particle-hole symmetry (\ie{} focusing on the $\lambda = 1$ case). 

\subsection{The spectral function of $\ket{\Psi^{\nu_+, \nu_-}_{\nu}}$ and its $\mathrm{U} \left( 4 \right) \times \mathrm{U} \left( 4 \right)$ rotations}\label{app:sec:spFuncSym:chiral}

In \cref{app:sec:spFuncSym:nonchiral}, we discussed the spectral function of the integer-filled states $\ket{\Psi_{\nu}}$ and their nonchiral-flat $\Uncf$ rotations. The corresponding spectral function is independent on the $\Uncf$ rotation in question and no $\rthree$ ordering emerges at the level of the SLG lattice, even in the presence of intervalley coherence. To better understand the reasons behind the \emph{exact} cancellation from \cref{app:sec:spFuncSym:nonchiral} leading to the absence of $\rthree$ ordering, we here derive the minimal set of conditions necessary (but not sufficient) for the \emph{emergence} of $\rthree$ ordering in the STM signal in a given TBG insulator. 

For simplicity, we will focus on the chiral limit~\cite{TAR19,BER21a}. This is because any $\rthree$ ordering emerging in the chiral limit for a given TBG insulator will necessarily persist when the chiral symmetry is broken. In other words, the absence of $\rthree$ ordering for a given insulator is reliant on the \emph{presence} rather than the \emph{absence} of certain symmetries, so breaking chiral symmetry in itself cannot remove the $\rthree$ ordering. On the other hand, as shown in \cref{app:eqn:param_C_r_+,app:eqn:param_C_r_-}, in the chiral limit, the charge-one excitation matrices are proportional to identity, which vastly simplifies all algebraic manipulations.

We consider the general ground state $\ket{\varphi}$ defined in \cref{app:eqn:chosen_ground_state} for a generic $\mathrm{U} \left( 4 \right) \times \mathrm{U} \left( 4 \right)$ rotation $\hU$, where
\begin{equation}
	\label{app:eqn:cf_rot_U}
	\hU = \exp \left( i \sum_{a,b} \theta_+^{ab} S_+^{ab} + \theta_-^{ab} S_-^{ab}  \right)
	\quad \text{and} \quad 
	U = \exp \left( i \sum_{a,b} \theta_+^{ab} \left(s_+^{ab} \right)^{T} + \theta_-^{ab} \left(s_-^{ab} \right)^{T}  \right)
\end{equation}
respectively denote the rotation operator and the corresponding eight-dimensional unitary matrix $U$ implementing it in the energy band basis. 
In \cref{app:eqn:cf_rot_U}, $\theta_{\pm}^{ab} \in \mathbb{R}$ (for $a,b=0,x,y,z$) denote the angles parameterizing the $\mathrm{U} \left( 4 \right) \times \mathrm{U} \left( 4 \right)$ rotation, while $S_{\pm}^{ab}$ and $s_{\pm}^{ab}$ are the corresponding generators in the operator and matrix form, as given in \cref{app:eqn:generatorsu4u4,app:eqn:generatorsu4u4:matrix}, respectively. In order to specify the filled Chern-band, valley, and spin flavors in the unrotated ground state $\ket{\Psi^{\nu_+,\nu_-}_{\nu}}$, we will find it useful to define
\begin{equation}
	\label{app:eqn:a_b_occupation}
	a_{\eta,s} = \frac{1}{2} \sum_{e_Y} \rho_{e_Y,\eta,s}, \qquad
	b_{\eta,s} = \frac{1}{2} \sum_{e_Y} e_Y \rho_{e_Y,\eta,s},
\end{equation}
where $\rho_{e_Y,\eta,s} = 1,0$ was introduced in \cref{app:eqn:chosen_ground_state_rho}. In \cref{app:eqn:a_b_occupation}, $a_{\eta,s}$ specifies whether the valley-spin flavor $\left\lbrace \eta,s \right\rbrace$ has zero ($a_{\eta,s}=0$), one ($a_{\eta,s}=1/2$), or two ($a_{\eta,s}=1$) filled Chern bands in $\ket{\Psi^{\nu_+,\nu_-}_{\nu}}$. In addition, $2b_{\eta,s} = -1,0,1$ denotes the total Chern number of the occupied bands in the valley-spin flavor $\left\lbrace \eta,s \right\rbrace$. As an example, the insulator $\ket{\Psi^{1,2}_{-1}} = \prod_{\vk} \left( \hat{d}^\dagger_{\vk,-1,-,\uparrow} \hat{d}^\dagger_{\vk,+1,-,\uparrow} \hat{d}^\dagger_{\vk,-1,-,\downarrow} \right) \ket{0}$ (also depicted in \cref{app:fig:rotatedbases} of \cref{app:sec:charge:specific}) corresponds to $a_{-,\uparrow} = 1$, $a_{-,\downarrow} = 1/2$, $a_{+,\uparrow /\downarrow} = 0$, and $b_{-,\downarrow} = -1/2$, $b_{-,\uparrow} = b_{+,\uparrow /\downarrow} = 0$. This notation is consistent with the one introduced in \cref{app:eqn:ncf_occupations}: the states considered in \cref{app:sec:spFuncSym:chiral} correspond to $a_{\eta,s} = 0,1$ and $b_{\eta,s} = 0$. 

\subsubsection{Parameterized forms of $\Upsilon \left( \vk, \omega \right)$ and $\tilde{\Upsilon} \left( \vk, \omega \right)$ }\label{app:sec:spFuncSym:chiral:upsilon}

We are interested in obtaining parameterizatizations of the spectral function matrices $\mathcal{M}_{\varphi}^{\pm} \left( \omega \right)$. Focusing on the electron contribution, $\mathcal{M}^{+}_{\varphi} \left( \omega \right)$ can be written in terms of the matrix $\Upsilon \left( \vk, \omega \right)$ introduced in \cref{app:eqn:spec_func_from_rMat_ups+}, according to \cref{app:eqn:def_upsilons+}. In turn, we have argued in \cref{app:sec:spFuncSym:nonchiral} that $\Upsilon \left( \vk, \omega \right)$ shares identical eigenvectors with the matrix $\Pi U \left( R \left( \vk \right) \otimes s^0 \right) U^{\dagger} \Pi$, as can be seen by comparing \cref{app:eqn:ch_exc_unrot_eigendecomp_+,app:eqn:def_upsilons+}. However, in the first chiral limit, the charge-one excitation matrices are proportional to identity, implying from \cref{app:eqn:param_C_r_+} that the electron excitations are degenerate, having energies given by $E^p_{\vk} = d_{2} \left( \vk \right)$, for any $1 \leq p \leq 4-\nu$, which allows us derive an even stronger relation 
\begin{equation}
	\label{app:eqn:param_upsilon_chiral+}
	\Upsilon \left( \vk, \omega \right) = \frac{\delta \left[ \omega - d_{2} \left( \vk \right) \right]}{d_{2} \left( \vk \right)} \Pi U \left( R \left( \vk \right) \otimes s^0 \right) U^{\dagger} \Pi = \Pi \beta \left(\vk, \omega \right),
\end{equation}
where $\beta \left(\vk, \omega \right) = \delta \left[ \omega - d_{2} \left( \vk \right) \right]$ is a real function which is even with respect to momentum inversion. In the last equality of \cref{app:eqn:param_upsilon_chiral+}, we have used the fact that $R \left( \vk \right) \otimes s^0$ is proportional to the identity matrix and that the projector $\Pi$ is idempotent (\ie{} $\Pi^{2} = \Pi$). A similar relation can be written for hole contribution
\begin{equation}
	\label{app:eqn:param_upsilon_chiral-}
	\tilde{\Upsilon} \left( \vk, \omega \right) = \frac{\delta \left[ \omega - \tilde{d}_{2} \left( \vk \right) \right]}{\tilde{d}_{2} \left( \vk \right)} \tilde{\Pi} U \left( \tilde{R} \left( \vk \right) \otimes s^0 \right) U^{\dagger} \tilde{\Pi} = \tilde{\Pi} \tilde{\beta} \left(\vk, \omega \right),
\end{equation}
where $\tilde{\beta} \left(\vk, \omega \right) = \delta \left[ \omega - \tilde{d}_{2} \left( \vk \right) \right]$ is also a real function which is even with respect to momentum inversion.

Using the notation in \cref{app:eqn:a_b_occupation}, we find that the two projector matrices are given by 
\begin{align}
	\Pi=&\frac{1}{4} \zeta ^0 \tau ^0 s^0\left(4-a_{-,\downarrow }-a_{-,\uparrow }-a_{+,\downarrow }-a_{+,\uparrow }\right)+\frac{1}{4} \zeta ^0 \tau ^0 s^z\left(a_{-,\downarrow }-a_{-,\uparrow }+a_{+,\downarrow }-a_{+,\uparrow }\right)\nonumber \\ 
 +&\frac{1}{4} \zeta ^0 \tau ^z s^0\left(a_{-,\downarrow }+a_{-,\uparrow }-a_{+,\downarrow }-a_{+,\uparrow }\right)+\frac{1}{4} \zeta ^0 \tau ^z s^z\left(-a_{-,\downarrow }+a_{-,\uparrow }+a_{+,\downarrow }-a_{+,\uparrow }\right)\nonumber \\ 
 +&\frac{1}{4} \zeta ^y \tau ^0 s^0\left(b_{-,\downarrow }+b_{-,\uparrow }+b_{+,\downarrow }+b_{+,\uparrow }\right)+\frac{1}{4} \zeta ^y \tau ^0 s^z\left(-b_{-,\downarrow }+b_{-,\uparrow }-b_{+,\downarrow }+b_{+,\uparrow }\right)\nonumber \\ 
 +&\frac{1}{4} \zeta ^y \tau ^z s^0\left(-b_{-,\downarrow }-b_{-,\uparrow }+b_{+,\downarrow }+b_{+,\uparrow }\right)+\frac{1}{4} \zeta ^y \tau ^z s^z\left(b_{-,\downarrow }-b_{-,\uparrow }-b_{+,\downarrow }+b_{+,\uparrow }\right) 	, \label{app:eqn:cf_projectors_+}\\
	\tilde{\Pi}=&\zeta^{0} \tau^{0} s^{0} - \Pi, \label{app:eqn:cf_projectors_-}
\end{align}
or, equivalently, in index notation by
\begin{align}
	\Pi_{n \eta s, n' \eta' s'} &= \delta_{\eta,\eta'} \delta_{s,s'} \left\lbrace \left[ \zeta^{0} \right]_{n n'} \left( 1 - a_{\eta,s} \right) - \left[ \zeta^{y} \right]_{n n'} b_{\eta,s} \right\rbrace, \\
	\tilde{\Pi}_{n \eta s, n' \eta' s'} &= \delta_{\eta,\eta'} \delta_{s,s'} \left\lbrace \left[ \zeta^{0} \right]_{n n'}  a_{\eta,s} + \left[ \zeta^{y} \right]_{n n'} b_{\eta,s} \right\rbrace.
\end{align}

\subsubsection{Parameterized form of $\mathcal{M}_{\varphi}^{S\pm}$ }\label{app:sec:spFuncSym:chiral:mSymmetric}

To obtain the spectral function matrices, we perform one additional $\mathrm{U} \left( 4 \right) \times \mathrm{U} \left( 4 \right)$ rotation on  \cref{app:eqn:param_upsilon_chiral+,app:eqn:param_upsilon_chiral-}, which affords 
\begin{align}
	Y \left( \vk, \omega \right) = U^{\dagger} \Upsilon \left( \vk, \omega \right) U = \beta \left(\vk, \omega \right) U^{\dagger} \Pi U &= \beta \left(\vk, \omega \right) \left[ \frac{1}{4} \zeta^0 \tau^0 s^0 \left(4- \sum_{\eta,s} a_{\eta,s} \right) + \frac{1}{4} \zeta^y \tau^0 s^0 \left( \sum_{\eta,s} b_{\eta,s} \right) + N \right], \label{app:eqn:param_cf_upsilonU+} \\
	\tilde{Y} \left( \vk, \omega \right) = U^{\dagger} \tilde{\Upsilon} \left( \vk, \omega \right) U = \tilde{\beta} \left(\vk, \omega \right) U^{\dagger} \tilde{\Pi} U &= \tilde{\beta} \left(\vk, \omega \right) \left[ \frac{1}{4} \zeta^0 \tau^0 s^0 \left( \sum_{\eta,s} a_{\eta,s} \right) - \frac{1}{4} \zeta^y \tau^0 s^0 \left( \sum_{\eta,s} b_{\eta,s} \right) + \tilde{N} \right], \label{app:eqn:param_cf_upsilonU-}
\end{align}
where the matrices $N$ and $\tilde{N}$ depend on the exact rotation $\hU$, as well as on the parameters $a_{\eta,s}$ and $b_{\eta,s}$. In what follows, we will work out the parameterization of $N$ and $\tilde{N}$ in different cases and derive properties of the corresponding spectral function matrices.

\Cref{app:eqn:param_cf_upsilonU+,app:eqn:param_cf_upsilonU-} were derived in analogy with \cref{app:eqn:param_upsilonU+}. Focusing on the electron contribution (the hole contribution follows in exactly the same manner), the matrices $\zeta^0 \tau^0 s^0$ and $\zeta^y \tau^0 s^0$ commute with all the generators of the $\mathrm{U} \left( 4 \right) \times \mathrm{U} \left( 4 \right)$ group from \cref{app:eqn:generatorsu4u4:matrix}: the corresponding terms in the expansion in \cref{app:eqn:cf_projectors_+} remain invariant under the transformation $U$. The remaining matrices in the expansion of $\Pi$ are \emph{not} invariant under general $\mathrm{U} \left( 4 \right) \times \mathrm{U} \left( 4 \right)$ transformations, but belong to the set $\mathcal{N}_{\mathrm{U} \left( 4 \right) \times \mathrm{U} \left( 4 \right)}$ defined in \cref{app:eqn:def_N_u4u4}. Since $\mathcal{N}_{\mathrm{U} \left( 4 \right) \times \mathrm{U} \left( 4 \right)}$ \emph{is} invariant (as a set) under $\mathrm{U} \left( 4 \right) \times \mathrm{U} \left( 4 \right)$ rotations (\ie{} for any $T \in \mathcal{N}_{\mathrm{U} \left( 4 \right) \times \mathrm{U} \left( 4 \right)}$, $U^{\dagger} T U \in \mathcal{N}_{\mathrm{U} \left( 4 \right) \times \mathrm{U} \left( 4 \right)}$), it follows that $N,\tilde{N} \in \mathcal{N}_{\mathrm{U} \left( 4 \right) \times \mathrm{U} \left( 4 \right)}$. 

Remembering that the functions $\beta \left( \vk, \omega \right)$ and $\tilde{\beta} \left( \vk, \omega \right)$ are even with respect to momentum inversion, we can derive the parameterized form of the symmetrized spectral function from \cref{app:eqn:symmetrized_m_tensor,app:eqn:param_cf_upsilonU+,app:eqn:param_cf_upsilonU-}
\begin{align}
	\left[\mathcal{M}_{\varphi}^{S +}\left(\omega\right) \right]_{\vk n \eta,\vk n' \eta'} &= \delta_{\vk,\vk'} \frac{1}{2} \left( \frac{4 - \nu}{2}  \delta_{n,n'} \delta_{\eta,\eta'} + \left[ \tr_s N \right]_{n\eta,n'\eta'} + \left[ \tr_s N \right]_{n'(-\eta'),n(-\eta)} \right) \beta \left( \vk, \omega \right), \label{app:eqn:expansion_spec_func_cf_sym+} \\
	\left[\mathcal{M}_{\varphi}^{S -}\left(\omega\right) \right]_{\vk n \eta,\vk n' \eta'} &= \delta_{\vk,\vk'} \frac{1}{2} \left( \frac{4 + \nu}{2} \delta_{n,n'} \delta_{\eta,\eta'} + \left[ \tr_s \tilde{N} \right]_{n\eta,n'\eta'} + \left[ \tr_s \tilde{N} \right]_{n'(-\eta'),n(-\eta)} \right) \tilde{\beta} \left( \vk, \omega \right), \label{app:eqn:expansion_spec_func_cf_sym-}
\end{align}
where, as in \cref{app:eqn:expansion_spec_func_ncf+}, $\tr_s$ denotes the trace over the spin degree of freedom.
Depending on the rotation $\hU$ and the parameters $a_{\eta,s}$ and $b_{\eta,s}$, more restrictive statements can be made about the matrices $N$ and $\tilde{N}$. Without being exhaustive, we can identify two distinct cases with particular relevance to the existence of $\rthree$ ordering:
\begin{enumerate}
\item~\label{app:cases:cf_sf:case_1}$ \hU \in \Uncf $ and $ b_{\eta,s} = 0 $ for all $\eta=\pm $ and $ s=\uparrow,\downarrow$, where $\Uncf$ denotes the nonchiral-flat subgroup of $\mathrm{U} \left( 4 \right)\times \mathrm{U} \left( 4 \right)$ whose generators are given in \cref{app:eqn:generatorsu4}.
	
	If all $b_{\eta,s}$ are zero, then all matrices from \cref{app:eqn:cf_projectors_+,app:eqn:cf_projectors_-} which are not proportional to $\zeta^0 \tau^0 s^0$ or $\zeta^{y} \tau^0 s^0$ belong to the set $\mathcal{N}_{\Uncf}$, defined in \cref{app:eqn:def_N_u4}. Because $\mathcal{N}_{\Uncf}$ (as a set) is invariant under any $\Uncf$ rotations, it follows that $N,\tilde{N} \in \mathcal{N}_{\Uncf}$, implying that $\tr_s N, \tr_s \tilde{N} \in \tr_s \mathcal{N}_{\Uncf}$, where 
	\begin{equation}
		\tr_s \mathcal{N}_{\Uncf} = \setDef*{ \sum_{i=1}^{3} \phi_i t_i }{\phi_i \in \mathbb{C}, t_1 = \zeta^y \tau^x, t_2 = \zeta^y \tau^y, t_3 = \zeta^0 \tau^z }.
	\end{equation}	 
	In this case, one obtains that 
	\begin{equation}
		\label{app:eqn:cf_cancel_u4}
		\left[ \tr_s N \right]_{n\eta,n'\eta'} + \left[ \tr_s N \right]_{n'(-\eta'),n(-\eta)} = \left[ \tr_s \tilde{N} \right]_{n\eta,n'\eta'} + \left[ \tr_s \tilde{N} \right]_{n'(-\eta'),n(-\eta)} = 0,
	\end{equation}
	which implies from \cref{app:eqn:expansion_spec_func_cf_sym+,app:eqn:expansion_spec_func_cf_sym-} that the spectral function matrices are diagonal in the valley and band indices and independent on the rotation $\hU$. Therefore, no $\rthree$ ordering emerges in this case and the corresponding spectral functions are independent on the particular rotation $\hU$. This result can also be seen as a consequence of the general case away from the chiral limit, derived in \cref{app:sec:spFuncSym:nonchiral}.
\item~\label{app:cases:cf_sf:case_2}Generic $\hU \in \left[ \mathrm{U} \left( 4 \right) \times \mathrm{U} \left( 4 \right) \right] \setminus \Uncf$ rotations and/or $b_{\eta,s} \neq 0$ for some $\eta$ and $s$. 
	
	For any choices of $\ket{\varphi}$, which do not otherwise belong to \cref{app:cases:cf_sf:case_1}, we have $N,\tilde{N} \in \mathcal{N}_{\mathrm{U} \left( 4 \right) \times \mathrm{U} \left( 4 \right)}$. Therefore, we find that $\tr_s N, \tr_s \tilde{N} \in \tr_s \mathcal{N}_{\mathrm{U} \left( 4 \right) \times \mathrm{U} \left( 4 \right)}$, where  
	\begin{equation}
		\tr_s \mathcal{N}_{\mathrm{U} \left( 4 \right) \times \mathrm{U} \left( 4 \right)} = \setDef*{ \sum_{i=1}^{7} \phi_i t_i }{\phi_i \in \mathbb{C}, t_1 = \zeta^y \tau^x, t_2 = \zeta^y \tau^y, t_3 = \zeta^0 \tau^z, t_4 = \zeta^y \tau^0, t_5 = \zeta^0 \tau^x, t_6 = \zeta^0 \tau^y, t_7 = \zeta^y \tau^z},
	\end{equation}
	implying that for generic rotations $\hU$ we expect to have
	\begin{equation}
		\label{app:eqn:cf_noncancel_general}
		\left[ \tr_s N \right]_{n\eta,n'\eta'} + \left[ \tr_s N \right]_{n'(-\eta'),n(-\eta)} \neq 0 
		\quad \text{and} \quad 
		\left[ \tr_s \tilde{N} \right]_{n\eta,n'\eta'} + \left[ \tr_s \tilde{N} \right]_{n'(-\eta'),n(-\eta)} \neq 0.
	\end{equation}
	A simple example illustrating this case corresponds to an intervalley-coherent $\mathcal{C} = 1$ insulator at $\nu = -3$ 
	\begin{equation}
		\ket{\varphi} = \prod_{\vk} \left( \hat{d}^\dagger_{\vk,+1,+,\uparrow} \cos \alpha - \hat{d}^\dagger_{\vk,+1,-,\uparrow} \sin \alpha \right) \ket{0},
	\end{equation}
	which can be obtained by setting $\theta_{+}^{y0} = \alpha$ and all other angles $\theta^{ab}_{\pm} = 0$ in \cref{app:eqn:cf_rot_U}, as well as $b_{+,\uparrow} = a_{+,\uparrow} = 1/2$ and all other $b_{\eta,s} = a_{\eta,s} = 0$. From \cref{app:eqn:param_cf_upsilonU+,app:eqn:param_cf_upsilonU-}, we find that 
	\begin{equation}
		\begin{split}
		\frac{1}{2} \left(	\left[ \tr_s N \right]_{n\eta,n'\eta'} + \left[ \tr_s N \right]_{n'(-\eta'),n(-\eta)} \right) &= \frac{1}{4} \left[ \left( \zeta^{0} \tau^{x} \right) \sin 2\alpha + \left( \zeta^{y} \tau^{z} \right) \cos 2\alpha \right]_{n\eta,n'\eta'} \\
		\frac{1}{2} \left(	\left[ \tr_s \tilde{N} \right]_{n\eta,n'\eta'} + \left[ \tr_s \tilde{N} \right]_{n'(-\eta'),n(-\eta)} \right) &= - \frac{1}{4} \left[ \left( \zeta^{0} \tau^{x} \right) \sin 2\alpha + \left( \zeta^{y} \tau^{z} \right) \cos 2\alpha \right]_{n\eta,n'\eta'}
		\end{split}
	\end{equation}	 
	which manifestly obeys \cref{app:eqn:cf_noncancel_general}. 
		
	In this case, the symmetrized spectral function matrices from \cref{app:eqn:expansion_spec_func_cf_sym+,app:eqn:expansion_spec_func_cf_sym-} \emph{can} generically have non-vanishing off-diagonal valley components (as can be seen in the above example for general values of $\alpha$), leading to the emergence of $\rthree$ ordering in the corresponding STM patterns. This does \emph{not} however imply that for any intervalley-coherent insulator not contained in \cref{app:cases:cf_sf:case_1}, $\rthree$ ordering emerges, but rather that there is no symmetry precluding it. In the following \cref{app:sec:spFuncSym:add_examples}, we present one additional example of intervalley-coherent insulators \emph{without} $\rthree$ ordering, which do not belong to \cref{app:cases:cf_sf:case_1}. 
\end{enumerate}	

The analysis from this section complements the conclusions of \cref{app:sec:spFuncSym:nonchiral}: for an intervalley-coherent $\mathrm{U} \left( 4 \right) \times \mathrm{U} \left( 4 \right)$ rotation of the state $\ket{\Psi^{\nu_+, \nu_-}_{\nu}}$ defined in \cref{app:eqn:chiralGS}, which is not part of the set of insulators discussed in \cref{app:sec:spFuncSym:nonchiral}, an $\rthree$ pattern \emph{can} generically arise in the STM signal. Moreover, as $\rthree$ ordering is a symmetry-breaking effect, we expect the conditions for its emergence to hold even as we move away from the chiral limit (\ie{} the vanishing of $\rthree$ ordering relies on the \emph{presence} rather than the \emph{absence} of certain symmetries). 

On the other hand, the insulators discussed in \cref{app:sec:spFuncSym:nonchiral} are \emph{not} the only example of intervalley-coherent insulators $\ket{\varphi}$ of the form in \cref{app:eqn:chosen_ground_state} without $\rthree$ symmetry-breaking. Far from being exhaustive, in \cref{app:sec:spFuncSym:add_examples}, we will show by direct construction that a different submanifold of intervalley-coherent TBG ground states without $\rthree$ ordering exists. While deriving all the necessary \emph{and sufficient} conditions for the emergence of $\rthree$ symmetry-breaking is beyond the scope of this work, we will nevertheless build a complete picture describing the presence or absence of $\rthree$ ordering in the case of \emph{maximally spin-polarized} TBG ground states of the form in \cref{app:eqn:chosen_ground_state}.

	\subsection{Additional examples of intervalley-coherent ground states without $\rthree$ ordering }\label{app:sec:spFuncSym:add_examples}
In \cref{app:sec:spFuncSym:chiral}, we argued that an intervalley-coherent $\mathrm{U} \left( 4 \right) \times \mathrm{U} \left( 4 \right)$ rotation of the state $\ket{\Psi^{\nu_+, \nu_-}_{\nu}}$ defined in \cref{app:eqn:chiralGS} does \emph{generically} show $\rthree$ ordering, unless the rotation is part of the $\Uncf$ group, and the state $\ket{\Psi^{\nu_+, \nu_-}_{\nu}}$ has only fully filled and fully empty valley-spin flavors. In this section, we construct a different submanifold of TBG ground states of the form in \cref{app:eqn:chosen_ground_state} that are intervalley-coherent, but do not exhibit $\rthree$ symmetry breaking. Coupled with the analysis in \cref{app:sec:spFuncSym:nonchiral}, this will enable us discuss the $\rthree$ symmetry-breaking in the \emph{maximally spin-polarized} insulators of the form in \cref{app:eqn:chosen_ground_state}.

	\subsubsection{$\mathcal{C} = 0$ intervalley-coherent insulators with opposite valley polarization per Chern sector}\label{app:sec:spFuncSym:add_examples:example}
We start by defining $\susv$ to be a subgroup of $\mathrm{U} \left( 4 \right) \times \mathrm{U} \left( 4 \right)$ corresponding to independent $\mathrm{SU} \left( 2 \right)$ rotations in both the valley (V) and spin (S) subspaces. The $\susv$ group is generated by the six operators
\begin{equation}
	\label{app:eqn:generator_susv}
		\begin{split}
			S_{\mathrm{V}}^{a} &= \sum_{\vk}\sum_{\eta, s, \eta', s', n} \left[ \tau^0 s^a  \right]_{\eta s, \eta' s'} \hat{c}^\dagger_{\vk,n,\eta,s} \hat{c}_{\vk,n,\eta',s'}, \quad \text{for} \quad a=x,y,z, \\
			S_{\mathrm{S}}^{a} &= \sum_{\vk}\sum_{\eta,s, \eta', s', n} \left[ \tau^a s^0  \right]_{\eta s, \eta' s'} \hat{c}^\dagger_{\vk,n,\eta,s} \hat{c}_{\vk,n,\eta',s'}, \quad \text{for} \quad a=x,y,z,
		\end{split}
\end{equation}
which are diagonal in the band subspace, and therefore act \emph{identically} within the two Chern sectors. Unlike the $\Uncf$ group, within the $\susv$ group, the valley rotations generated by $S_{\mathrm{V}}^{a}$ rotate the two Chern-band operators in the same direction. 
	
Using the notation from \cref{app:eqn:a_b_occupation}, we now focus on the case 
\begin{equation}
	\label{app:eqn:add_example_b_condition}
	b_{\eta,s} = -\sum_{s'} M_{ss'} b_{-\eta,s'}, \quad \text{for all } \eta=\pm \text{ and } s=\uparrow,\downarrow
\end{equation}
where we take either $M = s^0$ or $M = s^x$. Depending on the matrix $M$, this case corresponds to occupying the Chern bands of the unrotated state $\ket{\Psi^{\nu_+,\nu_-}_{\nu}}$ two at a time, with \emph{opposite} Chern numbers, \emph{opposite} valley indices and \emph{identical} ($M=s^0$) or \emph{opposite} ($M=s^x$) spin projections along the $\hat{\vec{z}}$ axis. To obtain the rotated state $\ket{\varphi}$ defined in \cref{app:eqn:chosen_ground_state}, we perform two successive $\mathrm{SU} \left( 2 \right)$ rotations in the valley (V) and spin (S) subspaces by letting $\hU = \huS \huV = \huV \huS$, where
\begin{equation}
	\label{app:eqn:susv_rot_hU}
	\huV = \exp \left( i \sum_{a \in \left\lbrace x,y,z \right\rbrace} \theta_{\mathrm{V}}^{a} S_{\mathrm{V}}^{a}  \right),
	\quad \text{and} \quad
	\huS = \exp \left( i \sum_{a \in \left\lbrace x,y,z \right\rbrace} \theta_{\mathrm{S}}^{a} S_{\mathrm{S}}^{a}  \right),
\end{equation}
with $\theta^{a}_{\mathrm{V}},\theta^{a}_{\mathrm{S}} \in \mathbb{R}$ denoting the angles parameterizing the two $\mathrm{SU} \left( 2 \right)$ rotations, and $S_{\mathrm{V}}$ and $S_{\mathrm{S}}$ being the $\susv$ generators introduced in \cref{app:eqn:generator_susv}. Within the energy-band basis, the transformation $\hU$ is implemented by the eight-dimensional unitary matrix $U = \uS \uV = \uV  \uS$, with 
\begin{equation}
	\label{app:eqn:susv_rot_U}
	\uV^{T} = \exp \left( i \sum_{a \in \left\lbrace x,y,z \right\rbrace} \theta_{\mathrm{V}}^{a} \zeta^{0} \tau^{a} s^{0}  \right),
	\quad \text{and} \quad
	\uS^{T} = \exp \left( i \sum_{a \in \left\lbrace x,y,z \right\rbrace} \theta_{\mathrm{S}}^{a} \zeta^{0} \tau^{0} s^{a}  \right),
\end{equation}
corresponding to the two individual $\mathrm{SU} \left( 2 \right)$ rotations in the valley and spin subspaces. As the transformation $\huV$ rotates the two Chern sectors identically and the bands of the \emph{unrotated} state $\ket{\Psi^{\nu_+, \nu_-}_{\nu}}$ are occupied two at a time within opposite Chern sectors and opposite valleys, it follows that in the state $\ket{\varphi}$, any bands with opposite Chern numbers have exactly opposite valley polarizations.
	
We will now prove that no $\rthree$ ordering emerges in the STM signal of $\ket{\varphi}$, despite it being generically intervalley-coherent.	We first note that for the particular choice of occupying the bands of $\ket{\Psi^{\nu_+,\nu_-}_{\nu}}$, the projectors from \cref{app:eqn:cf_projectors_+,app:eqn:cf_projectors_-} obey 
\begin{equation}
	\label{app:eqn:susv_prop_proj}
	\left( \zeta^{0} \tau^{x} M \right) \Pi^{*}  \left( \zeta^{0} \tau^{x} M \right)^{\dagger} = \Pi
	\quad \text{and} \quad 
	\left( \zeta^{0} \tau^{x} M \right) \tilde{\Pi}^{*} \left( \zeta^{0} \tau^{x} M \right)^{\dagger} = \tilde{\Pi},
\end{equation}
where the matrix $M$ was defined in the text surrounding \cref{app:eqn:add_example_b_condition}. 
In addition, it can checked that for any $\susv$ rotation as defined in \cref{app:eqn:susv_rot_hU}, we have that
\begin{equation}
	\label{app:eqn:susv_prop_rot}
	\left( \zeta^{0} \tau^{z} s^{0} \right) \uV \left( \zeta^{0} \tau^{z} s^{0} \right)^{\dagger} = \left( \zeta^{0} \tau^{x} M \right) U^{*}_{\mathrm{V}} \left( \zeta^{0} \tau^{x} M \right)^{\dagger}.
\end{equation}
Focusing on the electron contribution to the spectral function matrix and letting $Q \equiv \left( \zeta^{0} \tau^{0} M \right) \uS \left( \zeta^{0} \tau^{0} M \right)$, we can show from \cref{app:eqn:susv_prop_proj,app:eqn:susv_prop_rot} that the projector matrix from \cref{app:eqn:cf_projectors_+} obeys 
\begin{align}
	\left( \zeta^{0} \tau^{x} M \right) \left( U^{\dagger} \Pi U \right)^{*} \left( \zeta^{0} \tau^{x} M \right)^{\dagger} &= Q^{T} \left( \zeta^{0} \tau^{x} M \right) \uV^{T} \Pi^{*} \uV^{*} \left( \zeta^{0} \tau^{x} M \right)^{\dagger} Q^{*} \nonumber \\
	&= Q^{T} \left( \zeta^{0} \tau^{x} M \right) \uV^{T} \left( \zeta^{0} \tau^{x} M \right)^{\dagger} \Pi \left( \zeta^{0} \tau^{x} M \right) \uV^{*} \left( \zeta^{0} \tau^{x} M \right)^{\dagger} Q^{*} \nonumber \\
	&= Q^{T} \left( \zeta^{0} \tau^{z} s^{0} \right) \uV^{\dagger} \left( \zeta^{0} \tau^{z} s^{0} \right)^{\dagger} \Pi \left( \zeta^{0} \tau^{z} s^{0} \right) \uV \left( \zeta^{0} \tau^{z} s^{0} \right)^{\dagger} Q^{*} \nonumber \\
	&= \left( \zeta^{0} \tau^{z} s^{0} \right) Q^{T} \uS \left( U^{\dagger}  \Pi U \right) \uS^{\dagger} Q^{*} \left( \zeta^{0} \tau^{z} s^{0} \right)^{\dagger}. \label{app:eqn:susv_long_eq}
\end{align}
In deriving \cref{app:eqn:susv_long_eq}, we have also used $M^2 = s^0$, $\left[ \tau^{x}, Q \right] = \left[ \uS, \uV \right] = 0$. Additionally, using the parametrization of the electron excitation matrix from \cref{app:eqn:param_r_+}, it is straightforward to show that 
\begin{equation}
	\left( \zeta^{0} \tau^{x} M \right)  \left( R \left( \vk \right) \otimes s^0 \right)^{*} \left( \zeta^{0} \tau^{x} M \right)^{\dagger} =  \left( R \left( -\vk \right) \otimes s^0 \right). \label{app:eqn:susv_r_prop}
\end{equation}
Combining \cref{app:eqn:susv_long_eq,app:eqn:susv_r_prop}, we can derive a relation for the electron excitations wave functions
\begin{equation}
	\label{app:eqn:susv_r_sym}
	\left( \zeta^{0} \tau^{x} M \right) \left[ U^{\dagger} \Pi U \left( R \left( \vk \right) \otimes s^0 \right) U^{\dagger} \Pi U \right]^{*} \left( \zeta^{0} \tau^{x} M \right)^{\dagger} = 
	\left( \zeta^{0} \tau^{z} s^{0} \right) Q^{T} \uS \left[ U^{\dagger} \Pi U \left( R \left( - \vk \right) \otimes s^0 \right) U^{\dagger} \Pi U \right] \uS^{\dagger} Q^{*} \left( \zeta^{0} \tau^{z} s^{0} \right)^{\dagger}.
\end{equation}
As discussed at the end of \cref{app:sec:computingME:integer}, the matrix $Y \left( \vk, \omega \right)$ defined in \cref{app:eqn:def_ys+} and $U^{\dagger} \Pi U \left( R \left( \vk \right) \otimes s^0 \right) U^{\dagger} \Pi U$, whose spectral decomposition was given in \cref{app:eqn:ch_exc_eigendecomp_+}, share identical eigendecompositions, up to a rescaling of the eigenvalues. As such, $Y \left( \vk, \omega \right)$ satisfies a similar relation to \cref{app:eqn:susv_r_sym}, namely 
\begin{equation}
	\label{app:eqn:susv_Y_sym}
	\left( \zeta^{0} \tau^{x} M \right) \left[ Y \left( \vk, \omega \right) \right]^{*} \left( \zeta^{0} \tau^{x} M \right)^{\dagger} = 
	\left( \zeta^{0} \tau^{z} s^{0} \right) Q^{T} \uS \left[ Y \left( - \vk, \omega \right) \right] \uS^{\dagger} Q^{*} \left( \zeta^{0} \tau^{z} s^{0} \right)^{\dagger}
\end{equation}
Tracing over the spin degree of freedom in \cref{app:eqn:susv_Y_sym} and using the Hermiticity of $Y \left( \vk, \omega \right)$, we find that
\begin{equation}
	\left[ \tr_s Y\left(- \vk, \omega \right) \right]_{n' (-\eta'),n (-\eta)} = 
	\left[ \left( \zeta^{0} \tau^{z} \right) \tr_s Y\left(\vk, \omega \right) \left( \zeta^{0} \tau^{z} \right)^{\dagger} \right]_{n \eta, n' \eta'} = 
	\eta \eta' \left[\tr_s Y\left(\vk, \omega \right) \right]_{n \eta, n' \eta'}.
\end{equation}
At the same time, using \cref{app:eqn:symmetrized_m_tensor,app:eqn:spec_func_from_rMat_y+}, we can find the symmetrized spectral function matrix corresponding to electron excitations to be
\begin{equation}
	\label{app:eqn:susv_m_+}
	\left[\mathcal{M}_{\varphi}^{S +}\left(\omega\right) \right]_{\vk n \eta,\vk n' \eta'} = \frac{1}{2} \delta_{\vk,\vk'} \left(1 + \eta \eta' \right) \left[ \tr_s Y\left(\vk, \omega \right) \right]_{n \eta, n' \eta'}.
\end{equation}
In analogous manner, the hole symmetrized spectral function matrix is given by 
\begin{equation}
	\label{app:eqn:susv_m_-}
	\left[\mathcal{M}_{\varphi}^{S -}\left(\omega\right) \right]_{\vk n \eta,\vk n' \eta'} = \frac{1}{2} \delta_{\vk,\vk'} \left(1 + \eta \eta' \right) \left[ \tr_s \tilde{Y}\left(\vk, \omega \right) \right]_{n \eta, n' \eta'}.
\end{equation}	
Inspecting \cref{app:eqn:susv_m_+,app:eqn:susv_m_-} reveals that the symmetrized spectral function matrices are diagonal in valley subspace, thus proving that no $\rthree$ ordering emerges for the insulator $\ket{\varphi}$. 
	
The insulators $ \hU \ket{\Psi^{\nu_+, \nu_-}_{\nu}}$ with $\hU$ restricted to $\susv$ and the occupancy of the Chern bands of $\ket{\Psi^{\nu_+, \nu_-}_{\nu}}$ given in \cref{app:eqn:susv_rot_hU} constitute another example of intervalley-coherent states without $\rthree$ ordering, in addition to the set of ground states discussed already in \cref{app:sec:spFuncSym:nonchiral}. The key difference between the two cases lies in the symmetries enforcing the absence of $\rthree$ ordering. In \cref{app:sec:spFuncSym:nonchiral}, we proved that the vanishing of $\rthree$ ordering is a direct consequence of restricting to the insulators with fully occupied or fully empty spin-valley flavors and only considering nonchiral $\Uncf$ rotations thereof, rather than the more general $\mathrm{U} \left( 4 \right) \times \mathrm{U} \left( 4 \right)$ transformations. This particular submanifold of ground states over the more general states given by \cref{app:eqn:chiralGS} becomes energetically favored - in the absence of chiral symmetry - in comparison to the latter at even fillings~\cite{LIA21,XIE21}. On the other hand, the absence of $\rthree$ ordering for the insulators discussed in this section is \emph{fine tuned}: there is no \emph{a priori} reason for restricting to the $\susv$ subgroup of $\mathrm{U} \left( 4 \right) \times \mathrm{U} \left( 4 \right)$ or to the unrotated states described by \cref{app:eqn:add_example_b_condition}. Nevertheless, coupled with the conclusions of \cref{app:sec:spFuncSym:nonchiral}, this example allows us to provide an intuitive picture explaining the vanishing of $\rthree$ ordering in \emph{maximally spin-polarized} TBG insulators of the form in \cref{app:eqn:chiralGS}.

\subsubsection{An intuitive picture for the maximally spin-polarized insulators}\label{app:sec:spFuncSym:add_examples:intuitive}

\begin{figure}[!t]
	\captionsetup[subfloat]{farskip=0pt}\sbox\nsubbox{
		\resizebox{\textwidth}{!}
		{\includegraphics[height=6cm]{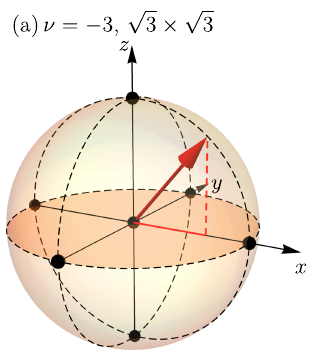}\includegraphics[height=6cm]{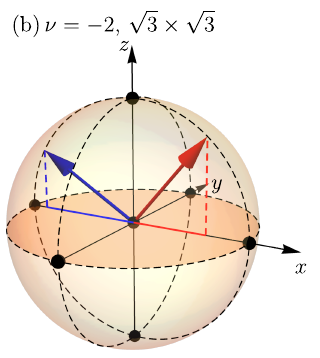}\includegraphics[height=6cm]{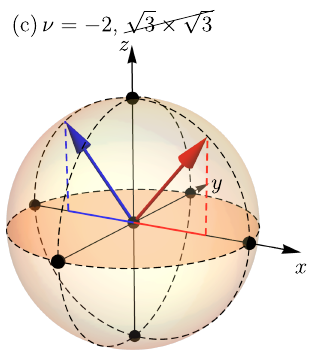}\includegraphics[height=6cm]{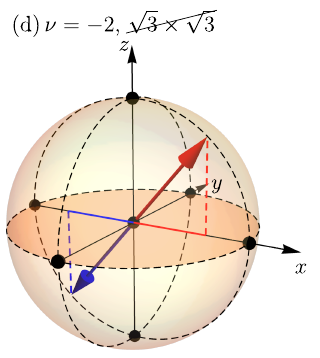}}}\setlength{\nsubht}{\ht\nsubbox}\centering\subfloat{\label{app:fig:blochMaxSpinPolarized:a}\includegraphics[height=\nsubht]{6a.pdf}}\subfloat{\label{app:fig:blochMaxSpinPolarized:b}\includegraphics[height=\nsubht]{6b.pdf}}\subfloat{\label{app:fig:blochMaxSpinPolarized:c}\includegraphics[height=\nsubht]{6c.pdf}}\subfloat{\label{app:fig:blochMaxSpinPolarized:d}\includegraphics[height=\nsubht]{6d.pdf}}
	\caption{The presence of $\rthree$ symmetry-breaking and the valley polarization Bloch sphere for the maximally spin-polarized TBG ground states. We show how the valley polarization of the $\mathcal{C} = +1$ (red) and $\mathcal{C} = -1$ (blue) bands for the maximally spin-polarized TBG insulators at $\nu = -3, -2$ determines the presence of $\rthree$ ordering. With only one Chern band filled [(a)], the intervalley-coherent, maximally spin-polarized $\nu = -3$ ground state will harbor $\rthree$ symmetry breaking. For the $\nu = -2$ and $\mathcal{C} = 0$ insulators [(b)-(d)], intervalley coherence does lead to the emergence of $\rthree$ ordering [(b)], unless the polarization vectors of the two Chern bands have opposite projections in the $xy$ plane [(c) and (d)]. The insulator in (c) corresponds to a $\Uncf$ rotation of a TBG ground state with a single fully filled valley-spin flavor, and thus, as shown in \cref{app:sec:spFuncSym:nonchiral}, does not display $\rthree$ ordering. The absence of $\rthree$ ordering in the insulator from (d) is due to the opposite valley polarizations of the two Chern bands, as shown in \cref{app:sec:spFuncSym:add_examples:example}. The vanishing of $\rthree$ ordering in (c) and (d) can be understood by noting that in both cases the $xy$-projections of the valley polarizations of the two Chern band exactly cancel.}
	\label{app:fig:blochMaxSpinPolarized}
\end{figure}

Intervalley-coherence is a necessary, but insufficient condition for the emergence of $\rthree$ patterns in the STM signals of the TBG insulators constructed in Ref.~\cite{LIA21}. This was illustrated in \cref{app:sec:spFuncSym:nonchiral,app:sec:spFuncSym:add_examples:example}, where we identified two distinct ground-state submanifolds that are manifestly intervalley-coherent, but do not harbor any $\rthree$ ordering. Due to the nature of the $\mathrm{U} \left( 4 \right) \times \mathrm{U} \left( 4 \right)$ group, whose rotations mix the valley and spin degrees of freedom, deriving \emph{all} the necessary \emph{and} sufficient conditions for the absence of $\rthree$ symmetry-breaking for the valley-coherent TBG insulators from \cref{app:eqn:chosen_ground_state} is beyond the scope of the current work. Nevertheless, an intuitive explanation for the absence of $\rthree$ ordering in some valley-coherent TBG insulators can be obtained by restricting to the subset of TBG ground states from \cref{app:eqn:chosen_ground_state}, which are \emph{maximally spin-polarized}.

The complete set of maximally spin-polarized (along the $\hat{\vec{z}}$ direction, without loss of generality) states $\ket{\varphi}$, as defined in \cref{app:eqn:chosen_ground_state}, can be obtained by letting $\hU$ be a product of $\mathrm{SU} \left( 2 \right)$ valley transformations in each of the two Chern sectors [\ie{} letting $\theta^{x0}_{\pm}$, $\theta^{y0}_{\pm}$, $\theta^{z0}_{\pm}$ be the only nonzero angles in \cref{app:eqn:cf_rot_U}], and occupying the spin-$\uparrow$  bands of the unrotated state $\ket{\Psi^{\nu_+, \nu_-}_{\nu}}$ first. For $\nu = -4$, TBG ground state is not intervalley-coherent, so no $\rthree$ ordering is expected. At $\nu = -3$, the only possibility is to occupy a single Chern band, and it was already shown in \cref{app:cases:cf_sf:case_2} of \cref{app:sec:spFuncSym:chiral:mSymmetric} that the resulting insulator does show $\rthree$ symmetry-breaking if it is intervalley-coherent (see \cref{app:fig:blochMaxSpinPolarized:a}). For $\nu = -2$, there are two choices. Filling two Chern bands with the same Chern number leads to a valley-polarized state, which does not display any $\rthree$ ordering. On the other hand filling two Chern bands with opposite Chern numbers will generically lead to the emergence of $\rthree$ ordering, unless $\ket{\varphi} = \hU \ket{\Psi_{\nu}}$, with $\hU \in \Uncf$ (\ie{} $\ket{\varphi}$ is one of the insulators discussed in \cref{app:sec:spFuncSym:nonchiral}), or $\ket{\varphi}$ has opposite valley polarizations per Chern sector (as discussed in \cref{app:sec:spFuncSym:add_examples:example}). As shown in \cref{app:fig:blochMaxSpinPolarized:b,app:fig:blochMaxSpinPolarized:c,app:fig:blochMaxSpinPolarized:d}, the maximally spin-polarized $\nu = -2$ and $\mathcal{C} = 0$ insulators $\ket{\varphi}$ generally harbor $\rthree$ symmetry-breaking, unless valley polarization projections of the two Chern bands in the $xy$ plane exactly cancel. 

At $\nu = -1$, the situation is similar to that at $\nu = -3$ for the maximally spin-polarized insulators. As both valleys are filled, one Chern sector will be valley-polarized. Assuming that the other Chern band is intervalley-coherent, $\rthree$ ordering will generally emerge for the insulator $\ket{\varphi}$. For $\nu = 0$, both valleys are fully filled for a given spin sector, thus precluding any $\rthree$ symmetry breaking. For any positive filling $\nu$, the conditions for the presence or absence of $\rthree$ ordering are identical to the ones at $-\nu$, as a consequence of the many-body charge-conjugation symmetry of TBG~\cite{BER21a}.

\subsection{Breaking the particle-hole symmetry}\label{app:sec:spFuncSym:breakingPH}

\begin{figure}[!t]
	\captionsetup[subfloat]{farskip=0pt}\sbox\nsubbox{
		\resizebox{0.75\textwidth}{!}
		{\includegraphics[height=6cm]{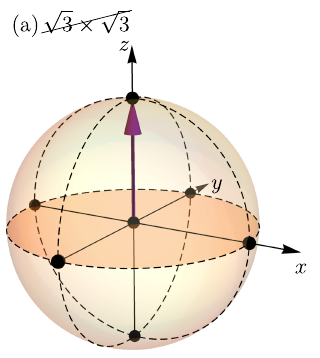}\includegraphics[height=6cm]{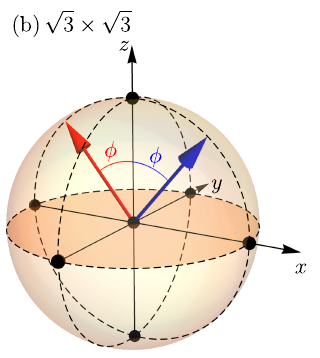}\includegraphics[height=6cm]{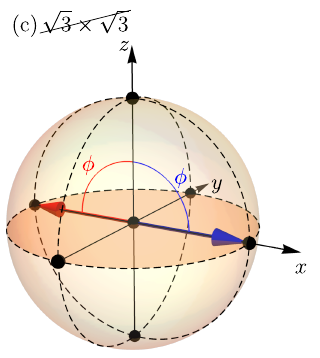}}}\setlength{\nsubht}{\ht\nsubbox}\centering\subfloat{\label{app:fig:phBreakBloch:a}\includegraphics[height=\nsubht]{7a.pdf}}\subfloat{\label{app:fig:phBreakBloch:b}\includegraphics[height=\nsubht]{7b.pdf}}\subfloat{\label{app:fig:phBreakBloch:c}\includegraphics[height=\nsubht]{7c.pdf}}
	\caption{Emergence of $\rthree$ ordering in the absence of exact particle-hole symmetry ($\lambda = 1$). We consider the $\nu = -2$ insulator from \cref{app:eqn:phBreak_state} for different rotation $\hU$ given by \cref{app:eqn:phBreak_rot_U_mat}, and show how the valley polarization of the $\mathcal{C} = +1$ (red) and $\mathcal{C} = -1$ (blue) bands determines the presence of $\rthree$ symmetry-breaking. The $\Uncf$ rotation from \cref{app:eqn:phBreak_rot_U_mat} rotates the polarizations of the two Chern bands by an angle $\phi$ in opposite directions within the $xz$ plane. As expected, the valley polarized case shown in (a) does not exhibit $\rthree$ ordering. Unlike the case with exact particle-hole symmetry ($\lambda = 0$), a general values of $\phi$ leads to the emergence of $\rthree$ ordering [as in (b)], unless the Chern bands are polarized in the $xy$ plane [as in (c)], when the $\rthree$ ordering vanishes even without \emph{exact} particle-hole symmetry.}
	\label{app:fig:phBreakBloch}
\end{figure}

In \cref{app:sec:spFuncSym:nonchiral}, we showed that the states $\hU \ket{\Psi_{\nu}}$, where $\hU \in \Uncf$ and the unrotated state $\ket{\Psi_{\nu}}$ defined in \cref{app:eqn:nonchiralGS} contains only fully filled or fully empty valley-spin flavors, do not harbor any $\rthree$ ordering \emph{even} in the presence of intervalley coherence. Among other reasons, we argued that the absence of $\rthree$ symmetry breaking is a direct result of the discrete symmetries of TBG: $C_{2z}$, $T$, and $P$. While the particle-hole symmetry is an excellent \emph{approximate} symmetry of TBG~\cite{SON19,SON21}, it is nevertheless not \emph{exact}. In this section, we investigate the effects of relaxing the assumption of \emph{exact} particle-hole symmetry by considering the unapproximated case $\lambda = 1$. 

For simplicity, we will focus on the insulator 
\begin{equation}
	\label{app:eqn:phBreak_state}
	\ket{\varphi} = \hU \ket{\Psi_{-2}} = \hU \left( \prod_{\vk} \hat{c}^\dagger_{\vk,+1,+,\uparrow} \hat{c}^\dagger_{\vk,-1,+,\uparrow} \right) \ket{0}
\end{equation} 
where the rotation $\hU$ is given in terms of the $\Uncf$ generators from \cref{app:eqn:generatorsu4} by 
\begin{equation}
	\label{app:eqn:phBreak_rot_U_mat}
	\hU = \exp\left(\frac{i\phi S^{y0}}{2}\right).
\end{equation}
The transformation $\hU$ rotates the valley polarization of the Chern sector $e_Y$, by an angle $e_Y \phi$ within the $xz$ plane (see \cref{app:fig:phBreakBloch}). Although $\hU$ is not the most general $\Uncf$ rotation, it is nevertheless sufficient for understanding the implications of breaking the exact particle-hole symmetry on the conclusions of \cref{app:sec:spFuncSym:nonchiral}. 

As in \cref{app:sec:spFuncSym:chiral}, to parameterize the spectral function matrix, we can leverage the fact that the matrix $Y \left( \vk, \omega \right)$ defined in \cref{app:eqn:def_ys+} and $U^{\dagger} \Pi U \left( R \left( \vk \right) \otimes s^0 \right) U^{\dagger} \Pi U$, whose spectral decomposition was given in \cref{app:eqn:ch_exc_eigendecomp_+}, share identical eigendecompositions, up to a rescaling of the eigenvalues (and similarly for the hole contribution). The matrices $U^{\dagger} \Pi U \left( R \left( \vk \right) \otimes s^0 \right) U^{\dagger} \Pi U$ and $U^{\dagger} \tilde{\Pi} U \left( \tilde{R} \left( \vk \right) \otimes s^0 \right) U^{\dagger} \tilde{\Pi} U$ can be obtained directly for $\ket{\varphi}$, by using the parameterizations of the charge-one excitation matrices in the $\lambda = 1$ case from \cref{app:eqn:param_r_+,app:eqn:param_r_-}
\begin{align}
	&U^{\dagger} \Pi U \left( R \left( \vec{k} \right) \otimes s^0 \right) U^{\dagger} \Pi U -=\frac{1}{4}\zeta ^0 \tau ^0 s^0 \left(-\cos (\phi ) d_1\left(\vec{k}\right)+3 d_2\left(\vec{k}\right)\right)\nonumber \\ 
-&\frac{1}{4}\zeta ^0 \tau ^0 s^z \left(-\cos (\phi ) d_1\left(\vec{k}\right)-d_2\left(\vec{k}\right)\right) -\frac{1}{8}\zeta ^0 \tau ^z s^0 \left((5+\cos (2 \phi )) d_1\left(\vec{k}\right)-2 \cos (\phi ) d_2\left(\vec{k}\right)\right)\nonumber \\ 
-&\frac{1}{8}\zeta ^0 \tau ^z s^z \left((-3+\cos (2 \phi )) d_1\left(\vec{k}\right)-2 \cos (\phi ) d_2\left(\vec{k}\right)\right)\nonumber \\ 
-&\frac{1}{8}\zeta ^x \tau ^0 s^0 \left(-2 \cos (\phi ) d_3\left(\vec{k}\right)+(5+\cos (2 \phi )) d_4\left(\vec{k}\right)\right)\nonumber \\ 
-&\frac{1}{8}\zeta ^x \tau ^0 s^z \left(-2 \cos (\phi ) d_3\left(\vec{k}\right)+(-3+\cos (2 \phi )) d_4\left(\vec{k}\right)\right) -\frac{1}{4}\zeta ^x \tau ^y s^0 \sin (\phi ) \left(d_5\left(\vec{k}\right)-\cos (\phi ) d_6\left(\vec{k}\right)\right)\nonumber \\ 
-&\frac{1}{4}\zeta ^x \tau ^y s^z \sin (\phi ) \left(d_5\left(\vec{k}\right)-\cos (\phi ) d_6\left(\vec{k}\right)\right) -\frac{1}{4}\zeta ^x \tau ^z s^0 \left(3 d_3\left(\vec{k}\right)-\cos (\phi ) d_4\left(\vec{k}\right)\right)\nonumber \\ 
-&\frac{1}{4}\zeta ^x \tau ^z s^z \left(-d_3\left(\vec{k}\right)-\cos (\phi ) d_4\left(\vec{k}\right)\right) -\frac{1}{4}\zeta ^y \tau ^x s^0 \sin (\phi ) \left(-\cos (\phi ) d_1\left(\vec{k}\right)+d_2\left(\vec{k}\right)\right)\nonumber \\ 
-&\frac{1}{4}\zeta ^y \tau ^x s^z \sin (\phi ) \left(-\cos (\phi ) d_1\left(\vec{k}\right)+d_2\left(\vec{k}\right)\right) -\frac{1}{8}\zeta ^z \tau ^0 s^0 \left(-2 \cos (\phi ) d_5\left(\vec{k}\right)+(5+\cos (2 \phi )) d_6\left(\vec{k}\right)\right)\nonumber \\ 
-&\frac{1}{8}\zeta ^z \tau ^0 s^z \left(-2 \cos (\phi ) d_5\left(\vec{k}\right)+(-3+\cos (2 \phi )) d_6\left(\vec{k}\right)\right) -\frac{1}{4}\zeta ^z \tau ^y s^0 \sin (\phi ) \left(-d_3\left(\vec{k}\right)+\cos (\phi ) d_4\left(\vec{k}\right)\right)\nonumber \\ 
-&\frac{1}{4}\zeta ^z \tau ^y s^z \sin (\phi ) \left(-d_3\left(\vec{k}\right)+\cos (\phi ) d_4\left(\vec{k}\right)\right) -\frac{1}{4}\zeta ^z \tau ^z s^0 \left(3 d_5\left(\vec{k}\right)-\cos (\phi ) d_6\left(\vec{k}\right)\right)\nonumber \\ 
-&\frac{1}{4}\zeta ^z \tau ^z s^z \left(-d_5\left(\vec{k}\right)-\cos (\phi ) d_6\left(\vec{k}\right)\right) 	, \label{app:eqn:paramPHBreak_1+} \\ 
	&U^{\dagger} \tilde{\Pi} U \left( \tilde{R} \left( \vec{k} \right) \otimes s^0 \right) U^{\dagger} \tilde{\Pi} U -=\frac{1}{4}\zeta ^0 \tau ^0 s^0 \left(\cos (\phi ) \tilde{d}_1\left(\vec{k}\right)+\tilde{d}_2\left(\vec{k}\right)\right) -\frac{1}{4}\zeta ^0 \tau ^0 s^z \left(\cos (\phi ) \tilde{d}_1\left(\vec{k}\right)+\tilde{d}_2\left(\vec{k}\right)\right)\nonumber \\ 
-&\frac{1}{4}\zeta ^0 \tau ^z s^0 \cos (\phi ) \left(\cos (\phi ) \tilde{d}_1\left(\vec{k}\right)+\tilde{d}_2\left(\vec{k}\right)\right) -\frac{1}{4}\zeta ^0 \tau ^z s^z \cos (\phi ) \left(\cos (\phi ) \tilde{d}_1\left(\vec{k}\right)+\tilde{d}_2\left(\vec{k}\right)\right)\nonumber \\ 
-&\frac{1}{4}\zeta ^x \tau ^0 s^0 \cos (\phi ) \left(\tilde{d}_3\left(\vec{k}\right)+\cos (\phi ) \tilde{d}_4\left(\vec{k}\right)\right) -\frac{1}{4}\zeta ^x \tau ^0 s^z \cos (\phi ) \left(\tilde{d}_3\left(\vec{k}\right)+\cos (\phi ) \tilde{d}_4\left(\vec{k}\right)\right)\nonumber \\ 
-&\frac{1}{4}\zeta ^x \tau ^y s^0 \sin (\phi ) \left(\tilde{d}_5\left(\vec{k}\right)+\cos (\phi ) \tilde{d}_6\left(\vec{k}\right)\right) -\frac{1}{4}\zeta ^x \tau ^y s^z \sin (\phi ) \left(\tilde{d}_5\left(\vec{k}\right)+\cos (\phi ) \tilde{d}_6\left(\vec{k}\right)\right)\nonumber \\ 
-&\frac{1}{4}\zeta ^x \tau ^z s^0 \left(\tilde{d}_3\left(\vec{k}\right)+\cos (\phi ) \tilde{d}_4\left(\vec{k}\right)\right) -\frac{1}{4}\zeta ^x \tau ^z s^z \left(\tilde{d}_3\left(\vec{k}\right)+\cos (\phi ) \tilde{d}_4\left(\vec{k}\right)\right)\nonumber \\ 
-&\frac{1}{4}\zeta ^y \tau ^x s^0 \sin (\phi ) \left(\cos (\phi ) \tilde{d}_1\left(\vec{k}\right)+\tilde{d}_2\left(\vec{k}\right)\right) -\frac{1}{4}\zeta ^y \tau ^x s^z \sin (\phi ) \left(\cos (\phi ) \tilde{d}_1\left(\vec{k}\right)+\tilde{d}_2\left(\vec{k}\right)\right)\nonumber \\ 
-&\frac{1}{4}\zeta ^z \tau ^0 s^0 \cos (\phi ) \left(\tilde{d}_5\left(\vec{k}\right)+\cos (\phi ) \tilde{d}_6\left(\vec{k}\right)\right) -\frac{1}{4}\zeta ^z \tau ^0 s^z \cos (\phi ) \left(\tilde{d}_5\left(\vec{k}\right)+\cos (\phi ) \tilde{d}_6\left(\vec{k}\right)\right)\nonumber \\ 
-&\frac{1}{4}\zeta ^z \tau ^y s^0 \sin (\phi ) \left(\tilde{d}_3\left(\vec{k}\right)+\cos (\phi ) \tilde{d}_4\left(\vec{k}\right)\right) -\frac{1}{4}\zeta ^z \tau ^y s^z \sin (\phi ) \left(\tilde{d}_3\left(\vec{k}\right)+\cos (\phi ) \tilde{d}_4\left(\vec{k}\right)\right)\nonumber \\ 
-&\frac{1}{4}\zeta ^z \tau ^z s^0 \left(\tilde{d}_5\left(\vec{k}\right)+\cos (\phi ) \tilde{d}_6\left(\vec{k}\right)\right) -\frac{1}{4}\zeta ^z \tau ^z s^z \left(\tilde{d}_5\left(\vec{k}\right)+\cos (\phi ) \tilde{d}_6\left(\vec{k}\right)\right) 	, \label{app:eqn:paramPHBreak_1-}
\end{align}
with the parity of the functions $d_i \left( \vk \right)$ and $\tilde{d}_i \left( \vk \right)$ (for $1 \leq i \leq 6$) being given in \cref{app:eqn:parity_d_c2z}. For general values of $\phi$, none of the terms in the matrix expansions from \cref{app:eqn:paramPHBreak_1+,app:eqn:paramPHBreak_1-} has definite momentum parity, implying that the matrices $U^{\dagger} \Pi U \left( R \left( \vk \right) \otimes s^0 \right) U^{\dagger} \Pi U$ and $U^{\dagger} \tilde{\Pi} U \left( \tilde{R} \left( \vk \right) \otimes s^0 \right) U^{\dagger} \tilde{\Pi} U$ at opposite momenta cannot be related through a similarity transformation. As a result, $Y \left( \vk , \omega \right)$ and $\tilde{Y} \left( \vk, \omega \right)$ defined in \cref{app:eqn:def_ys+,app:eqn:def_ys-}, which share identical eigenvectors with the matrices from \cref{app:eqn:paramPHBreak_1+,app:eqn:paramPHBreak_1-}, are not similar with the matrices $Y \left( - \vk , \omega \right)$ and $\tilde{Y} \left( - \vk, \omega \right)$, respectively. At the same time, for general values of $\phi$, the matrices $U^{\dagger} \Pi U \left( R \left( \vk \right) \otimes s^0 \right) U^{\dagger} \Pi U$ and $U^{\dagger} \tilde{\Pi} U \left( \tilde{R} \left( \vk \right) \otimes s^0 \right) U^{\dagger} \tilde{\Pi} U$ have non-vanishing valley-off-diagonal terms, implying that $Y \left( \vk , \omega \right)$ and $\tilde{Y} \left( \vk, \omega \right)$ are not necessarily diagonal in the valley subspace. Since $Y \left( \vk , \omega \right)$ and $\tilde{Y} \left( \vk, \omega \right)$ have different eigenvalues at opposite momenta, no \emph{exact} cancellation of the valley-off-diagonal matrix elements can occur when computing the symmetrized spectral function matrices. As a result, $\rthree$ ordering generally emerges in the STM signal of $\ket{\varphi}$. 

In addition to the cases $\phi = 0, \pi$, which correspond to valley-polarized insulators, there is one other notable exception for which $\rthree$ ordering vanishes \emph{exactly} even in the $\lambda = 1$ case: $\phi =\pi/2$. For $\phi = \pi /2$ the valley polarizations of the two Chern bands lie in the $xy$ plane, a configuration which was shown by Ref.~\cite{BUL20,LIA21} to be energetically favored when the effects of kinetic energy are considered (\ie{} the K-IVC state discussed in the main paper). To show that $\rthree$ order vanishes in this case, despite $\ket{\varphi}$ being valley-coherent, we note that for $\phi = \pi/2$, the matrices from \cref{app:eqn:paramPHBreak_1+,app:eqn:paramPHBreak_1-} are given by 
\begin{align}
	&U^{\dagger} \Pi U \left( R \left( \vk \right) \otimes s^0 \right) U^{\dagger} \Pi U = \nonumber \\ &\zeta ^0 \tau ^0 s^0\frac{3 d_2\left(\vec{k}\right)}{4} -\frac{1}{4}\zeta ^0 \tau ^0 s^z d_2\left(\vec{k}\right) -\zeta ^0 \tau ^z s^0\frac{d_1\left(\vec{k}\right)}{2} -\frac{1}{2}\zeta ^0 \tau ^z s^z d_1\left(\vec{k}\right) -\zeta ^x \tau ^0 s^0\frac{d_4\left(\vec{k}\right)}{2} -\frac{1}{2}\zeta ^x \tau ^0 s^z d_4\left(\vec{k}\right)\nonumber \\ 
-&\zeta ^x \tau ^y s^0\frac{d_5\left(\vec{k}\right)}{4} -\zeta ^x \tau ^y s^z\frac{d_5\left(\vec{k}\right)}{4} -\zeta ^x \tau ^z s^0\frac{3 d_3\left(\vec{k}\right)}{4} -\frac{1}{4}\zeta ^x \tau ^z s^z d_3\left(\vec{k}\right) -\zeta ^y \tau ^x s^0\frac{d_2\left(\vec{k}\right)}{4} -\zeta ^y \tau ^x s^z\frac{d_2\left(\vec{k}\right)}{4}\nonumber \\ 
-&\zeta ^z \tau ^0 s^0\frac{d_6\left(\vec{k}\right)}{2} -\frac{1}{2}\zeta ^z \tau ^0 s^z d_6\left(\vec{k}\right) -\frac{1}{4}\zeta ^z \tau ^y s^0 d_3\left(\vec{k}\right) -\frac{1}{4}\zeta ^z \tau ^y s^z d_3\left(\vec{k}\right) -\zeta ^z \tau ^z s^0\frac{3 d_5\left(\vec{k}\right)}{4} -\frac{1}{4}\zeta ^z \tau ^z s^z d_5\left(\vec{k}\right) 	, \label{app:eqn:paramPHBreak_2+} \\ 
	&U^{\dagger} \tilde{\Pi} U \left( \tilde{R} \left( \vk \right) \otimes s^0 \right) U^{\dagger} \tilde{\Pi} U = \nonumber \\
	&\frac{1}{4}\zeta ^0 \tau ^0 s^0 \tilde{d}_2\left(\vec{k}\right) -\frac{1}{4}\zeta ^0 \tau ^0 s^z \tilde{d}_2\left(\vec{k}\right) -\frac{1}{4}\zeta ^x \tau ^y s^0 \tilde{d}_5\left(\vec{k}\right) -\frac{1}{4}\zeta ^x \tau ^y s^z \tilde{d}_5\left(\vec{k}\right) -\frac{1}{4}\zeta ^x \tau ^z s^0 \tilde{d}_3\left(\vec{k}\right) -\frac{1}{4}\zeta ^x \tau ^z s^z \tilde{d}_3\left(\vec{k}\right)\nonumber \\ 
-&\frac{1}{4}\zeta ^y \tau ^x s^0 \tilde{d}_2\left(\vec{k}\right) -\frac{1}{4}\zeta ^y \tau ^x s^z \tilde{d}_2\left(\vec{k}\right) -\frac{1}{4}\zeta ^z \tau ^y s^0 \tilde{d}_3\left(\vec{k}\right) -\frac{1}{4}\zeta ^z \tau ^y s^z \tilde{d}_3\left(\vec{k}\right) -\frac{1}{4}\zeta ^z \tau ^z s^0 \tilde{d}_5\left(\vec{k}\right)\nonumber \\ 
-&\frac{1}{4}\zeta ^z \tau ^z s^z \tilde{d}_5\left(\vec{k}\right) 	. \label{app:eqn:paramPHBreak_2-}
\end{align}
Using the parity and the reality of the functions $d_i \left( \vk \right)$ and $\tilde{d}_i \left( \vk \right)$ (for $1 \leq i \leq 6$) from \cref{app:eqn:parity_d_c2z}, we find that
\begin{align}
	\left( \zeta^{0} \tau^{y} s^{0} \right) U^{\dagger} \Pi U \left( R \left( \vk \right) \otimes s^0 \right) U^{\dagger} \Pi U \left( \zeta^{0} \tau^{y} s^{0} \right)^{\dagger} =& 
	U^{\dagger} \Pi U \left( R \left( - \vk \right) \otimes s^0 \right)^* U^{\dagger} \Pi U, \\
	\left( \zeta^{0} \tau^{y} s^{0} \right) U^{\dagger} \tilde{\Pi} U \left( \tilde{R} \left( \vk \right) \otimes s^0 \right) U^{\dagger} \tilde{\Pi} U \left( \zeta^{0} \tau^{y} s^{0} \right)^{\dagger} =& 
	U^{\dagger} \tilde{\Pi} U \left( \tilde{R} \left( - \vk \right) \otimes s^0 \right)^* U^{\dagger} \tilde{\Pi} U, 
\end{align}
which implies (as a result of sharing identical eigenvectors) that the $Y \left( \vk, \omega \right)$ and $\tilde{Y} \left( \vk, \omega \right)$ matrices defined in \cref{app:eqn:def_ys+,app:eqn:def_ys-} obey an identical property
\begin{align}
	\left( \zeta^{0} \tau^{y} s^{0} \right) Y \left( \vec{k}, \omega \right) \left( \zeta^{0} \tau^{y} s^{0} \right)^{\dagger} =& 
	Y^* \left( - \vec{k}, \omega \right), \\
	\left( \zeta^{0} \tau^{y} s^{0} \right) \tilde{Y} \left( \vec{k}, \omega \right) \left( \zeta^{0} \tau^{y} s^{0} \right)^{\dagger} =& 
	\tilde{Y}^* \left( - \vec{k}, \omega \right). 
\end{align}
Focusing on the electron contribution, tracing over the spin degree of freedom of $Y \left( \vec{k}, \omega \right)$ and using its Hermiticity, we obtain that 
\begin{equation}
	\eta \eta' \left[ \tr_s Y\left(- \vk, \omega \right) \right]_{n' (-\eta'),n (-\eta)} = 
	\left[\tr_s Y\left(\vk, \omega \right) \right]_{n \eta, n' \eta'}.
\end{equation}
At the same time, using \cref{app:eqn:symmetrized_m_tensor,app:eqn:spec_func_from_rMat_y+}, we can find the symmetrized spectral function matrix corresponding to electron excitations to be
\begin{equation}
	\label{app:eqn:PHBreak_m_+}
	\left[\mathcal{M}_{\varphi}^{S +}\left(\omega\right) \right]_{\vk n \eta,\vk n' \eta'} = \frac{1}{2} \delta_{\vk,\vk'} \left(1 + \eta \eta' \right) \left[ \tr_s Y\left(\vk, \omega \right) \right]_{n \eta, n' \eta'}.
\end{equation}
In analogous manner, the hole symmetrized spectral function matrix is given by 
\begin{equation}
	\label{app:eqn:PHBreak_m_-}
	\left[\mathcal{M}_{\varphi}^{S -}\left(\omega\right) \right]_{\vk n \eta,\vk n' \eta'} = \frac{1}{2} \delta_{\vk,\vk'} \left(1 + \eta \eta' \right) \left[ \tr_s \tilde{Y}\left(\vk, \omega \right) \right]_{n \eta, n' \eta'}.
\end{equation}	
Inspecting \cref{app:eqn:PHBreak_m_+,app:eqn:PHBreak_m_-} reveals that the symmetrized spectral function matrices are manifestly valley-diagonal, thus proving that no $\rthree$ ordering emerges for the insulator $\ket{\varphi}$ for $\phi = \pi/2$. 

In conclusion, breaking the exact particle-hole symmetry generally leads to the emergence of $\rthree$ ordering in $\Uncf$ rotations of the insulators $\ket{\Psi_{\nu}}$. Nevertheless, because the particle-hole symmetry-breaking terms in \cref{app:eqn:paramPHBreak_1+,app:eqn:paramPHBreak_1-} (\ie{} $d_{i} \left( \vk \right)$ and $\tilde{d}_{i} \left( \vk \right)$ for $i=1,4,6$) are relatively small, as shown in \cref{app:eqn:parity_d_p_approx}, we expect the amplitude of the $\rthree$ ordering signal to be small in an STM measurement, implying that the conclusions of \cref{app:sec:spFuncSym:nonchiral} hold \emph{approximately} even in the $\lambda = 1$ case. Nevertheless, we note that when the valley polarizations of the two Chern bands of $\ket{\varphi}$ are in the $xy$ plane, a configuration which was shown to be energetically favorable when kinetic energy effects are included~\cite{LIA21}, the $\rthree$ ordering vanishes \emph{exactly}, even in the absence of \emph{exact} particle-hole symmetry.    

\FloatBarrier

\newpage{}
\section{Experimental method}\label{app:sec:experimental}

\begin{figure}[!t]
    \centering
    \includegraphics[width=\textwidth]{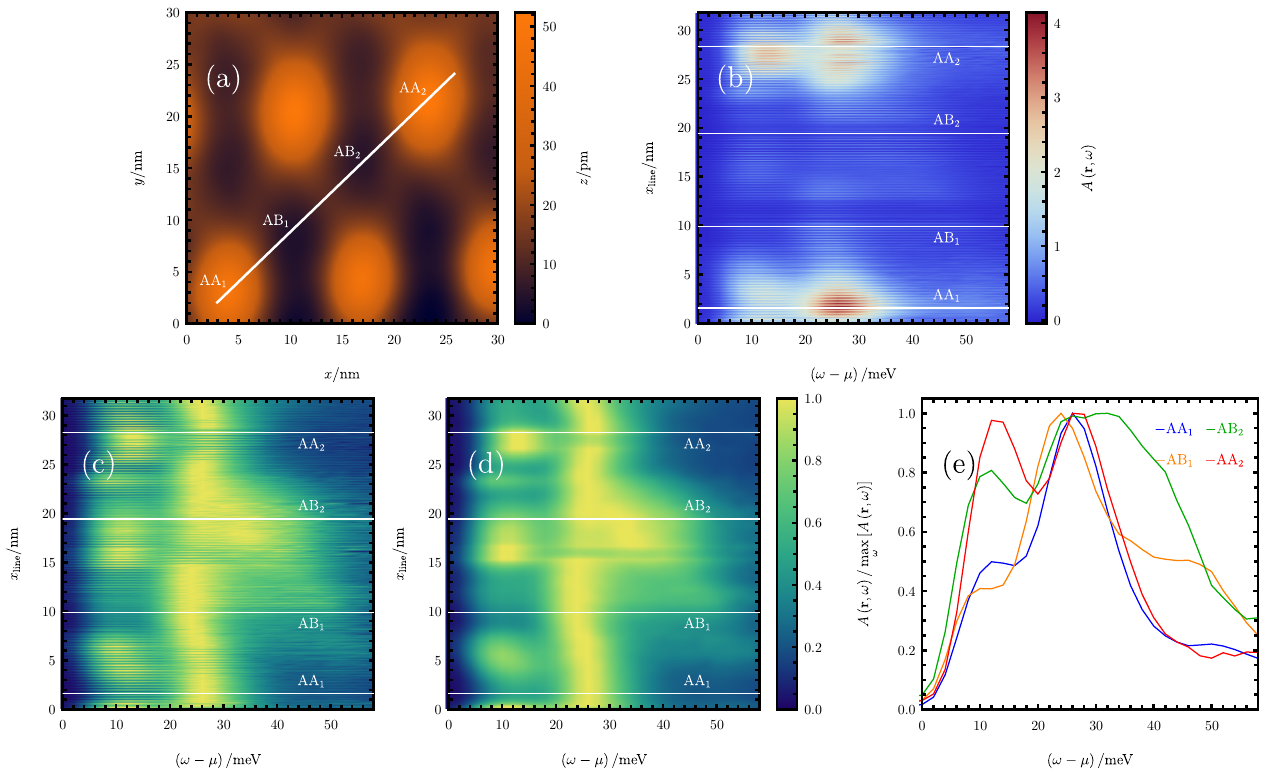}\subfloat{\label{app:fig:STM_linecut_data1:a}}\subfloat{\label{app:fig:STM_linecut_data1:b}}\subfloat{\label{app:fig:STM_linecut_data1:c}}\subfloat{\label{app:fig:STM_linecut_data1:d}}\subfloat{\label{app:fig:STM_linecut_data1:e}}\caption{STM signal along a line connecting two AA stacking sites of TBG for the $\nu = -4$ band insulator with a gate voltage $V_g = \SI{-43}{\volt}$. In (a), we show the topography map of the sample together with the line connecting the two AA regions, passing through one AB and one BA regions. Because the AB and BA regions of TBG are equivalent, we label the centers of the staking regions along the line by $\mathrm{AA}_1$, $\mathrm{AB}_1$, $\mathrm{AB}_2$, and $\mathrm{AA}_2$. The (unnormalized) differential conductance is measured along the line and shown in (b) as a function of the bias ($\omega- \mu$), as well as the distance along the line $x_{\mathrm{line}}$, with the four stacking regions marked by vertical lines. As explained in \cref{app:sec:spectralf}, for each value of $x_{\mathrm{line}}$, we normalize the differential conductance such that the maximum value is 1 and show the resulting data in (c). There is significant variability in the differential conductance on the atomic graphene scale. For that reason, we convolve the signals \emph{along} the line with a Gaussian with standard deviation $\sigma = \SI{0.8}{\nano\meter}$, resulting in the data shown in (d). In (e), we show the resulting four STM signals at $\mathrm{AA}_1$, $\mathrm{AB}_1$, $\mathrm{AB}_2$, and $\mathrm{AA}_2$, of which the $\mathrm{AA}_2$ and $\mathrm{AB}_1$ are showcased in \cref{fig:expVStheory} of the main text.}
    \label{app:fig:STM_linecut_data1}
\end{figure}

\begin{figure}[!t]
    \centering
    \includegraphics[width=\textwidth]{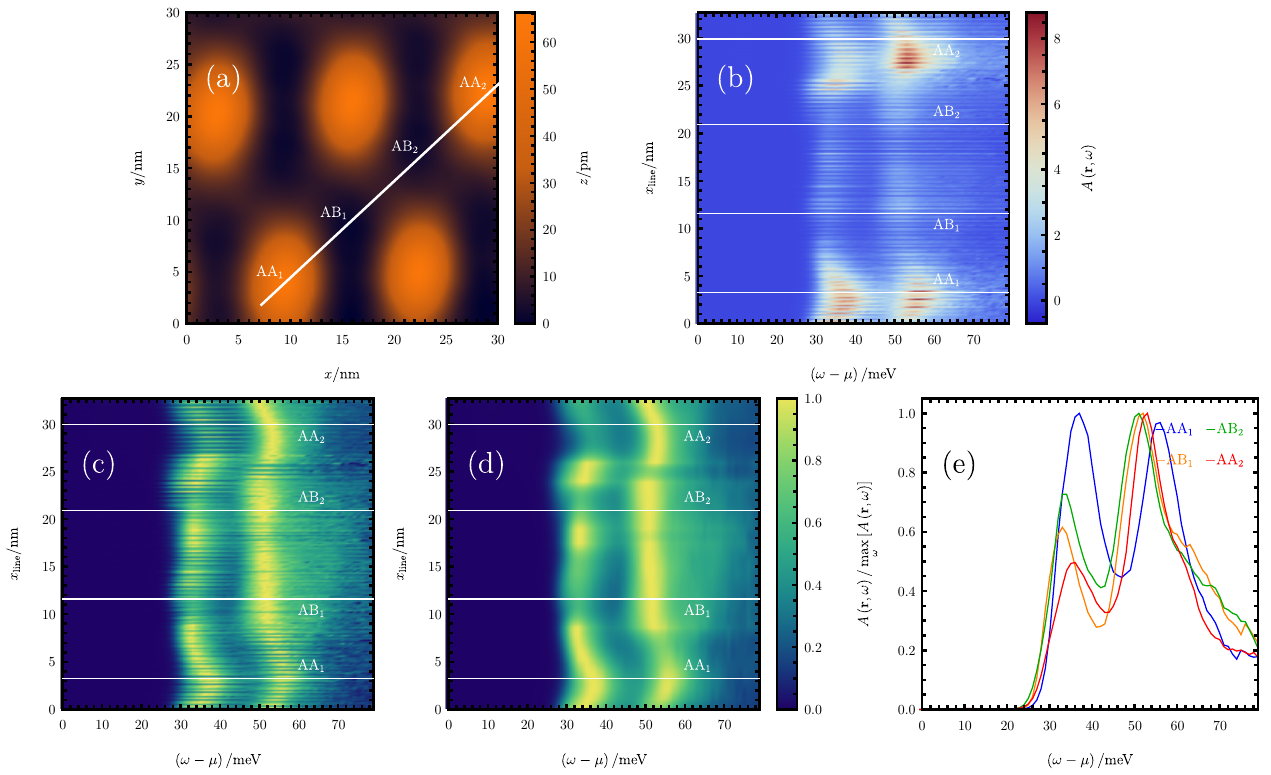}\subfloat{\label{app:fig:STM_linecut_data2:a}}\subfloat{\label{app:fig:STM_linecut_data2:b}}\subfloat{\label{app:fig:STM_linecut_data2:c}}\subfloat{\label{app:fig:STM_linecut_data2:d}}\subfloat{\label{app:fig:STM_linecut_data2:e}}\caption{STM signal along a line connecting two AA sites of TBG for the $\nu = -4$ band insulator with a gate voltage $V_g = \SI{-45}{\volt}$ at a different location than \cref{app:fig:STM_linecut_data1} (but the same sample). The panels have the same meaning as in \cref{app:fig:STM_linecut_data1}.}
    \label{app:fig:STM_linecut_data2}
\end{figure}

In this \siSection{}, we provide additional information about the experimental data discussed in the article. Tunneling measurements were performed on a homebuilt, ultra-high vacuum (UHV) STM~\cite{WON20a} using tungsten tips calibrated against the Cu$(111)$ Shockley surface state. Tungsten and copper were chosen to match the work function of graphene. TBG devices, which consist of TBG/hexagonal boron nitride (hBN)/SiO$_2$/Si, were electrically contacted through pre-patterned gold electrodes. The TBG was biased relative to the tip, which was referenced to ground, and a gate voltage was applied to Si to tune the carrier density to $\nu=\pm 4$. The differential conductance $\dd I/\dd V$ (\ie{} the STM signal) was obtained via lock-in detection of the AC tunnel current induced by an AC modulation voltage added to the sample bias $V_{\mathrm{bias}}$. As shown in Ref.~\cite{COL15}, the differential conductance as a function of the the bias voltage is proportional to the spectral function of the sample
\begin{equation}
	\label{app:eqn:prop_dIdV_A}
	\dv{I}{V} \left( V_{\mathrm{bias}} \right) \propto A \left( \vec{r}, \omega \right), \qq{where} eV_{\mathrm{bias}} = \omega - \mu.
\end{equation}
The proportionality constant in \cref{app:eqn:prop_dIdV_A} depends on the specifics of the tip~\cite{COL15}, but is eliminated for our choice of normalization. We can directly compare the experimentally-obtained differential conductance and the theoretically-computed spectral function by normalizing both quantities by their maxima at the location of interest since
\begin{equation}
	\frac{\dv{I}{V} \left( V_{\mathrm{bias}} \right)}{\max_{V_{\mathrm{bias}}} \left[ \dv{I}{V} \left( V_{\mathrm{bias}} \right) \right] } = \frac{A \left(\vec{r}, \omega \right)}{\max_{\omega} \left[ A \left(\vec{r}, \omega \right) \right]}, \qq{where} eV_{\mathrm{bias}} = \omega - \mu.
\end{equation}
As a result of this and to ensure the consistency of notation between the measurements and theoretical prediction, we will present the experimental results in terms of normalized spectral function $A \left(\vec{r}, \omega \right) / \max_{\omega} \left[ A \left(\vec{r}, \omega \right) \right]$ and tunneling energy $\omega - \mu$.

The energy resolution of the measured STM signal is influenced by a variety of factors the most important of which are the thermal noise, the modulation voltage from the amplifier~\cite{SON10}, as well as the finite capacitance of the tip (which depends on the tip shape)~\cite{AST16}. For the temperature ($T = \SI{6}{\kelvin}$) at which we perform our measurements, the effect of tip capacitance on the energy resolution is significantly smaller than the thermal broadening and can thus be ignored~\cite{AST16}. The energy resolution $\delta \omega$ of the STM signal is thus given by the thermal broadening added in quadrature to the amplifier modulation voltage noise ($V_{\mathrm{mod}} = \SI{1}{\milli\volt}$)~\cite{SON10}
\begin{equation}
	\delta \omega \approx \sqrt{\left( 3.5 k_B T \right)^2 + \left(2.5 e V_{\mathrm{mod}} \right)^2} \approx \SI{3}{\milli\eV}.
\end{equation}

TBG devices were fabricated through a ``tear-and-stack'' method \cite{KIM16a} whereby monolayer graphene was torn into two pieces via pickup on a sacrificial polyvinyl alcohol (PVA) structure. One monolayer was rotated relative to the other and was stacked on top of the other. As explained in greater detail in Ref.~\cite{WON20a}, the PVA structure was dissolved through water injection, and the subsequent heterostructure was annealed at 400$^{\circ}$C.

In the main paper, we present STM data at $\nu = \pm 4$ which fits with the overall features of the TBG spectral function in the strong-coupling picture. We now give a broader overview of the data at $\nu = -4$ in \cref{app:fig:STM_linecut_data1,app:fig:STM_linecut_data2}. The two figures are obtained from the same sample but at two different locations, and we will detail \cref{app:fig:STM_linecut_data1} knowing that a similar analysis can be done for \cref{app:fig:STM_linecut_data2}. We measure the STM signal along a line connecting two AA regions and passing through an AB and a BA region: as depicted in \cref{app:fig:STM_linecut_data1:a}, the measurement starts at the lower-left endpoint and ends approximately at the upper-right corner (with a small amount of drift). The resulting differential conductance from \cref{app:fig:STM_linecut_data1:b} is then normalized with respect to its maximum in \cref{app:fig:STM_linecut_data1:c}. The data shows a significant degree of variability in the STM signal on the atomic graphene scale. We circumvent this issue by applying a Gaussian filter with standard deviation $\sigma = \SI{0.8}{\nano\meter}$ \emph{before normalization}, resulting in \cref{app:fig:STM_linecut_data1:d}. Finally, we extract the normalized STM signals corresponding to the two AA and two AB/BA regions and illustrated them in \cref{app:fig:STM_linecut_data1:e}. Note that the normalized STM signals shown in \cref{fig:expVStheory:a} of the main text are the $\mathrm{AA}_2$ and $\mathrm{AB}_1$ plots of \cref{app:fig:STM_linecut_data1:e}. It is also worth noting that the exact position of the peaks in terms of bias voltage (or tunneling energy) seems to change slightly in both \cref{app:fig:STM_linecut_data1,app:fig:STM_linecut_data2} by up to approximately $\SI{8}{\milli\eV}$. This is due to the STM tip acting as a top gate -- a phenomenon known as \emph{tip-induced band bending}~\cite{XIE19}: empirically, the amount of tip-induced band bending that occurs when the tip is above an AA site compared to when it is above an AB site appears to differ slightly. Although not a quantitative argument, this could be thought of as being due to the greater local density of states at the AA site compared to the one at the AB site. As a consequence, the chemical potential might shift due to tip gating when moving from an AA to AB region.

The results shown in both \cref{app:fig:STM_linecut_data1:e,app:fig:STM_linecut_data2:e} indicate a fair amount of variability between different AA and AB/BA sites of the same sample, primarily in how the ratio between the two peaks changes upon moving from the AA to the AB/BA site. Nevertheless, as explained in the main paper, the signal broadening is consistent with the interaction energy scale as opposed to the bandwidth of the TBG bands. Thus, it validates the strong coupling approach in describing the physics of TBG.

\newpage
\section{Additional numerical results}\label{app:sec:results}

In this \siSection{} we provide numerical results for 14 different TBG insulator states (listed in \cref{app:tab:model_insulators}), as well as additional data for the six insulators considered in \cref{fig:realSpaceExamples} of the main text. Firstly, by focusing on the insulating states shown in \cref{fig:realSpaceExamples} of the main text, we present and discuss the real-space features of the TBG spectral function across the \emph{entire} moir\'e unit cell (as opposed to restricting to the AA stacking center, as was done in the main). We find that the spectral function near the AB/BA stacking centers reveals no further distinguishing features, compared to the AA-centered spectral function. Next, for each insulator listed in \cref{app:tab:model_insulators}, we consider four parameter values $\left( w_0/w_1, \lambda \right) = \left( 0.0, 0 \right)$, $\left( w_0/w_1, \lambda \right) = \left( 0.4, 0 \right)$, $\left( w_0/w_1, \lambda \right) = \left( 0.8, 0 \right)$, $\left( w_0/w_1, \lambda \right) = \left( 0.8, 1 \right)$. In each case, we show the spectral function at the AA and AB stacking centers, together with the charge-one excitation dispersion plotted along the high-symmetry lines of the MBZ. For each insulator and parameter choice, we also plot the spatial structure of the TBG spectral function in both real space (around the AA site) and momentum space for two values of $\omega - \mu$ (corresponding roughly to the maxima of the AA STM signal for hole or electron tunneling, respectively). Due the decay of the TBG spectral function in momentum space $A \left( \vec{q} ,\omega \right)$ with $\abs{\vec{q}}$, in all the plots of this work, we plot $\abs{A \left( \vec{q} ,\omega \right)}^{1/3}$ (as opposed to $\abs{A \left( \vec{q} ,\omega \right)}$ directly), thus effectively enhancing all the features of the spectral function. 

\begin{table}[t]
\centering
\begin{tabular}{c||l||c|p{5 cm}|p{0.5 cm}|p{0.5 cm}|p{0.5 cm}|p{0.5 cm}|p{0.5 cm}|p{0.5 cm}}
\hline
\multirow{ 3}{*}{$\nu$} & \multirow{ 3}{*}{$\ket{\varphi}$} & \multirow{ 3}{*}{\SiSection{}} & \multirow{ 3}{*}{Description} & \multicolumn{3}{c|}{$w_0/w_1$} \\
\cline{5-7}
& & & & $ 0.0 $ & $ 0.4 $ & $ 0.8 $\\
\hline 
\hline
$-4$ & $ \displaystyle \ket{\varphi^{(0)}} = \ket{0} $ & \cref{app:sec:example:stateZero} & \tiny Insulator with all TBG active bands empty & \ding{51} & \ding{51} & \ding{51} \\
\hline
\hline
$-3$ & $ \displaystyle \ket{\varphi^{(1)}}  = \prod_{\vec{k}} \hat{d}^\dagger_{\vec{k},+1,+,\uparrow} \ket{0} $ & \cref{app:sec:example:stateOne} & \tiny $\mathcal{C}=1$, valley polarized & \ding{51} & \ding{51} & \ding{55}\\
\hline
$-3$ & $ \displaystyle \ket{\varphi^{(2)}}  = \prod_{\vec{k}} \frac{\hat{d}^\dagger_{\vec{k},+1,+,\uparrow} + \hat{d}^\dagger_{\vec{k},+1,-,\uparrow} }{\sqrt{2}} \ket{0} $ & \cref{app:sec:example:stateTwo} & \tiny $\mathcal{C}=1$, intervalley-coherent & \ding{51} & \ding{51} & \ding{55}\\
\hline
\hline
$-2$ & $ \displaystyle \ket{\varphi^{(3)}}  = \prod_{\vec{k}} \prod_{e_Y = \pm 1} \frac{\hat{d}^\dagger_{\vec{k},e_Y,+,\uparrow} + e_Y \hat{d}^\dagger_{\vec{k},e_Y,-,\uparrow} }{\sqrt{2}} \ket{0}  $ & \cref{app:sec:example:stateThree} & \tiny $\mathcal{C}=0$, K-IVC  & \ding{51} & \ding{51} & \ding{51}\\
\hline
$-2$ & $ \displaystyle \ket{\varphi^{(4)}}  = \prod_{\vec{k}} \prod_{e_Y = \pm 1} \frac{\hat{d}^\dagger_{\vec{k},e_Y,+,\uparrow} + \hat{d}^\dagger_{\vec{k},e_Y,-,\uparrow} }{\sqrt{2}} \ket{0} $ & \cref{app:sec:example:stateFour} & \tiny $\mathcal{C}=0$, T-IVC & \ding{51} & \ding{51} & \ding{51}\\
\hline
$-2$ & $ \displaystyle \ket{\varphi^{(5)}}  = \prod_{\vec{k}} \hat{d}^\dagger_{\vec{k},+1,+,\uparrow} \hat{d}^\dagger_{\vec{k},-1,+,\uparrow} \ket{0} $ & \cref{app:sec:example:stateFive} & \tiny $\mathcal{C}=0$, valley polarized & \ding{51} & \ding{51} & \ding{51} \\
\hline
$-2$ & $ \displaystyle \ket{\varphi^{(6)}}  = \prod_{\vec{k}} \hat{d}^\dagger_{\vec{k},+1,+,\uparrow} \hat{d}^\dagger_{\vec{k},+1,+,\downarrow} \ket{0} $ & \cref{app:sec:example:stateSix} & \tiny $\mathcal{C}=2$, valley polarized & \ding{51} & \ding{51} & \ding{51} \\
\hline
$-2$ & $ \displaystyle \ket{\varphi^{(7)}}  = \prod_{\vec{k}} \prod_{s=\uparrow,\downarrow} \frac{\hat{d}^\dagger_{\vec{k},+1,+,s} + \hat{d}^\dagger_{\vec{k},+1,-,s} }{\sqrt{2}} \ket{0} $ & \cref{app:sec:example:stateSeven} & \tiny $\mathcal{C}=2$, intervalley-coherent & \ding{51} & \ding{51} & \ding{51} \\
\hline
\hline
$-1$ & $ \displaystyle \ket{\varphi^{(8)}}  = \prod_{\vec{k}} \hat{d}^\dagger_{\vec{k},+1,+,\uparrow} \hat{d}^\dagger_{\vec{k},+1,+,\downarrow} \hat{d}^\dagger_{\vec{k},+1,-,\downarrow} \ket{0} $ & \cref{app:sec:example:stateEight} & \tiny $\mathcal{C}=3$, valley polarized & \ding{51} & \ding{51} & \ding{51} \\
\hline
$-1$ & $ \displaystyle \ket{\varphi^{(9)}}  = \prod_{\vec{k}} \hat{d}^\dagger_{\vec{k},+1,-,\downarrow} \prod_{e_Y = \pm 1} \frac{\hat{d}^\dagger_{\vec{k},e_Y,+,\uparrow} + e_Y \hat{d}^\dagger_{\vec{k},e_Y,-,\uparrow} }{\sqrt{2}} \ket{0} $ & \cref{app:sec:example:stateNine} & \tiny $\mathcal{C}=1$, two occupied K-IVC bands and a valley polarized band & \ding{51} & \ding{51} & \ding{51} \\
\hline
\hline
$0$ & $ \displaystyle \ket{\varphi^{(10)}}  = \prod_{\vec{k}} \prod_{e_Y = \pm 1} \prod_{s = \uparrow,\downarrow} \frac{\hat{d}^\dagger_{\vec{k},e_Y,+,s} + e_Y \hat{d}^\dagger_{\vec{k},e_Y,-,s} }{\sqrt{2}} \ket{0} $ & \cref{app:sec:example:stateTen} & \tiny $\mathcal{C}=0$, K-IVC & \ding{51} & \ding{51} & \ding{51} \\
\hline
$0$ & $ \displaystyle \ket{\varphi^{(11)}}  = \prod_{\vec{k}} \prod_{e_Y = \pm 1} \prod_{s = \uparrow,\downarrow} \hat{d}^\dagger_{\vec{k},e_Y,+,s} \ket{0} $ & \cref{app:sec:example:stateEleven} & \tiny $\mathcal{C}=0$, valley polarized & \ding{51} & \ding{51} & \ding{51} \\
\hline
$0$ & $ \displaystyle \ket{\varphi^{(12)}}  = \prod_{\vec{k}} \prod_{\eta = \pm} \prod_{s = \uparrow,\downarrow} \hat{d}^\dagger_{\vec{k},+1,\eta,s} \ket{0} $ & \cref{app:sec:example:stateTwelve} & \tiny $\mathcal{C}=4$, valley polarized & \ding{51} & \ding{51} & \ding{51} \\
\hline
\hline
$+4$ & $ \displaystyle \ket{\varphi^{(13)}}  = \prod_{\vec{k}} \prod_{\eta = \pm} \prod_{s = \uparrow,\downarrow} \prod_{e_Y = \pm 1} \hat{d}^\dagger_{\vec{k},e_Y,\eta,s} \ket{0} $ & \cref{app:sec:example:stateThirteen} & \tiny Insulator with all TBG active bands filled & \ding{51} & \ding{51} & \ding{51} \\
\hline
\end{tabular}
\caption{Model insulator states of TBG, for which we study the spectral function. Each model insulator is investigated for multiple values of the $w_0 / w_1$. For each value of $w_0 / w_1$ and type of potential, we indicate whether the insulator shows only positive-energy charge-one and charge neutral excitations (\ding{51}) or not (\ding{55}). If an insulator has negative-energy excitations, then we do not compute the spectral function (as this implies the insulators is not a ground state of $H_I$).}
\label{app:tab:model_insulators}
\end{table}

\subsection{The real-space features of the TBG spectral function around the AB/BA sites}\label{app:sec:results:AB_site}

\begin{figure}[!t]
	\includegraphics[width=\textwidth]{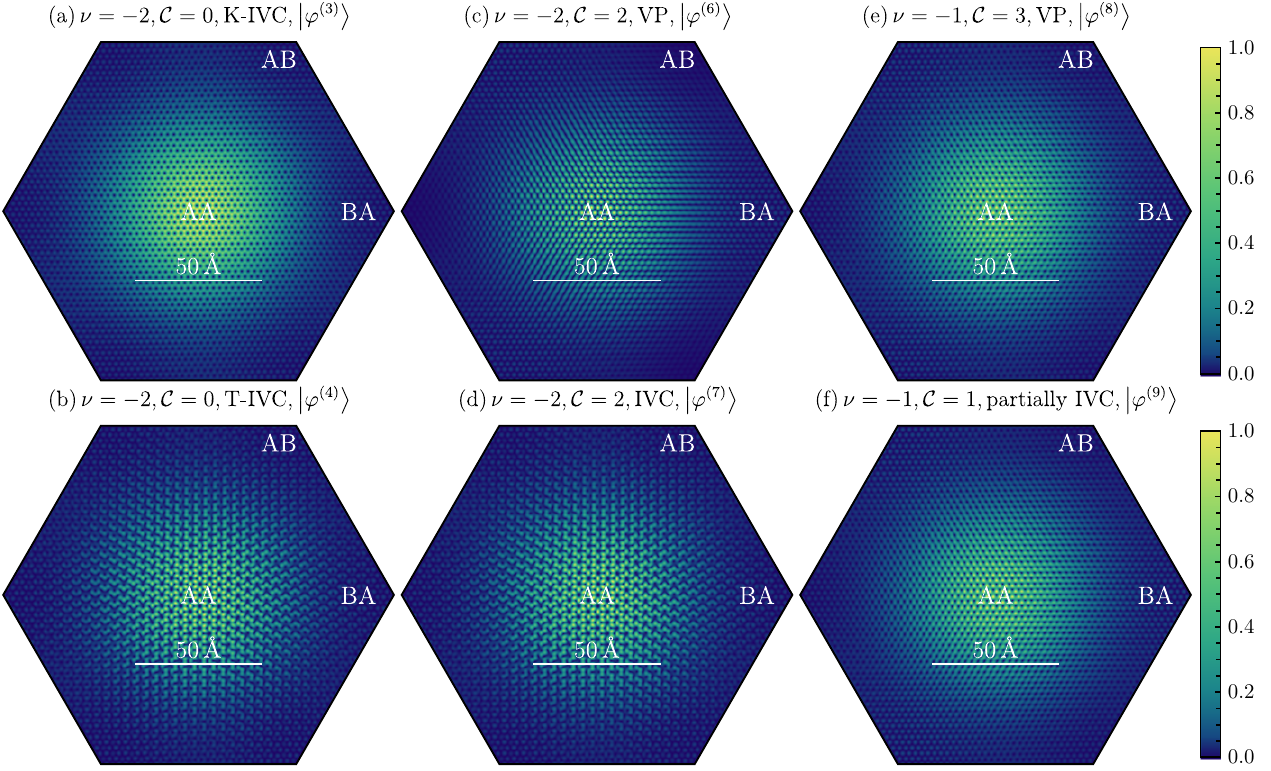}\subfloat{\label{app:fig:realSpaceExamples:a}}\subfloat{\label{app:fig:realSpaceExamples:b}}\subfloat{\label{app:fig:realSpaceExamples:c}}\subfloat{\label{app:fig:realSpaceExamples:d}}\subfloat{\label{app:fig:realSpaceExamples:e}}\subfloat{\label{app:fig:realSpaceExamples:f}}\caption{The real-space features of the TBG spectral function for an entire moir\'e unit cell. We focus on the same insulators as in \cref{fig:realSpaceExamples} of the main text (employing the same conventions for numbering the panels), such that (a)-(f) correspond respectively to $\displaystyle \ket{\varphi^{(3)}}$, $ \displaystyle \ket{\varphi^{(4)}}$, $ \displaystyle \ket{\varphi^{(6)}}$, $ \displaystyle \ket{\varphi^{(7)}}$, $ \displaystyle \ket{\varphi^{(8)}}$, and $ \displaystyle \ket{\varphi^{(9)}}$, in the notation of \cref{app:tab:model_insulators}. We consider the $\nu = -2$, $\mathcal{C} = 0$, K-IVC [(a)] and T-IVC [(b)] states, the $\nu = -2$, $\mathcal{C} = 2$, valley polarized [(c)] and intervalley-coherent [(d)] Chern insulators, as well as the $\nu = -1$, $\mathcal{C} = 3$ fully valley polarized [(e)] and $\nu = -1$, $\mathcal{C} = 1$ partially intervalley-coherent [(f)] Chern insulators. Similar to \cref{fig:realSpaceExamples}, the presence of Kekul\'e distortion in (b) and (d) appears as threefold enlargement of the SLG unit cell manifested across the entire moir\'e unit cell.}
	\label{app:fig:realSpaceExamples}
\end{figure}

\begin{figure}[!t]
	\includegraphics[width=\textwidth]{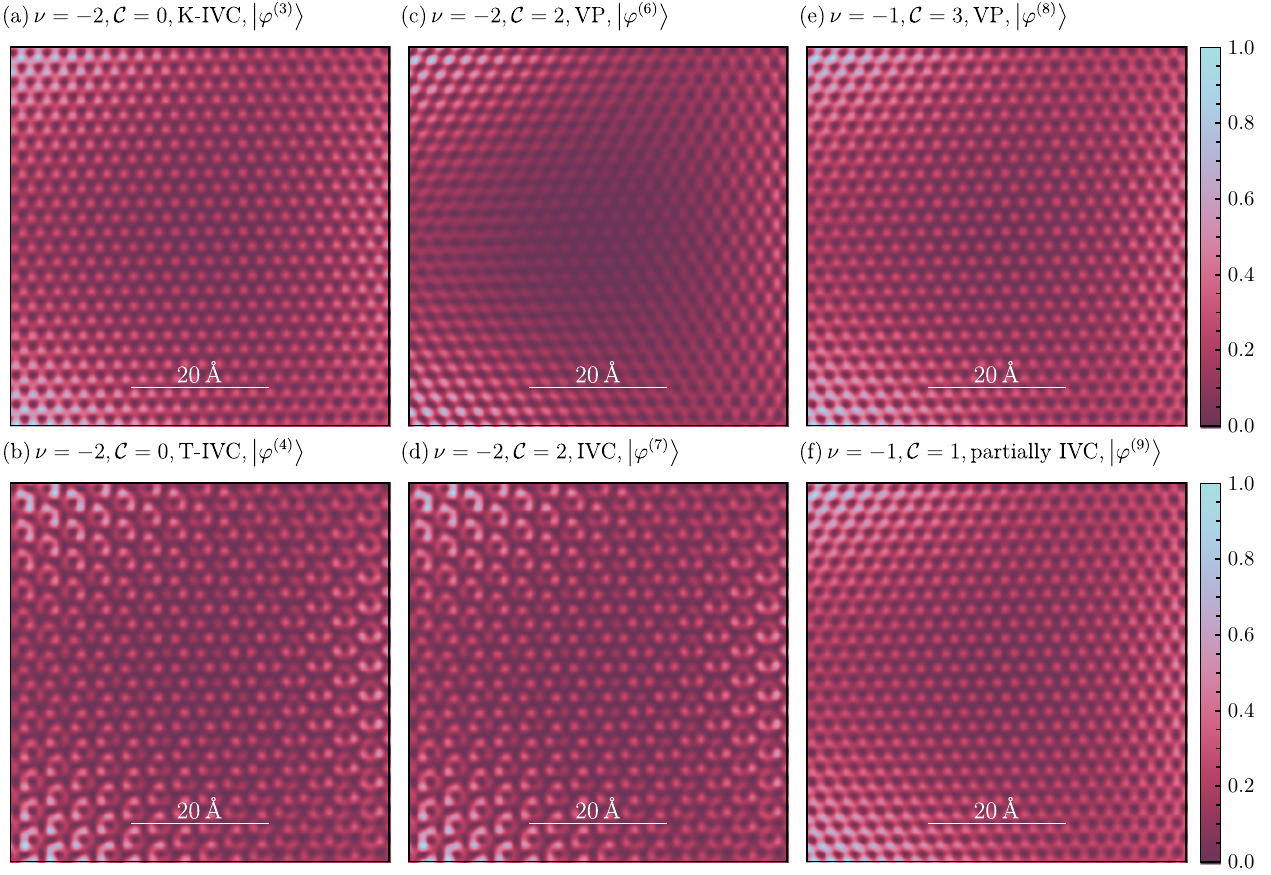}\subfloat{\label{app:fig:realSpaceExamples_AB:a}}\subfloat{\label{app:fig:realSpaceExamples_AB:b}}\subfloat{\label{app:fig:realSpaceExamples_AB:c}}\subfloat{\label{app:fig:realSpaceExamples_AB:d}}\subfloat{\label{app:fig:realSpaceExamples_AB:e}}\subfloat{\label{app:fig:realSpaceExamples_AB:f}}\caption{The real-space features of the TBG spectral function around the AB stacking center. We focus on the same insulators as in \cref{fig:realSpaceExamples} of the main text and in \cref{app:fig:realSpaceExamples} (employing the same conventions for numbering the panels). Note that the color scale is normalized to one with respect to the \emph{maximum} value of the spectral function in the region considered in each panel. To avoid potential confusion, we have thus used a different color scheme than in \cref{app:fig:realSpaceExamples}. For each insulator [(a)-(f)], we show the real-space spectral function centered at the AB site. Note that the TBG spectral function is approximately A-sublattice polarized near the AB stacking center. As a consequence, the spectral function of the $ \displaystyle \ket{\varphi^{(3)}}$ state (which is B-sublattice polarized in the AA region), becomes negligible at the AB stacking center. Similar to \cref{fig:realSpaceExamples} of the main text, the presence of Kekul\'e distortion in (b) and (d) appears as threefold enlargement of the SLG unit cell.}
	\label{app:fig:realSpaceExamples_AB}
\end{figure}

In \cref{fig:realSpaceExamples} of the main text, we have shown the spatial features of the TBG spectral function for six insulators (namely $ \displaystyle \ket{\varphi^{(3)}}$, $ \displaystyle \ket{\varphi^{(4)}}$, $ \displaystyle \ket{\varphi^{(6)}}$, $ \displaystyle \ket{\varphi^{(7)}}$, $ \displaystyle \ket{\varphi^{(8)}}$, and $ \displaystyle \ket{\varphi^{(9)}}$, in the notation of \cref{app:tab:model_insulators}) around the AA stacking site. This allowed us to identify distinguishing spectral features for each insulator (such as the presence or absence of Kekul\'e distortion, or sublattice polarization). We now present additional numerical data for the same insulating states, but instead consider the spectral function across the \emph{entire} moir\'e unit cell in \cref{app:fig:realSpaceExamples} and around the AB stacking centers in \cref{app:fig:realSpaceExamples_AB}. 

Similarly to the AA-centered spectral function presented in \cref{fig:realSpaceExamples} of the main text, the presence of Kekul\'e distortion is manifested for the $\nu = -2$, $\mathcal{C} = 0$, T-IVC and the $\nu = -2$, $\mathcal{C} = 2$, intervalley-coherent insulators over the whole moir\'e unit cell as an apparent threefold enlargement of the graphene unit cell. Focusing on the AB stacking center, we note however that the exact $\rthree{}$ patterns shown in \cref{app:fig:realSpaceExamples_AB:b,app:fig:realSpaceExamples_AB:d} differ from the corresponding ones near the AA sites (shown in \cref{fig:realSpaceExamples} of the main text): the spectral function instead changes continuously across the entire moir\'e unit cell (as a consequence of the long-wavelength moir\'e{}-scale modulation of the spectral function). 

Moreover, it can be seen that the TBG spectral function near the AB/BA stacking centers is primarily supported on a single graphene sublattice of the top graphene layer. This is because only one of the two carbon atoms in each unit cell of the top graphene layer is located directly above a carbon atom from the bottom layer and can thus have a non-negligible contribution to the active TBG bands renormalized by interactions. This should be contrasted with the AA sites where both graphene sublattices of the top layer are located approximately above carbon atoms from the bottom layer and can thus contribute to the active TBG band wave functions. Therefore, the corresponding TBG spectral function always appears A-sublattice (B-sublattice) polarized at the AB (BA) stacking center, as can be seen in \cref{app:fig:realSpaceExamples} and in the AB-centered spectral function real-space plots from \cref{app:fig:realSpaceExamples_AB}. For example, the spectral function for the $ \displaystyle \ket{\varphi^{(3)}}$ state shown in \cref{app:fig:realSpaceExamples:a,app:fig:realSpaceExamples_AB:a} which displays no sublattice polarization at the AA sites, is sublattice-polarized at the AB/BA stacking centers. Moreover, when the TBG spectral function is sublattice-polarized at the AA site (due to the nature of the corresponding insulating state), the spectral function may even become negligible near the AB/BA stacking centers. This is exactly the case for for the $ \displaystyle \ket{\varphi^{(6)}}$ insulator shown in \cref{app:fig:realSpaceExamples:c}, which is B-sublattice polarized at the AA site: the corresponding spectral function, shown in more detail in \cref{app:fig:realSpaceExamples_AB:c}, approximately vanishes at the AB site. As a consequence, sublattice polarization is a generic feature of the TBG spectral function at the AB or BA stacking centers, and cannot be used to reliably distinguish between different insulating candidate states.

In summary, the spatial variation of the TBG spectral function around the AB (or BA) stacking centers does not reveal any further qualitative features compared to the AA-centered spectral function considered in the main text. Coupled with the fact that the electron density of the TBG active band wave functions is mostly concentrated around the AA sites, this justifies our choice of solely focusing on the AA sites when plotting the real-space TBG spectral function. 
\newpage
\clearpage

\FloatBarrier
\subsection{$\nu=-4$, insulator with all TBG active bands empty{}: $\ket{0}$}\label{app:sec:example:stateZero}

\begin{figure}[!h]
\centering
\includegraphics[width=0.95\textwidth]{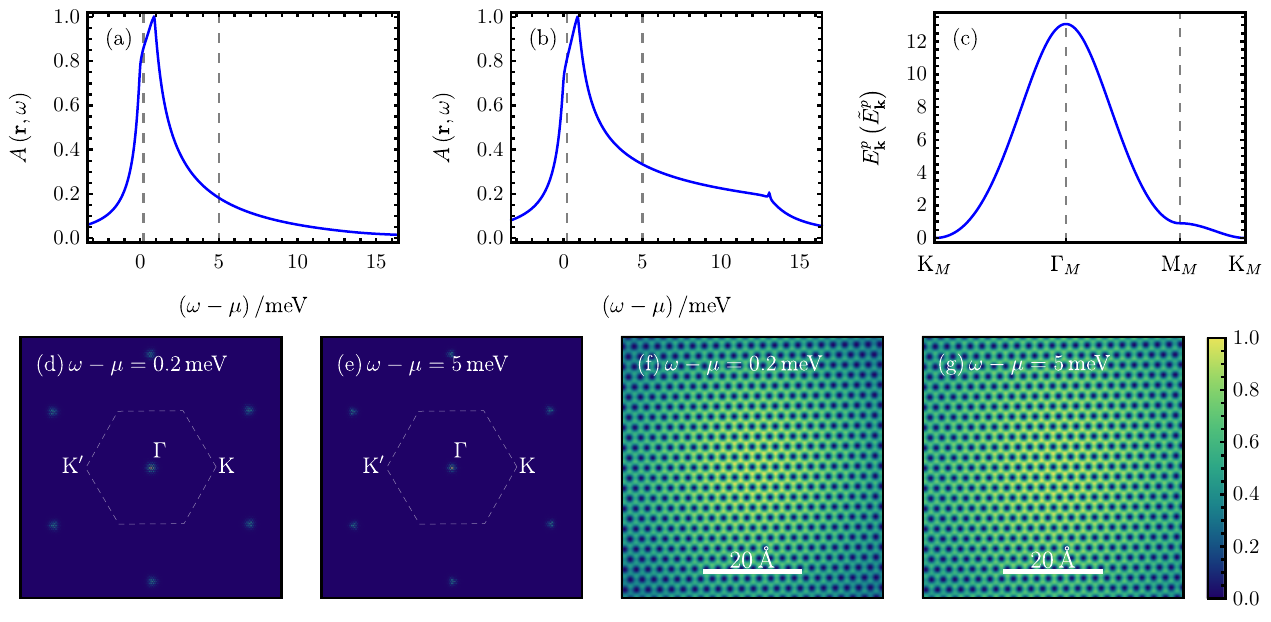}
\caption{{\it $\nu=-4$, insulator with all TBG active bands empty{}}: The real-space spectral function averaged over three graphene unit cells at the AA site [(a)] and at the AB site [(b)]. The electron (blue) and hole (orange) dispersion is shown in (c). In (d)-(g), we illustrate the spatial variation of the spectral function in momentum space [(d) and (e)], as well as real space [(f) and (g)] for the two energy choices depicted by dashed gray lines in (a) and (b). Here, we focus on $\left(w_0/w_1, \lambda \right) = \left(0.0, 0 \right)$.}\label{app:fig:realspaceZeroWZeroLambdaZero}
\end{figure}

\begin{figure}[!h]
\centering
\includegraphics[width=0.95\textwidth]{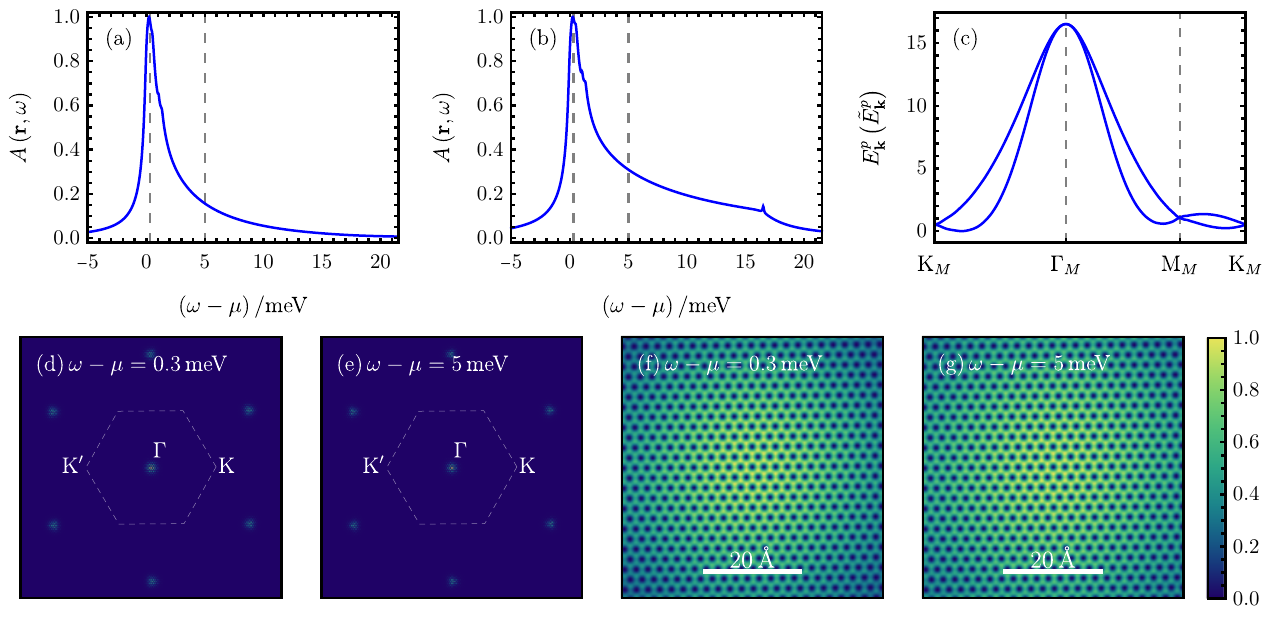}
\caption{{\it $\nu=-4$, insulator with all TBG active bands empty{}}: The real-space spectral function averaged over three graphene unit cells at the AA site [(a)] and at the AB site [(b)]. The electron (blue) and hole (orange) dispersion is shown in (c). In (d)-(g), we illustrate the spatial variation of the spectral function in momentum space [(d) and (e)], as well as real space [(f) and (g)] for the two energy choices depicted by dashed gray lines in (a) and (b). Here, we focus on $\left(w_0/w_1, \lambda \right) = \left(0.4, 0 \right)$.}\label{app:fig:realspaceZeroWFourLambdaZero}
\end{figure}

\begin{figure}[!h]
\centering
\includegraphics[width=0.95\textwidth]{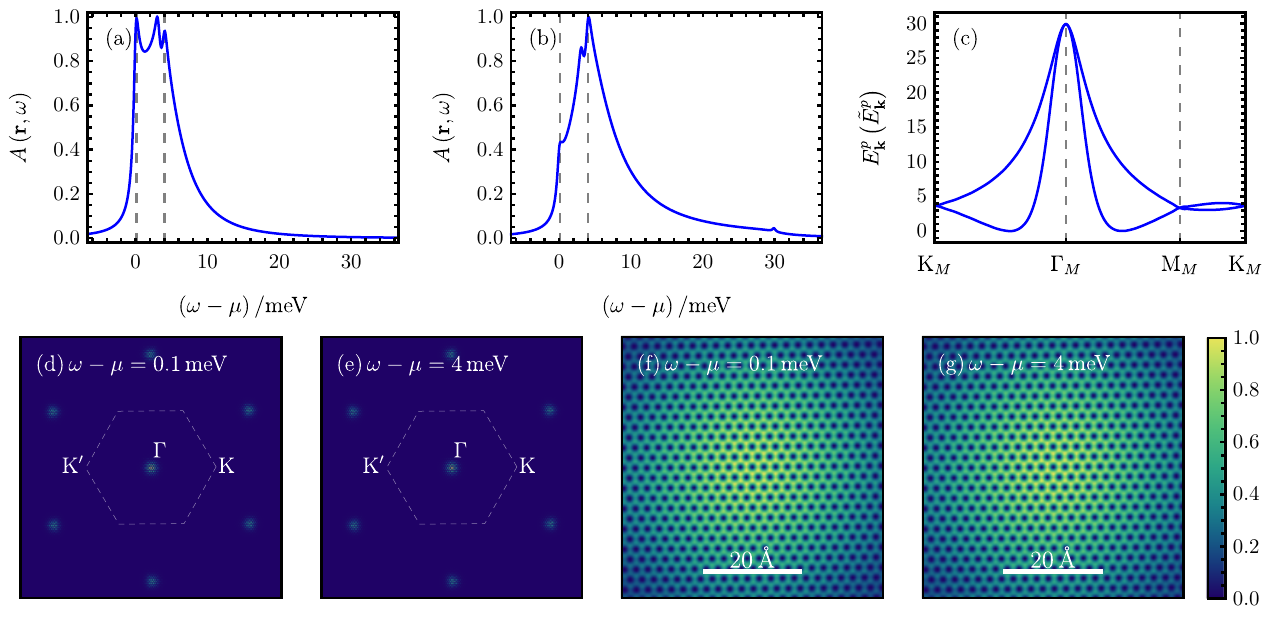}
\caption{{\it $\nu=-4$, insulator with all TBG active bands empty{}}: The real-space spectral function averaged over three graphene unit cells at the AA site [(a)] and at the AB site [(b)]. The electron (blue) and hole (orange) dispersion is shown in (c). In (d)-(g), we illustrate the spatial variation of the spectral function in momentum space [(d) and (e)], as well as real space [(f) and (g)] for the two energy choices depicted by dashed gray lines in (a) and (b). Here, we focus on $\left(w_0/w_1, \lambda \right) = \left(0.8, 0 \right)$.}\label{app:fig:realspaceZeroWEightLambdaZero}
\end{figure}

\begin{figure}[!h]
\centering
\includegraphics[width=0.95\textwidth]{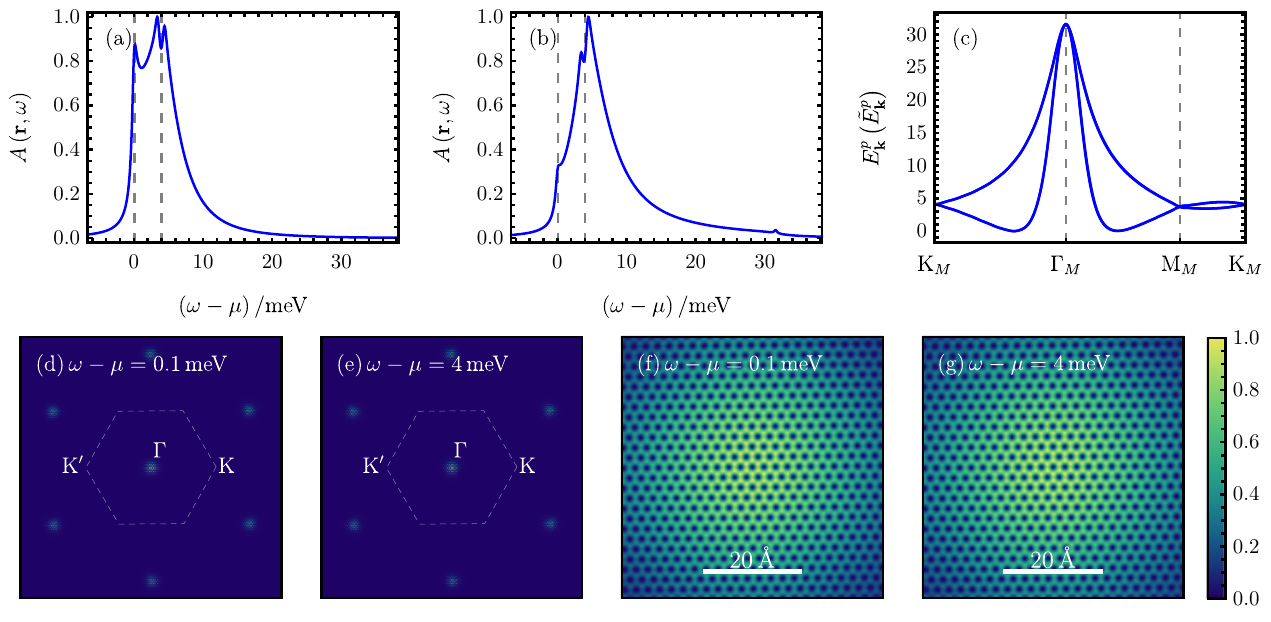}
\caption{{\it $\nu=-4$, insulator with all TBG active bands empty{}}: The real-space spectral function averaged over three graphene unit cells at the AA site [(a)] and at the AB site [(b)]. The electron (blue) and hole (orange) dispersion is shown in (c). In (d)-(g), we illustrate the spatial variation of the spectral function in momentum space [(d) and (e)], as well as real space [(f) and (g)] for the two energy choices depicted by dashed gray lines in (a) and (b). Here, we focus on $\left(w_0/w_1, \lambda \right) = \left(0.8, 1 \right)$.}\label{app:fig:realspaceZeroWEightLambdaOne}
\end{figure}

\newpage

\clearpage

\FloatBarrier
\subsection{$\nu=-3$, $\mathcal{C}=1$, valley polarized{}: $\prod_{\vec{k}} \hat{d}^\dagger_{\vec{k},+1,+,\uparrow} \ket{0}$}\label{app:sec:example:stateOne}

\begin{figure}[!h]
\centering
\includegraphics[width=0.95\textwidth]{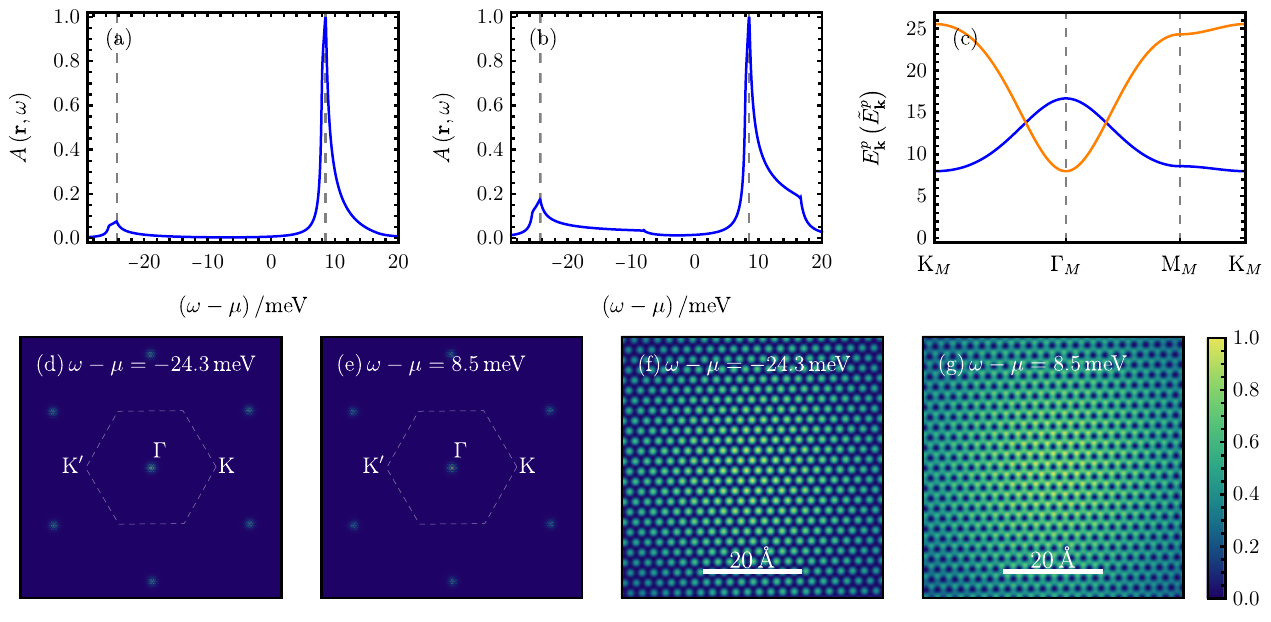}
\caption{{\it $\nu=-3$, $\mathcal{C}=1$, valley polarized{}}: The real-space spectral function averaged over three graphene unit cells at the AA site [(a)] and at the AB site [(b)]. The electron (blue) and hole (orange) dispersion is shown in (c). In (d)-(g), we illustrate the spatial variation of the spectral function in momentum space [(d) and (e)], as well as real space [(f) and (g)] for the two energy choices depicted by dashed gray lines in (a) and (b). Here, we focus on $\left(w_0/w_1, \lambda \right) = \left(0.0, 0 \right)$.}\label{app:fig:realspaceOneWZeroLambdaZero}
\end{figure}

\begin{figure}[!h]
\centering
\includegraphics[width=0.95\textwidth]{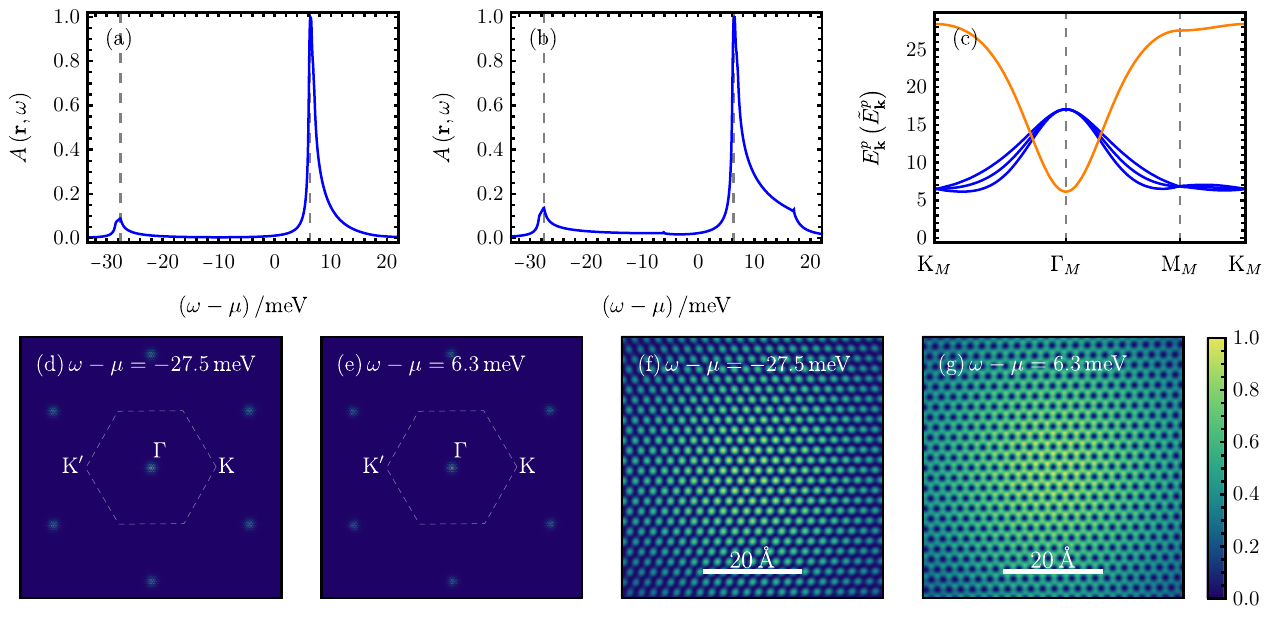}
\caption{{\it $\nu=-3$, $\mathcal{C}=1$, valley polarized{}}: The real-space spectral function averaged over three graphene unit cells at the AA site [(a)] and at the AB site [(b)]. The electron (blue) and hole (orange) dispersion is shown in (c). In (d)-(g), we illustrate the spatial variation of the spectral function in momentum space [(d) and (e)], as well as real space [(f) and (g)] for the two energy choices depicted by dashed gray lines in (a) and (b). Here, we focus on $\left(w_0/w_1, \lambda \right) = \left(0.4, 0 \right)$.}\label{app:fig:realspaceOneWFourLambdaZero}
\end{figure}

\begin{figure}[!h]
\centering
\includegraphics[width=0.95\textwidth]{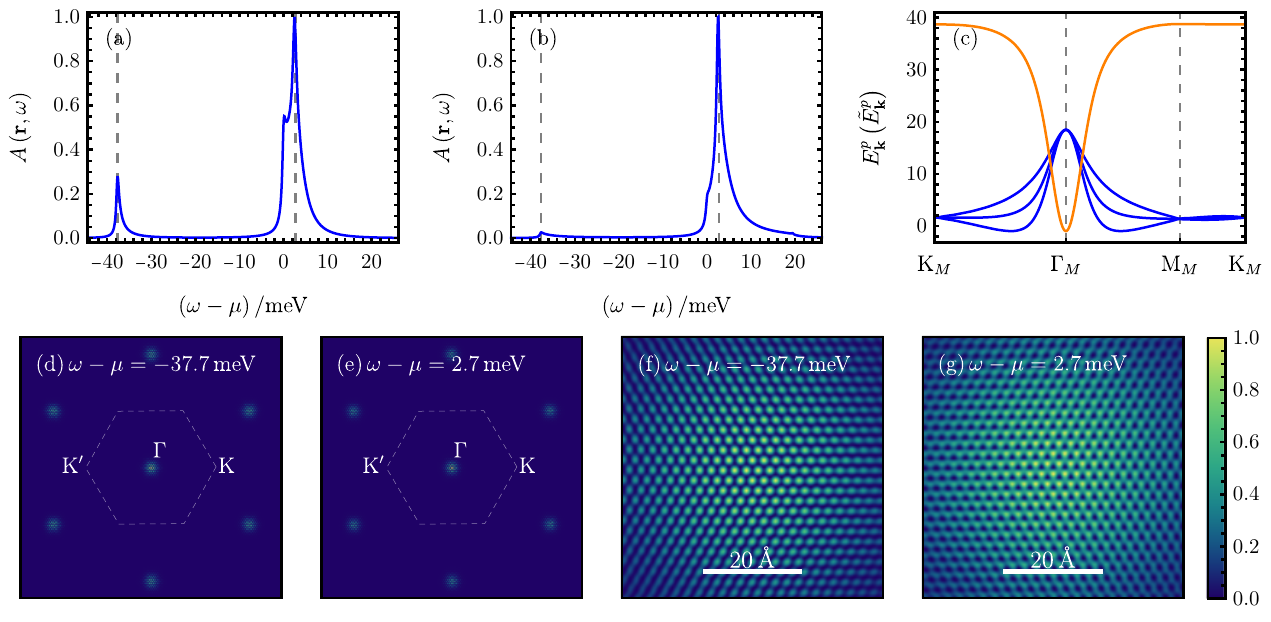}
\caption{{\it $\nu=-3$, $\mathcal{C}=1$, valley polarized{}}: The real-space spectral function averaged over three graphene unit cells at the AA site [(a)] and at the AB site [(b)]. The electron (blue) and hole (orange) dispersion is shown in (c). In (d)-(g), we illustrate the spatial variation of the spectral function in momentum space [(d) and (e)], as well as real space [(f) and (g)] for the two energy choices depicted by dashed gray lines in (a) and (b). Here, we focus on $\left(w_0/w_1, \lambda \right) = \left(0.8, 0 \right)$.}\label{app:fig:realspaceOneWEightLambdaZero}
\end{figure}

\begin{figure}[!h]
\centering
\includegraphics[width=0.95\textwidth]{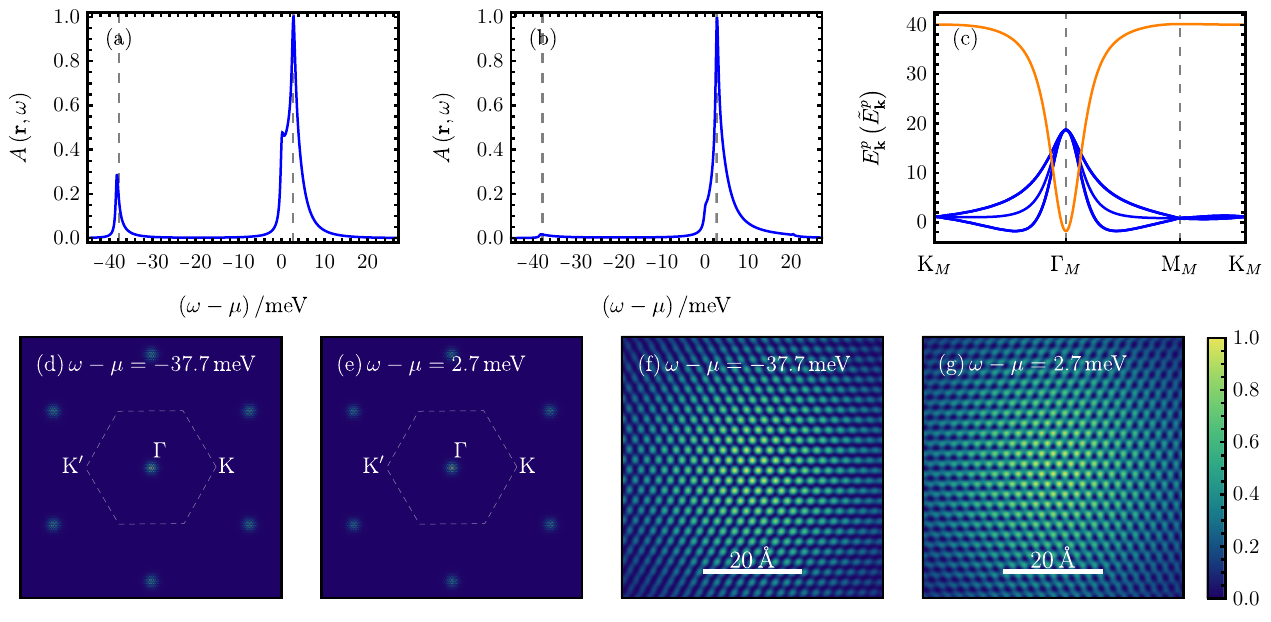}
\caption{{\it $\nu=-3$, $\mathcal{C}=1$, valley polarized{}}: The real-space spectral function averaged over three graphene unit cells at the AA site [(a)] and at the AB site [(b)]. The electron (blue) and hole (orange) dispersion is shown in (c). In (d)-(g), we illustrate the spatial variation of the spectral function in momentum space [(d) and (e)], as well as real space [(f) and (g)] for the two energy choices depicted by dashed gray lines in (a) and (b). Here, we focus on $\left(w_0/w_1, \lambda \right) = \left(0.8, 1 \right)$.}\label{app:fig:realspaceOneWEightLambdaOne}
\end{figure}

\newpage

\clearpage

\FloatBarrier
\subsection{$\nu=-3$, $\mathcal{C}=1$, valley coherent{}: $\prod_{\vec{k}} \frac{\hat{d}^\dagger_{\vec{k},+1,+,\uparrow} + \hat{d}^\dagger_{\vec{k},+1,-,\uparrow} }{\sqrt{2}} \ket{0}$}\label{app:sec:example:stateTwo}

\begin{figure}[!h]
\centering
\includegraphics[width=0.95\textwidth]{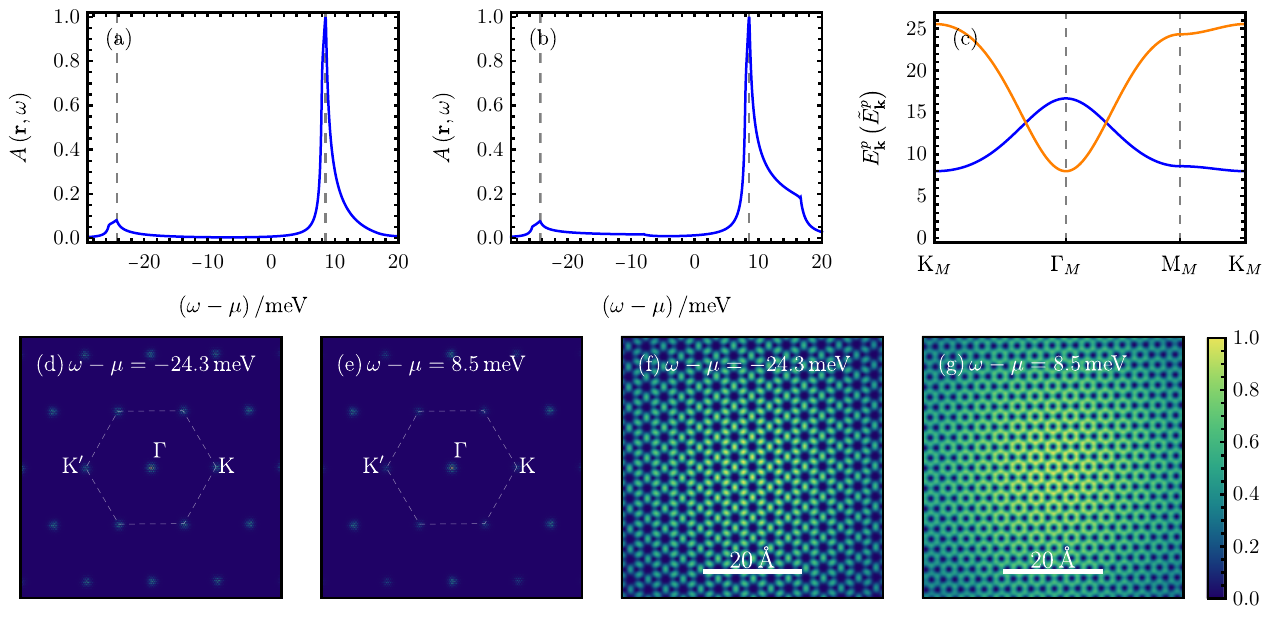}
\caption{{\it $\nu=-3$, $\mathcal{C}=1$, valley coherent{}}: The real-space spectral function averaged over three graphene unit cells at the AA site [(a)] and at the AB site [(b)]. The electron (blue) and hole (orange) dispersion is shown in (c). In (d)-(g), we illustrate the spatial variation of the spectral function in momentum space [(d) and (e)], as well as real space [(f) and (g)] for the two energy choices depicted by dashed gray lines in (a) and (b). Here, we focus on $\left(w_0/w_1, \lambda \right) = \left(0.0, 0 \right)$.}\label{app:fig:realspaceTwoWZeroLambdaZero}
\end{figure}

\begin{figure}[!h]
\centering
\includegraphics[width=0.95\textwidth]{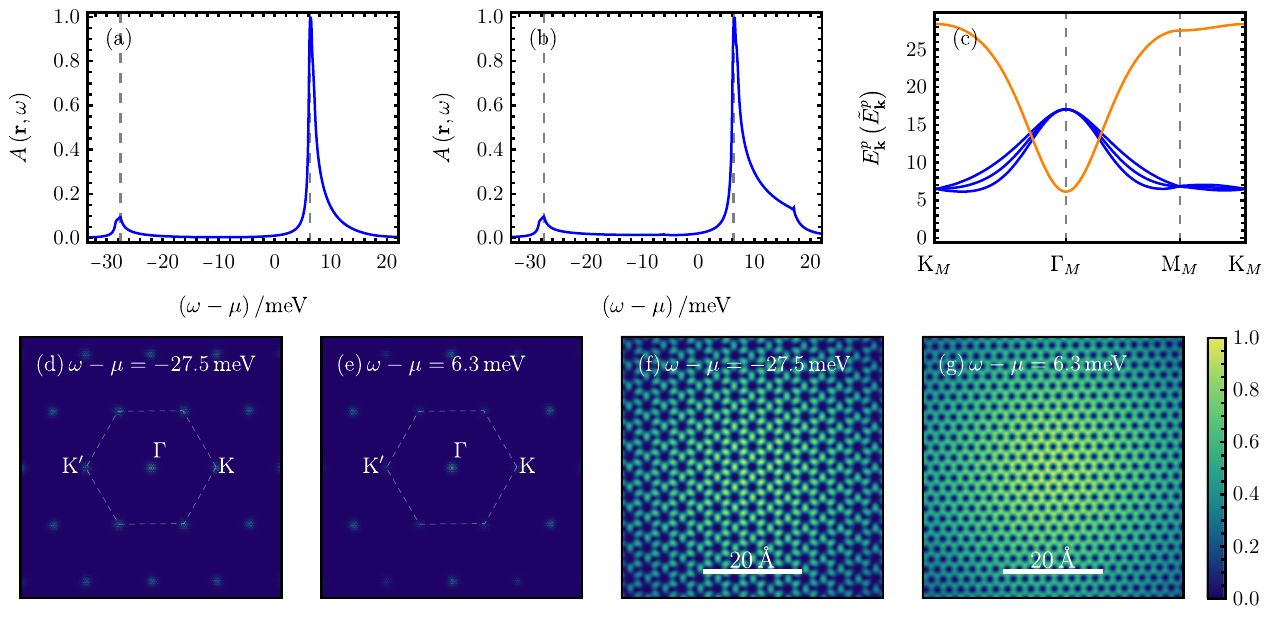}
\caption{{\it $\nu=-3$, $\mathcal{C}=1$, valley coherent{}}: The real-space spectral function averaged over three graphene unit cells at the AA site [(a)] and at the AB site [(b)]. The electron (blue) and hole (orange) dispersion is shown in (c). In (d)-(g), we illustrate the spatial variation of the spectral function in momentum space [(d) and (e)], as well as real space [(f) and (g)] for the two energy choices depicted by dashed gray lines in (a) and (b). Here, we focus on $\left(w_0/w_1, \lambda \right) = \left(0.4, 0 \right)$.}\label{app:fig:realspaceTwoWFourLambdaZero}
\end{figure}

\begin{figure}[!h]
\centering
\includegraphics[width=0.95\textwidth]{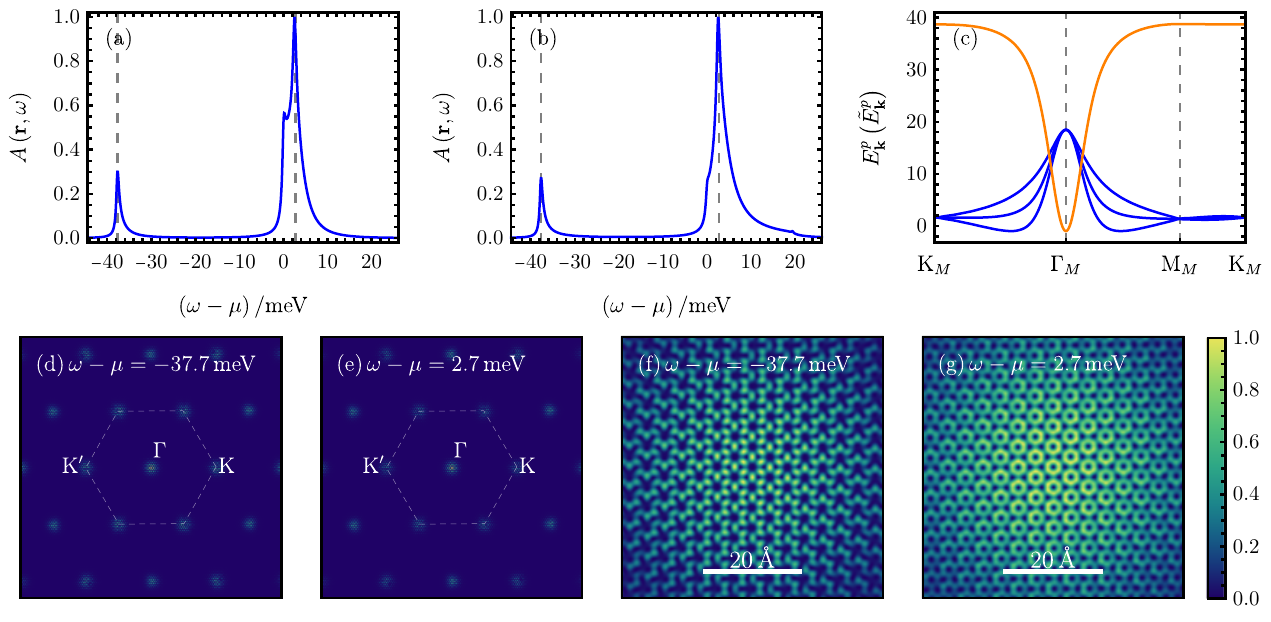}
\caption{{\it $\nu=-3$, $\mathcal{C}=1$, valley coherent{}}: The real-space spectral function averaged over three graphene unit cells at the AA site [(a)] and at the AB site [(b)]. The electron (blue) and hole (orange) dispersion is shown in (c). In (d)-(g), we illustrate the spatial variation of the spectral function in momentum space [(d) and (e)], as well as real space [(f) and (g)] for the two energy choices depicted by dashed gray lines in (a) and (b). Here, we focus on $\left(w_0/w_1, \lambda \right) = \left(0.8, 0 \right)$.}\label{app:fig:realspaceTwoWEightLambdaZero}
\end{figure}

\begin{figure}[!h]
\centering
\includegraphics[width=0.95\textwidth]{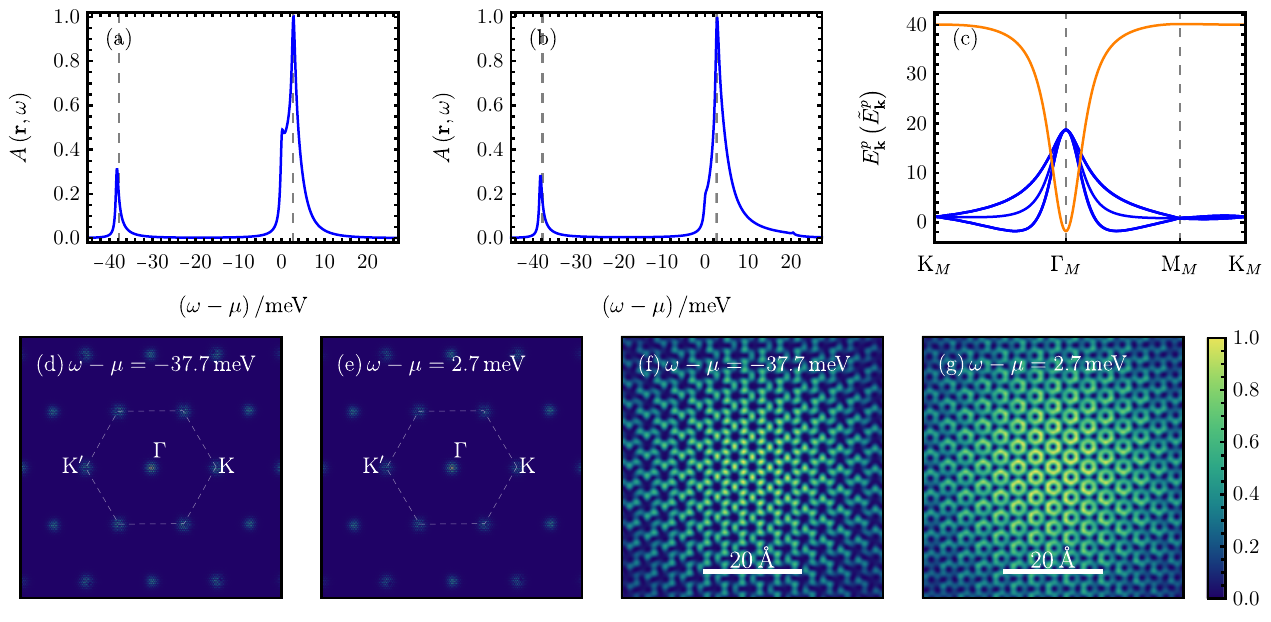}
\caption{{\it $\nu=-3$, $\mathcal{C}=1$, valley coherent{}}: The real-space spectral function averaged over three graphene unit cells at the AA site [(a)] and at the AB site [(b)]. The electron (blue) and hole (orange) dispersion is shown in (c). In (d)-(g), we illustrate the spatial variation of the spectral function in momentum space [(d) and (e)], as well as real space [(f) and (g)] for the two energy choices depicted by dashed gray lines in (a) and (b). Here, we focus on $\left(w_0/w_1, \lambda \right) = \left(0.8, 1 \right)$.}\label{app:fig:realspaceTwoWEightLambdaOne}
\end{figure}

\newpage

\clearpage

\FloatBarrier
\subsection{$\nu=-2$, $\mathcal{C}=0$, K-IVC{}: $\prod_{\vec{k}} \prod_{e_Y = \pm 1} \frac{\hat{d}^\dagger_{\vec{k},e_Y,+,\uparrow} + e_Y \hat{d}^\dagger_{\vec{k},e_Y,-,\uparrow} }{\sqrt{2}} \ket{0} $}\label{app:sec:example:stateThree}

\begin{figure}[!h]
\centering
\includegraphics[width=0.95\textwidth]{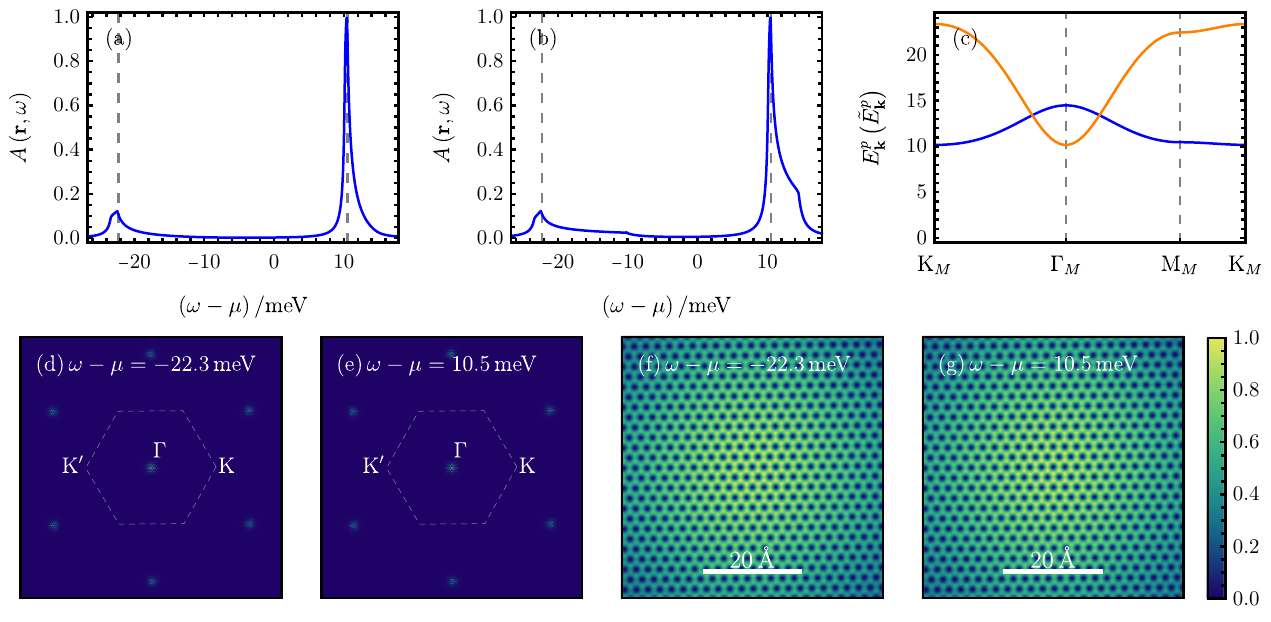}
\caption{{\it $\nu=-2$, $\mathcal{C}=0$, K-IVC{}}: The real-space spectral function averaged over three graphene unit cells at the AA site [(a)] and at the AB site [(b)]. The electron (blue) and hole (orange) dispersion is shown in (c). In (d)-(g), we illustrate the spatial variation of the spectral function in momentum space [(d) and (e)], as well as real space [(f) and (g)] for the two energy choices depicted by dashed gray lines in (a) and (b). Here, we focus on $\left(w_0/w_1, \lambda \right) = \left(0.0, 0 \right)$.}\label{app:fig:realspaceThreeWZeroLambdaZero}
\end{figure}

\begin{figure}[!h]
\centering
\includegraphics[width=0.95\textwidth]{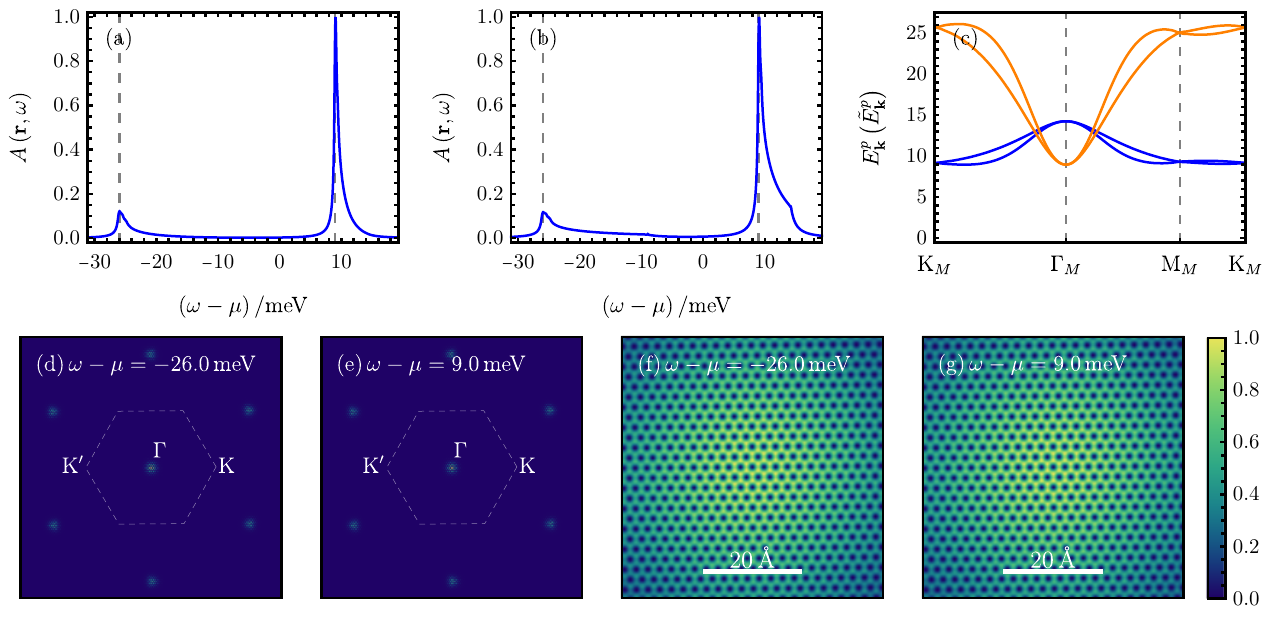}
\caption{{\it $\nu=-2$, $\mathcal{C}=0$, K-IVC{}}: The real-space spectral function averaged over three graphene unit cells at the AA site [(a)] and at the AB site [(b)]. The electron (blue) and hole (orange) dispersion is shown in (c). In (d)-(g), we illustrate the spatial variation of the spectral function in momentum space [(d) and (e)], as well as real space [(f) and (g)] for the two energy choices depicted by dashed gray lines in (a) and (b). Here, we focus on $\left(w_0/w_1, \lambda \right) = \left(0.4, 0 \right)$.}\label{app:fig:realspaceThreeWFourLambdaZero}
\end{figure}

\begin{figure}[!h]
\centering
\includegraphics[width=0.95\textwidth]{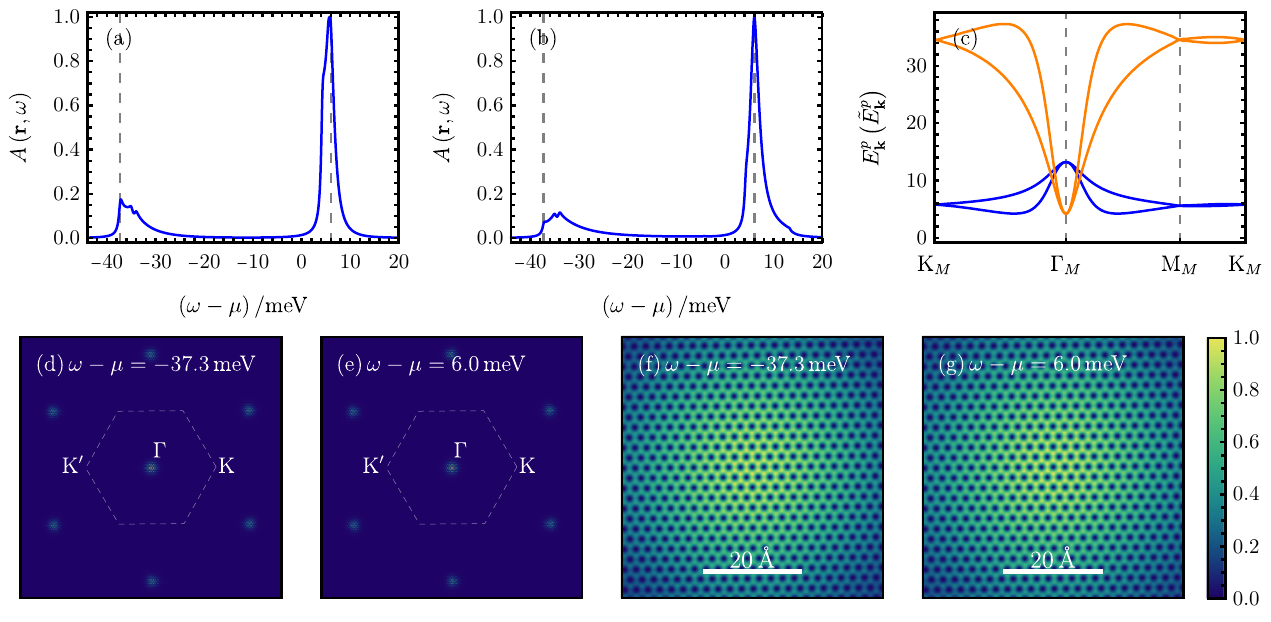}
\caption{{\it $\nu=-2$, $\mathcal{C}=0$, K-IVC{}}: The real-space spectral function averaged over three graphene unit cells at the AA site [(a)] and at the AB site [(b)]. The electron (blue) and hole (orange) dispersion is shown in (c). In (d)-(g), we illustrate the spatial variation of the spectral function in momentum space [(d) and (e)], as well as real space [(f) and (g)] for the two energy choices depicted by dashed gray lines in (a) and (b). Here, we focus on $\left(w_0/w_1, \lambda \right) = \left(0.8, 0 \right)$.}\label{app:fig:realspaceThreeWEightLambdaZero}
\end{figure}

\begin{figure}[!h]
\centering
\includegraphics[width=0.95\textwidth]{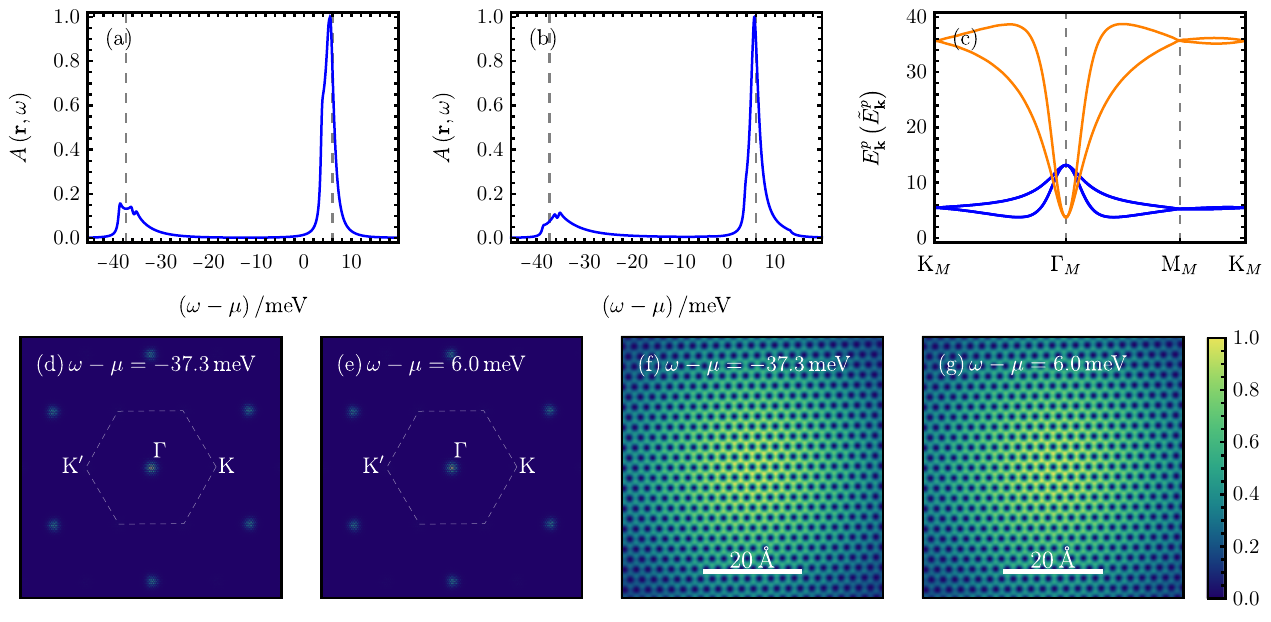}
\caption{{\it $\nu=-2$, $\mathcal{C}=0$, K-IVC{}}: The real-space spectral function averaged over three graphene unit cells at the AA site [(a)] and at the AB site [(b)]. The electron (blue) and hole (orange) dispersion is shown in (c). In (d)-(g), we illustrate the spatial variation of the spectral function in momentum space [(d) and (e)], as well as real space [(f) and (g)] for the two energy choices depicted by dashed gray lines in (a) and (b). Here, we focus on $\left(w_0/w_1, \lambda \right) = \left(0.8, 1 \right)$.}\label{app:fig:realspaceThreeWEightLambdaOne}
\end{figure}

\newpage

\clearpage

\FloatBarrier
\subsection{$\nu=-2$, $\mathcal{C}=0$, T-IVC{}: $\prod_{\vec{k}} \prod_{e_Y = \pm 1} \frac{\hat{d}^\dagger_{\vec{k},e_Y,+,\uparrow} + \hat{d}^\dagger_{\vec{k},e_Y,-,\uparrow} }{\sqrt{2}} \ket{0}$}\label{app:sec:example:stateFour}

\begin{figure}[!h]
\centering
\includegraphics[width=0.95\textwidth]{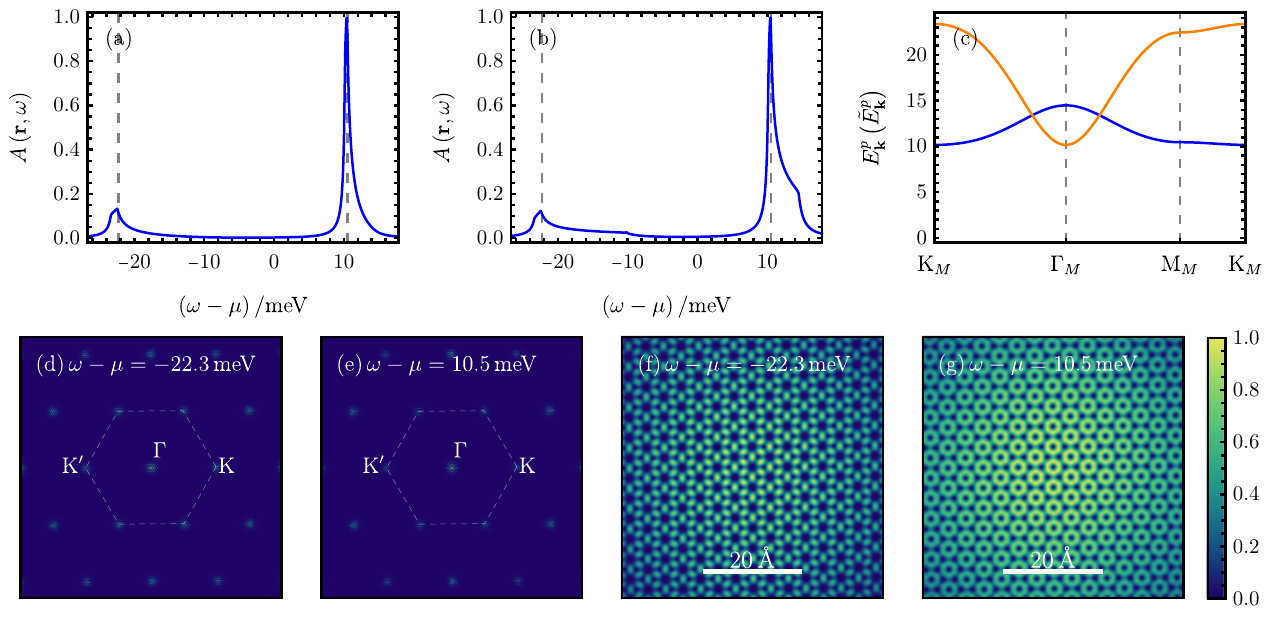}
\caption{{\it $\nu=-2$, $\mathcal{C}=0$, T-IVC{}}: The real-space spectral function averaged over three graphene unit cells at the AA site [(a)] and at the AB site [(b)]. The electron (blue) and hole (orange) dispersion is shown in (c). In (d)-(g), we illustrate the spatial variation of the spectral function in momentum space [(d) and (e)], as well as real space [(f) and (g)] for the two energy choices depicted by dashed gray lines in (a) and (b). Here, we focus on $\left(w_0/w_1, \lambda \right) = \left(0.0, 0 \right)$.}\label{app:fig:realspaceFourWZeroLambdaZero}
\end{figure}

\begin{figure}[!h]
\centering
\includegraphics[width=0.95\textwidth]{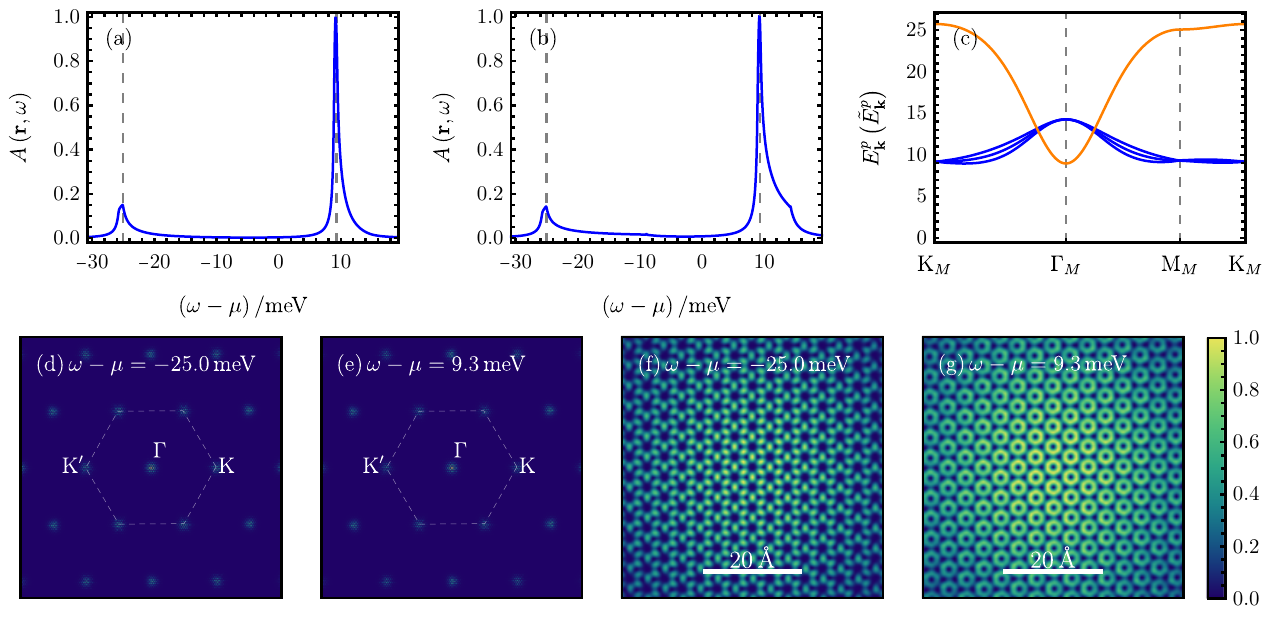}
\caption{{\it $\nu=-2$, $\mathcal{C}=0$, T-IVC{}}: The real-space spectral function averaged over three graphene unit cells at the AA site [(a)] and at the AB site [(b)]. The electron (blue) and hole (orange) dispersion is shown in (c). In (d)-(g), we illustrate the spatial variation of the spectral function in momentum space [(d) and (e)], as well as real space [(f) and (g)] for the two energy choices depicted by dashed gray lines in (a) and (b). Here, we focus on $\left(w_0/w_1, \lambda \right) = \left(0.4, 0 \right)$.}\label{app:fig:realspaceFourWFourLambdaZero}
\end{figure}

\begin{figure}[!h]
\centering
\includegraphics[width=0.95\textwidth]{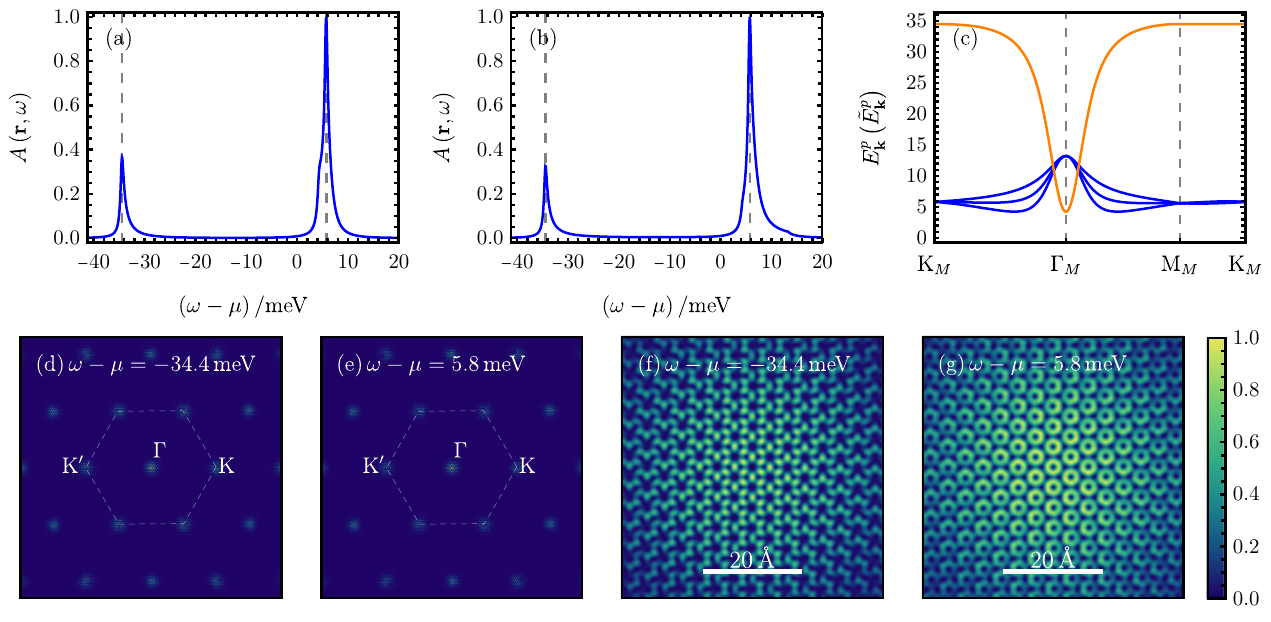}
\caption{{\it $\nu=-2$, $\mathcal{C}=0$, T-IVC{}}: The real-space spectral function averaged over three graphene unit cells at the AA site [(a)] and at the AB site [(b)]. The electron (blue) and hole (orange) dispersion is shown in (c). In (d)-(g), we illustrate the spatial variation of the spectral function in momentum space [(d) and (e)], as well as real space [(f) and (g)] for the two energy choices depicted by dashed gray lines in (a) and (b). Here, we focus on $\left(w_0/w_1, \lambda \right) = \left(0.8, 0 \right)$.}\label{app:fig:realspaceFourWEightLambdaZero}
\end{figure}

\begin{figure}[!h]
\centering
\includegraphics[width=0.95\textwidth]{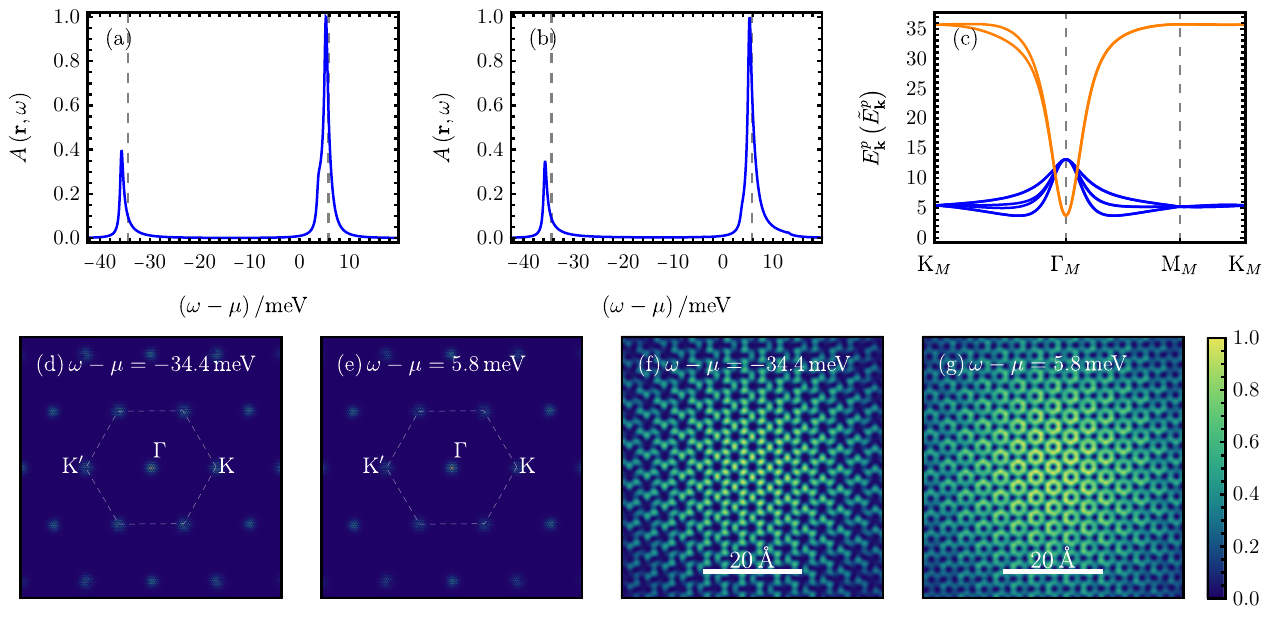}
\caption{{\it $\nu=-2$, $\mathcal{C}=0$, T-IVC{}}: The real-space spectral function averaged over three graphene unit cells at the AA site [(a)] and at the AB site [(b)]. The electron (blue) and hole (orange) dispersion is shown in (c). In (d)-(g), we illustrate the spatial variation of the spectral function in momentum space [(d) and (e)], as well as real space [(f) and (g)] for the two energy choices depicted by dashed gray lines in (a) and (b). Here, we focus on $\left(w_0/w_1, \lambda \right) = \left(0.8, 1 \right)$.}\label{app:fig:realspaceFourWEightLambdaOne}
\end{figure}

\newpage

\clearpage

\FloatBarrier
\subsection{$\nu=-2$, $\mathcal{C}=0$, valley polarized
{}: $\prod_{\vec{k}} \hat{d}^\dagger_{\vec{k},+1,+,\uparrow} \hat{d}^\dagger_{\vec{k},-1,+,\uparrow} \ket{0}$}\label{app:sec:example:stateFive}

\begin{figure}[!h]
\centering
\includegraphics[width=0.95\textwidth]{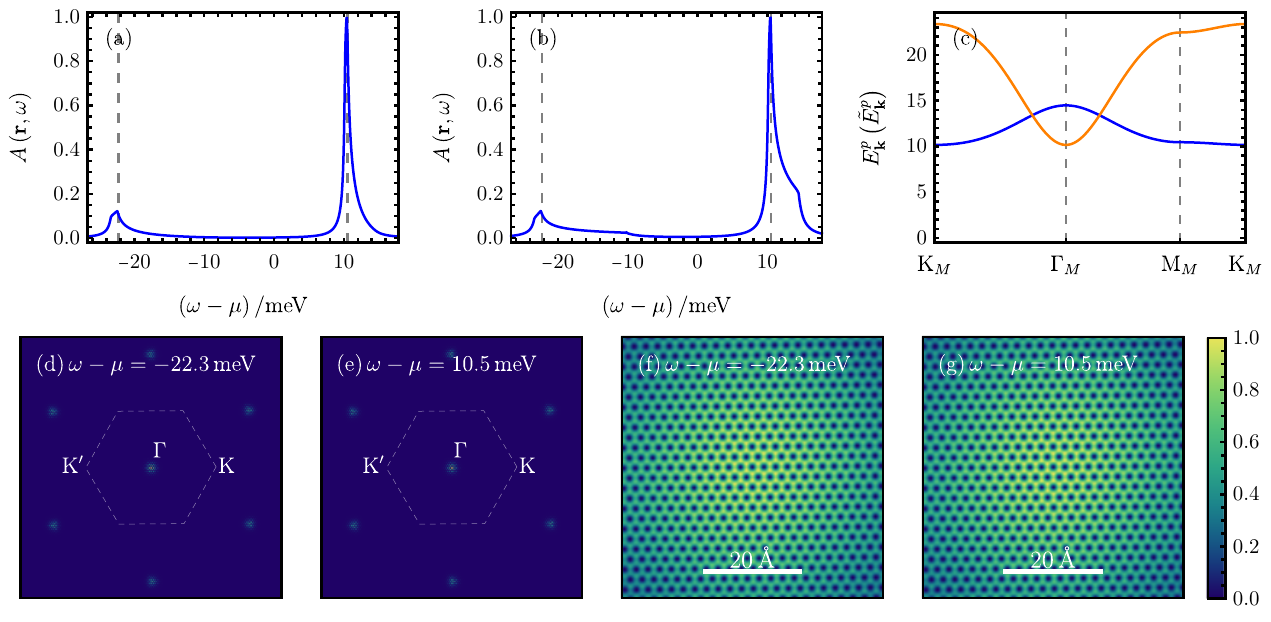}
\caption{{\it $\nu=-2$, $\mathcal{C}=0$, valley polarized
{}}: The real-space spectral function averaged over three graphene unit cells at the AA site [(a)] and at the AB site [(b)]. The electron (blue) and hole (orange) dispersion is shown in (c). In (d)-(g), we illustrate the spatial variation of the spectral function in momentum space [(d) and (e)], as well as real space [(f) and (g)] for the two energy choices depicted by dashed gray lines in (a) and (b). Here, we focus on $\left(w_0/w_1, \lambda \right) = \left(0.0, 0 \right)$.}\label{app:fig:realspaceFiveWZeroLambdaZero}
\end{figure}

\begin{figure}[!h]
\centering
\includegraphics[width=0.95\textwidth]{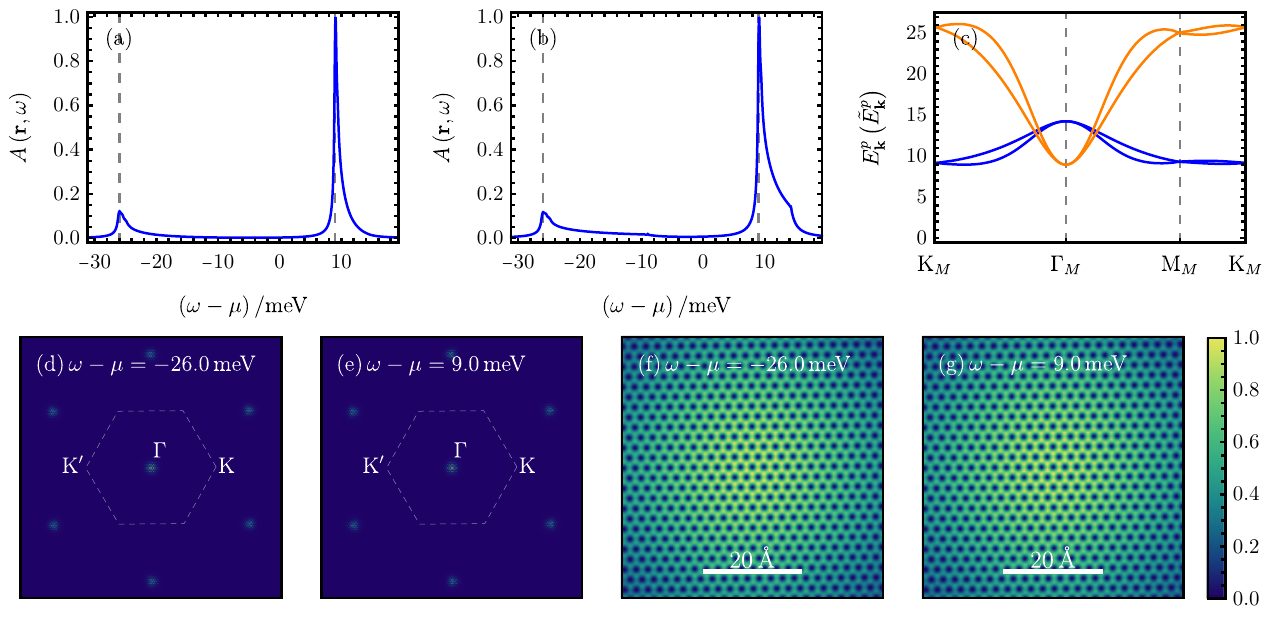}
\caption{{\it $\nu=-2$, $\mathcal{C}=0$, valley polarized
{}}: The real-space spectral function averaged over three graphene unit cells at the AA site [(a)] and at the AB site [(b)]. The electron (blue) and hole (orange) dispersion is shown in (c). In (d)-(g), we illustrate the spatial variation of the spectral function in momentum space [(d) and (e)], as well as real space [(f) and (g)] for the two energy choices depicted by dashed gray lines in (a) and (b). Here, we focus on $\left(w_0/w_1, \lambda \right) = \left(0.4, 0 \right)$.}\label{app:fig:realspaceFiveWFourLambdaZero}
\end{figure}

\begin{figure}[!h]
\centering
\includegraphics[width=0.95\textwidth]{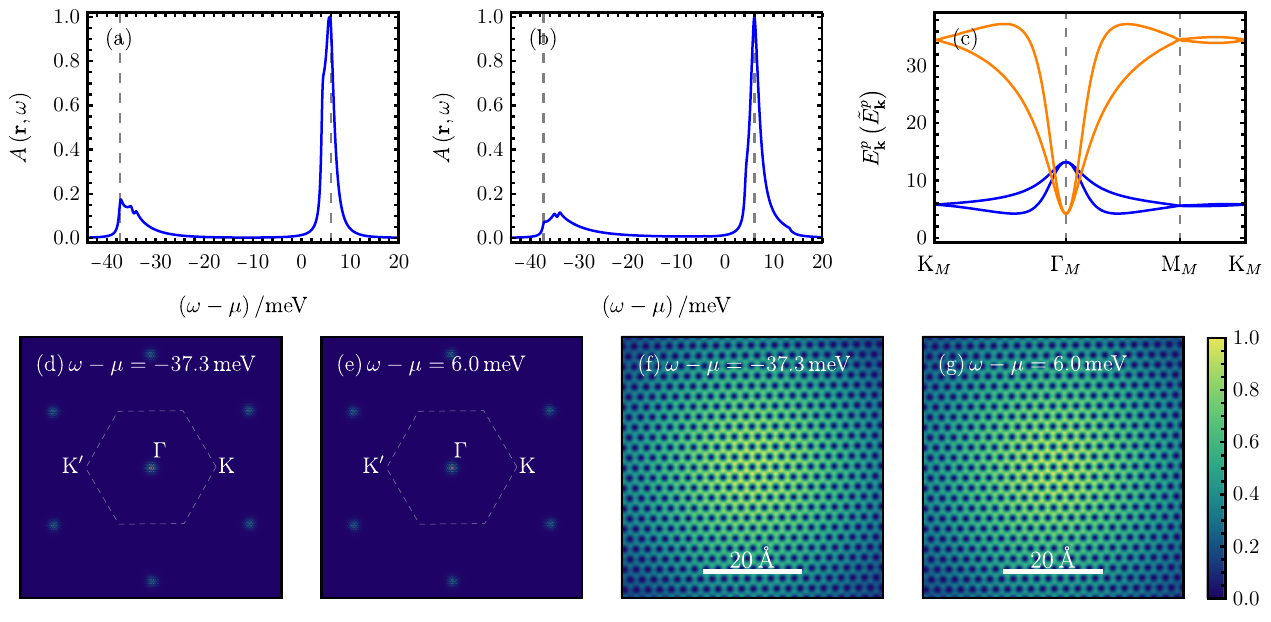}
\caption{{\it $\nu=-2$, $\mathcal{C}=0$, valley polarized
{}}: The real-space spectral function averaged over three graphene unit cells at the AA site [(a)] and at the AB site [(b)]. The electron (blue) and hole (orange) dispersion is shown in (c). In (d)-(g), we illustrate the spatial variation of the spectral function in momentum space [(d) and (e)], as well as real space [(f) and (g)] for the two energy choices depicted by dashed gray lines in (a) and (b). Here, we focus on $\left(w_0/w_1, \lambda \right) = \left(0.8, 0 \right)$.}\label{app:fig:realspaceFiveWEightLambdaZero}
\end{figure}

\begin{figure}[!h]
\centering
\includegraphics[width=0.95\textwidth]{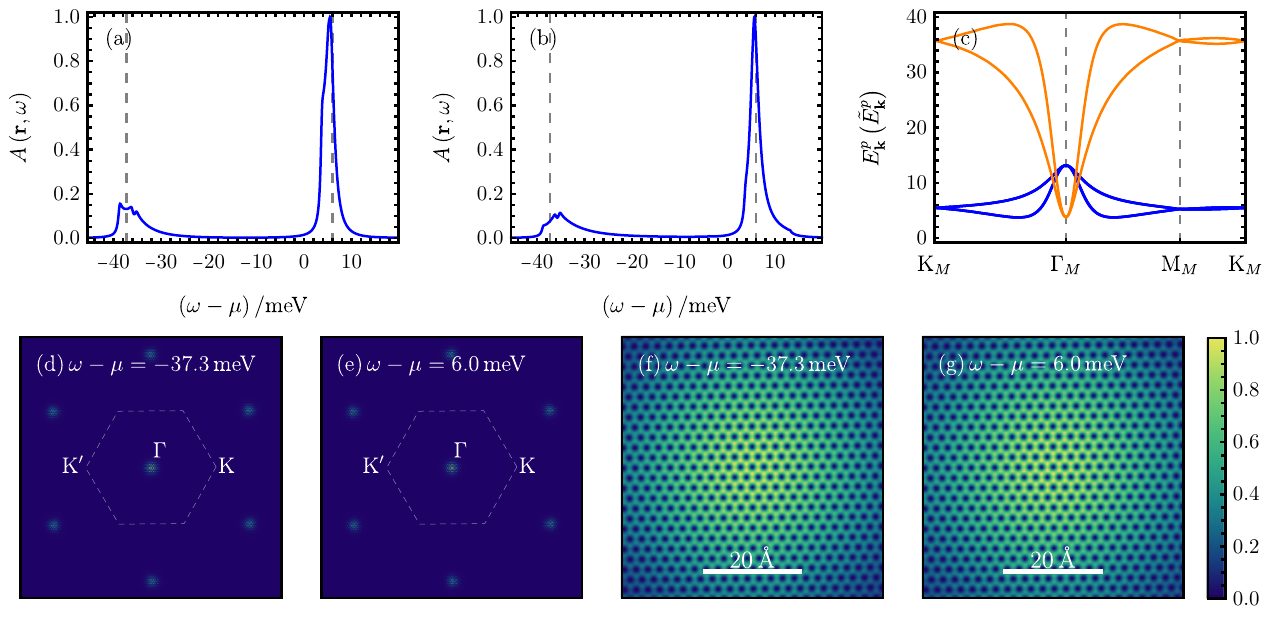}
\caption{{\it $\nu=-2$, $\mathcal{C}=0$, valley polarized
{}}: The real-space spectral function averaged over three graphene unit cells at the AA site [(a)] and at the AB site [(b)]. The electron (blue) and hole (orange) dispersion is shown in (c). In (d)-(g), we illustrate the spatial variation of the spectral function in momentum space [(d) and (e)], as well as real space [(f) and (g)] for the two energy choices depicted by dashed gray lines in (a) and (b). Here, we focus on $\left(w_0/w_1, \lambda \right) = \left(0.8, 1 \right)$.}\label{app:fig:realspaceFiveWEightLambdaOne}
\end{figure}

\newpage

\clearpage

\FloatBarrier
\subsection{$\nu=-2$, $\mathcal{C}=2$, valley polarized
{}: $\prod_{\vec{k}} \hat{d}^\dagger_{\vec{k},+1,+,\uparrow} \hat{d}^\dagger_{\vec{k},+1,+,\downarrow} \ket{0}$}\label{app:sec:example:stateSix}

\begin{figure}[!h]
\centering
\includegraphics[width=0.95\textwidth]{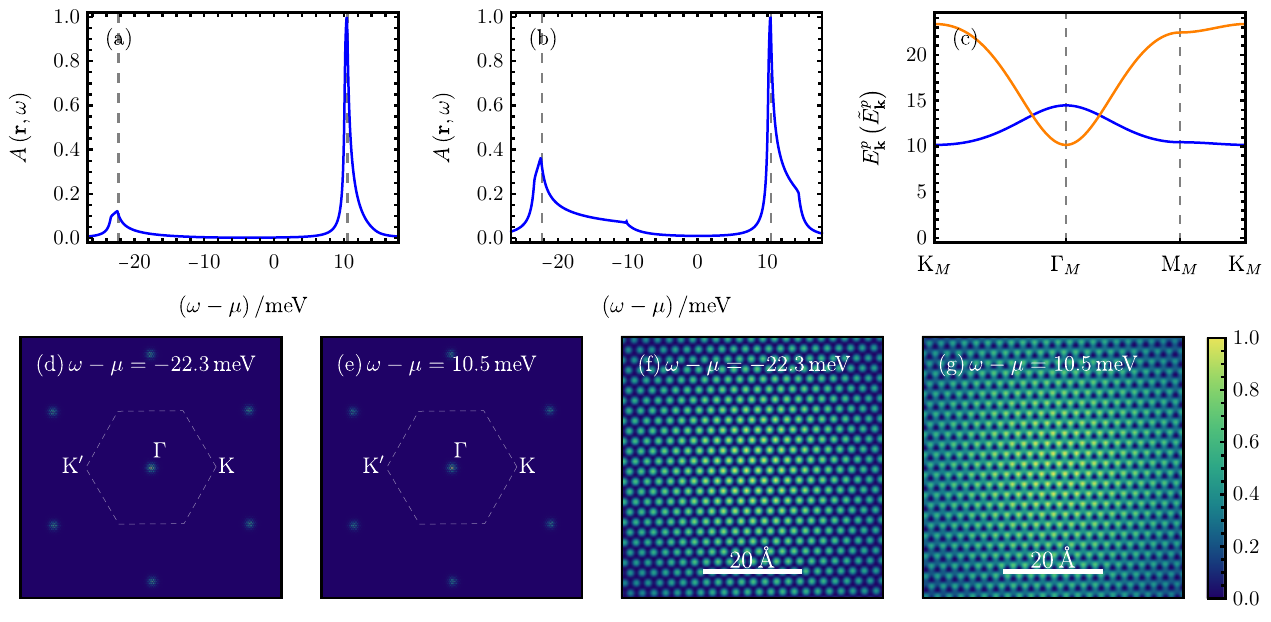}
\caption{{\it $\nu=-2$, $\mathcal{C}=2$, valley polarized
{}}: The real-space spectral function averaged over three graphene unit cells at the AA site [(a)] and at the AB site [(b)]. The electron (blue) and hole (orange) dispersion is shown in (c). In (d)-(g), we illustrate the spatial variation of the spectral function in momentum space [(d) and (e)], as well as real space [(f) and (g)] for the two energy choices depicted by dashed gray lines in (a) and (b). Here, we focus on $\left(w_0/w_1, \lambda \right) = \left(0.0, 0 \right)$.}\label{app:fig:realspaceSixWZeroLambdaZero}
\end{figure}

\begin{figure}[!h]
\centering
\includegraphics[width=0.95\textwidth]{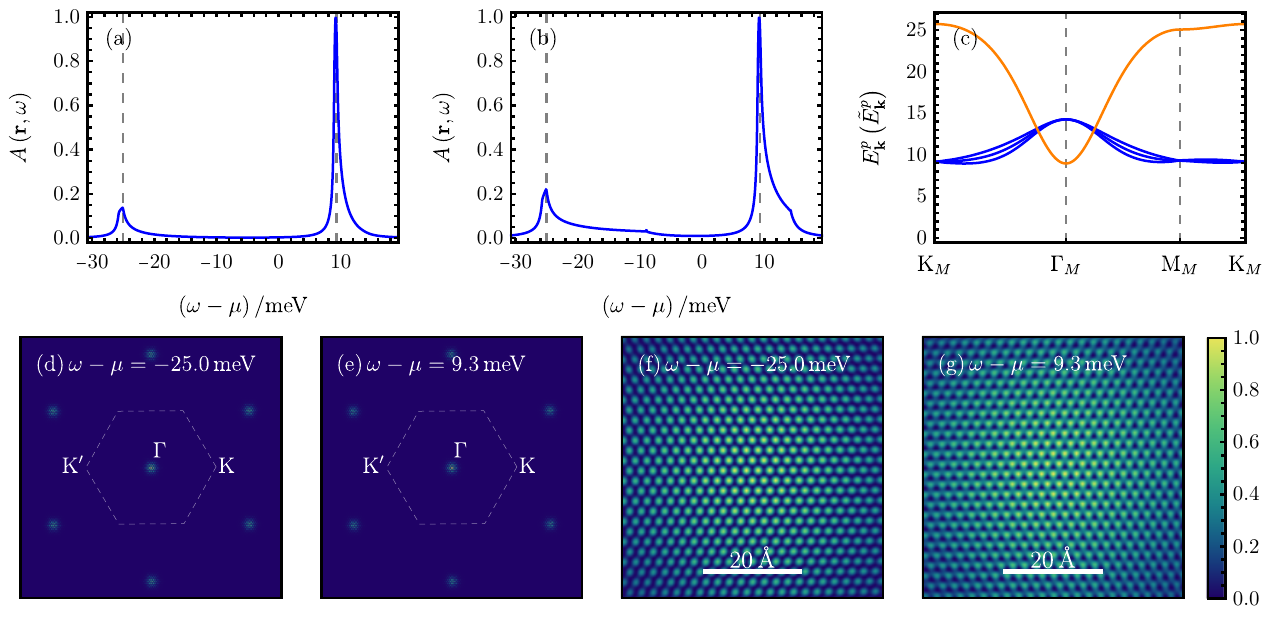}
\caption{{\it $\nu=-2$, $\mathcal{C}=2$, valley polarized
{}}: The real-space spectral function averaged over three graphene unit cells at the AA site [(a)] and at the AB site [(b)]. The electron (blue) and hole (orange) dispersion is shown in (c). In (d)-(g), we illustrate the spatial variation of the spectral function in momentum space [(d) and (e)], as well as real space [(f) and (g)] for the two energy choices depicted by dashed gray lines in (a) and (b). Here, we focus on $\left(w_0/w_1, \lambda \right) = \left(0.4, 0 \right)$.}\label{app:fig:realspaceSixWFourLambdaZero}
\end{figure}

\begin{figure}[!h]
\centering
\includegraphics[width=0.95\textwidth]{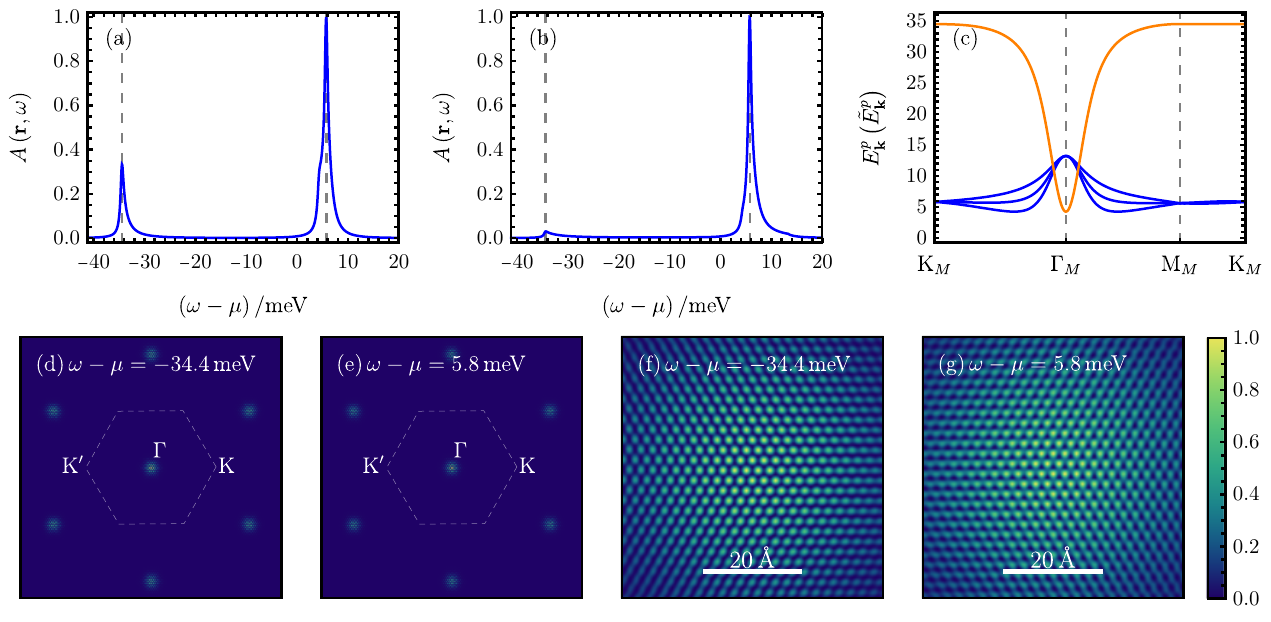}
\caption{{\it $\nu=-2$, $\mathcal{C}=2$, valley polarized
{}}: The real-space spectral function averaged over three graphene unit cells at the AA site [(a)] and at the AB site [(b)]. The electron (blue) and hole (orange) dispersion is shown in (c). In (d)-(g), we illustrate the spatial variation of the spectral function in momentum space [(d) and (e)], as well as real space [(f) and (g)] for the two energy choices depicted by dashed gray lines in (a) and (b). Here, we focus on $\left(w_0/w_1, \lambda \right) = \left(0.8, 0 \right)$.}\label{app:fig:realspaceSixWEightLambdaZero}
\end{figure}

\begin{figure}[!h]
\centering
\includegraphics[width=0.95\textwidth]{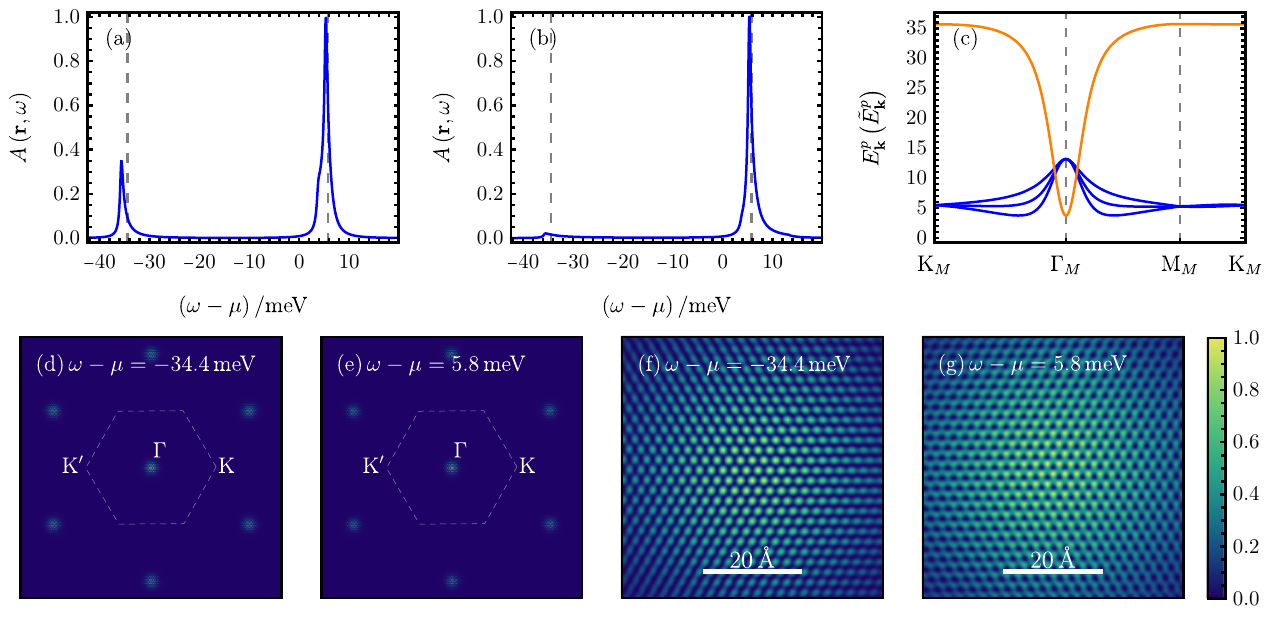}
\caption{{\it $\nu=-2$, $\mathcal{C}=2$, valley polarized
{}}: The real-space spectral function averaged over three graphene unit cells at the AA site [(a)] and at the AB site [(b)]. The electron (blue) and hole (orange) dispersion is shown in (c). In (d)-(g), we illustrate the spatial variation of the spectral function in momentum space [(d) and (e)], as well as real space [(f) and (g)] for the two energy choices depicted by dashed gray lines in (a) and (b). Here, we focus on $\left(w_0/w_1, \lambda \right) = \left(0.8, 1 \right)$.}\label{app:fig:realspaceSixWEightLambdaOne}
\end{figure}

\newpage

\clearpage

\FloatBarrier
\subsection{$\nu=-2$, $\mathcal{C}=2$, valley coherent
{}: $\prod_{\vec{k}} \prod_{s=\uparrow,\downarrow} \frac{\hat{d}^\dagger_{\vec{k},+1,+,s} + \hat{d}^\dagger_{\vec{k},+1,-,s} }{\sqrt{2}} \ket{0}$}\label{app:sec:example:stateSeven}

\begin{figure}[!h]
\centering
\includegraphics[width=0.95\textwidth]{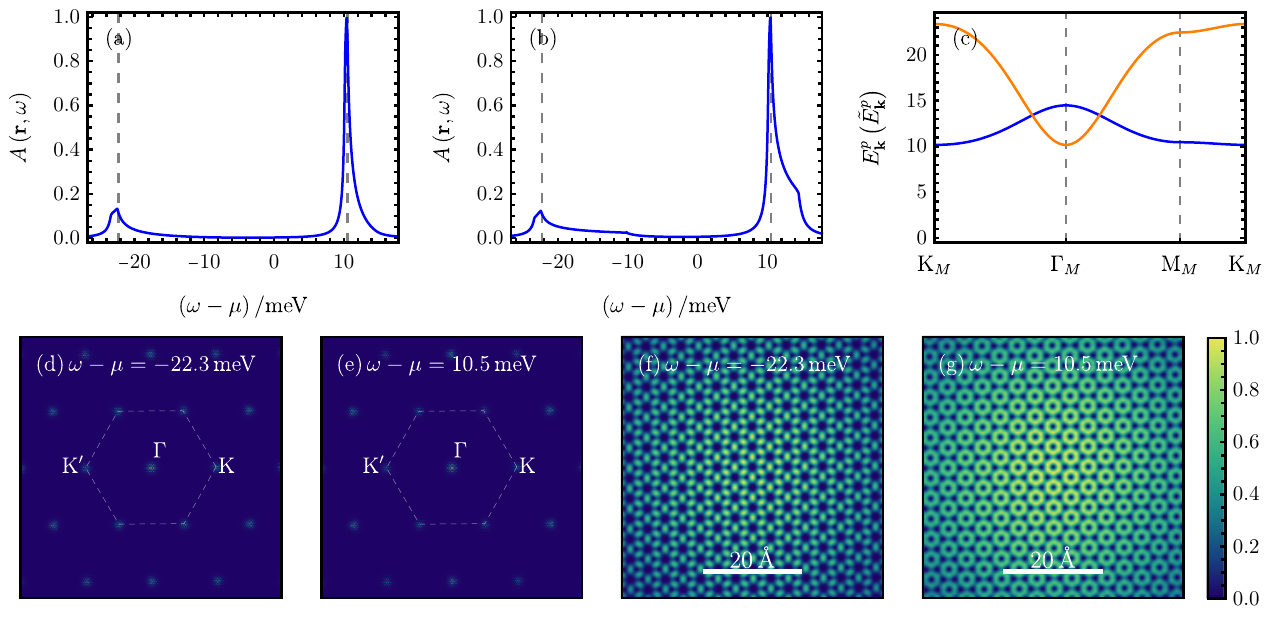}
\caption{{\it $\nu=-2$, $\mathcal{C}=2$, valley coherent
{}}: The real-space spectral function averaged over three graphene unit cells at the AA site [(a)] and at the AB site [(b)]. The electron (blue) and hole (orange) dispersion is shown in (c). In (d)-(g), we illustrate the spatial variation of the spectral function in momentum space [(d) and (e)], as well as real space [(f) and (g)] for the two energy choices depicted by dashed gray lines in (a) and (b). Here, we focus on $\left(w_0/w_1, \lambda \right) = \left(0.0, 0 \right)$.}\label{app:fig:realspaceSevenWZeroLambdaZero}
\end{figure}

\begin{figure}[!h]
\centering
\includegraphics[width=0.95\textwidth]{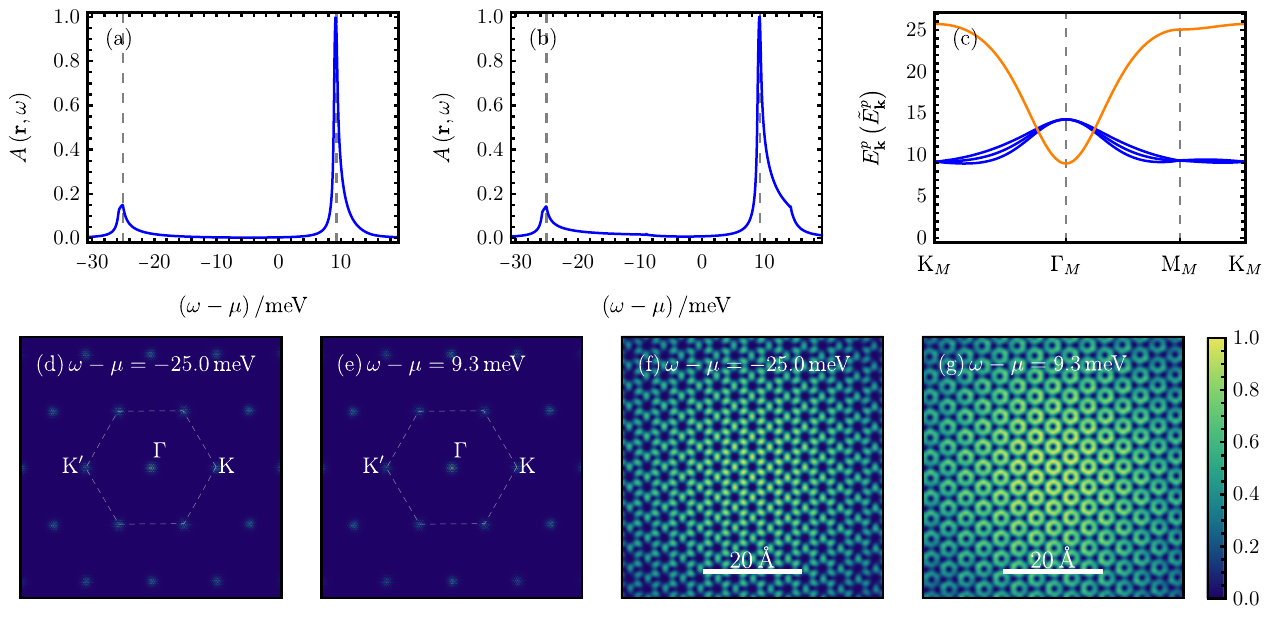}
\caption{{\it $\nu=-2$, $\mathcal{C}=2$, valley coherent
{}}: The real-space spectral function averaged over three graphene unit cells at the AA site [(a)] and at the AB site [(b)]. The electron (blue) and hole (orange) dispersion is shown in (c). In (d)-(g), we illustrate the spatial variation of the spectral function in momentum space [(d) and (e)], as well as real space [(f) and (g)] for the two energy choices depicted by dashed gray lines in (a) and (b). Here, we focus on $\left(w_0/w_1, \lambda \right) = \left(0.4, 0 \right)$.}\label{app:fig:realspaceSevenWFourLambdaZero}
\end{figure}

\begin{figure}[!h]
\centering
\includegraphics[width=0.95\textwidth]{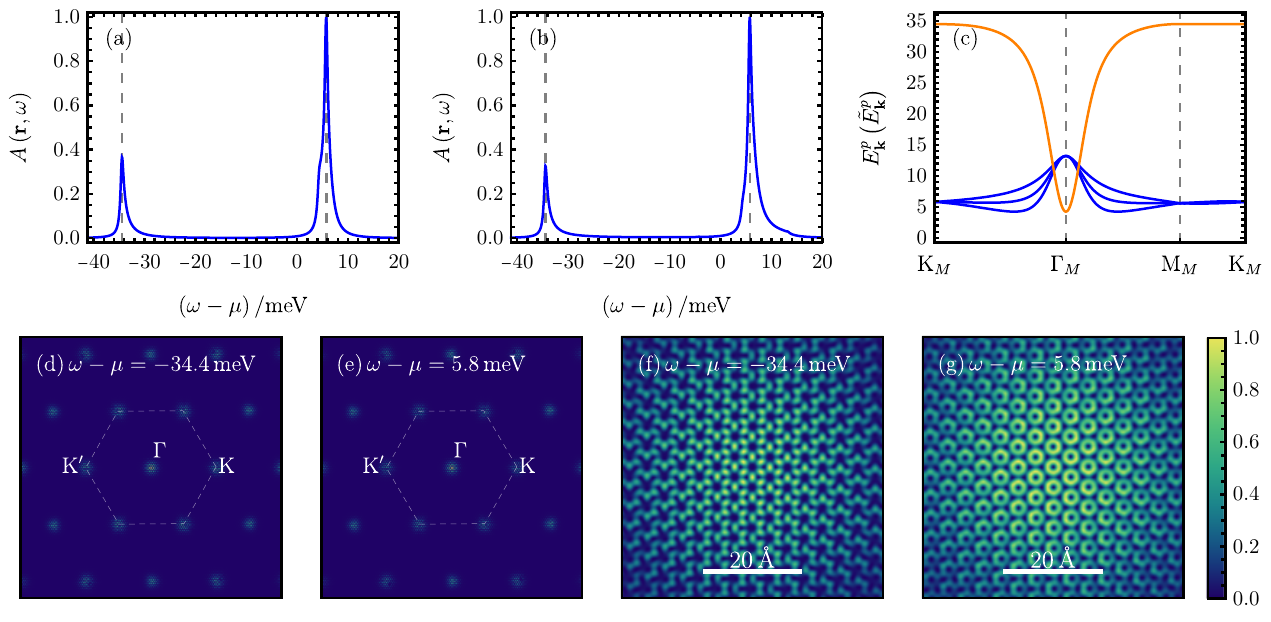}
\caption{{\it $\nu=-2$, $\mathcal{C}=2$, valley coherent
{}}: The real-space spectral function averaged over three graphene unit cells at the AA site [(a)] and at the AB site [(b)]. The electron (blue) and hole (orange) dispersion is shown in (c). In (d)-(g), we illustrate the spatial variation of the spectral function in momentum space [(d) and (e)], as well as real space [(f) and (g)] for the two energy choices depicted by dashed gray lines in (a) and (b). Here, we focus on $\left(w_0/w_1, \lambda \right) = \left(0.8, 0 \right)$.}\label{app:fig:realspaceSevenWEightLambdaZero}
\end{figure}

\begin{figure}[!h]
\centering
\includegraphics[width=0.95\textwidth]{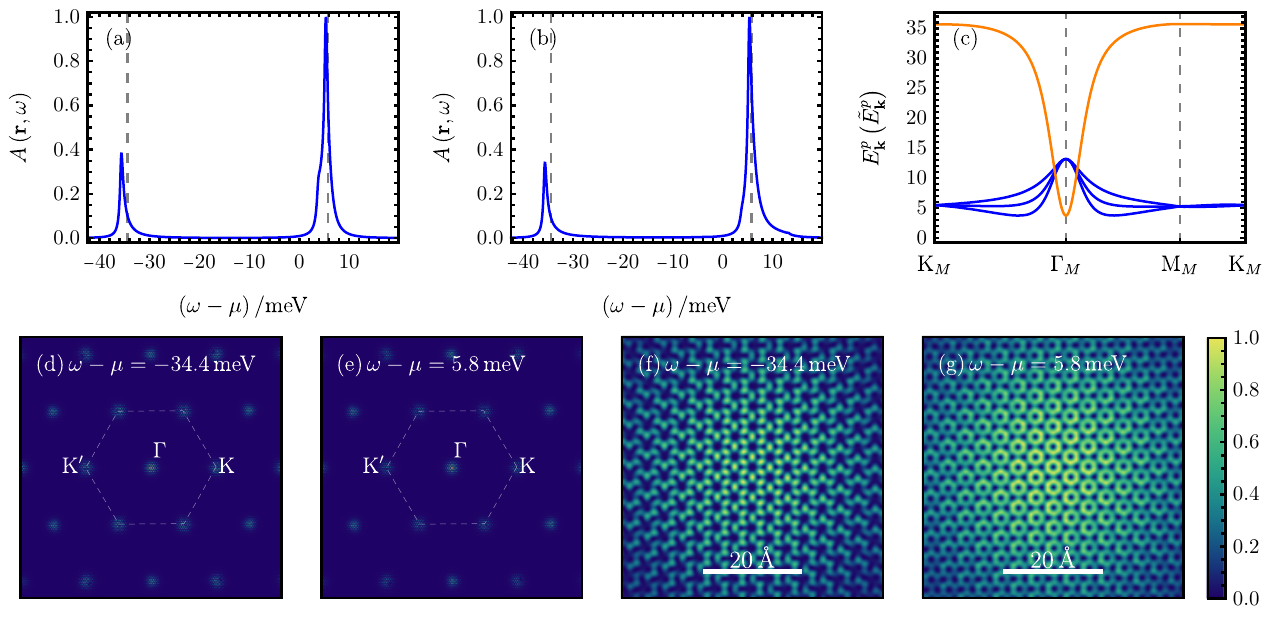}
\caption{{\it $\nu=-2$, $\mathcal{C}=2$, valley coherent
{}}: The real-space spectral function averaged over three graphene unit cells at the AA site [(a)] and at the AB site [(b)]. The electron (blue) and hole (orange) dispersion is shown in (c). In (d)-(g), we illustrate the spatial variation of the spectral function in momentum space [(d) and (e)], as well as real space [(f) and (g)] for the two energy choices depicted by dashed gray lines in (a) and (b). Here, we focus on $\left(w_0/w_1, \lambda \right) = \left(0.8, 1 \right)$.}\label{app:fig:realspaceSevenWEightLambdaOne}
\end{figure}

\newpage

\clearpage

\FloatBarrier
\subsection{$\nu=-1$, $\mathcal{C}=3$, valley polarized
{}: $\prod_{\vec{k}} \hat{d}^\dagger_{\vec{k},+1,+,\uparrow} \hat{d}^\dagger_{\vec{k},+1,+,\downarrow} \hat{d}^\dagger_{\vec{k},+1,-,\downarrow} \ket{0}$}\label{app:sec:example:stateEight}

\begin{figure}[!h]
\centering
\includegraphics[width=0.95\textwidth]{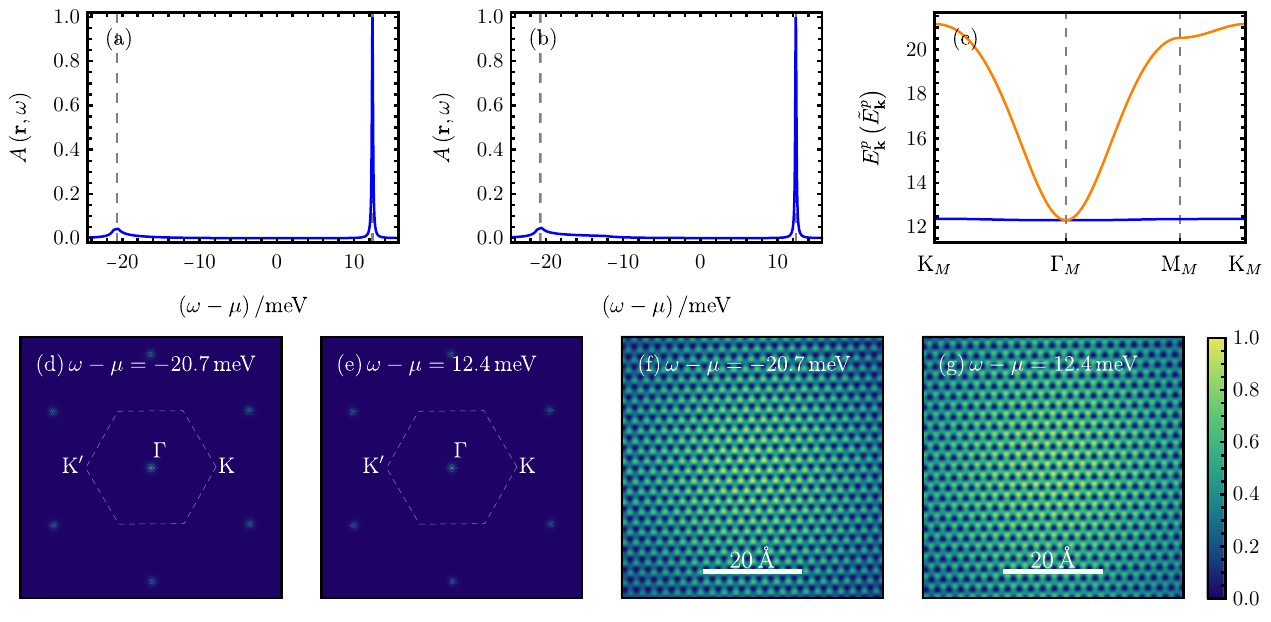}
\caption{{\it $\nu=-1$, $\mathcal{C}=3$, valley polarized
{}}: The real-space spectral function averaged over three graphene unit cells at the AA site [(a)] and at the AB site [(b)]. The electron (blue) and hole (orange) dispersion is shown in (c). In (d)-(g), we illustrate the spatial variation of the spectral function in momentum space [(d) and (e)], as well as real space [(f) and (g)] for the two energy choices depicted by dashed gray lines in (a) and (b). Here, we focus on $\left(w_0/w_1, \lambda \right) = \left(0.0, 0 \right)$.}\label{app:fig:realspaceEightWZeroLambdaZero}
\end{figure}

\begin{figure}[!h]
\centering
\includegraphics[width=0.95\textwidth]{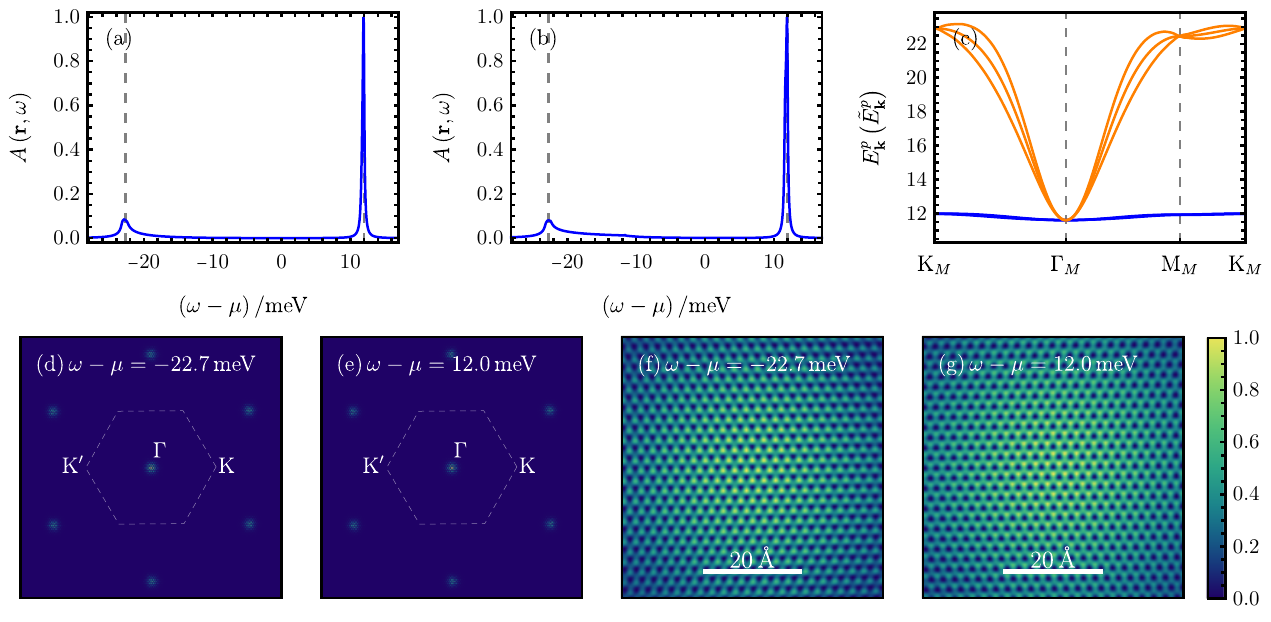}
\caption{{\it $\nu=-1$, $\mathcal{C}=3$, valley polarized
{}}: The real-space spectral function averaged over three graphene unit cells at the AA site [(a)] and at the AB site [(b)]. The electron (blue) and hole (orange) dispersion is shown in (c). In (d)-(g), we illustrate the spatial variation of the spectral function in momentum space [(d) and (e)], as well as real space [(f) and (g)] for the two energy choices depicted by dashed gray lines in (a) and (b). Here, we focus on $\left(w_0/w_1, \lambda \right) = \left(0.4, 0 \right)$.}\label{app:fig:realspaceEightWFourLambdaZero}
\end{figure}

\begin{figure}[!h]
\centering
\includegraphics[width=0.95\textwidth]{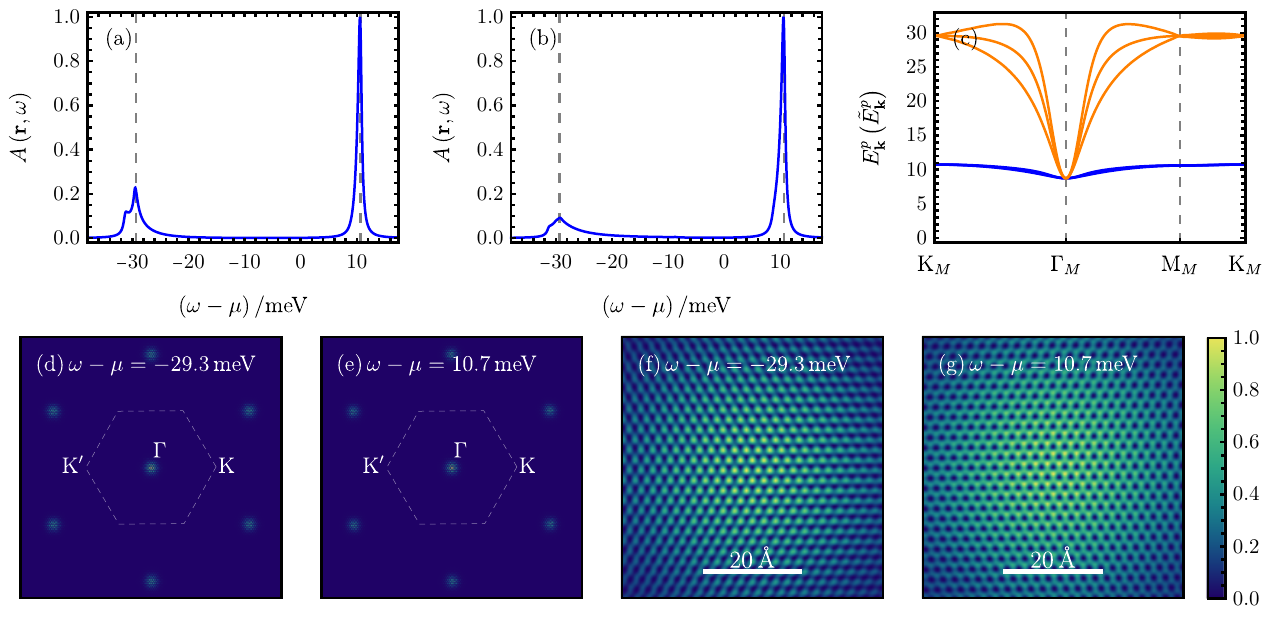}
\caption{{\it $\nu=-1$, $\mathcal{C}=3$, valley polarized
{}}: The real-space spectral function averaged over three graphene unit cells at the AA site [(a)] and at the AB site [(b)]. The electron (blue) and hole (orange) dispersion is shown in (c). In (d)-(g), we illustrate the spatial variation of the spectral function in momentum space [(d) and (e)], as well as real space [(f) and (g)] for the two energy choices depicted by dashed gray lines in (a) and (b). Here, we focus on $\left(w_0/w_1, \lambda \right) = \left(0.8, 0 \right)$.}\label{app:fig:realspaceEightWEightLambdaZero}
\end{figure}

\begin{figure}[!h]
\centering
\includegraphics[width=0.95\textwidth]{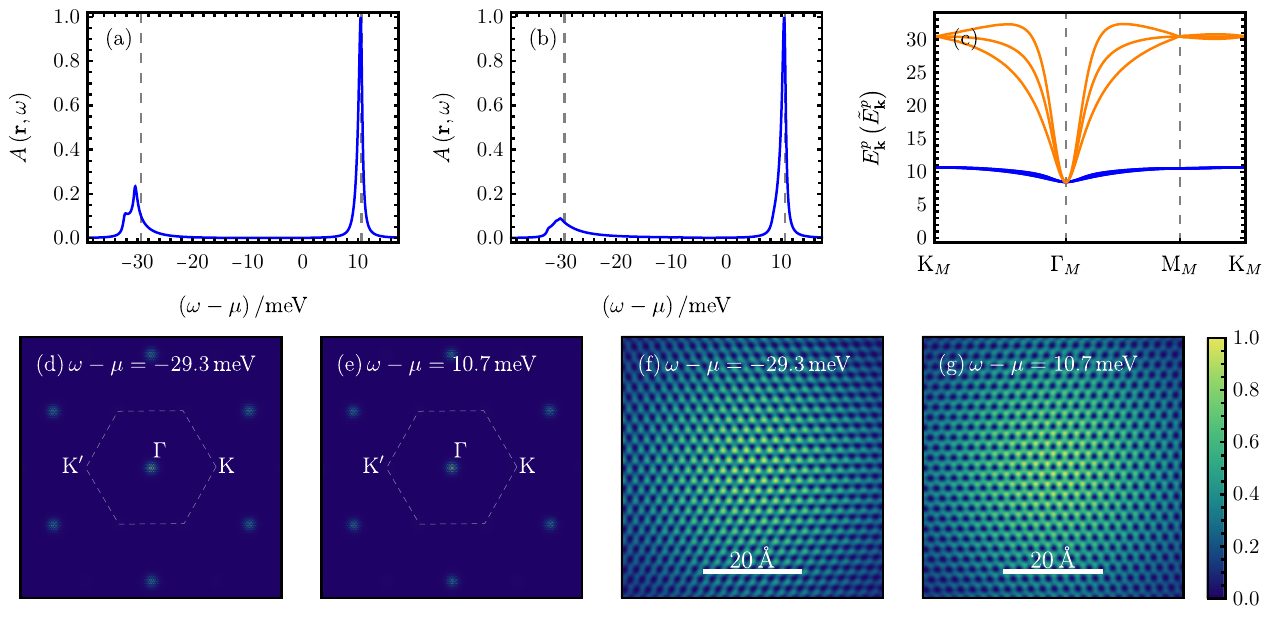}
\caption{{\it $\nu=-1$, $\mathcal{C}=3$, valley polarized
{}}: The real-space spectral function averaged over three graphene unit cells at the AA site [(a)] and at the AB site [(b)]. The electron (blue) and hole (orange) dispersion is shown in (c). In (d)-(g), we illustrate the spatial variation of the spectral function in momentum space [(d) and (e)], as well as real space [(f) and (g)] for the two energy choices depicted by dashed gray lines in (a) and (b). Here, we focus on $\left(w_0/w_1, \lambda \right) = \left(0.8, 1 \right)$.}\label{app:fig:realspaceEightWEightLambdaOne}
\end{figure}

\newpage

\clearpage

\FloatBarrier
\subsection{$\nu=-1$, $\mathcal{C}=1$, two occupied K-IVC bands and a valley polarized band{}: $\prod_{\vec{k}} \hat{d}^\dagger_{\vec{k},+1,-,\downarrow} \prod_{e_Y = \pm 1} \frac{\hat{d}^\dagger_{\vec{k},e_Y,+,\uparrow} + e_Y \hat{d}^\dagger_{\vec{k},e_Y,-,\uparrow} }{\sqrt{2}} \ket{0}$}\label{app:sec:example:stateNine}
\vspace{-2 em}
\begin{figure}[!h]
\centering
\includegraphics[width=0.95\textwidth]{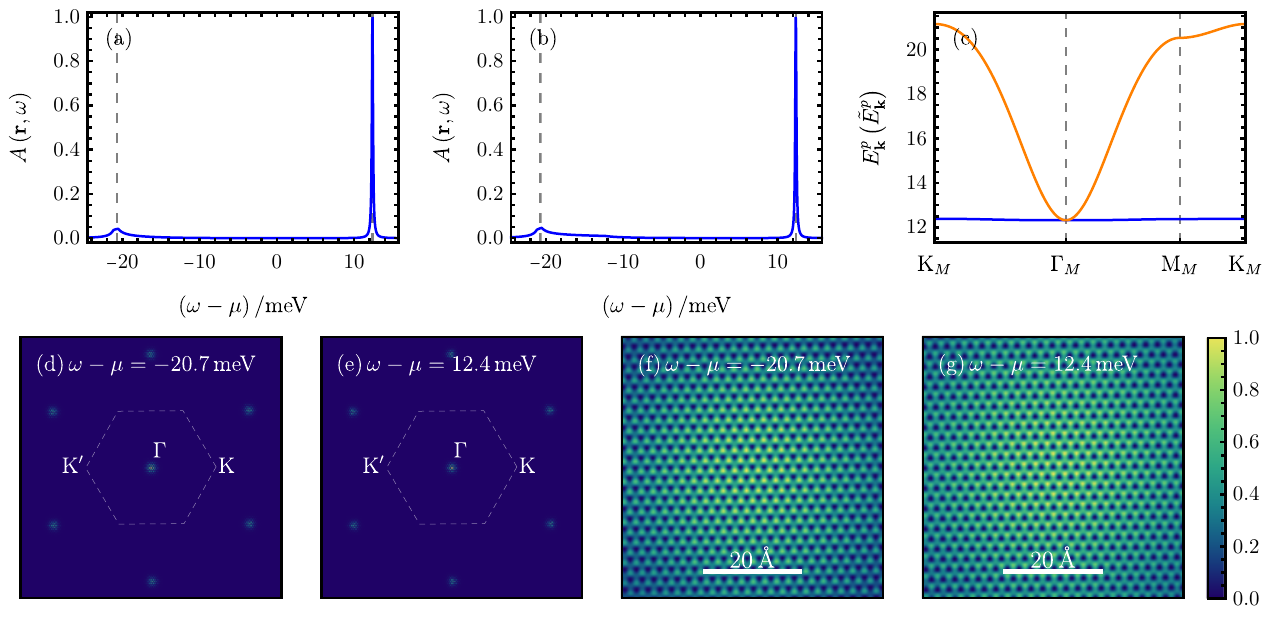}
\caption{{\it $\nu=-1$, $\mathcal{C}=1$, two occupied K-IVC bands and a valley polarized band{}}: The real-space spectral function averaged over three graphene unit cells at the AA site [(a)] and at the AB site [(b)]. The electron (blue) and hole (orange) dispersion is shown in (c). In (d)-(g), we illustrate the spatial variation of the spectral function in momentum space [(d) and (e)], as well as real space [(f) and (g)] for the two energy choices depicted by dashed gray lines in (a) and (b). Here, we focus on $\left(w_0/w_1, \lambda \right) = \left(0.0, 0 \right)$.}\label{app:fig:realspaceNineWZeroLambdaZero}
\end{figure}

\begin{figure}[!h]
\centering
\includegraphics[width=0.95\textwidth]{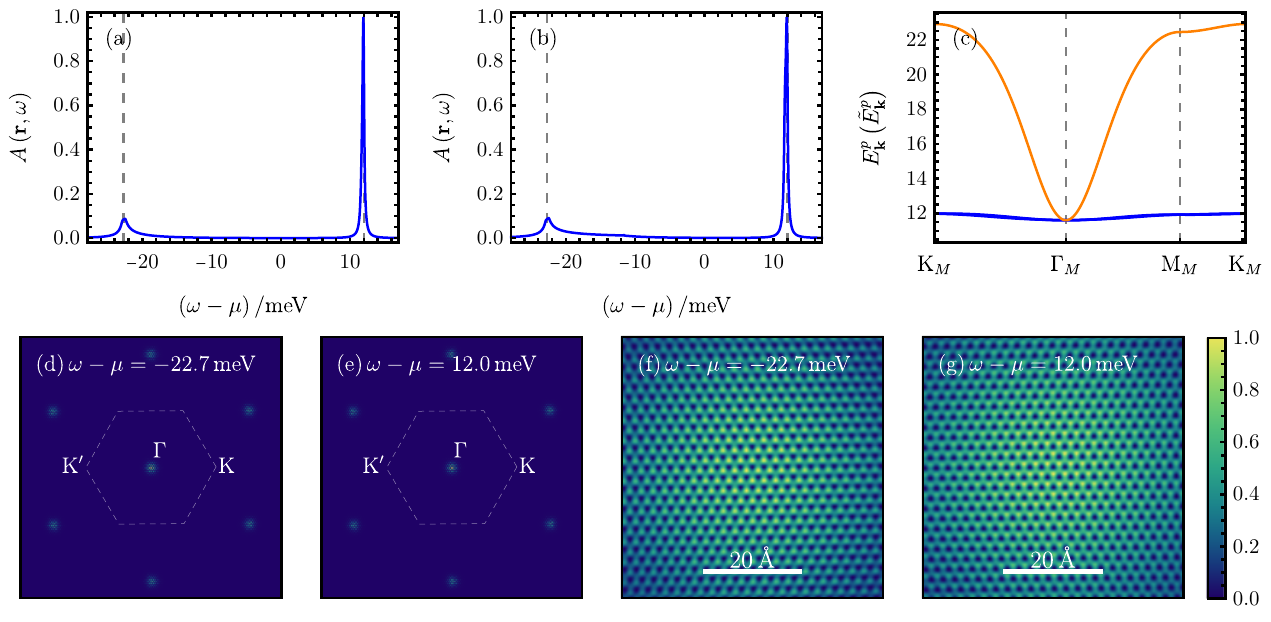}
\caption{{\it $\nu=-1$, $\mathcal{C}=1$, two occupied K-IVC bands and a valley polarized band{}}: The real-space spectral function averaged over three graphene unit cells at the AA site [(a)] and at the AB site [(b)]. The electron (blue) and hole (orange) dispersion is shown in (c). In (d)-(g), we illustrate the spatial variation of the spectral function in momentum space [(d) and (e)], as well as real space [(f) and (g)] for the two energy choices depicted by dashed gray lines in (a) and (b). Here, we focus on $\left(w_0/w_1, \lambda \right) = \left(0.4, 0 \right)$.}\label{app:fig:realspaceNineWFourLambdaZero}
\end{figure}

\begin{figure}[!h]
\centering
\includegraphics[width=0.95\textwidth]{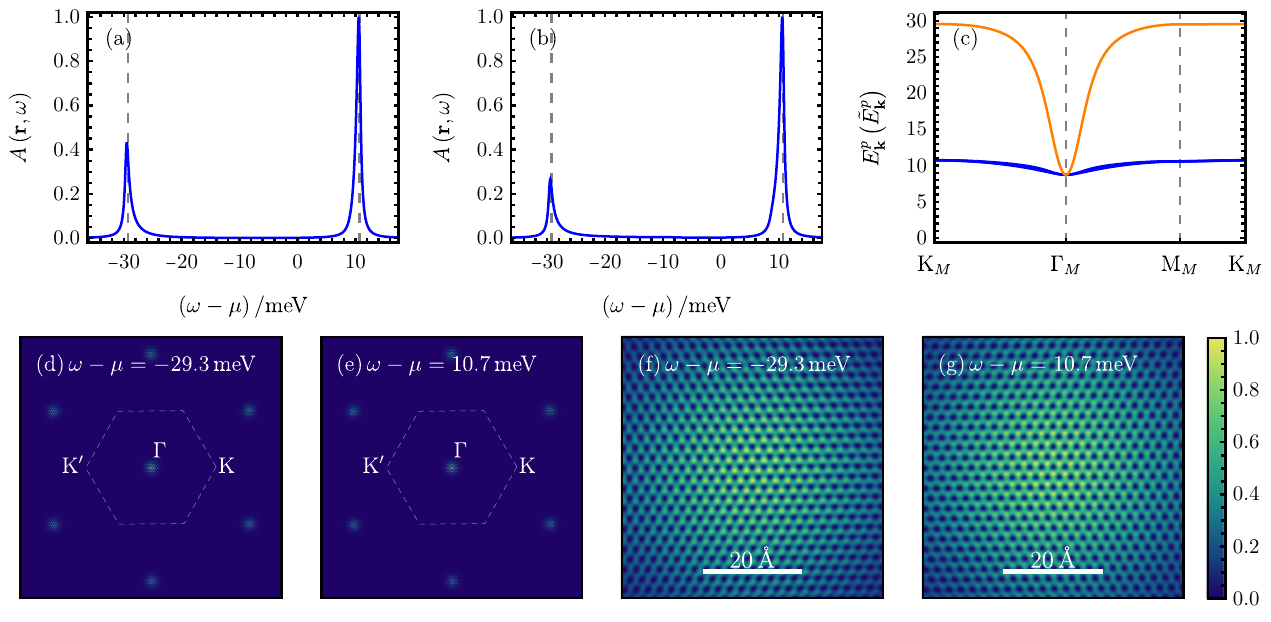}
\caption{{\it $\nu=-1$, $\mathcal{C}=1$, two occupied K-IVC bands and a valley polarized band{}}: The real-space spectral function averaged over three graphene unit cells at the AA site [(a)] and at the AB site [(b)]. The electron (blue) and hole (orange) dispersion is shown in (c). In (d)-(g), we illustrate the spatial variation of the spectral function in momentum space [(d) and (e)], as well as real space [(f) and (g)] for the two energy choices depicted by dashed gray lines in (a) and (b). Here, we focus on $\left(w_0/w_1, \lambda \right) = \left(0.8, 0 \right)$.}\label{app:fig:realspaceNineWEightLambdaZero}
\end{figure}

\begin{figure}[!h]
\centering
\includegraphics[width=0.95\textwidth]{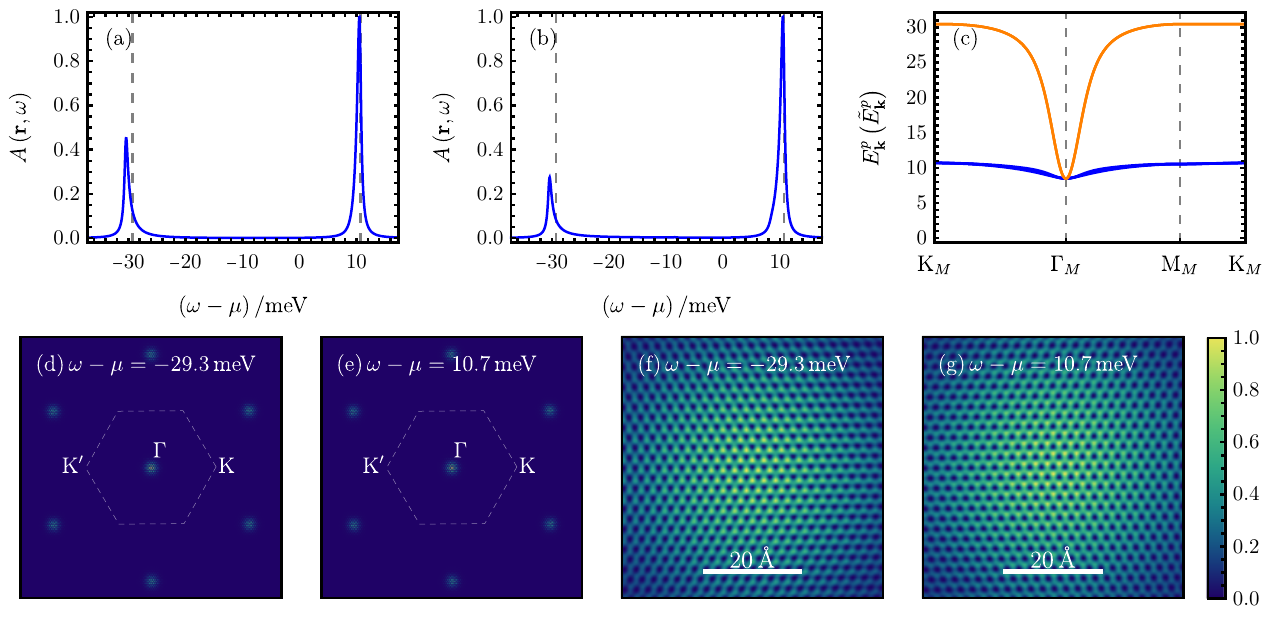}
\caption{{\it $\nu=-1$, $\mathcal{C}=1$, two occupied K-IVC bands and a valley polarized band{}}: The real-space spectral function averaged over three graphene unit cells at the AA site [(a)] and at the AB site [(b)]. The electron (blue) and hole (orange) dispersion is shown in (c). In (d)-(g), we illustrate the spatial variation of the spectral function in momentum space [(d) and (e)], as well as real space [(f) and (g)] for the two energy choices depicted by dashed gray lines in (a) and (b). Here, we focus on $\left(w_0/w_1, \lambda \right) = \left(0.8, 1 \right)$.}\label{app:fig:realspaceNineWEightLambdaOne}
\end{figure}

\newpage

\clearpage

\FloatBarrier
\subsection{$\nu=0$, $\mathcal{C}=0$, K-IVC{}: $\prod_{\vec{k}} \prod_{e_Y = \pm 1} \prod_{s = \uparrow,\downarrow} \frac{\hat{d}^\dagger_{\vec{k},e_Y,+,s} + e_Y \hat{d}^\dagger_{\vec{k},e_Y,-,s} }{\sqrt{2}} \ket{0}$}\label{app:sec:example:stateTen}

\begin{figure}[!h]
\centering
\includegraphics[width=0.95\textwidth]{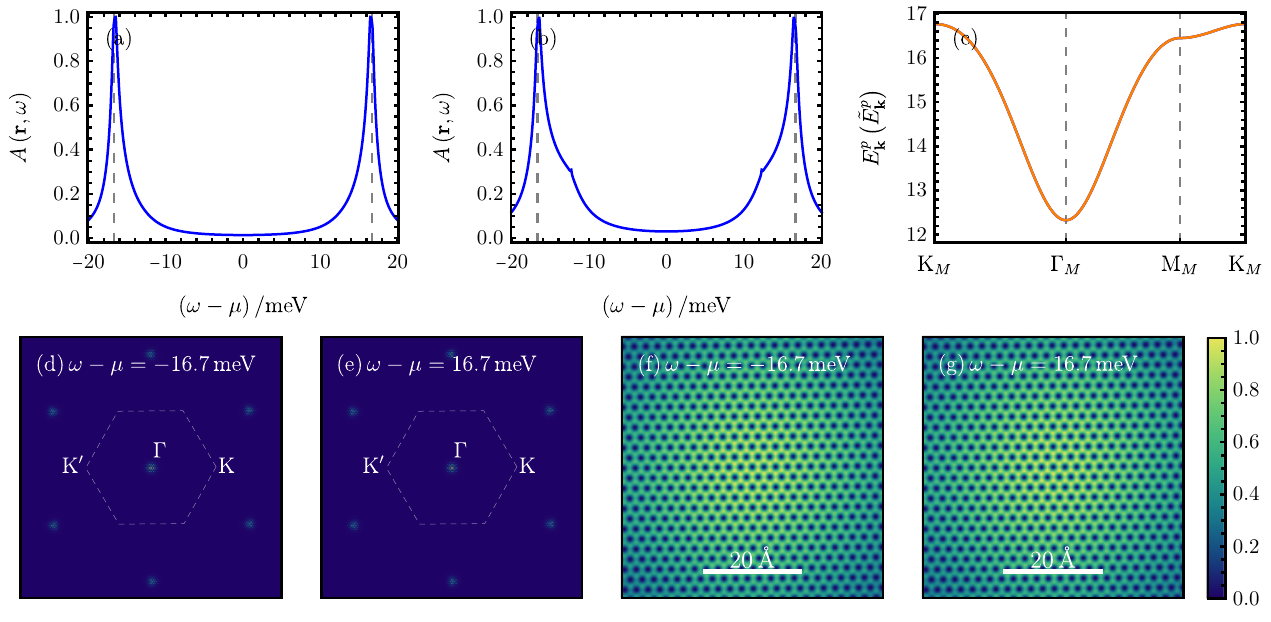}
\caption{{\it $\nu=0$, $\mathcal{C}=0$, K-IVC{}}: The real-space spectral function averaged over three graphene unit cells at the AA site [(a)] and at the AB site [(b)]. The electron (blue) and hole (orange) dispersion is shown in (c). In (d)-(g), we illustrate the spatial variation of the spectral function in momentum space [(d) and (e)], as well as real space [(f) and (g)] for the two energy choices depicted by dashed gray lines in (a) and (b). Here, we focus on $\left(w_0/w_1, \lambda \right) = \left(0.0, 0 \right)$.}\label{app:fig:realspaceTenWZeroLambdaZero}
\end{figure}

\begin{figure}[!h]
\centering
\includegraphics[width=0.95\textwidth]{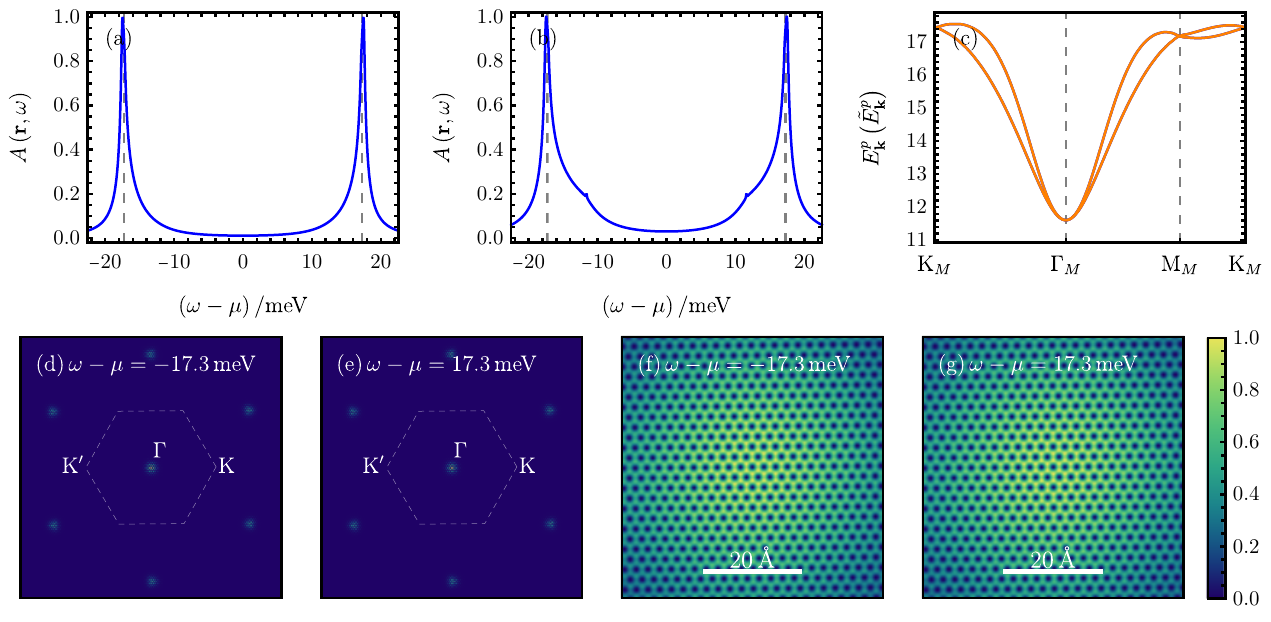}
\caption{{\it $\nu=0$, $\mathcal{C}=0$, K-IVC{}}: The real-space spectral function averaged over three graphene unit cells at the AA site [(a)] and at the AB site [(b)]. The electron (blue) and hole (orange) dispersion is shown in (c). In (d)-(g), we illustrate the spatial variation of the spectral function in momentum space [(d) and (e)], as well as real space [(f) and (g)] for the two energy choices depicted by dashed gray lines in (a) and (b). Here, we focus on $\left(w_0/w_1, \lambda \right) = \left(0.4, 0 \right)$.}\label{app:fig:realspaceTenWFourLambdaZero}
\end{figure}

\begin{figure}[!h]
\centering
\includegraphics[width=0.95\textwidth]{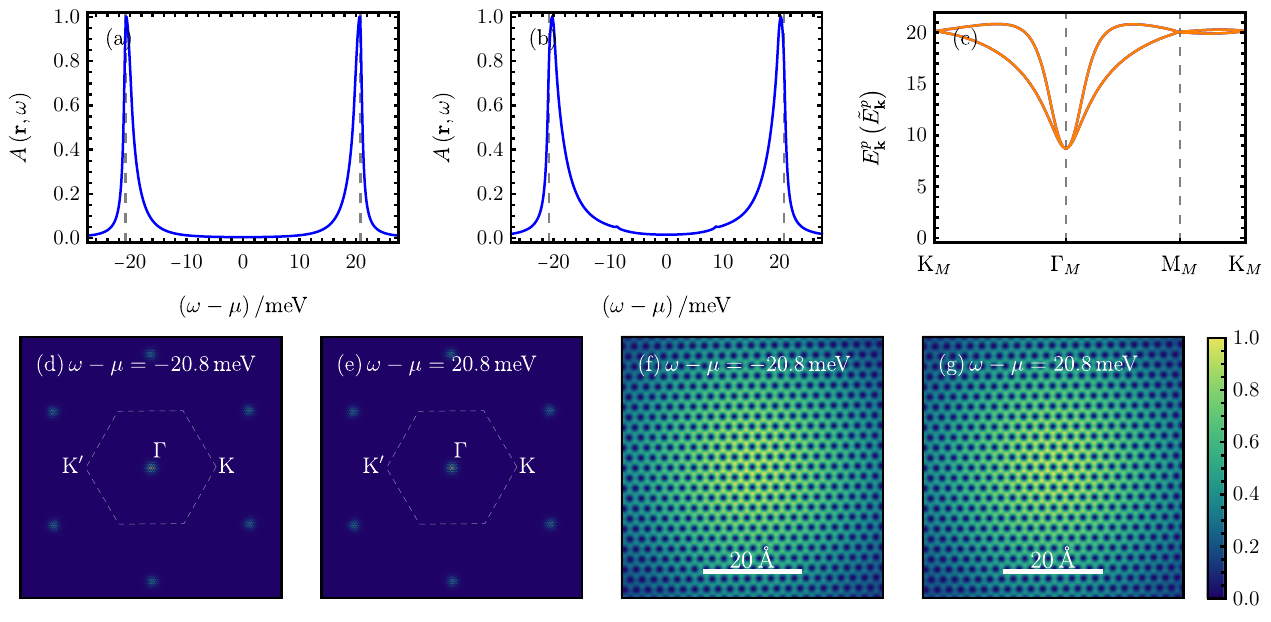}
\caption{{\it $\nu=0$, $\mathcal{C}=0$, K-IVC{}}: The real-space spectral function averaged over three graphene unit cells at the AA site [(a)] and at the AB site [(b)]. The electron (blue) and hole (orange) dispersion is shown in (c). In (d)-(g), we illustrate the spatial variation of the spectral function in momentum space [(d) and (e)], as well as real space [(f) and (g)] for the two energy choices depicted by dashed gray lines in (a) and (b). Here, we focus on $\left(w_0/w_1, \lambda \right) = \left(0.8, 0 \right)$.}\label{app:fig:realspaceTenWEightLambdaZero}
\end{figure}

\begin{figure}[!h]
\centering
\includegraphics[width=0.95\textwidth]{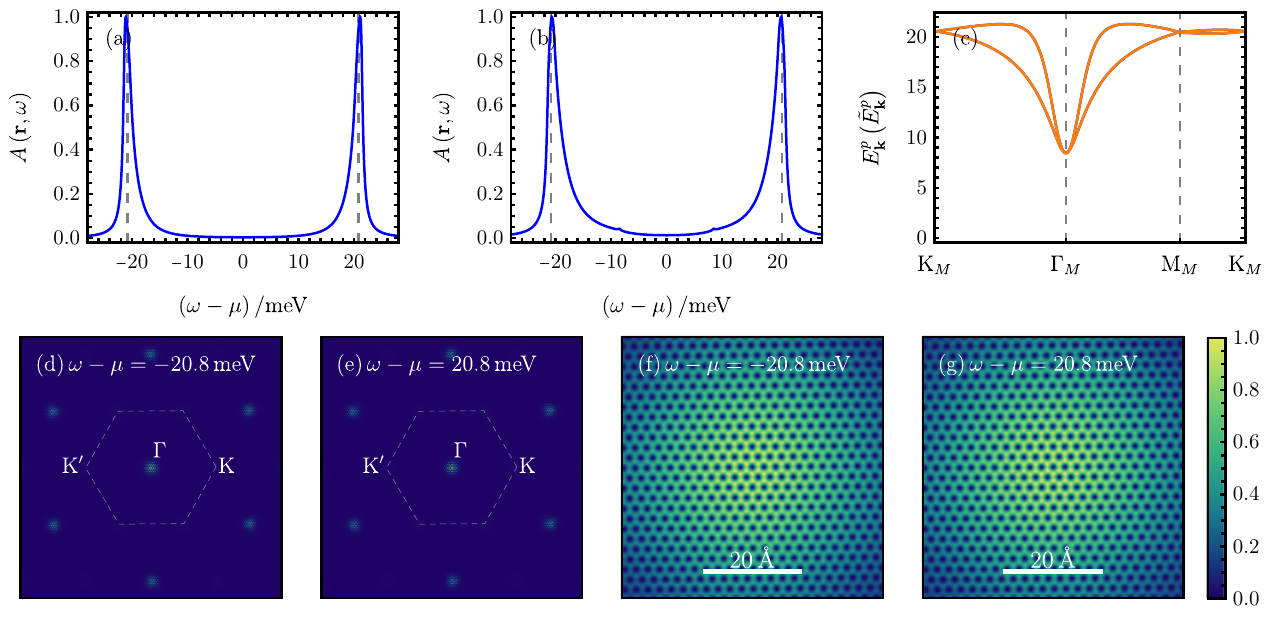}
\caption{{\it $\nu=0$, $\mathcal{C}=0$, K-IVC{}}: The real-space spectral function averaged over three graphene unit cells at the AA site [(a)] and at the AB site [(b)]. The electron (blue) and hole (orange) dispersion is shown in (c). In (d)-(g), we illustrate the spatial variation of the spectral function in momentum space [(d) and (e)], as well as real space [(f) and (g)] for the two energy choices depicted by dashed gray lines in (a) and (b). Here, we focus on $\left(w_0/w_1, \lambda \right) = \left(0.8, 1 \right)$.}\label{app:fig:realspaceTenWEightLambdaOne}
\end{figure}

\newpage

\clearpage

\FloatBarrier
\subsection{$\nu=0$, $\mathcal{C}=0$, valley polarized{}: $\prod_{\vec{k}} \prod_{e_Y = \pm 1} \prod_{s = \uparrow,\downarrow} \hat{d}^\dagger_{\vec{k},e_Y,+,s} \ket{0}$}\label{app:sec:example:stateEleven}

\begin{figure}[!h]
\centering
\includegraphics[width=0.95\textwidth]{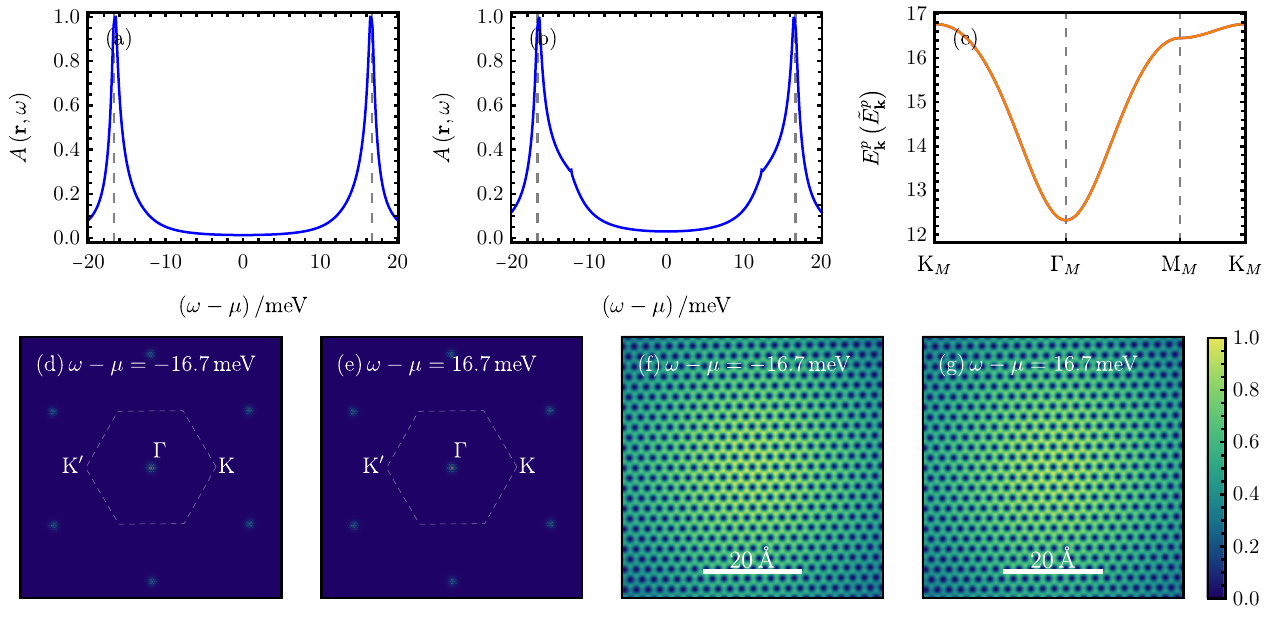}
\caption{{\it $\nu=0$, $\mathcal{C}=0$, valley polarized{}}: The real-space spectral function averaged over three graphene unit cells at the AA site [(a)] and at the AB site [(b)]. The electron (blue) and hole (orange) dispersion is shown in (c). In (d)-(g), we illustrate the spatial variation of the spectral function in momentum space [(d) and (e)], as well as real space [(f) and (g)] for the two energy choices depicted by dashed gray lines in (a) and (b). Here, we focus on $\left(w_0/w_1, \lambda \right) = \left(0.0, 0 \right)$.}\label{app:fig:realspaceElevenWZeroLambdaZero}
\end{figure}

\begin{figure}[!h]
\centering
\includegraphics[width=0.95\textwidth]{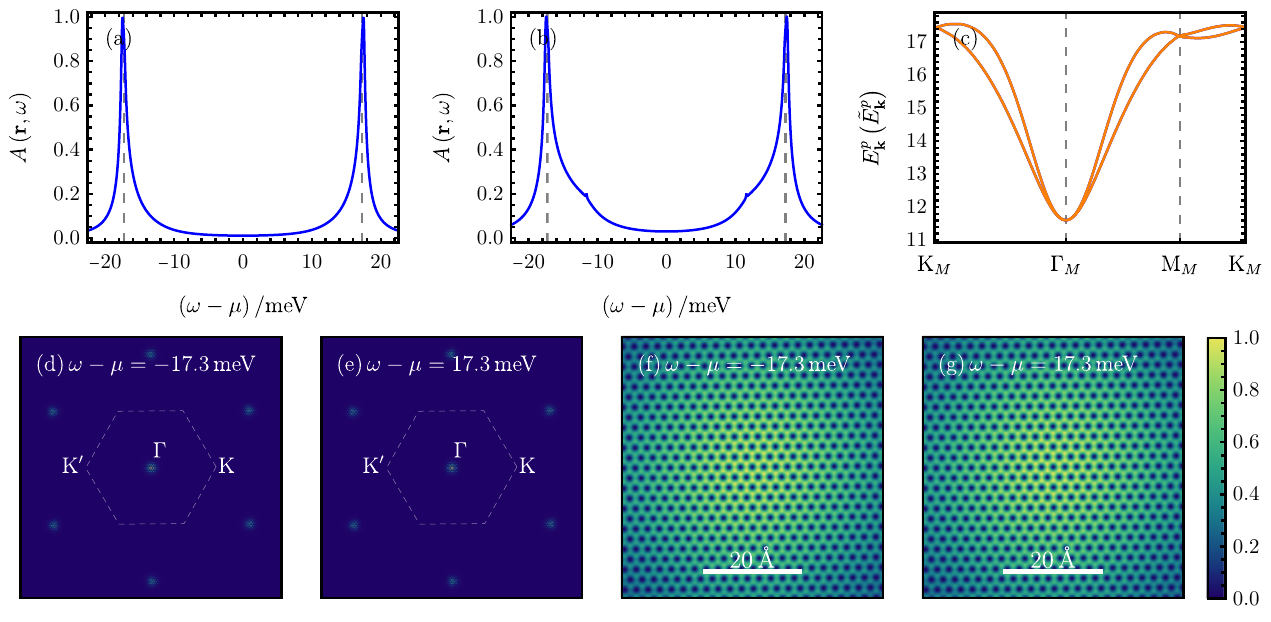}
\caption{{\it $\nu=0$, $\mathcal{C}=0$, valley polarized{}}: The real-space spectral function averaged over three graphene unit cells at the AA site [(a)] and at the AB site [(b)]. The electron (blue) and hole (orange) dispersion is shown in (c). In (d)-(g), we illustrate the spatial variation of the spectral function in momentum space [(d) and (e)], as well as real space [(f) and (g)] for the two energy choices depicted by dashed gray lines in (a) and (b). Here, we focus on $\left(w_0/w_1, \lambda \right) = \left(0.4, 0 \right)$.}\label{app:fig:realspaceElevenWFourLambdaZero}
\end{figure}

\begin{figure}[!h]
\centering
\includegraphics[width=0.95\textwidth]{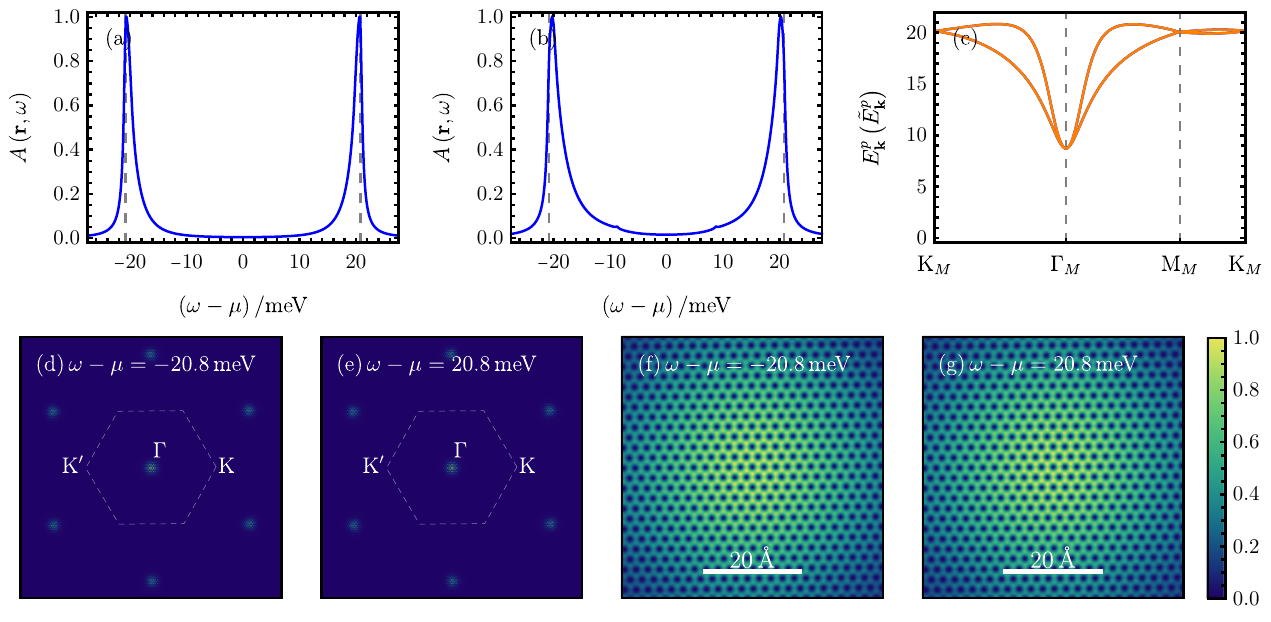}
\caption{{\it $\nu=0$, $\mathcal{C}=0$, valley polarized{}}: The real-space spectral function averaged over three graphene unit cells at the AA site [(a)] and at the AB site [(b)]. The electron (blue) and hole (orange) dispersion is shown in (c). In (d)-(g), we illustrate the spatial variation of the spectral function in momentum space [(d) and (e)], as well as real space [(f) and (g)] for the two energy choices depicted by dashed gray lines in (a) and (b). Here, we focus on $\left(w_0/w_1, \lambda \right) = \left(0.8, 0 \right)$.}\label{app:fig:realspaceElevenWEightLambdaZero}
\end{figure}

\begin{figure}[!h]
\centering
\includegraphics[width=0.95\textwidth]{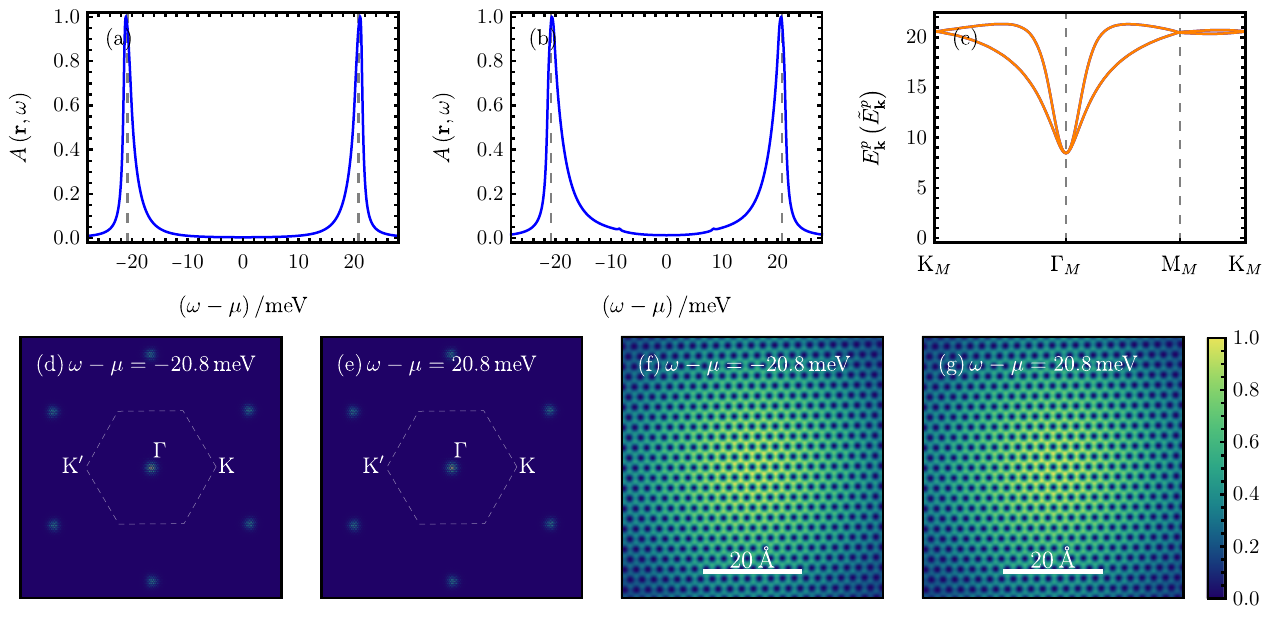}
\caption{{\it $\nu=0$, $\mathcal{C}=0$, valley polarized{}}: The real-space spectral function averaged over three graphene unit cells at the AA site [(a)] and at the AB site [(b)]. The electron (blue) and hole (orange) dispersion is shown in (c). In (d)-(g), we illustrate the spatial variation of the spectral function in momentum space [(d) and (e)], as well as real space [(f) and (g)] for the two energy choices depicted by dashed gray lines in (a) and (b). Here, we focus on $\left(w_0/w_1, \lambda \right) = \left(0.8, 1 \right)$.}\label{app:fig:realspaceElevenWEightLambdaOne}
\end{figure}

\newpage

\clearpage

\FloatBarrier
\subsection{$\nu=0$, $\mathcal{C}=4$, valley polarized{}: $\prod_{\vec{k}} \prod_{\eta = \pm} \prod_{s = \uparrow,\downarrow} \hat{d}^\dagger_{\vec{k},+1,\eta,s} \ket{0}$}\label{app:sec:example:stateTwelve}

\begin{figure}[!h]
\centering
\includegraphics[width=0.95\textwidth]{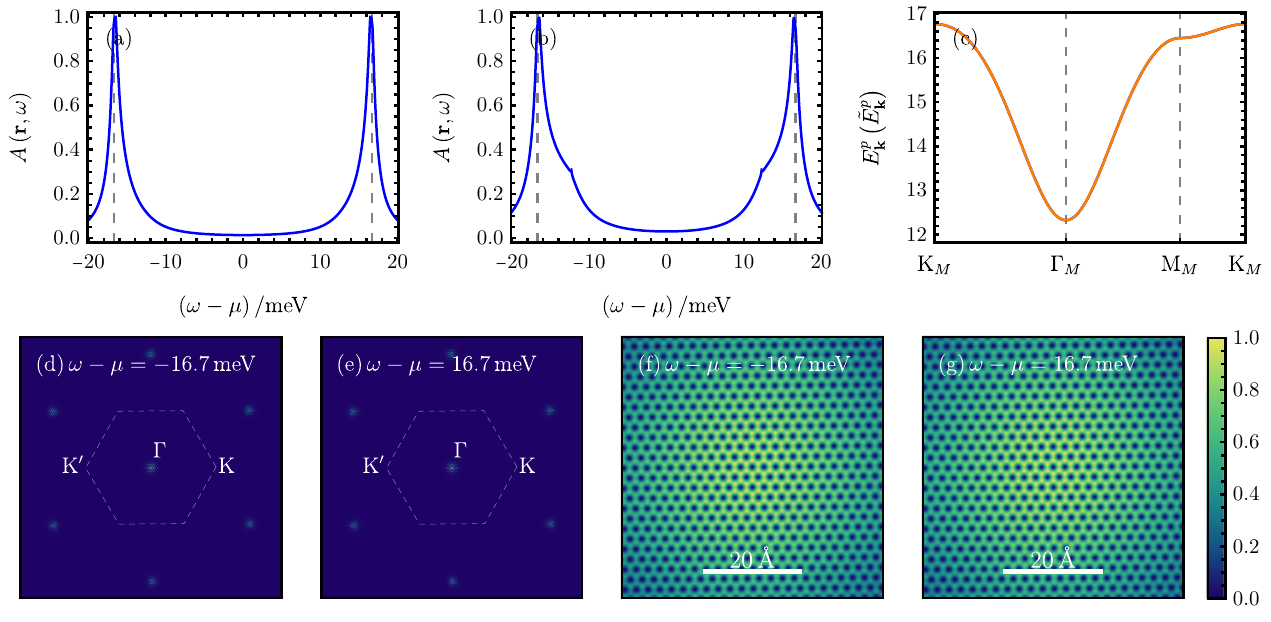}
\caption{{\it $\nu=0$, $\mathcal{C}=4$, valley polarized{}}: The real-space spectral function averaged over three graphene unit cells at the AA site [(a)] and at the AB site [(b)]. The electron (blue) and hole (orange) dispersion is shown in (c). In (d)-(g), we illustrate the spatial variation of the spectral function in momentum space [(d) and (e)], as well as real space [(f) and (g)] for the two energy choices depicted by dashed gray lines in (a) and (b). Here, we focus on $\left(w_0/w_1, \lambda \right) = \left(0.0, 0 \right)$.}\label{app:fig:realspaceTwelveWZeroLambdaZero}
\end{figure}

\begin{figure}[!h]
\centering
\includegraphics[width=0.95\textwidth]{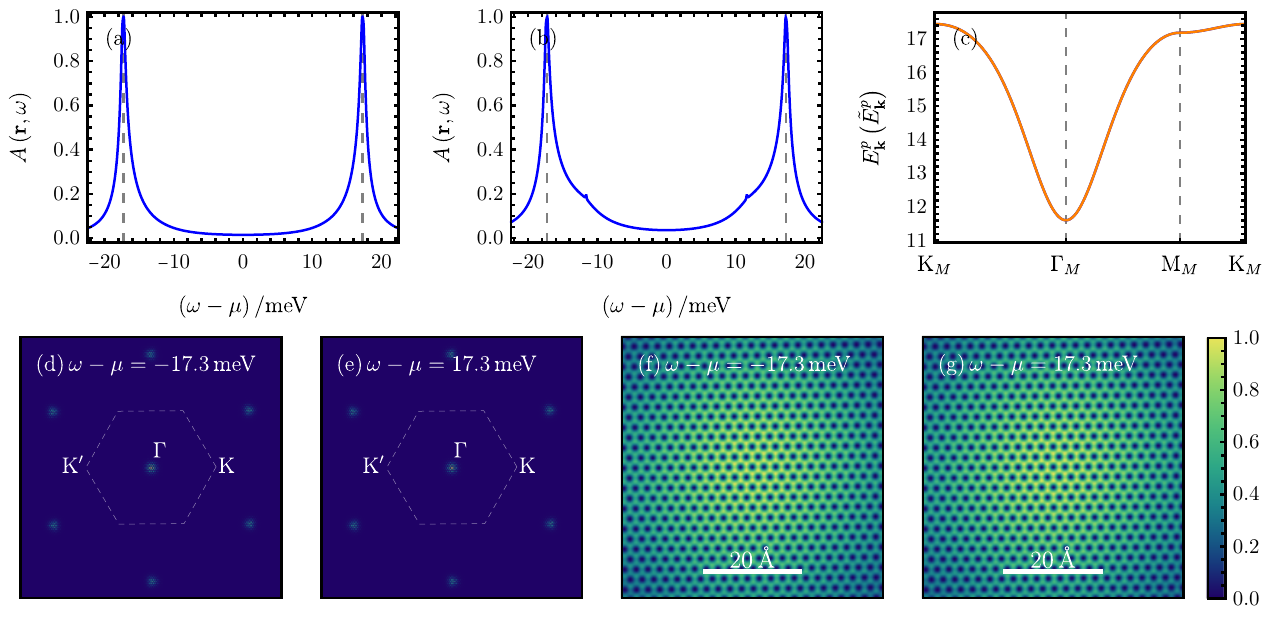}
\caption{{\it $\nu=0$, $\mathcal{C}=4$, valley polarized{}}: The real-space spectral function averaged over three graphene unit cells at the AA site [(a)] and at the AB site [(b)]. The electron (blue) and hole (orange) dispersion is shown in (c). In (d)-(g), we illustrate the spatial variation of the spectral function in momentum space [(d) and (e)], as well as real space [(f) and (g)] for the two energy choices depicted by dashed gray lines in (a) and (b). Here, we focus on $\left(w_0/w_1, \lambda \right) = \left(0.4, 0 \right)$.}\label{app:fig:realspaceTwelveWFourLambdaZero}
\end{figure}

\begin{figure}[!h]
\centering
\includegraphics[width=0.95\textwidth]{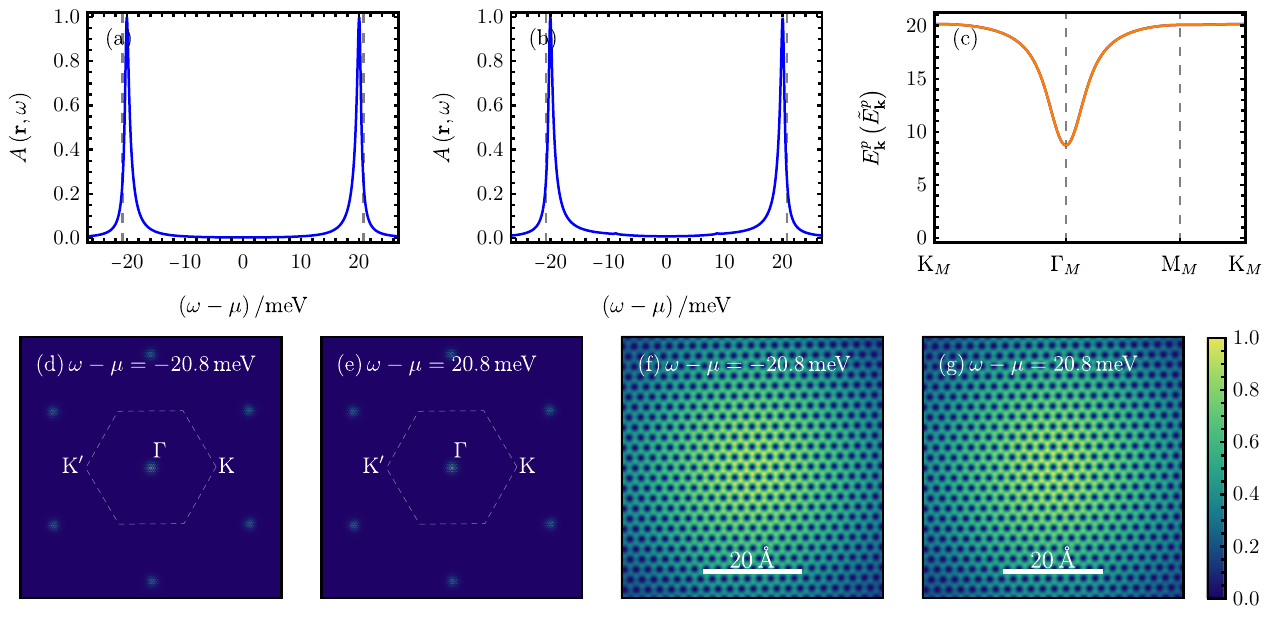}
\caption{{\it $\nu=0$, $\mathcal{C}=4$, valley polarized{}}: The real-space spectral function averaged over three graphene unit cells at the AA site [(a)] and at the AB site [(b)]. The electron (blue) and hole (orange) dispersion is shown in (c). In (d)-(g), we illustrate the spatial variation of the spectral function in momentum space [(d) and (e)], as well as real space [(f) and (g)] for the two energy choices depicted by dashed gray lines in (a) and (b). Here, we focus on $\left(w_0/w_1, \lambda \right) = \left(0.8, 0 \right)$.}\label{app:fig:realspaceTwelveWEightLambdaZero}
\end{figure}

\begin{figure}[!h]
\centering
\includegraphics[width=0.95\textwidth]{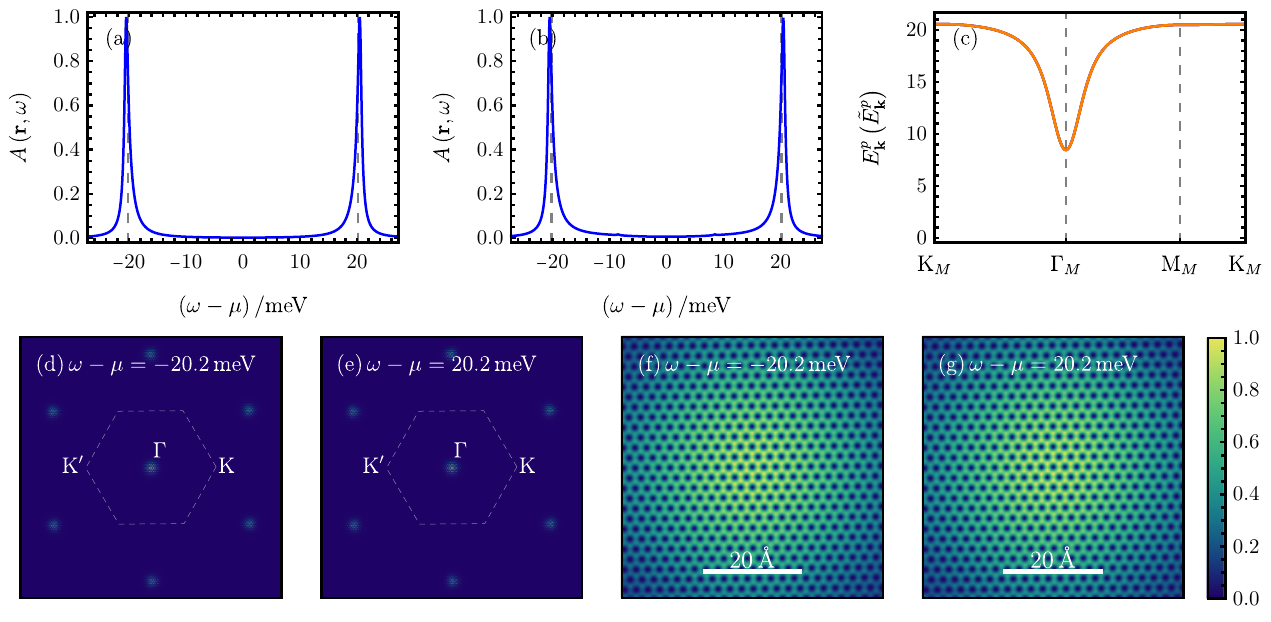}
\caption{{\it $\nu=0$, $\mathcal{C}=4$, valley polarized{}}: The real-space spectral function averaged over three graphene unit cells at the AA site [(a)] and at the AB site [(b)]. The electron (blue) and hole (orange) dispersion is shown in (c). In (d)-(g), we illustrate the spatial variation of the spectral function in momentum space [(d) and (e)], as well as real space [(f) and (g)] for the two energy choices depicted by dashed gray lines in (a) and (b). Here, we focus on $\left(w_0/w_1, \lambda \right) = \left(0.8, 1 \right)$.}\label{app:fig:realspaceTwelveWEightLambdaOne}
\end{figure}

\newpage

\clearpage

\FloatBarrier
\subsection{$\nu=+4$, insulator with all TBG active bands filled{}: $\prod_{\vec{k}} \prod_{\eta = \pm} \prod_{s = \uparrow,\downarrow} \prod_{e_Y = \pm 1} \hat{d}^\dagger_{\vec{k},e_Y,\eta,s} \ket{0}$}\label{app:sec:example:stateThirteen}

\begin{figure}[!h]
\centering
\includegraphics[width=0.95\textwidth]{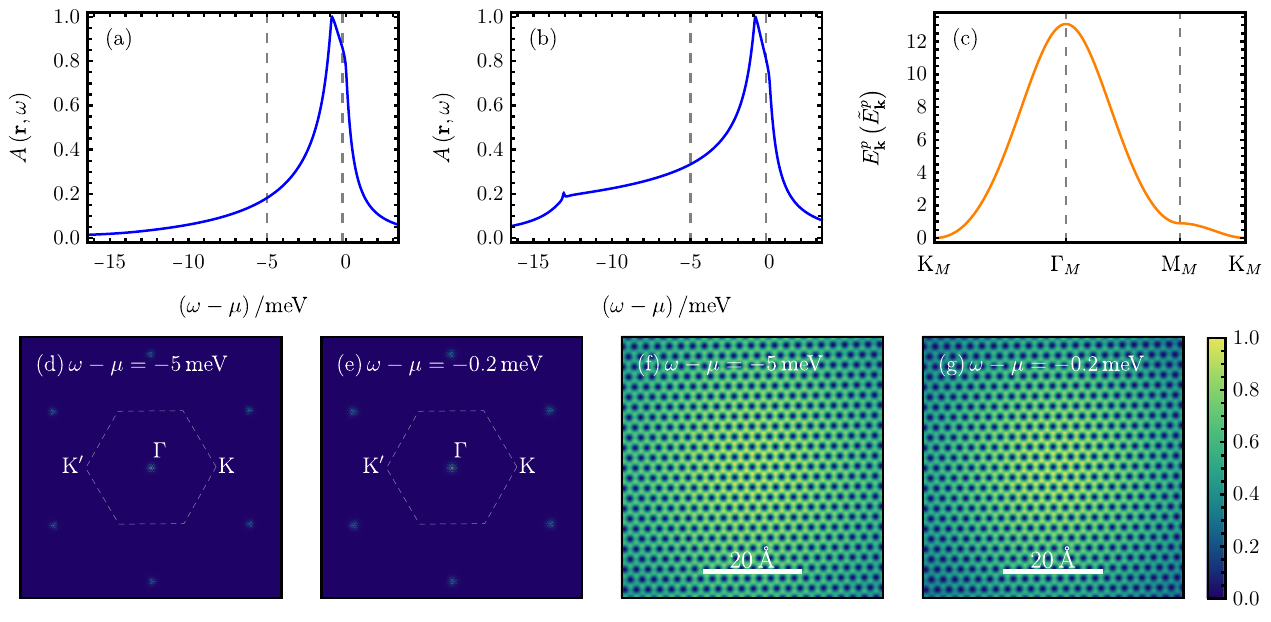}
\caption{{\it $\nu=+4$, insulator with all TBG active bands filled{}}: The real-space spectral function averaged over three graphene unit cells at the AA site [(a)] and at the AB site [(b)]. The electron (blue) and hole (orange) dispersion is shown in (c). In (d)-(g), we illustrate the spatial variation of the spectral function in momentum space [(d) and (e)], as well as real space [(f) and (g)] for the two energy choices depicted by dashed gray lines in (a) and (b). Here, we focus on $\left(w_0/w_1, \lambda \right) = \left(0.0, 0 \right)$.}\label{app:fig:realspaceThirteenWZeroLambdaZero}
\end{figure}

\begin{figure}[!h]
\centering
\includegraphics[width=0.95\textwidth]{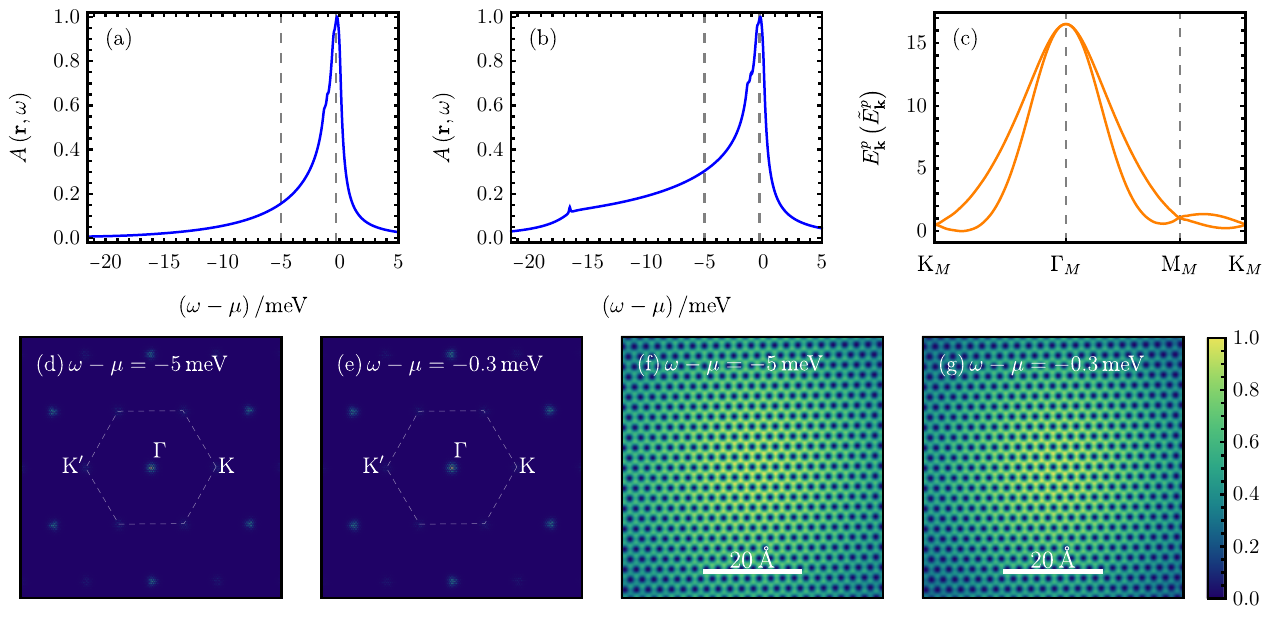}
\caption{{\it $\nu=+4$, insulator with all TBG active bands filled{}}: The real-space spectral function averaged over three graphene unit cells at the AA site [(a)] and at the AB site [(b)]. The electron (blue) and hole (orange) dispersion is shown in (c). In (d)-(g), we illustrate the spatial variation of the spectral function in momentum space [(d) and (e)], as well as real space [(f) and (g)] for the two energy choices depicted by dashed gray lines in (a) and (b). Here, we focus on $\left(w_0/w_1, \lambda \right) = \left(0.4, 0 \right)$.}\label{app:fig:realspaceThirteenWFourLambdaZero}
\end{figure}

\begin{figure}[!h]
\centering
\includegraphics[width=0.95\textwidth]{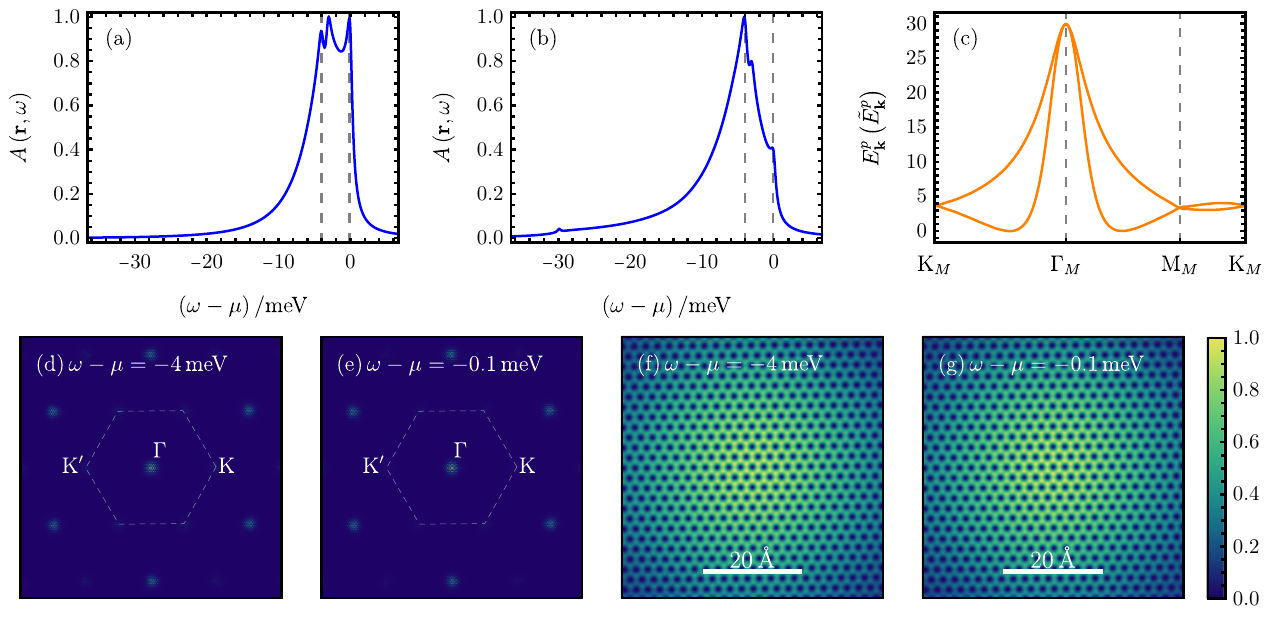}
\caption{{\it $\nu=+4$, insulator with all TBG active bands filled{}}: The real-space spectral function averaged over three graphene unit cells at the AA site [(a)] and at the AB site [(b)]. The electron (blue) and hole (orange) dispersion is shown in (c). In (d)-(g), we illustrate the spatial variation of the spectral function in momentum space [(d) and (e)], as well as real space [(f) and (g)] for the two energy choices depicted by dashed gray lines in (a) and (b). Here, we focus on $\left(w_0/w_1, \lambda \right) = \left(0.8, 0 \right)$.}\label{app:fig:realspaceThirteenWEightLambdaZero}
\end{figure}

\begin{figure}[!h]
\centering
\includegraphics[width=0.95\textwidth]{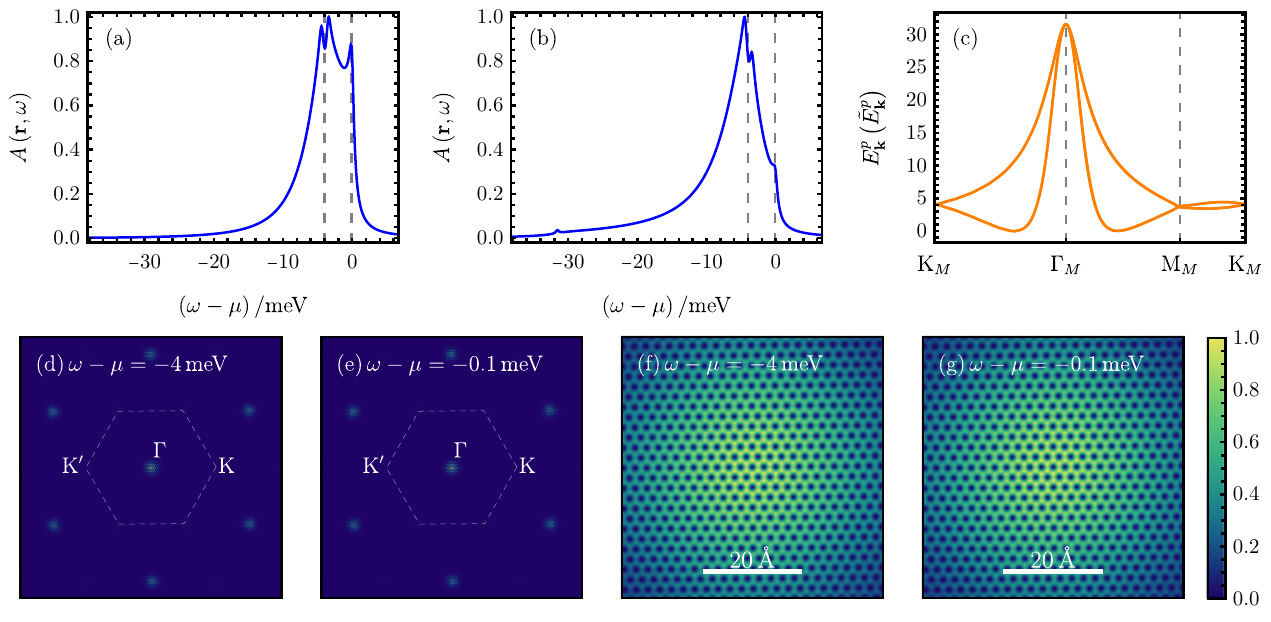}
\caption{{\it $\nu=+4$, insulator with all TBG active bands filled{}}: The real-space spectral function averaged over three graphene unit cells at the AA site [(a)] and at the AB site [(b)]. The electron (blue) and hole (orange) dispersion is shown in (c). In (d)-(g), we illustrate the spatial variation of the spectral function in momentum space [(d) and (e)], as well as real space [(f) and (g)] for the two energy choices depicted by dashed gray lines in (a) and (b). Here, we focus on $\left(w_0/w_1, \lambda \right) = \left(0.8, 1 \right)$.}\label{app:fig:realspaceThirteenWEightLambdaOne}
\end{figure}

\end{document}